 \documentclass[twocolumn]{aastex62}

\shorttitle{Light Curve Analysis of Novae}
\shortauthors{Hachisu \& Kato}

\begin{document}

\title{The $UBV$ Color Evolution of Classical Novae. III. 
Time-Stretched Color-Magnitude Diagram of Novae in Outburst}


\author[0000-0002-0884-7404]{Izumi Hachisu}
\affil{Department of Earth Science and Astronomy, 
College of Arts and Sciences, The University of Tokyo,
3-8-1 Komaba, Meguro-ku, Tokyo 153-8902, Japan} 
\email{hachisu@ea.c.u-tokyo.ac.jp}


\author[0000-0002-8522-8033]{Mariko Kato}
\affil{Department of Astronomy, Keio University, 
Hiyoshi, Kouhoku-ku, Yokohama 223-8521, Japan} 

%


%
%



\begin{abstract}
We propose a modified color-magnitude diagram for novae in outburst, i.e., 
$(B-V)_0$ versus $(M_V-2.5 \log f_{\rm s})$, where $f_{\rm s}$ is 
the timescaling factor of a (target) nova against a comparison
(template) nova, $(B-V)_0$ is the intrinsic $B-V$ color, and $M_V$ is 
the absolute $V$ magnitude.  We dub it the time-stretched color-magnitude
diagram.  We carefully reanalyzed 20 novae based on the time-stretching
method and revised their extinctions $E(B-V)$, distance moduli 
in the $V$ band $(m-M)_V$, distances $d$, and timescaling
factors $f_{\rm s}$ against the template nova LV~Vul.
We have found that these 20 nova outburst tracks
broadly follow one of the two template tracks, LV~Vul/V1668~Cyg or 
V1500~Cyg/V1974~Cyg group, in the time-stretched color-magnitude diagram.
In addition, we estimate the white dwarf masses and $(m-M)_V$ of the novae
by directly fitting the absolute $V$ model light curves ($M_V$)
with observational apparent $V$ magnitudes ($m_V$).  
A good agreement in the two estimates of $(m-M)_V$ confirms 
the consistency of the time-stretched color-magnitude diagram.
Our distance estimates are in good agreement with the results 
of Gaia Data Release 2.
\end{abstract}


\keywords{novae, cataclysmic variables --- stars: individual 
(V574~Pup, V679~Car, V1369~Cen, V5666~Sgr) --- stars: winds}


\section{Introduction}
\label{introduction}
A nova is a thermonuclear explosion event on a mass-accreting 
white dwarf (WD) in a close binary.  When the mass of the hydrogen-rich
envelope on the WD reaches a critical value, hydrogen ignites 
at the bottom of the hydrogen-rich envelope to trigger a nova explosion.
Then, the envelope expands to a red-giant size and emits strong winds 
\citep{kat94h}.  
After the maximum expansion of the photosphere (when the maximum wind
mass-loss rate is attained), the photospheric radius begins to shrink and 
the photospheric temperature rises with time.
The main emitting wavelength region thus shifts from optical to 
UV and supersoft X-ray \citep[see, e.g., Figure 23 of][for 
V1974 Cyg]{hac06kb}. 
A large part of the hydrogen-rich envelope is blown in the wind.
The wind stops and the nova enters a supersoft X-ray source (SSS) phase.
The envelope mass decreases further by nuclear burning.  
When it decreases to below the critical mass
(minimum mass for steady hydrogen shell-burning),
nuclear burning extinguishes.  The WD cools down and the nova outburst ends.
These evolutions of novae are briefly summarized in \citet{hac17k}. 

Evolutions of nova outbursts can be followed by the optically thick
wind theory. \citet{kat94h} calculated nova evolution for various WD masses
(from $0.5~M_\sun$ to $1.38~M_\sun$) and showed that the main parameter 
which determines the speed class of novae is the WD mass.
A nova on a more massive WD evolves faster because the ignition mass
(envelope mass at the epoch when hydrogen ignites) 
is much smaller and is blown off in the wind in a much shorter time.

The acceleration mechanism of the wind is the so-called continuum 
radiation-driven \citep{fri66}.  The matter in the envelope is
accelerated deep inside the photosphere, in other words, at/near
the iron peak of the radiative opacity ($\log T$ (K) $\sim 5.2$). 
So, it is called the optically thick winds. 
This mechanism is common among all the WD masses ranging from
$\sim0.5~M_\sun$ to $1.38~M_\sun$ \citep{kat94h}. 


Nova spectra in the decay phase are dominated by free-free emission 
\citep{gal76, enn77, kra84, nai09} 
until the nova entered the nebular phase where strong emission 
lines such as [\ion{O}{3}] are dominant. 
Free-free emission of novae is originating from the expanding hot and
optically thin gaseous plasma, that is, outside the photosphere of
the optically thick winds. 
Free-free emission light curves of novae are theoretically calculated
from the wind-mass loss rate $\dot M_{\rm wind}$, photospheric velocity
$v_{\rm ph}$, and photospheric radius $R_{\rm ph}$, that is,
$F_\nu \propto \dot M^2_{\rm wind} / v^2_{\rm ph} R_{\rm ph}$
\citep{hac06kb},
where $F_\nu$ is the flux of free-free emission at the frequency
$\nu$.   It should be noted that the flux of free-free emission $F_\nu$
is almost independent of the frequency $\nu$ in the optical and
near-infrared (NIR) region. 

\citet{hac06kb} calculated free-free emission light curves 
for various WD masses (from $0.5~M_\sun$ to $1.38~M_\sun$) 
and envelope chemical compositions. 
Since the flux of free-free emission $F_\nu$ is almost independent of
the frequency $\nu$, the light curve shapes are very similar among
the optical and NIR bands until the nebular phase or dust formation started. 
For example, the $V$, $J$, $H$, $2.3\mu$m, and $3.6 \mu$m band light
curves of V1668 Cyg show an almost identical light curve shape,
which are well reproduced by a theoretical light curve 
until the optically-thin dust shell formation makes a deviation from
the theoretical one \citep[see, e.g., Figure 18 of][]{hac06kb}.  

\citet{hac06kb} further found that the theoretical free-free emission
light curves and UV (1455 \AA) light curves show a scaling law
among various WD masses and chemical compositions.  
The UV~1455\AA\  band is a narrow (1445--1465\AA) and emission
line-free band invented by \citet{cas02} and represents continuum flux
of UV.  All the UV~1455\AA\  model light curves also follow the same
timescaling law as that of free-free emission light curves.  In other
words, if we properly squeeze the timescales of novae, these theoretical
light curves converge into one \citep[see Figure 18 of][]{hac06kb}. 
The main part of theoretical free-free emission light curve decays 
as $F_\nu \propto t^{-1.75}$.   \citet{hac06kb} called this property 
the universal decline law of classical novae.  

Because the decline rate of optical light curve is a good indicator of the 
WD mass, we can estimate the WD mass from comparison of the theoretical 
light curve with the observed optical one.   If multiwavelength light
curves are available, we are able to determine the WD mass more accurately. 
Using multiwavelength light curves in optical, UV~1455\AA, and
supersoft X-ray turn-on/turnoff times, \citet{hac06kb, hac10k, hac16k} 
determined the WD masses of V1974 Cyg and V1668 Cyg.
The UV~1455\AA\  band flux represents the photospheric emission
in the wind phase while the supersoft X-ray flux does in the post-wind phase.
These multiwavelength data are so useful because their dependences
on the WD mass and chemical composition are different. 
It should be noted that the soft X-ray light curves do not obey
the timescaling law \citep[see, e.g., Figure 6 of][]{hac10k}.
We are further able to discriminate these two parameters from multiwavelength
light curve fitting.  Thus, \citet{hac15k, hac16k} determined the WD masses
within error of $\pm0.01~M_\sun$ by simultaneously reproducing
the observed multiwavelength light curves for a number of novae.  

\citet{hac10k} introduced the timescaling factor $f_{\rm s}$,
which is the timescale ratio of a target nova against a template nova.
Using this $f_{\rm s}$, we can compare various nova light curves
in a quantitative way.  They showed that, if a large part of the nova
light/color curves overlap each other by properly squeezing/stretching
their timescales, the absolute brightness ($M_V$) of the target nova
is related to the absolute brightness ($M'_V$) of the other (template)
nova by $M_V = M'_V + 2.5 \log f_{\rm s}$.
This means that the time-stretched absolute light curves of the two novae
overlap each other in the $(t/f_{\rm s})$-$(M_V-2.5 \log f_{\rm s})$ plane.
They adopted V1668~Cyg and V1974~Cyg as template novae, of which 
the distances and extinctions are known.  Using this property, 
\citet{hac10k, hac15k, hac16k} determined the absolute magnitude $M_V$
for each free-free emission light curve of different WD mass and 
chemical composition.  They called this method the time-stretching method.  
By directly fitting the theoretical light curves of $M_V$ with the observed
$m_V$, we can also estimate the WD mass $M_{\rm WD}$ and distance
modulus in the $V$ band of $\mu_V\equiv (m-M)_V$ for a target nova.
\citet{hac18kb} confirmed that the time-stretching method is
applicable to fast and very fast novae in our Galaxy, LMC, and M31.
 
There are also common properties in the color evolution in nova 
outbursts.  \citet[][hereafter Paper I]{hac14k} found that many novae
in outburst evolve on a common  path in the $(B-V)_0$-$(U-B)_0$
color-color diagram.  Here, $(B-V)_0$ and $(U-B)_0$ are the intrinsic
colors of $B-V$ and $U-B$, respectively.  The evolution path starts
from the nova-giant sequence and then moves to the free-free emission
phase followed by the emission line dominant nebular phase.  
(See, e.g., their Figure 28.)

\citet[][Paper II]{hac16kb} found that many novae follow similar
evolution path in the $(B-V)_0$-$M_V$ color-magnitude diagram. 
Here, $M_V$ is the absolute $V$ magnitude in the $V$ band.
They extensively examined the color-magnitude tracks of 40 novae
in outburst and found several distinct ``template tracks''
in the $(B-V)_0$-$M_V$ diagram.  They proposed six template tracks
of novae and dubbed them the V1500~Cyg, V1668~Cyg, V1974~Cyg, LV~Vul,
FH~Ser, and PU~Vul type tracks.  The difference among the groups
may be related to the envelope mass, which is closely related to
the nova evolution speed, but was not fully clarified. 
They categorized 40 novae into one of these six subgroups, 
depending on their similarity of the tracks in the
$(B-V)_0$-$M_V$ diagram.  They are not fully successful, however, to
explain why each nova has a similar $M_V$ in the $(B-V)_0$-$M_V$ diagram.

Nova light curves show a similar property in the 
$(t/f_{\rm s})$-$(M_V-2.5 \log f_{\rm s})$ plane.
The color-magnitude evolution should also be treated in the same 
way, that is, based on the time-stretching method. 
In the present work, we propose the $(B-V)_0$-$(M_V-2.5\log f_{\rm s})$
color magnitude diagram.
We call this ``time-stretched color-magnitude diagram'' in order to
distinguish it from the usual $(B-V)_0$-$M_V$ diagram.
The time-stretching method is useful because many nova light curves
follow a universal decline law.  Once their absolute $V$ magnitude 
is time-stretched, nova light curves overlap each other 
in the $(t/f_{\rm s})$-$(M_V-2.5\log f_{\rm s})$ light curve \citep{hac10k}.
Here, we stretch/squeeze the timescale of a target nova
as $t'=t/f_{\rm s}$ and overlap its light curve to the comparison (template)
nova.  Then, we have the relation of
\begin{eqnarray}
\left( M_V[t] \right)_{\rm template} 
&=& \left( M'_V[t'] \right)_{\rm target} \cr
&=& \left( M_V[t/f_{\rm s}]-2.5\log f_{\rm s} \right)_{\rm target}
\label{time-stretching_general}
\end{eqnarray}
between $(t', M'_V)$ and $(t, M_V)$ coordinates systems
\citep[see Appendix A of][]{hac16k}, where $M_V[t]$ means that
$M_V$ is a function of time $t$.  
The color $(B-V)_0$ is not affected by this time-stretch
because the both fluxes of $B$ and $V$ are
shifted by the same quantities, $-2.5\log f_{\rm s}$ \citep{hac18k}.
We thus examine each nova evolution 
in the $(B-V)_0$-$(M_V-2.5\log f_{\rm s})$ diagram.
The above two reasons, Equation (\ref{time-stretching_general}) and
that the color $(B-V)_0$ is not affected by time-stretch, are 
the theoretical background for the property that each nova track
overlaps in the time-stretched color-magnitude diagram.

To summarize the theoretical background, 
both the colors $(B-V)_0$ and $(U-B)_0$ are not affected by time-stretch.
Therefore, the $(t/f_{\rm s})$-$(B-V)_0$ and $(t/f_{\rm s})$-$(U-B)_0$
color curves overlap each other after time-stretch. 
As a result, many novae in outburst evolve on a common  path 
in the usual $(B-V)_0$-$(U-B)_0$ color-color diagram.  This is the
reason why \citet{hac14k} found a common track in the $(B-V)_0$-$(U-B)_0$
diagram.   On the other hand, the $(t/f_{\rm s})$-$M_V$ light curves
do not overlap because the absolute $M_V$ magnitudes are different 
among novae.  This is the reason why \citet{hac16kb} did not find
common tracks in the $(B-V)_0$-$M_V$ diagram.  If we use
the $(B-V)_0$-$(M_V-2.5\log f_{\rm s})$ diagram, we could find
common tracks because the $(t/f_{\rm s})$-$(M_V-2.5\log f_{\rm s})$
light curves overlap each other after time-stretch.

In this work, we examine nova evolution in the time-stretched
color-magnitude diagram.  We adopt total 20 novae, including 12
novae that were already analyzed in Paper II, but reanalyzed
here in a new light of the time-stretched color-magnitude diagram.
Our aim is to provide a uniform set of 20 novae analyzed by 
the same time-stretched color-magnitude diagram method.   

Our paper is organized as follows.  First we describe our method
for four well-studied novae in Section \ref{method_example}.  Then,
we apply this method to four recent novae V574~Pup, V679~Car, V1369~Cen,
and V5666~Sgr to determine their various physical parameters
in Sections \ref{v574_pup_cmd}, \ref{v679_car_cmd}, \ref{v1369_cen_cmd},
and \ref{v5666_sgr_cmd}, respectively.  In Section \ref{twelve_novae},
we reanalyze twelve novae, which were studied in Paper II
on the usual color-magnitude diagram.  We confirm that all the novae
follow one of the two template tracks, V1500~Cyg or LV~Vul,
in the time-stretched color-magnitude diagram.  In Section \ref{discussion},
we compared our distance estimates with the Gaia Data Release 2.
Conclusions follows in Section \ref{conclusions}.


\begin{deluxetable*}{llllrrll}
\tabletypesize{\scriptsize}
\tablecaption{Extinctions, distance moduli, and distances for selected novae
\label{extinction_various_novae}}
\tablewidth{0pt}
\tablehead{
\colhead{Object} & \colhead{Outburst} & \colhead{$E(B-V)$} 
& \colhead{$(m-M)_V$} & \colhead{$d$} 
& \colhead{$\log f_{\rm s}$\tablenotemark{a}} &
\colhead{$(m-M')_V$} & \colhead{Ref.\tablenotemark{b}} \\
  & year &  &  &  (kpc) &  &  & 
} 
\startdata
CI~Aql & 2000 & 1.0 & 15.7 & 3.3 & $-0.22$ & 15.15 & 3 \\
V1419~Aql & 1993 & 0.52 & 15.0 & 4.7 & $+0.15$ & 15.35 & 4 \\
V679~Car & 2008 & 0.69 & 16.1 & 6.2 & $+0.0$ & 16.1 & 4 \\
V705~Cas & 1993 & 0.45 & 13.45 & 2.6 & $+0.45$ & 14.55 & 4 \\
V1065~Cen & 2007 & 0.45 & 15.0 & 5.3 & $+0.0$ & 15.0 & 2 \\
V1369~Cen & 2013 & 0.11 & 10.25 & 0.96 & $+0.17$ & 10.65 & 4 \\
IV~Cep & 1971 & 0.65 & 14.5 & 3.1 & $+0.0$ & 14.5 & 2 \\
T~CrB & 1946 & 0.056 & 10.1 & 0.96 & $-1.32$ & 13.4 & 3 \\
V407~Cyg & 2010 & 1.0 & 16.1 & 3.9 & $-0.37$ & 15.2 & 3 \\
V1500~Cyg & 1975 & 0.45 & 12.3 & 1.5 & $-0.22$ & 11.75 & 1 \\
V1668~Cyg & 1978 & 0.30 & 14.6 & 5.4 & $+0.0$ & 14.6 & 1 \\
V1974~Cyg & 1992 & 0.30 & 12.2 & 1.8 & $+0.03$ & 12.3 & 1 \\
V2362~Cyg & 2006 & 0.60 & 15.4 & 5.1 & $+0.25$ & 16.05 & 4 \\
V2468~Cyg & 2008 & 0.65 & 16.2 & 6.9 & $+0.38$ & 17.15 & 4 \\
V2491~Cyg & 2008 & 0.45 & 17.4 & 15.9 & $-0.34$ & 16.55 & 4 \\
YY~Dor & 2010 & 0.12 & 18.9 & 50.0 & $-0.72$ & 17.1 & 3 \\
V446~Her & 1960 & 0.40 & 11.95 & 1.38 & $+0.0$ & 11.95 & 4 \\
V533~Her & 1963 & 0.038 & 10.65 & 1.28 & $+0.08$ & 10.85 & 4 \\
V838~Her & 1991 & 0.53 & 13.7 & 2.6 & $-1.22$ & 10.65 & 3 \\
V959~Mon & 2012 & 0.38 & 13.15 & 2.5 & $+0.14$ & 13.5 & 2 \\
RS~Oph  & 2006 & 0.65 & 12.8 & 1.4 & $-1.02$ & 10.25 & 3 \\
V2615~Oph  & 2007 & 0.90 & 15.95 & 4.3 & $+0.20$ & 16.45 & 4 \\
V574~Pup  & 2004 & 0.45 & 15.0 & 5.3 & $+0.10$ & 15.25 & 4 \\
U~Sco & 2010 & 0.26 & 16.3 & 12.6 & $-1.32$ & 13.0 & 3 \\
V745~Sco & 2014 & 0.70 & 16.6 & 7.8 & $-1.32$ & 13.3 & 3 \\
V1534~Sco & 2014 & 0.93 & 17.6 & 8.8 & $-1.22$ & 14.55 & 3 \\
V496~Sct & 2009 & 0.45 & 13.7 & 2.9 & $+0.30$ & 14.45 & 4 \\
V5114~Sgr & 2004 & 0.47 & 16.65 & 10.9 & $-0.12$ & 16.35 & 4 \\
V5666~Sgr & 2014 & 0.50 & 15.4 & 5.8 & $+0.25$ & 16.0 & 4 \\
V382~Vel & 1999 & 0.25 & 11.5 & 1.4 & $-0.29$ & 10.75 & 4 \\
LV~Vul & 1968 & 0.60 & 11.85 & 1.0 & $+0.0$ & 11.85 & 1,2,4 \\
PW~Vul & 1984 & 0.57 & 13.0 & 1.8 & $+0.35$ & 13.85 & 4 \\
LMCN~2009a & 2014 & 0.12 & 18.9 & 50.0 & $-0.52$ & 17.6 & 3 \\
LMCN~2012a & 2012 & 0.12 & 18.9 & 50.0 & $-1.22$ & 15.85 & 3 \\
LMCN~2013 & 2013 & 0.12 & 18.9 & 50.0 & $-0.42$ & 17.85 & 3 \\
SMCN~2016 & 2016 & 0.08 & 16.8 & 20.4 & $-0.72$ & 15.0 & 3 \\
M31N~2008-12a & 2015 & 0.30 & 24.8 & 780 & $-1.32$ & 21.5 & 3 \\
\enddata
\tablenotetext{a}{
$f_{\rm s}$ is the timescale against that of LV~Vul.}
\tablenotetext{b}{
1 - \citet{hac16kb},
2 - \citet{hac18k},
3 - \citet{hac18kb},
4 - present paper,
} 
\end{deluxetable*}


\begin{deluxetable*}{lrlllll}
\tabletypesize{\scriptsize}
\tablecaption{White dwarf masses of selected novae
\label{wd_mass_novae}}
\tablewidth{0pt}
\tablehead{
\colhead{Object} & \colhead{$\log f_{\rm s}$}
& \colhead{$M_{\rm WD}$}
& \colhead{$M_{\rm WD}$}
& \colhead{$M_{\rm WD}$}
& \colhead{$M_{\rm WD}$}
& \colhead{Chem.comp.}
\\
\colhead{} & \colhead{}
& \colhead{$f_{\rm s}$\tablenotemark{a}}
& \colhead{UV~1455~\AA\tablenotemark{b}}
& \colhead{$t_{\rm SSS-on}$\tablenotemark{c}}
& \colhead{$t_{\rm SSS-off}$\tablenotemark{d}}
&
\\
\colhead{} & \colhead{} & \colhead{($M_\sun$)} & \colhead{($M_\sun$)}
& \colhead{($M_\sun$)} & \colhead{($M_\sun$)} &
}
\startdata
CI~Aql   & $-0.22$ & 1.18  & --- & --- & --- & interp.\tablenotemark{e}\\
V1419~Aql & $+0.15$ & 0.90 & --- & --- & --- & CO3 \\
V679~Car & $+0.0$ & 0.98 & --- & --- & --- & CO3 \\
V705~Cas & $+0.45$ & 0.78 & 0.78 & --- & --- & CO4 \\
V1065~Cen & $+0.0$ & 0.98 & --- & --- & --- & CO3 \\
V1369~Cen & $+0.17$ & 0.90 & --- & --- & --- & CO3 \\
IV~Cep & $+0.0$ & 0.98 & --- & --- & --- & CO3 \\
T~CrB   & $-1.32$ & 1.38  & --- & --- & --- & interp.\\
V407~Cyg   & $-0.37$ & 1.22  & --- & --- & --- & interp.\\
V1500~Cyg & $-0.22$ & 1.20 & --- & --- & --- & Ne2 \\
V1668~Cyg & $+0.0$ & 0.98 & 0.98 & --- & --- & CO3 \\
V1974~Cyg & $+0.03$ & 0.98 & 0.98 & 0.98 & 0.98 & CO3 \\
V2362~Cyg & $+0.25$ & 0.85 & --- & --- & --- & interp. \\
V2468~Cyg & $+0.38$ & 0.85 & --- & --- & --- & CO4 \\
V2491~Cyg & $-0.34$ & 1.35 & --- & 1.35 & 1.35 & Ne2 \\
YY~Dor  & $-0.72$ & 1.29  & ---  & --- & --- & interp. \\
V446~Her & $+0.0$ & 0.98 & --- & --- & --- & CO3 \\
V533~Her & $+0.08$ & 1.03 & --- & --- & --- & Ne2 \\
V838~Her & $-1.22$ & 1.35 & 1.35 & --- & --- & Ne2 \\
V838~Her & $-1.22$ & 1.37 & 1.37 & --- & --- & Ne3 \\
V959~Mon & $+0.14$ & 0.95 & --- & 0.95 & --- & CO3 \\
V959~Mon & $+0.14$ & 1.05 & --- & 1.05 & --- & Ne2 \\
V959~Mon & $+0.14$ & 1.1 & --- & 1.10 & --- & Ne3 \\
RS~Oph   & $-1.02$ & 1.35 & 1.35 & 1.35 & 1.35 & evol.\tablenotemark{f} \\
V2615~Oph  & $+0.20$ & 0.90 & --- & --- & --- & CO3 \\
V574~Pup  & $+0.10$ & 1.05 & --- & 1.05 & 1.05 & Ne2 \\
U~Sco    & $-1.32$ & 1.37 & --- & 1.37 & 1.37 & evol. \\
V745~Sco & $-1.32$ & 1.38 & --- & 1.385 & 1.385 & evol. \\
V1534~Sco & $-1.22$ & 1.37 & --- & --- & --- & interp.\\
V496~Sct & $+0.30$ & 0.85 & --- & --- & ---0 & CO3 \\
V5114~Sgr & $-0.12$ & 1.15 & --- & --- & --- & Ne2 \\
V5666~Sgr & $+0.25$ & 0.85 & --- & --- & --- & CO3 \\
V382~Vel & $-0.29$ & 1.23 & --- & --- & 1.23 & Ne2 \\
LV~Vul   &  $+0.0$  & 0.98 & --- & --- & --- & CO3 \\ 
PW~Vul & $+0.35$ & 0.83 & 0.83 & --- & --- & CO4 \\
LMC~N~2009a & $-0.52$ & 1.25 & --- & 1.25 & 1.25 & Ne3 \\
LMC~N~2012a & $-1.22$ & 1.37 & --- & --- & --- & interp.\\
LMC~N~2013 & $-0.42$ & 1.23 & --- & --- & --- & interp. \\
SMC~N~2016 & $-0.72$ & 1.29 & --- & --- & 1.3 & Ne3 \\
SMC~N~2016 & $-0.72$ & 1.29 & --- & --- & 1.25 & Ne2 \\
M31N~2008-12a & $-1.32$ & 1.38 & --- & 1.38 & 1.38 & evol. \\
\enddata
\tablenotetext{a}{WD mass estimated from the $f_{\rm s}$ timescale.}
\tablenotetext{b}{WD mass estimated from the UV~1455~\AA\  fit.}
\tablenotetext{c}{WD mass estimated from the $t_{\rm SSS-on}$ fit.}
\tablenotetext{d}{WD mass estimated 
from the $t_{\rm SSS-off}$ fit.}
\tablenotetext{e}{``interp.'': WD mass estimated from a linear interpolation
of $\log f_{\rm s}$ vs. WD mass relation \citep[see][]{hac18kb}.}
\tablenotetext{f}{``evol.'': WD mass estimated from the time-evolution
calculation with a Henyey type code \citep[see, e.g.,][]{hac18kb}.}
\end{deluxetable*}

\section{Method and Template Novae}
\label{method_example}
\subsection{Method}
\label{reddening_distance_method}
First, we need to know the intrinsic color and time-stretching
factor $f_{\rm s}$ in order to analyze a particular nova light curve.  
The procedure is summarized as follows:\\
{\bf 1.} Calculate the intrinsic colors of $(B-V)_0$ and $(U-B)_0$ as
\begin{equation}
(B-V)_0 = (B-V) - E(B-V),
\label{dereddening_eq_bv}
\end{equation}
and
\begin{equation}
(U-B)_0 = (U-B) - 0.64 E(B-V),
\label{dereddening_eq_ub}
\end{equation}
where the factor of $0.64$ is taken from \citet{rie85}. 
The color excess $E(B-V)$ is taken from literature,
if available.  \\
{\bf 2.} Determine the time-stretching factor $f_{\rm s}$ of 
a target nova from a comparison with a template nova in the light/color
curves (see, e.g., Figure \ref{v574_pup_v1974_cyg_v_bv_logscale}).
We adopt the template nova from one of LV~Vul, V1500~Cyg, V1668~Cyg,
and V1974~Cyg, of which the distance modulus $\mu_V\equiv (m-M)_V$
and extinction $E(B-V)$ are well determined. 
We use LV~Vul as the template nova unless otherwise specified.
The timescaling factor $f_{\rm s}$ is measured against the timescale
of LV~Vul.
In the light/color curves, we stretch/squeeze the timescale of the
target nova (horizontal shift in a logarithmic timescale)
and also shift the magnitude in the vertical direction.
The light curve of the target nova overlaps
the template nova by stretching horizontally 
(in the time-direction) by a factor of $f_{\rm s}$
and by shifting vertically by $\Delta V$,
we estimate the distance modulus of the target nova,
\begin{equation}
(m-M)_{V,\rm target} 
= \left( (m-M)_V + \Delta V \right)_{\rm template} - 2.5 \log f_{\rm s}.
\label{distance_modulus_general_temp}
\end{equation}
\citet{hac10k} called this method the time-stretching method.
We usually increase/decrease the horizontal shift by a step of 
$\delta (\Delta \log t) = \delta (\log f_{\rm s}) = 0.01$ 
and the vertical shift by a step of 
$\delta (\Delta V)= 0.1$~mag, and determine the best one by eye.
The model light curves obey the universal decline law.  
This phase corresponds to the optically thick wind phase
before the nebular phase started.  In fact, many novae show a good
agreement with the model light curve in the optically thick phase.
Moreover, we find that the $V$ light curve of a target nova often
show a good agreement with that of a template nova even in the nebular
(optically thin) phase.  In such a case, we use all the phases
to determine the timescaling factor (see, e.g., Figure
\ref{lv_vul_yy_dor_lmcn_2009a_v_b_logscale_2fig}).
\\
{\bf 3.} Determine the WD mass and absolute brightness from
the comparison with model light curves calculated by 
\citet{hac10k, hac15k, hac16k}.  These light curves are obtained
for various WD masses and chemical compositions.
The distance modulus in the $V$ band, $(m-M)_V$, is determined 
independently of procedure No.2.  \\
{\bf 4.} The above procedures in No.2 and No.3 are independent methods.
Thus, good agreement of the two $(m-M)_V$ confirms our method.
We obtain the distance $d$ to the target nova from the relation of
\begin{eqnarray}
(m-M)_V = 3.1 E(B-V) + 5 \log (d/10~{\rm pc}),
\label{distance_modulus_rv}
\end{eqnarray}
where the factor $R_V=A_V/E(B-V)=3.1$ is the ratio of total to selective
extinction \citep[e.g.,][]{rie85}.\\
{\bf 5.}
We apply the same time-stretching method to the $U$, $B$, $I_{\rm C}$,
and $K_{\rm s}$ bands, if they are available, that is,
\begin{equation}
(m-M)_{U,\rm target} = ((m-M)_U + \Delta U
- 2.5 \log f_{\rm s})_{\rm template},
\label{distance_modulus_general_temp_u}
\end{equation}
\begin{equation}
(m-M)_{B,\rm target} = ((m-M)_B + \Delta B
- 2.5 \log f_{\rm s})_{\rm template},
\label{distance_modulus_general_temp_b}
\end{equation}
\begin{equation}
(m-M)_{I,\rm target} = ((m-M)_I + \Delta I_{\rm C}
- 2.5 \log f_{\rm s})_{\rm template},
\label{distance_modulus_general_temp_i}
\end{equation}
\begin{equation}
(m-M)_{K,\rm target} = ((m-M)_K + \Delta K_{\rm s}
- 2.5 \log f_{\rm s})_{\rm template},
\label{distance_modulus_general_temp_k}
\end{equation}
where the timescaling factor $f_{\rm s}$ is unique and the same as that in
Equation (\ref{distance_modulus_general_temp}).  Thus, we independently
obtain the five distance moduli in the $U$, $B$, $V$, $I_{\rm C}$, and
$K_{\rm s}$ bands
(for example, Figure \ref{v574_pup_yy_dor_lmcn_2009a_b_v_i_k_logscale_4fig}).
Then, we obtain the relation between the distance $d$ and color excess
$E(B-V)$ to the target nova
where we use the relation of
\begin{eqnarray}
(m-M)_U = 4.75 E(B-V) + 5 \log (d/10~{\rm pc}),
\label{distance_modulus_ru}
\end{eqnarray}
\begin{eqnarray}
(m-M)_B = 4.1 E(B-V) + 5 \log (d/10~{\rm pc}),
\label{distance_modulus_rb}
\end{eqnarray}
\begin{eqnarray}
(m-M)_I = 1.5 E(B-V) + 5 \log (d/10~{\rm pc}),
\label{distance_modulus_ri}
\end{eqnarray}
\begin{eqnarray}
(m-M)_K = 0.35 E(B-V) + 5 \log (d/10~{\rm pc}),
\label{distance_modulus_rk}
\end{eqnarray}
and $A_U/ E(B-V)=4.75$, $A_B/E(B-V)=4.1$, $A_I/E(B-V)= 1.5$, and
$A_K/E(B-V)= 0.35$ are taken from \citet{rie85}.
We plot these relations of Equations
(\ref{distance_modulus_ru}),
(\ref{distance_modulus_rb}),
(\ref{distance_modulus_rv}), 
(\ref{distance_modulus_ri}), and
(\ref{distance_modulus_rk}) in the reddening-distance plane.
If these lines cross at the same point, this point gives
correct values of reddening $E(B-V)$ and distance $d$ (for example,
Figure \ref{distance_reddening_v574_pup_v679_car_v1369_cen_v5666_sgr}(a)).\\
{\bf 6.} Using $E(B-V)$, $(m-M)_V$, and $f_{\rm s}$ obtained in the above
procedures, we plot the $(B-V)_0$-$M_V$ color-magnitude diagram 
and the $(B-V)_0$-$(M_V - 2.5\log f_{\rm s})$ 
time-stretched color-magnitude diagram of the target nova
(for example, Figures
\ref{hr_diagram_v574_pup_v679_car_v1369_cen_v5666_sgr_outburst_mv}
and \ref{hr_diagram_v574_pup_v679_car_v1369_cen_v5666_sgr_outburst},
respectively).
We derive the relation of
\begin{eqnarray}
(m-M')_{V,\rm target} & \equiv & \left( m_V-(M_V - 
2.5\log f_{\rm s})\right)_{\rm target} \cr
&=& ((m-M)_V + \Delta V)_{\rm template},
\label{distance_modulus_general_dot}
\end{eqnarray}
from Equations (\ref{time-stretching_general}) and
(\ref{distance_modulus_general_temp}).\\
{\bf 7.} If the track of the target nova overlaps to the LV~Vul
(or V1500~Cyg, V1668~Cyg, V1974~Cyg) template track
in the time-stretched color-magnitude diagram,
we regard that our set of $f_{\rm s}$, $E(B-V)$, $(m-M)_V$, and 
distance $d$ are reasonable.


\begin{figure*}
\plotone{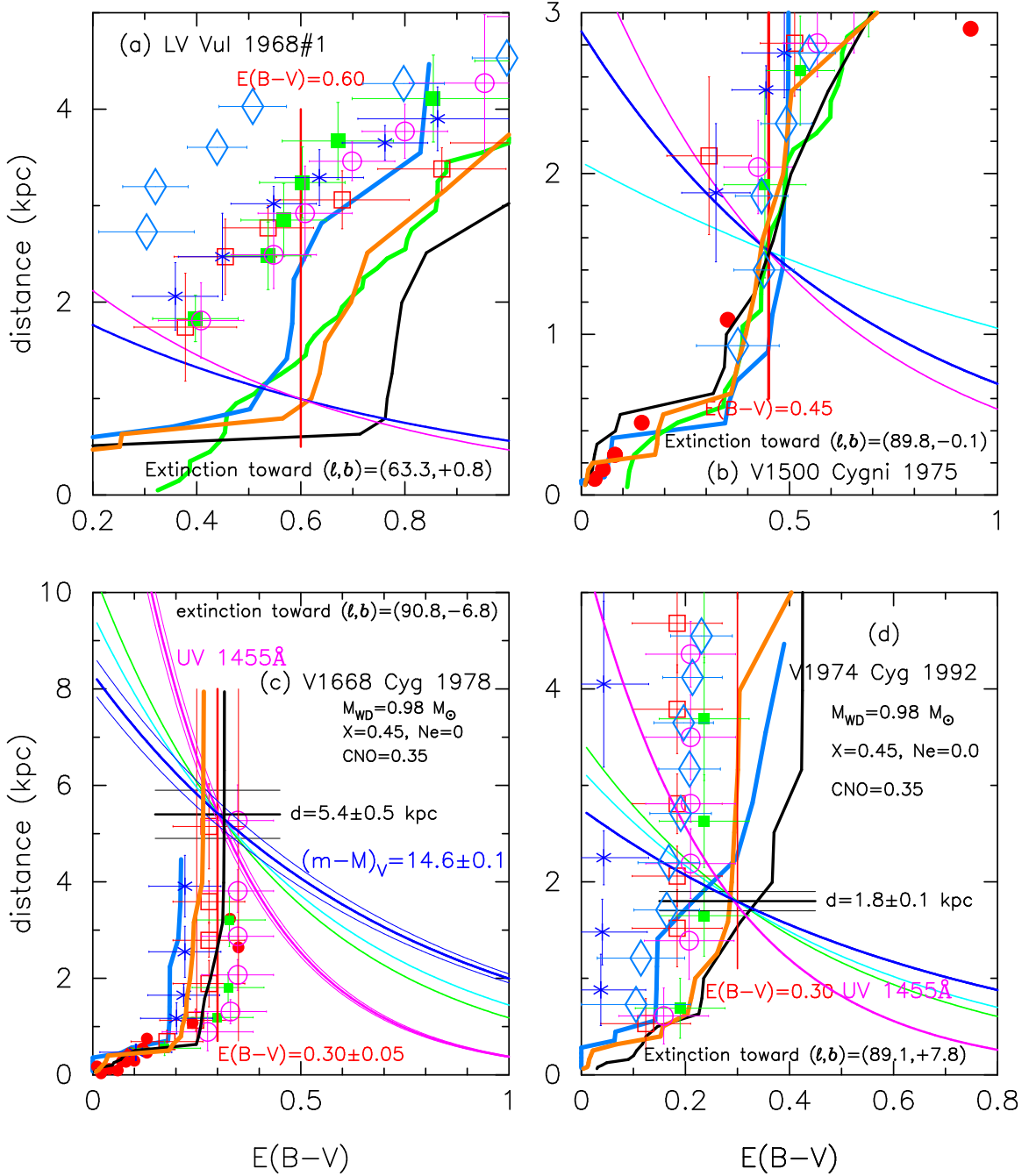}
\caption{
Distance-reddening relations toward our template novae:
(a) LV~Vul, (b) V1500~Cyg, (c) V1668~Cyg, and (d) V1974~Cyg.
The thick solid blue lines denote
(a) $(m-M)_V=11.85$, (b) $(m-M)_V=12.3$, (c) $(m-M)_V=14.6$,
and (d) $(m-M)_V=12.2$.
The vertical solid red lines represent the color excess of each nova.
The open cyan-blue diamonds denote the distance-reddening relations given
by \citet{ozd16}.  The solid black/orange lines represent the 
distance-reddening relations given by \citet{gre15, gre18}, respectively.
The thick solid cyan-blue lines denote the relations given by \citet{chen18}.
The four sets (open red squares, filled green squares, blue asterisks,
and open magenta circles) of data with error bars show 
the distance-reddening relations of \citet{mar06} 
in four nearby directions of each nova. In panels (a) and (b),
the green solid lines indicate the distance-reddening relations
given by \citet{sal14}.  In panels (a) and (b), the thin solid magenta line
denotes the $B$ band distance modulus.  In panel (b), the solid 
cyan line represents the $I$ band distance modulus.
In panels (b) and (c), the filled red circles denote the distances
and reddenings of nearby stars given by \citet{you76} and \citet{slo79},
respectively.  In panels (c) and (d), the solid magenta lines represent
the distance-reddening relation from the UV~1455\AA\  light curve fitting,
and the thin solid green/cyan lines depict the distance moduli of
$U$/$B$ bands, respectively.
See text for more details.
\label{distance_reddening_lv_vul_v1500_cyg_v1668_cyg_v1974_cyg}}
\end{figure*}

\subsection{Reddening and distance}
\label{distance_comparison}
In what follows, we use LV Vul, V1500 Cyg, V1668 Cyg, and V1974 Cyg
as template novae, because their distances and extinctions have been
well determined.  
Figure \ref{distance_reddening_lv_vul_v1500_cyg_v1668_cyg_v1974_cyg} 
shows the distance-reddening relations toward the four template novae,
(a) LV Vul, (b) V1500 Cyg, (c) V1668 Cyg, and (d) V1974 Cyg.
We have already analyzed them in our previous papers 
\citep[e.g.,][]{hac16k, hac16kb}, but here we reanalyze them because
new distance-reddening relations are recently available 
\citep[e.g.,][]{gre18, chen18}.  
To summarize, we use five results: (1) 
\citet{mar06} published a three-dimensional (3D) extinction map
of our Galaxy.  The range is $-100\fdg0 \le l \le 100\fdg0$
and $-10\fdg0 \le b \le +10\fdg0$ and the resolution of grids is
$\Delta l=0\fdg25$ and $\Delta b=0\fdg25$, where
$(l, b)$ is the galactic coordinates.
(2) \citet{sal14} presented a 3D reddening map with
the region of $30\arcdeg \le l < 215\arcdeg$ and $|b|<5\arcdeg$.
Their data are based on the INT Photometric H-Alpha Survey (IPHAS) photometry.
(3) \citet{gre15} published a 3D reddening map with
a wider range of three quarters of the sky and with much finer grids of
3\farcm4 to 13\farcm7 whose distance resolution is 25\%.  
The data of distance-reddening relation were recently revised
by \citet{gre18}.
(4) \citet{ozd16} obtained distance-reddening relations toward 46 novae
based on the unique position of the red clump giants
in the color-magnitude diagram.  Recently, \citet{ozd18} added
the data of recent novae.
(5) \citet{chen18} presented 3D interstellar dust reddening maps
of the galactic plane in three colors, $E(G - K_{\rm s})$, 
$E(G_{\rm BP}-G_{\rm RP})$,  and $E(H - K_{\rm s})$. 
We use the conversion of $E(B-V)= 0.75 E(G_{\rm BP}-G_{\rm RP})$
from \citet{chen18}. 
The maps have a spatial angular resolution of 6 arcmin and covers 
the galactic longitude $0 < l < 360^\circ$ and latitude $|b| < 10^\circ$.
The maps are constructed from parallax estimates from the Gaia Data Release 2
(Gaia DR2) combined with the photometry from the Gaia DR2
and the infrared photometry from the 2MASS and WISE surveys.

These five 3D dust maps often show large discrepancies (see, e.g., Figure
\ref{distance_reddening_lv_vul_v1500_cyg_v1668_cyg_v1974_cyg}(a)).
The 3D dust maps essentially give an averaged value
of a relatively broad region and thus, the pinpoint reddening could be
different from it, because
the resolutions of these dust maps are considerably larger
than molecular cloud structures observed in the interstellar medium.
For example, Figure
\ref{distance_reddening_lv_vul_v1500_cyg_v1668_cyg_v1974_cyg}(a) shows
a large discrepancy among the various relations.  The orange line
\citep{gre18} is quite consistent with our values, i.e., 
the crossing point of $(m-M)_V=11.85$ (blue line)
and $E(B-V)=0.60$ (vertical solid red line).
This will be discussed in detail in Section \ref{lv_vul}.


\begin{figure}
\plotone{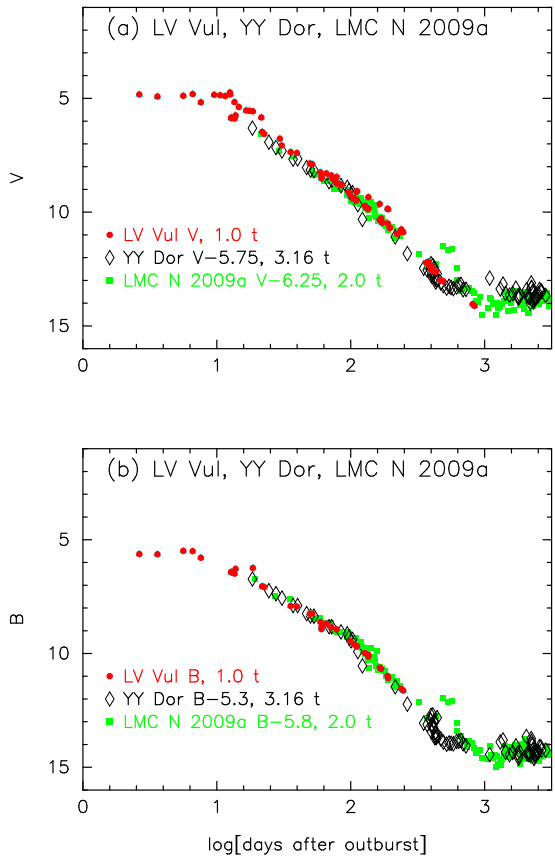}
\caption{
The (a) $V$ and (b) $B$ light curves of LV Vul,
YY Dor (2004), and LMC N 2009a on logarithmic timescales.
Each light curve is horizontally moved by $\Delta \log t = \log f_{\rm s}$
and vertically shifted by $\Delta V$ (or $\Delta B$)
with respect to that of LV Vul,
as indicated in the figure by, for example, 
``YY Dor V$-5.75$, 3.16 t,'' (``YY Dor B$-5.3$, 3.16 t,'' ),
where $\Delta V= -5.75$ ($\Delta B= -5.3$) and $f_{\rm s}= 3.16$. 
\label{lv_vul_yy_dor_lmcn_2009a_v_b_logscale_2fig}}
\end{figure}

\subsection{Template nova light curves}
\label{template_novae}

Here, we reexamine the four template novae LV~Vul, V1500~Cyg, V1668~Cyg,
and V1974~Cyg, with the new distance-reddening relations
\citep{gre18, chen18} and fix the distances and extinctions for later use.

\subsubsection{LV~Vul 1968\#1}
\label{lv_vul}

The reddening toward LV~Vul was estimated by \citet{fer69} to be
$E(B-V)=0.6\pm 0.2$ from the color excesses of 14 B stars near the line
of sight.  \citet{tem72} obtained $E(B-V)=0.55$ from the color 
at optical maximum; i.e., $E(B-V)=(B-V)_{\rm max} - (B-V)_{0, \rm max}= 
0.9 - 0.35 = 0.55$.  He adopted $(B-V)_{0, \rm max}= + 0.35$ 
\citep{sch57} instead of $(B-V)_{0, \rm max}=+0.23$ \citep{van87}.
\citet{hac14k} obtained $E(B-V)=0.60\pm0.05$ by fitting with the
typical color-color evolution track of nova outbursts.
\citet{sla95} obtained the distance toward LV~Vul to be 
$d=0.92\pm 0.08$~kpc by a nebular expansion parallax method.

Figure \ref{lv_vul_yy_dor_lmcn_2009a_v_b_logscale_2fig} shows the (a) 
$V$ and (b) $B$ light curves of LV Vul as well as YY Dor (2004) 
and LMC N 2009a.  The LV~Vul data are the same as those in Figure 4 of
Paper II.  The light curves of YY Dor and LMC N 2009a are 
time-stretched against LV Vul.  These two novae are analyzed 
by \citet{hac18kb} and
the various properties are summarized in their Tables 1, 2, and 3.
They determined the timescaling factors of YY~Dor and LMC~N~2009a
against LV~Vul, which are plotted in Figure 
\ref{lv_vul_yy_dor_lmcn_2009a_v_b_logscale_2fig}.
In this case, we shift up/down the YY~Dor and LMC~N~2009a light curves
in a step of $\delta(\Delta V)= 0.1$ mag (or $\delta(\Delta B)= 0.1$ mag)
and find the best fit one by eye.  We use all the part of the light curve
if they overlap each other.  If not, we use only the part of the
optically thick wind phase before the nebular (or dust blackout) phase
started.  Fortunately, the three light curves of LV Vul, YY Dor,
and LMC N 2009a broadly overlap each other even during the nebular phase
as shown in Figure \ref{lv_vul_yy_dor_lmcn_2009a_v_b_logscale_2fig}.

YY Dor and LMC N 2009a belong to the Large Magellanic Cloud (LMC)
and their distances are well constrained.
We adopt $\mu_{0,\rm LMC}=18.493\pm0.048$ \citep{pie13}.
The reddenings toward YY Dor is assumed to be $E(B-V)=0.12$, which is
the typical reddening toward the LMC \citep{imara07}.
The distance modulus in the $V$ band
is $(m-M)_V= \mu_0 + A_V= 18.49 + 3.1\times 0.12 = 18.86$ 
toward the LMC novae.

Applying the obtained $f_{\rm s}$ and $\Delta V$ to Equation
(\ref{distance_modulus_general_temp}), we have the relation of
\begin{eqnarray}
(m&-&M)_{V, \rm LV~Vul} \cr 
&=& \left( (m-M)_V + \Delta V\right)_{\rm YY~Dor} - 2.5 \log 3.16 \cr 
&=& 18.86 - 5.75\pm0.2 - 1.25 = 11.86\pm0.2 \cr
&=& \left( (m-M)_V + \Delta V\right)_{\rm LMCN~2009a} - 2.5 \log 2.0 \cr 
&=& 18.86 - 6.25\pm0.2 - 0.75 = 11.86\pm0.2.
\label{distance_modulus_lv_vul_yy_dor_lmcn_2009a_v}
\end{eqnarray}
Thus, we obtain $(m-M)_V=11.86\pm0.1$ for LV~Vul, being consistent with
our previous results of $(m-M)_V=11.85\pm0.1$ \citep{hac18kb}.

In the same way, we obtain $\Delta B$ from Figure
\ref{lv_vul_yy_dor_lmcn_2009a_v_b_logscale_2fig}(b).
Using this $\Delta B$ and $f_{\rm s}$, we obtain
\begin{eqnarray}
(m&-&M)_{B, \rm LV~Vul} \cr 
&=& \left( (m-M)_B + \Delta B\right)_{\rm YY~Dor} - 2.5 \log 3.16 \cr 
&=& 18.98 - 5.3\pm0.2 - 1.25 = 12.43\pm0.2 \cr
&=& \left( (m-M)_B + \Delta B\right)_{\rm LMCN~2009a} - 2.5 \log 2.0 \cr 
&=& 18.98 - 5.8\pm0.2 - 0.75 = 12.43\pm0.2,
\label{distance_modulus_lv_vul_yy_dor_lmcn_2009a_b}
\end{eqnarray}
from Equation (\ref{distance_modulus_general_temp_b}).
Here, we adopt $(m-M)_B = 18.49 + 4.1\times 0.12= 18.98$ for the LMC novae.
Thus, we obtain $(m-M)_B=12.43\pm0.1$ for LV~Vul.

We determine the stretching factor $f_{\rm s}$ by horizontally shifting
the light and color curves.  Its step is $\delta \log f_{\rm s} = 0.01$
as explained in Section \ref{reddening_distance_method}.  Usually we use
three $BVI_{\rm C}$ light curves and $(B-V)_0$ color curve to overlap
them with those of the template novae against a unique $f_{\rm s}$.
The error of $\log f_{\rm s}$ depends on how well these light/color
curves overlap.  Typically we have $0.03-0.05$ for the error
of $\log f_{\rm s}$.
This error is not independent but propagates to the error of
vertical fit of $\Delta V$.  We estimate the typical total error of
$\epsilon (\Delta V - 2.5 \log f_{\rm s})= (\pm 0.2) + (\pm 0.1)$.
In the case of LV~Vul, $\log f_{\rm s}$ is rather well determined
against YY~Dor and LMC~N~2009a, that is, the error of
$\epsilon( \log f_{\rm s}) = \pm0.02$ \citep[see Section 5.1 of][]{hac18kb}.
Therefore, the total error is about 
$\epsilon (\Delta V - 2.5 \log f_{\rm s})= (\pm 0.2) + (\pm 0.05)$.
In what follows, we include the error of the $f_{\rm s}$ determination
in the error of vertical fit $\Delta V$, because the vertical fit
always reflect the error of $f_{\rm s}$ determination.

We plot these two distance-reddening relations of Equations
(\ref{distance_modulus_rv}) and (\ref{distance_modulus_rb}) in
Figure \ref{distance_reddening_lv_vul_v1500_cyg_v1668_cyg_v1974_cyg}(a)
by the thin solid blue and magenta lines, respectively.
The crossing point of the blue, magenta, orange, and vertical red lines,
i.e., $d=1.0$~kpc and $E(B-V)=0.60$, is in reasonable
agreement with $d=0.92\pm 0.08$~kpc \citep{sla95} and $E(B-V)=0.60\pm0.05$
\citep{hac14k, hac16kb}.  Thus, we confirm that the distance modulus
in the $V$ band is $(m-M)_V=11.85\pm0.1$, the reddening is
$E(B-V)=0.60\pm0.05$, and the distance is $d=1.0\pm0.2$~kpc.
These values are listed in Table \ref{extinction_various_novae}.


\begin{figure}
\plotone{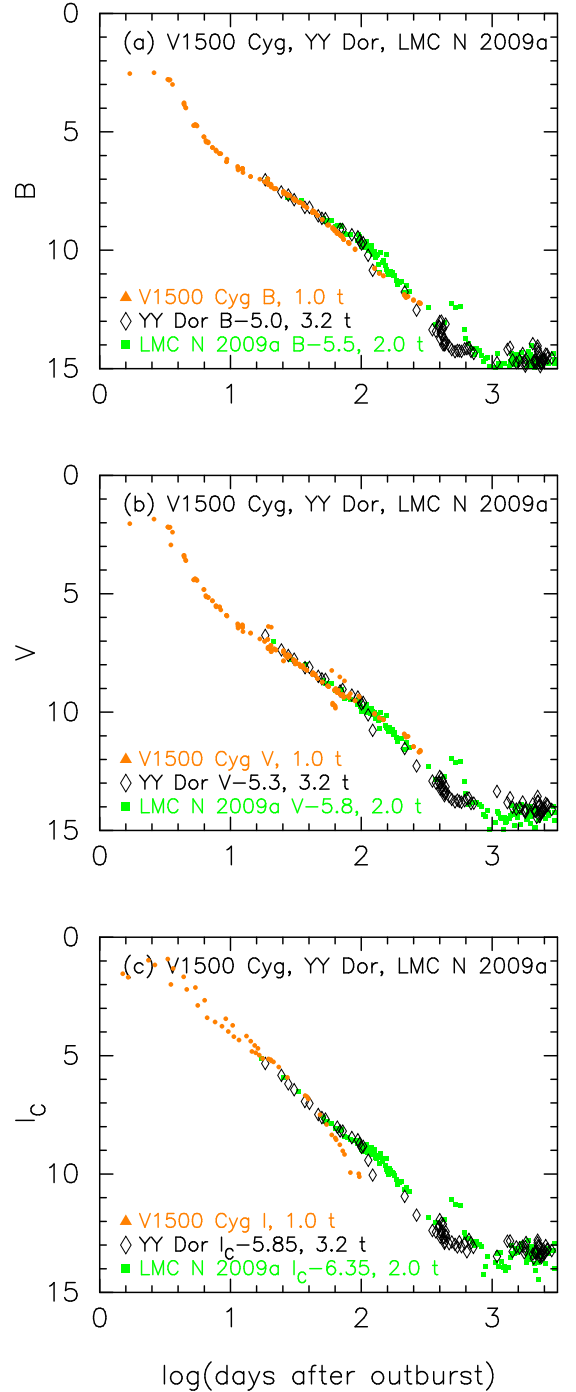}
\caption{
The (a) $B$, (b) $V$, and (c) $I$ 
light curves of V1500 Cyg as well as YY Dor and LMC N 2009a.  
The $BV$ data of V1500~Cyg are taken from \citet{wil77}, 
\citet{tem79}, \citet{pfa76}, and \citet{ark76}.
The $I$ data of V1500~Cyg are taken from \citet{bel77} and 
\citet{the76}.  We plot the $I_{\rm C}$ data for YY Dor and LMC N 2009a
because no $I$ data are available.  
\label{v1500_cyg_yy_dor_lmcn_2009a_b_v_i_logscale_3fig}}
\end{figure}

\subsubsection{V1500~Cyg 1975}
\label{v1500_cyg}

The distance to V1500~Cyg was discussed by many authors.  \citet{you76}
estimated the distance to be $1.4 \pm 0.1$~kpc for $E(B-V)= 0.45$
\citep{tom76} from the distance-reddening law toward the nova (depicted
by the filled red circles in Figure
\ref{distance_reddening_lv_vul_v1500_cyg_v1668_cyg_v1974_cyg}(b)).
We add Marshall et al.'s (2006) relations
for four directions close to V1500~Cyg,
that is, $(l, b)=(89\fdg75, 0\fdg00)$ (red open squares),
$(90\fdg00,  0\fdg00)$ (green filled squares),
$(89\fdg75, -0\fdg25)$ (blue asterisks), and
$(90\fdg00, -0\fdg25)$ (magenta open circles).
We further add the relations of \citet{gre15, gre18}, \citet{sal14},
\citet{ozd16}, and \citet{chen18}.
Green et al.'s and Sale et al.'s
relations (black and green lines)
give $d=1.5$~kpc for $E(B-V)=0.45$.

A firm upper limit to the apparent distance modulus was obtained as 
$(m-M)_V \le 12.5$ by \citet{and76} from the galactic rotational 
velocities of interstellar H and K absorption lines.  
The nebular expansion parallax method also gives an estimate of the distance.
\citet{bec80} first imaged an expanding
nebular ($0\farcs 25$~yr$^{-1}$) of V1500~Cyg and estimated a distance
of 1350 pc from the expansion velocity of $v_{\rm exp} = 1600$~km~s$^{-1}$.
However, \citet{wad91} resolved an expanding nebula and
obtained a much lower expansion rate of $0\farcs 16$~yr$^{-1}$; they
estimated the distance to be 1.56~kpc, with a much smaller expansion
velocity of $v_{\rm exp} = 1180$~km~s$^{-1}$
\citep{coh85}.  \citet{sla95} obtained a similar expanding
angular velocity of the nebula ($0\farcs16$~yr$^{-1}$)
and obtained a distance of 1550 pc from $v_{\rm exp} = 1180$~km~s$^{-1}$.
Therefore, the distance is reasonably determined to be $d=1.5\pm0.1$~kpc
from the nebular expansion parallax method.

\citet{hac14k} obtained $(m-M)_V=12.3$ for V1500 Cyg
by the time-stretching method together with the light curves of GK Per
and V603 Aql (see Figure 40 and Equation (A1) of Paper I).
We plot $(m-M)_V=12.3$ (blue solid line) and $E(B-V)=0.45$ 
(vertical red solid line) in Figure
\ref{distance_reddening_lv_vul_v1500_cyg_v1668_cyg_v1974_cyg}(b).
The crossing point of these two lines is consistent with
the distance-reddening relations of \citet{gre15, gre18},
\citet{sal14}, \citet{ozd16}, and \citet{chen18} except \citet{mar06}.
Thus, we confirm $d=1.5\pm0.1$~kpc,
$E(B-V)=0.45\pm0.05$, and $(m-M)_V=12.3\pm0.1$.
The timescaling factor of V1500~Cyg is determined to be 
$f_{\rm s}=0.60$ against LV~Vul.  
We list these values in Table \ref{extinction_various_novae}.

We check the distance moduli of V1500~Cyg with the LMC novae,
YY~Dor and LMC~N~2009a, based on the time-stretching method.
Figure \ref{v1500_cyg_yy_dor_lmcn_2009a_b_v_i_logscale_3fig} shows
the $BVI$ light curves of V1500~Cyg as well as YY~Dor and
LMC~N~2009a.  Here we use the $I$ band data of V1500~Cyg because
no $I_{\rm C}$ data are available.
We apply Equation (\ref{distance_modulus_general_temp_b}) 
for the $B$ band to Figure 
\ref{v1500_cyg_yy_dor_lmcn_2009a_b_v_i_logscale_3fig}(a)
and obtain
\begin{eqnarray}
(m&-&M)_{B, \rm V1500~Cyg} \cr
&=& ((m - M)_B + \Delta B)_{\rm YY~Dor} - 2.5 \log 3.2 \cr
&=& 18.98 - 5.0\pm0.2 - 1.25 = 12.73\pm0.2 \cr
&=& ((m - M)_B + \Delta B)_{\rm LMC~N~2009a} - 2.5 \log 2.0 \cr
&=& 18.98 - 5.5\pm0.2 - 0.75 = 12.73\pm0.2, 
\label{distance_modulus_b_v1500_cyg_yy_dor_lmcn2009a}
\end{eqnarray}
where we adopt $(m-M)_B = 18.49 + 4.1\times 0.12=18.98$ for the LMC novae.
Thus, we obtain $(m-M)_{B, \rm V1500~Cyg}= 12.73\pm0.1$.

For the $V$ light curves in Figure 
\ref{v1500_cyg_yy_dor_lmcn_2009a_b_v_i_logscale_3fig}(b), we similarly obtain
\begin{eqnarray}
(m&-&M)_{V, \rm V1500~Cyg} \cr
&=& ((m - M)_V + \Delta V)_{\rm YY~Dor} - 2.5 \log 3.2 \cr
&=& 18.86 - 5.3\pm0.2 - 1.25 = 12.31\pm0.2 \cr
&=& ((m - M)_V + \Delta V)_{\rm LMC~N~2009a} - 2.5 \log 2.0 \cr
&=& 18.86 - 5.8\pm0.2 - 0.75 = 12.31\pm0.2,
\label{distance_modulus_v_v1500_cyg_yy_dor_lmcn2009a}
\end{eqnarray}
where we adopt $(m-M)_V= 18.49 + 3.1\times 0.12 = 18.86$ for the LMC novae.
Thus, we obtain $(m-M)_{V, \rm V1500~Cyg}= 12.31\pm0.1$.

We apply Equation (\ref{distance_modulus_general_temp_i}) to Figure
\ref{v1500_cyg_yy_dor_lmcn_2009a_b_v_i_logscale_3fig}(c) and obtain
\begin{eqnarray}
(m&-&M)_{I, \rm V1500~Cyg} \cr
&=& ((m - M)_I + \Delta I_C)_{\rm YY~Dor} - 2.5 \log 3.2 \cr
&=& 18.66 - 5.85\pm0.3 - 1.25 = 11.56\pm 0.3 \cr
&=& ((m - M)_I + \Delta I_C)_{\rm LMC~N~2009a} - 2.5 \log 2.0 \cr
&=& 18.66 - 6.35\pm0.3 - 0.75 = 11.56\pm 0.3, 
\label{distance_modulus_i_v1500_cyg_yy_dor_lmcn2009a}
\end{eqnarray}
where we adopt 
$(m-M)_I = 18.49 + 1.5\times 0.12=18.66$ for the LMC novae.
Thus, we obtain $(m-M)_{I, \rm V1500~Cyg}= 11.56\pm0.2$.

We plot three distance moduli in the $B$, $V$, and $I_{\rm C}$
bands in Figure 
\ref{distance_reddening_lv_vul_v1500_cyg_v1668_cyg_v1974_cyg}(b)
by the magenta, blue, and cyan lines, that is,
$(m-M)_B= 12.73$, $(m-M)_V= 12.31$, and $(m-M)_I= 11.56$,
together with Equations (\ref{distance_modulus_rb}), 
(\ref{distance_modulus_rv}), and (\ref{distance_modulus_ri}),
respectively.
These three lines cross at $d=1.5$~kpc and $E(B-V)=0.45$.
This crossing point is an independent confirmation of
the distance and reddening.


\begin{figure}
\plotone{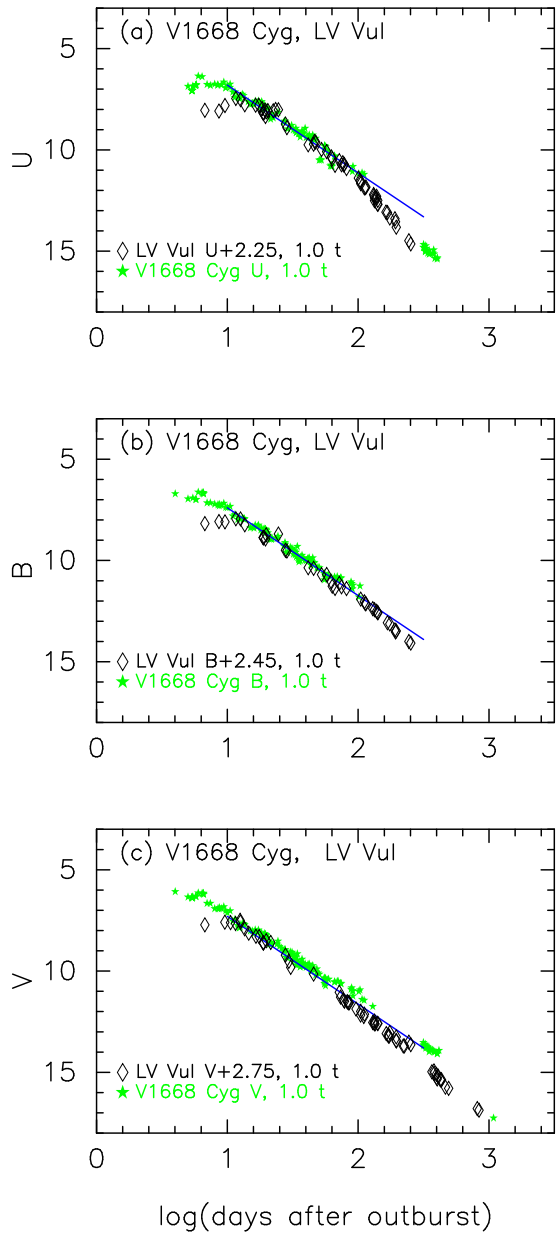}
\caption{
The (a) $U$, (b) $B$,  and (c) $V$ light curves of V1668~Cyg and LV~Vul
on a logarithmic timescale.
Each light curve is vertically shifted by $\Delta U$ (or $\Delta B$
or $\Delta V$) with respect to that of V1668~Cyg as indicated 
in the figure.  The timescales of the two novae are the same.  
The solid blue lines denote the slope of $F_\nu\propto t^{-1.75}$,
which represents well the optically thick wind phase \citep{hac06kb}. 
The $UBV$ data of V1668~Cyg and LV~Vul are the same as
Figures 1 and 4 of \citet{hac16kb}, respectively.
\label{v1668_cyg_lv_vul_u_b_v_logscale_3fig}}
\end{figure}

\subsubsection{V1668~Cyg 1978}
\label{v1668_cyg}
Figure \ref{v1668_cyg_lv_vul_u_b_v_logscale_3fig} shows the $UBV$
light curves of V1668~Cyg and LV~Vul.  The timescale of V1668~Cyg
is the same as LV~Vul, that is, $f_{\rm s}=1.0$ against LV~Vul.
If the chemical composition of V1668~Cyg is similar to that of LV~Vul,
both the WD masses should be similar.  The peak of V1668~Cyg
is slightly brighter than that of LV~Vul.  In general, the peak
brightness depends mainly on the WD mass and initial envelope mass.
The peak magnitude is brighter for the more massive ignition mass
even if the WD masses are the same.
Such a tendency was discussed in detail in \citet{hac10k}
(see their Figure 2 for the physical explanation).
The initial envelope mass, that is, the ignition mass, is closely related
to the mass accretion rate on the WD.  The larger the mass accretion rate, 
the smaller the ignition mass \citep[see, e.g., Figure 3 of][]{kat14shn}.
We suppose that the mass accretion rate is smaller in V1668~Cyg than 
in LV~Vul even if the two WD masses are similar.  

These two nova light curves well overlap except for the peak
brightnesses.  The distance moduli of LV~Vul are well determined 
in Section \ref{lv_vul}, so the distance moduli of V1668~Cyg are
calculated to be
$(m-M)_U = 12.84 + 2.25\pm0.1 = 15.1\pm0.1$,
$(m-M)_B = 12.45 + 2.45\pm0.1 = 14.9\pm0.1$, and
$(m-M)_V = 11.85 + 2.75\pm0.1 = 14.6\pm0.1$
from the time-stretching method.
These three distance moduli cross at $d= 5.4$~kpc and $E(B-V)= 0.30$
in the distance-reddening plane of Figure
\ref{distance_reddening_lv_vul_v1500_cyg_v1668_cyg_v1974_cyg}(c).

\citet{hac06kb, hac16k} calculated free-free emission model light curves 
based on the optically thick wind model \citep{kat94h} for
various WD masses and chemical compositions and fitted
their free-free emission model light curves with the $V$ and $y$
light curves of V1668~Cyg.  They further
fitted their blackbody emission UV~1455\AA\  model light curves
with the {\it International Ultraviolet Explore (IUE)} 
UV~1455\AA\ observation, and estimated the WD mass.
This is a narrow-band (1445--1465~\AA) flux that represents well
the continuum UV fluxes of novae \citep{cas02}.

Assuming the chemical composition of CO nova 3 for
V1668~Cyg, \citet{hac16k} obtained the best fit model of $0.98~M_\sun$ WD
with both the $V$ and UV~1455~\AA\  light curves.
(Such model light curve fittings are also plotted in Figure
\ref{v679_car_lv_vul_v1668_cyg_v_bv_ub_logscale} in Appendix
\ref{v679_car}.)
We plot the distance-reddening relation of $(m-M)_V=14.6\pm0.1$ 
(thick solid blue line flanked by two thin blue lines in Figure
\ref{distance_reddening_lv_vul_v1500_cyg_v1668_cyg_v1974_cyg}(c))
and another distance-reddening relation of the UV~1455\AA\  fit
(thick solid magenta line flanked by two thin solid magenta lines) of
\begin{eqnarray}
& & 2.5 \log F_{1455}^{\rm mod}
- 2.5 \log F_{1455}^{\rm obs} \cr
&=& 5 \log \left({d \over {10\mbox{~kpc}}} \right)  + 8.3 \times E(B-V),
\label{qu_vul_uv1455_fit_eq2}
\end{eqnarray}
where $F_{1455}^{\rm mod}$ is the model flux at the distance of
10~kpc and $F_{1455}^{\rm obs}$ the observed flux.  
Here we assume an absorption of
$A_\lambda= 8.3 \times E(B-V)$ at $\lambda= 1455$~\AA\  \citep{sea79}.

We further plot other distance-reddening relations toward V1668~Cyg in 
Figure \ref{distance_reddening_lv_vul_v1500_cyg_v1668_cyg_v1974_cyg}(c):
given by \citet{slo79} (filled red circles), by \citet{mar06}, 
by \citet{gre15, gre18}, and by \citet{chen18}.
Marshall et al.'s are for four nearby directions:
$(l, b)= (90\fdg75,-6\fdg75)$ (red open squares),
$(91\fdg00,-6\fdg75)$ (green filled squares),
$(90\fdg75,-7\fdg00)$ (blue asterisks), and
$(91\fdg00,-7\fdg00)$ (magenta open circles).
These trends/lines broadly cross at $d=5.4\pm0.5$~kpc and
$E(B-V)= 0.30\pm0.05$ except that of \citet{chen18}.
The extinction almost saturates 
at the distance of $d\gtrsim 3$~kpc as shown in
Figure \ref{distance_reddening_lv_vul_v1500_cyg_v1668_cyg_v1974_cyg}(c).
This is consistent with the galactic 2D dust absorption map
of $E(B-V)=0.29 \pm 0.02$ in the direction toward V1668~Cyg
at the NASA/IPAC Infrared Science
Archive\footnote{http://irsa.ipac.caltech.edu/applications/DUST/},
which is calculated on the basis of data from \citet{schl11}.
Thus, we confirm $E(B-V)= 0.30\pm0.05$.


\begin{figure}
\plotone{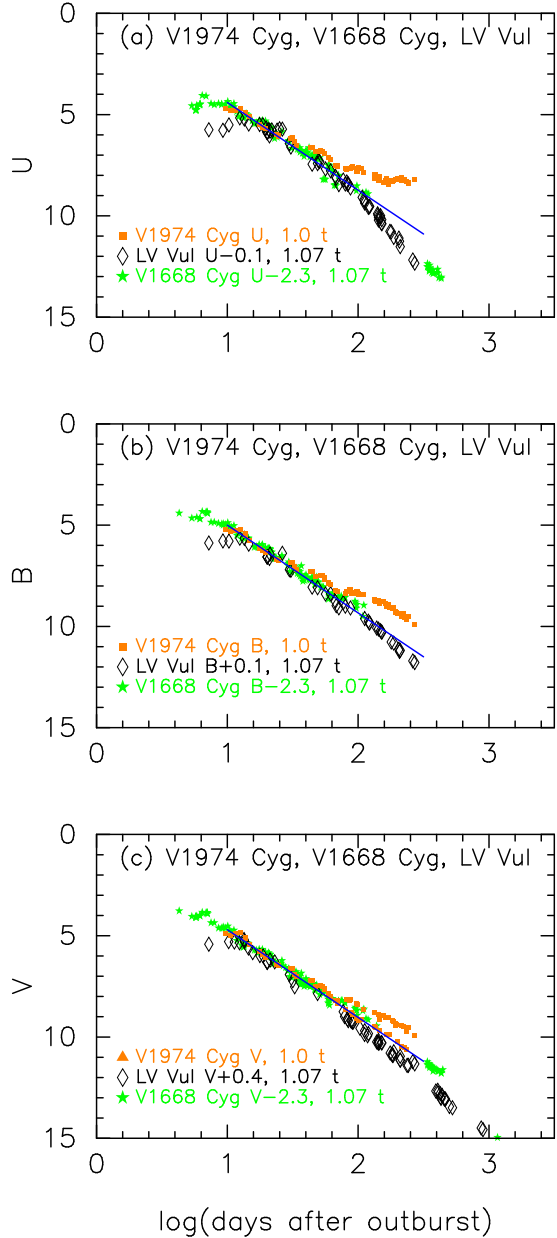}
\caption{
The $UBV$ light curves of V1974~Cyg are plotted together with
those of LV~Vul and V1668~Cyg.  
The solid blue lines denote the slope of $F_\nu\propto t^{-1.75}$,
which represents well the optically thick wind phase \citep{hac06kb}. 
The $UBV$ data of V1974~Cyg are taken from \citet{cho93}.
The $UBV$ data of LV~Vul and V1668~Cyg are the same as those
in Figure \ref{v1668_cyg_lv_vul_u_b_v_logscale_3fig}.
\label{v1974_cyg_v1668_cyg_lv_vul_u_b_v_logscale_3fig}}
\end{figure}

\subsubsection{V1974~Cyg 1992}
\label{v1974_cyg}
We obtain the reddening and distance from the time-stretching method.
We plot the $UBV$ light curves of V1974~Cyg together
with those of LV~Vul and V1668~Cyg in Figure 
\ref{v1974_cyg_v1668_cyg_lv_vul_u_b_v_logscale_3fig}.
The $UBV$ light curves well overlap until the nebular phase started
on Day $\sim 60$.  We use only the part of optically thick wind 
(or optically thick ejecta) phase, which are represented by the decline
trend of $F_\nu \propto t^{-1.75}$ \citep{hac06kb}.
The timescaling factor of $f_{\rm s}= 1.07$ against LV~Vul
is well determined from the fitting of UV~1455\AA\  
light curve between V1668~Cyg and V1974~Cyg (see, e.g., Figure
\ref{v382_vel_lv_vul_v1668_cyg_v1974_cyg_v_bv_ub_color_logscale}(a)
in Appendix \ref{v382_vel}).

We apply Equation (\ref{distance_modulus_general_temp_u}) for the $U$ band to
Figure \ref{v1974_cyg_v1668_cyg_lv_vul_u_b_v_logscale_3fig}(a) and obtain
\begin{eqnarray}
(m&-&M)_{U, \rm V1974~Cyg} \cr
&=& ((m - M)_U + \Delta U)_{\rm LV~Vul} - 2.5 \log 1.07 \cr
&=& 12.85 - 0.1\pm0.2 - 0.08 = 12.67\pm0.2 \cr
&=& ((m - M)_U + \Delta U)_{\rm V1668~Cyg} - 2.5 \log 1.07 \cr
&=& 15.1 - 2.3\pm0.2 - 0.08 = 12.72\pm0.2, 
\label{distance_modulus_u_v1974_cyg_lv_vul_v1668_cyg}
\end{eqnarray}
where we adopt $(m-M)_{U, \rm LV~Vul}=11.85 + (4.75-3.1) \times 0.60
=12.85$, and $(m-M)_{U, \rm V1668~Cyg}=14.6 + (4.75-3.1) \times 0.30
=15.10$.
Thus, we obtain $(m-M)_{U, \rm V1974~Cyg}= 12.7\pm0.1$.
For the $B$ light curves in Figure 
\ref{v1974_cyg_v1668_cyg_lv_vul_u_b_v_logscale_3fig}(b),
we similarly obtain
\begin{eqnarray}
(m&-&M)_{B, \rm V1974~Cyg} \cr
&=& ((m - M)_B + \Delta B)_{\rm LV~Vul} - 2.5 \log 1.07 \cr
&=& 12.45 + 0.1\pm0.2 - 0.08 = 12.47\pm0.2 \cr
&=& ((m - M)_B + \Delta B)_{\rm V1668~Cyg} - 2.5 \log 1.07 \cr
&=& 14.9 - 2.3\pm0.2 - 0.08 = 12.52\pm0.2, 
\label{distance_modulus_b_v1974_cyg_lv_vul_v1668_cyg}
\end{eqnarray}
where we adopt $(m-M)_{B, \rm LV~Vul}= 11.85 + 1.0\times 0.6 =12.45$ 
and $(m-M)_{B, \rm V1668~Cyg}= 14.6 + 1.0\times 0.3 =14.9$.
Thus, we obtain $(m-M)_{B, \rm V1974~Cyg}= 12.5\pm0.1$.
We apply Equation (\ref{distance_modulus_general_temp}) to Figure
\ref{v1974_cyg_v1668_cyg_lv_vul_u_b_v_logscale_3fig}(c) and obtain
\begin{eqnarray}
(m&-&M)_{V, \rm V1974~Cyg} \cr
&=& ((m - M)_V + \Delta V)_{\rm LV~Vul} - 2.5 \log 1.07 \cr
&=& 11.85 + 0.4\pm0.2 - 0.08 = 12.17\pm0.2 \cr
&=& ((m - M)_V + \Delta V)_{\rm V1668~Cyg} - 2.5 \log 1.07 \cr
&=& 14.6 - 2.3\pm0.2 - 0.08 = 12.22\pm0.2.
\label{distance_modulus_v_v1974_cyg_lv_vul_v1668_cyg}
\end{eqnarray}
Thus, we obtain $(m-M)_{V, \rm V1974~Cyg}= 12.2\pm0.1$.
These three distance moduli, that is, 
$(m-M)_U= 12.7$, $(m-M)_B= 12.5$, and $(m-M)_V= 12.2$ 
together with Equations (\ref{distance_modulus_ru}),
(\ref{distance_modulus_rb}), and
(\ref{distance_modulus_rv}),
cross at $d=1.8$~kpc and $E(B-V)=0.30$ as shown in Figure
\ref{distance_reddening_lv_vul_v1500_cyg_v1668_cyg_v1974_cyg}(d).

\citet{hac16k} modeled the $V$, UV~1455\AA, and supersoft X-ray
light curves and fitted a $0.98~M_\sun$ WD model with the observed
$V$, UV~1455\AA, and X-ray fluxes 
assuming the chemical composition of CO nova 3
(see, e.g., Figure \ref{v533_her_lv_vul_v1974_cyg_v_bv_ub_logscale}(a)
for such a model light curve fit).  
The supersoft X-ray flux was calculated assuming that the photospheric
emission is approximated by blackbody emission in the supersoft X-ray band.
The $V$ model light curve gives a distance modulus
in the $V$ band of $(m-M)_V=12.2$ (blue line) as shown in Figure 
\ref{distance_reddening_lv_vul_v1500_cyg_v1668_cyg_v1974_cyg}(d), while
the UV~1455~\AA\  light curve fit gives a relation of magenta line.

These two lines (blue and magenta lines) cross at $d\approx 1.8$~kpc and
$E(B-V)\approx 0.29$ as shown in Figure
\ref{distance_reddening_lv_vul_v1500_cyg_v1668_cyg_v1974_cyg}(d).
The figure also depicts other distance-reddening relations
toward V1974~Cyg, whose galactic coordinates are
$(l, b)=(89\fdg1338, 7\fdg8193)$.  
The four sets of data points with error bars correspond to
the distance-reddening relations in four directions close to V1974~Cyg:
$(l, b)=(89\fdg00,7\fdg75)$ (red open squares),
$(89\fdg25, 7\fdg75)$ (green filled squares),
$(89\fdg00,  8\fdg00)$ (blue asterisks),
and $(89\fdg25,  8\fdg00)$ (magenta open circles),
the data for which are taken from \citet{mar06}.
We also add distance-reddening relations given by \citet{gre15, gre18}
(black/orange lines), \citet{ozd16} (open cyan-blue diamonds
with error bars), and \citet{chen18} (cyan-blue line).
The orange line is consistent with the crossing point 
at $d\approx 1.8$~kpc and $E(B-V)\approx 0.30$.

\citet{cho97a} estimated the distance to V1974~Cyg as
$d=1.77\pm 0.11$~kpc by a nebular expansion parallax method.
This distance is consistent with our value of $d=1.8\pm 0.1$~kpc.
The reddening was also estimated by many researchers.  For example,
\citet{aus96} obtained the reddening toward V1974~Cyg mainly on the basis
of the UV and optical line ratios for days 200 through 500, i.e.,
$E(B-V)=0.3 \pm 0.1$.  The NASA/IPAC galactic 2D dust absorption map gives
$E(B-V)=0.35 \pm 0.01$ in the direction toward V1974~Cyg.  This slightly
larger value of reddening indicates that the reddening toward V1974~Cyg
does not saturate yet at the distance of 1.8 kpc, as shown in Figure 
\ref{distance_reddening_lv_vul_v1500_cyg_v1668_cyg_v1974_cyg}(d).
These are all consistent with our estimates, that is, 
$(m-M)_V=12.2\pm0.1$, $d=1.8\pm 0.1$~kpc, and $E(B-V)=0.3 \pm 0.05$.
These values are listed in Table
\ref{extinction_various_novae}.


\begin{figure}
\plotone{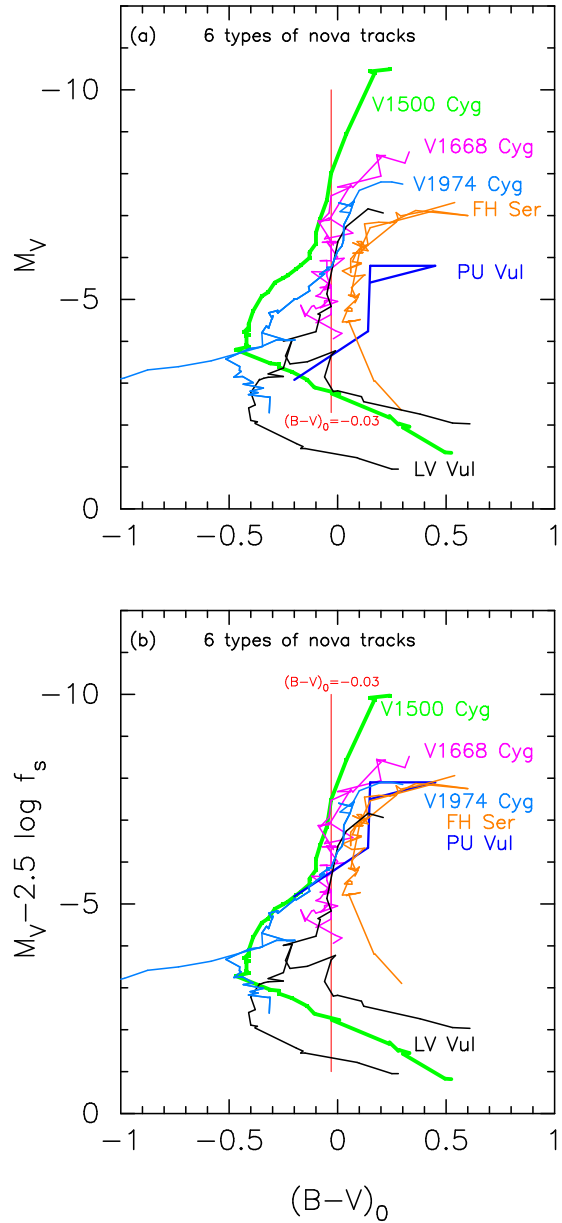}
\caption{
(a) Six typical nova tracks in the color-magnitude diagram:
from left to right:
V1500~Cyg (thick solid green lines), V1668~Cyg (solid magenta lines),
V1974~Cyg (solid sky-blue lines), LV~Vul (solid black lines),
FH~Ser (solid orange lines), and PU~Vul (thick solid blue lines).
The vertical solid red line indicates $(B-V)_0=-0.03$, the color
of optically thick free-free emission.
(b) The time-stretched color-magnitude diagram of six typical nova tracks.
Here, $f_{\rm s}$ is the timescaling factor against that of LV~Vul.
\label{hr_diagram_6types_novae_stretch.epsi}}
\end{figure}

\subsection{Template nova tracks in the time-stretched color-magnitude
diagram}
\label{template_tracks}
\citet{hac16kb} proposed six template tracks of novae with well-determined 
distance modulus in the $V$ band $(m-M)_V$ and color excess $E(B-V)$.
They dubbed them the V1500~Cyg, V1668~Cyg, V1974~Cyg, LV~Vul, FH~Ser,
and PU~Vul tracks as shown in Figure 
\ref{hr_diagram_6types_novae_stretch.epsi}(a).
Here, we adopt 
$E(B-V)=0.60$ and $(m-M)_V=11.85$ for LV~Vul from Section \ref{lv_vul},
$E(B-V)=0.45$ and $(m-M)_V=12.3$ for V1500~Cyg from Section \ref{v1500_cyg}, 
$E(B-V)=0.30$ and $(m-M)_V=14.6$ for V1668~Cyg from Section \ref{v1668_cyg},
$E(B-V)=0.30$ and $(m-M)_V=12.2$ for V1974~Cyg from Section \ref{v1974_cyg},
$E(B-V)=0.60$ and $(m-M)_V=11.7$ for FH~Ser and
$E(B-V)=0.30$ and $(m-M)_V=14.3$ for PU~Vul both from \citet{hac16kb}.
They categorized 40 novae into one of
these six subgroups, depending on their similarity in the
$(B-V)_0$-$M_V$ diagram.

The LV~Vul and V1974~Cyg tracks split into two branches just after
the nebular phase started.  As already discussed in our previous
papers \citep[see, e.g., Figure 10 and Section 3.3 of][]{hac14k},
strong emission lines such as [\ion{O}{3}] contribute to the blue
edge of $V$ filter.  A small difference in the $V$ filter response
function makes a large difference in the $V$ magnitude and results
in a large difference in the $B-V$ color.  This is the
reason why the track splits into two (or three or even four) 
branches among various observatories.

In the present work, we propose
the $(B-V)_0$-$(M_V-2.5\log f_{\rm s})$ diagram, that is,
Figure \ref{hr_diagram_6types_novae_stretch.epsi}(b).
In this new color-magnitude diagram,
the above six nova tracks are shifted up or down.
Here, we show the above six tracks of V1500~Cyg, V1974~Cyg, 
V1668~Cyg, LV~Vul, FH~Ser, and PU~Vul, adopting
$(m-M')_V=11.75$ for V1500~Cyg,
$(m-M')_V=12.3$ for V1974~Cyg, 
$(m-M')_V=11.85$ for LV~Vul.
$(m-M')_V=14.6$ for V1668~Cyg,
$(m-M')_V=12.45$ for FH~Ser, and 
$(m-M')_V=16.4$ for PU~Vul.

The tracks of V1500~Cyg and V1974~Cyg are different from
each other in the early phase but almost overlap in the middle part
of the tracks.  They turn from toward the blue to toward the red
at the similar place, that is, the turning corner.  
Therefore, we regard that V1974~Cyg belongs
to the V1500~Cyg type in the time-stretched color-magnitude diagram.

Also, the V1668~Cyg track almost follows the template track of LV~Vul. 
Thus, these two template tracks can be merged into the same group.
The track of FH~Ser is close to that of V1668~Cyg and LV~Vul.
The track of PU~Vul is close to that of V1500~Cyg.
Thus, the six tracks seem to closely converge into one of the two groups,
LV~Vul/V1668~Cyg or V1500~Cyg/V1974~Cyg, in the time-stretched
color-magnitude diagram.

The main difference between them is that the V1668~Cyg/LV~Vul group
novae evolve almost straight down along $(B-V)_0\sim -0.03$ during
$(M_V-2.5\log f_{\rm s})= -7$ and $-4$, while
the V1500~Cyg/V1974~Cyg group novae evolve blueward up to
$(B-V)_0\sim -0.5$ during $(M_V-2.5\log f_{\rm s})= -5$ and $-3$.
This tendency is clear, e.g., in the case of V2362~Cyg (Figure
\ref{hr_diagram_v705_cas_v382_vel_v5114_sgr_v2362_cyg_outburst}(d)).


\begin{figure*}
\plotone{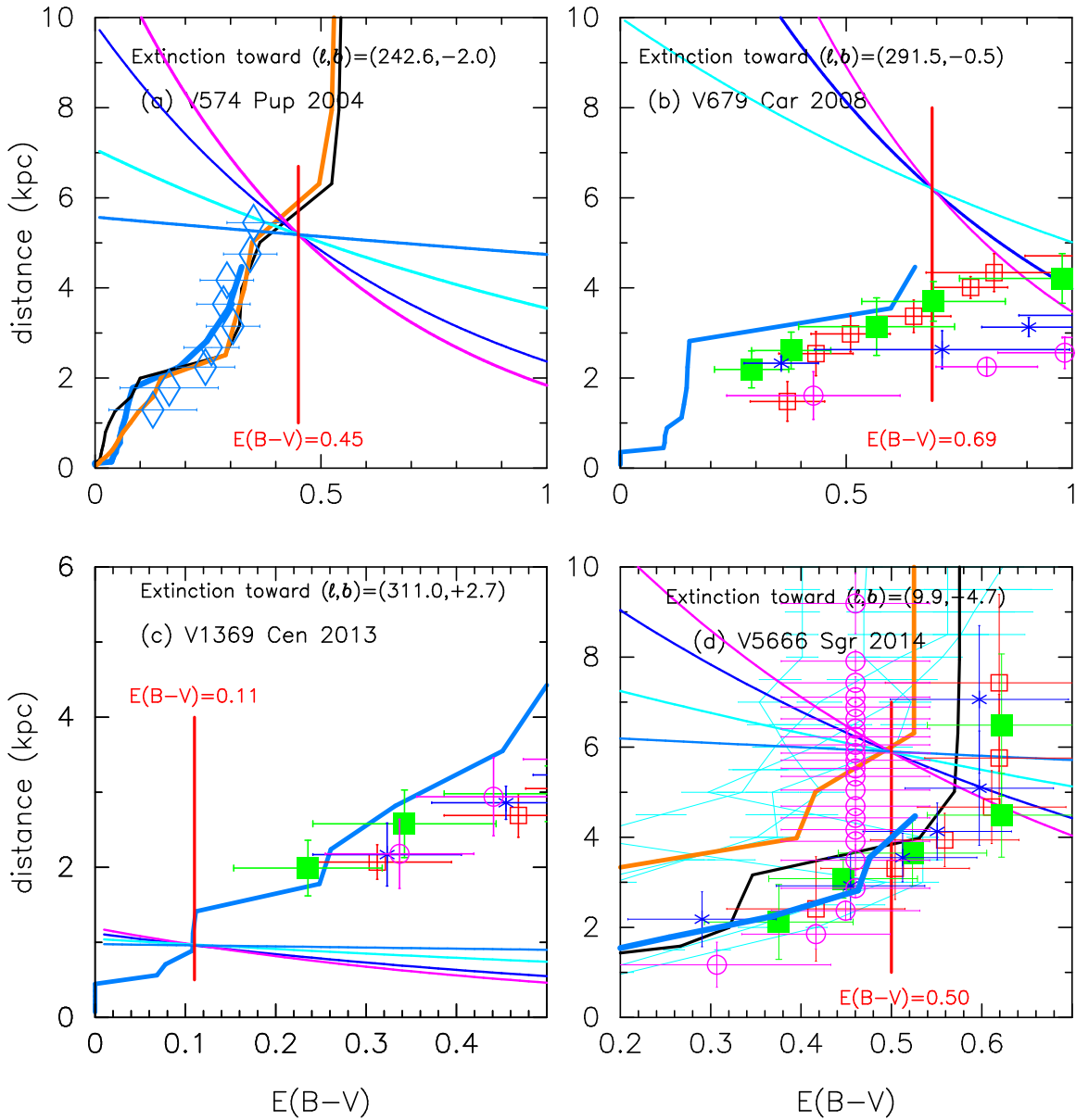}
\caption{
Same as Figure \ref{distance_reddening_lv_vul_v1500_cyg_v1668_cyg_v1974_cyg},
but for (a) V574~Pup, (b) V679~Car, (c) V1369~Cen, and (d) V5666~Sgr.
In panel (d), the four very thin solid cyan lines with error bars indicate
the distance-reddening relations given by \citet{schu14}. 
\label{distance_reddening_v574_pup_v679_car_v1369_cen_v5666_sgr}}
\end{figure*}


\begin{figure*}
\plotone{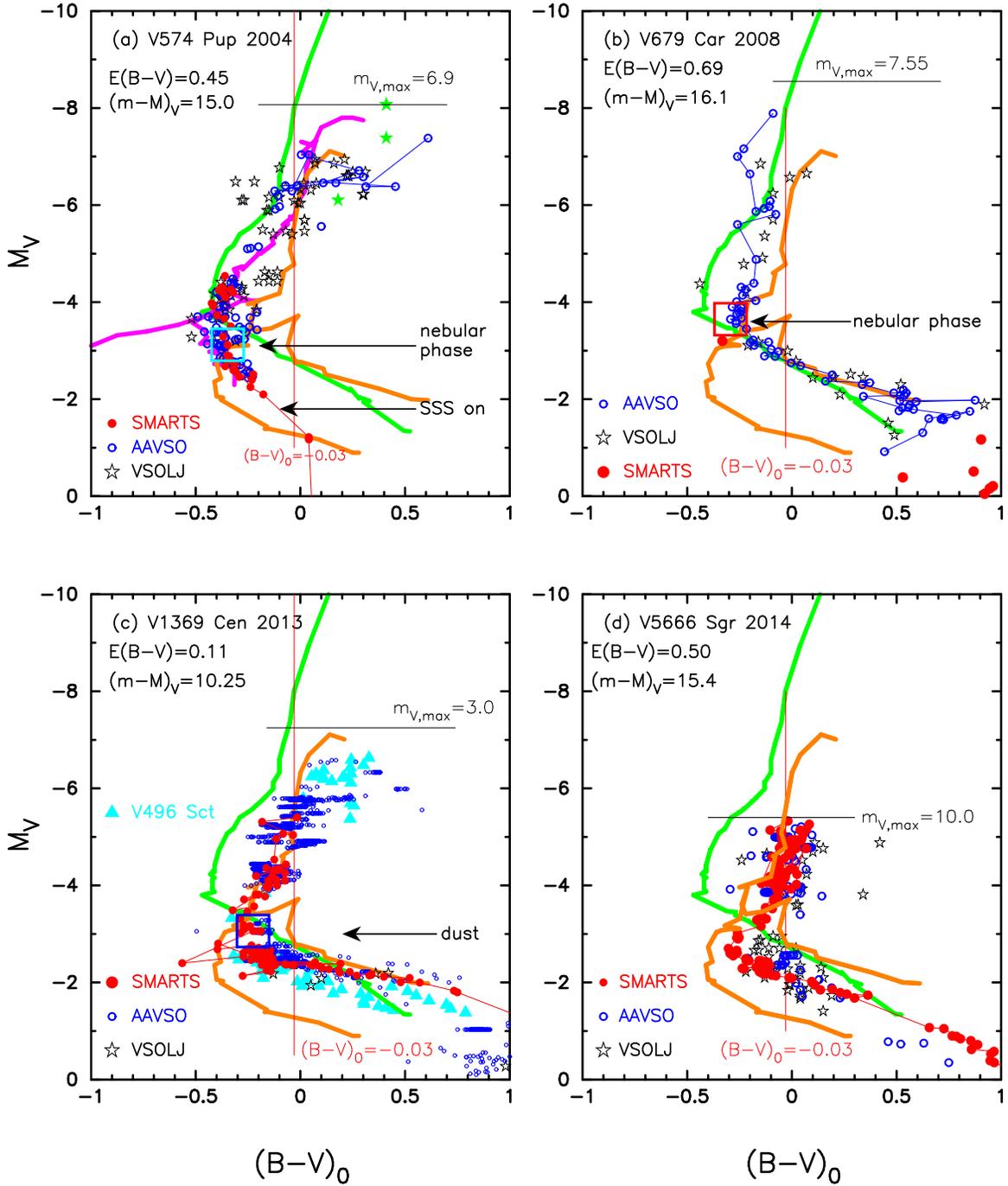}
\caption{
Color-magnitude diagram for
(a) V574~Pup, (b) V679~Car, (c) V1369~Cen, and (d) V5666~Sgr.  
The thick solid green, magenta, and orange lines denote the template
tracks of V1500~Cyg, V1974~Cyg, and LV~Vul, respectively.
In panel (a), we show the start of the nebular phase by
the large open cyan square, while in panel (b), the red square.
In panel (a), we add the start of 
the supersoft X-ray source (SSS) phase by ``SSS on.''    
In panel (c), we add the data of V496~Sct (filled cyan triangles), which
are taken from Section \ref{v496_sct_cmd} and Appendix \ref{v496_sct}.
We also show the start of the dust black out phase by
the large open blue square.  
\label{hr_diagram_v574_pup_v679_car_v1369_cen_v5666_sgr_outburst_mv}}
\end{figure*}


\begin{figure*}
\plotone{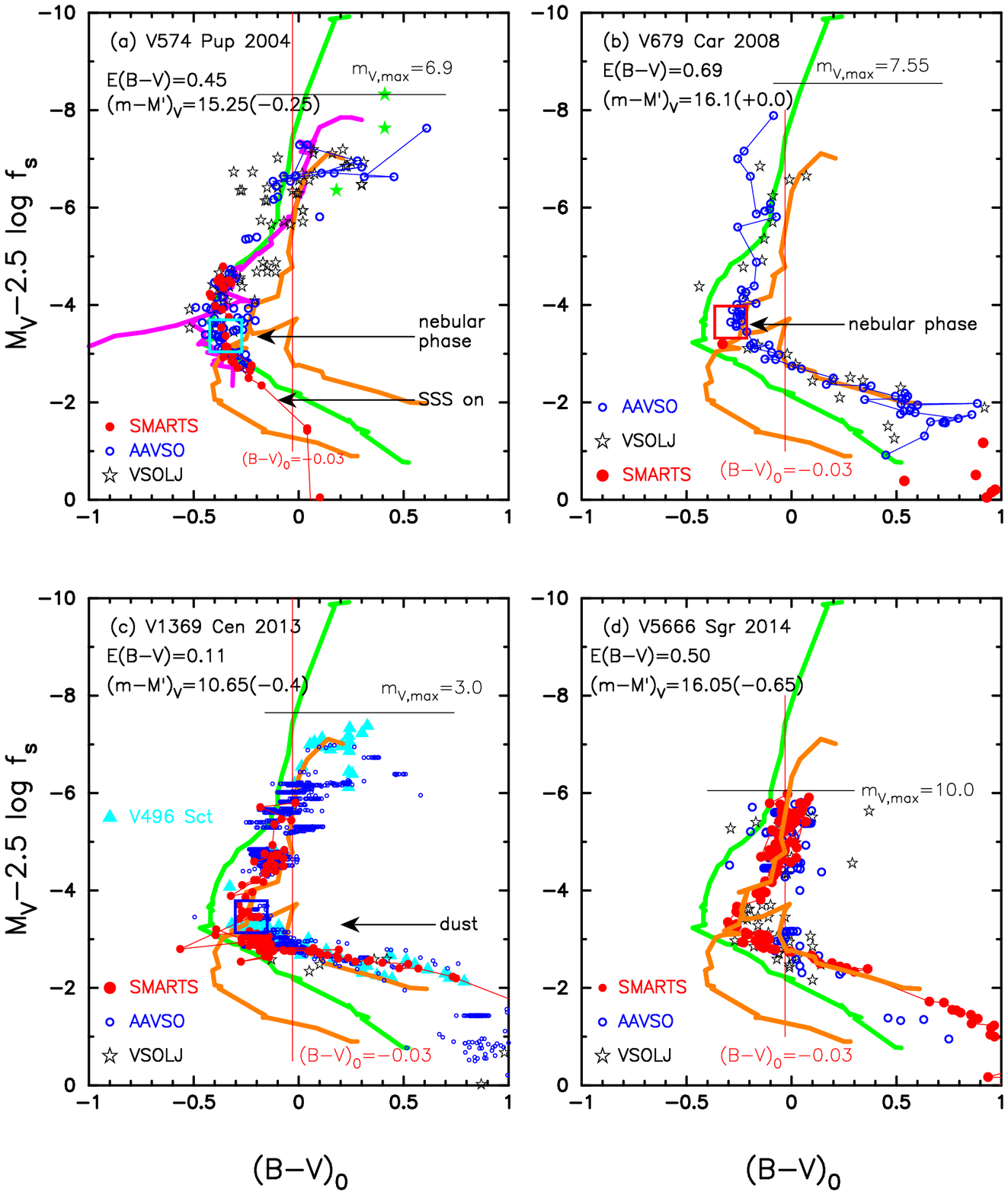}
\caption{
Time-stretched color-magnitude diagram for
(a) V574~Pup, (b) V679~Car, (c) V1369~Cen, and (d) V5666~Sgr.  
Here, $f_{\rm s}$ is the timescaling factor against that of LV~Vul.
The green, magenta, and orange lines denote the template
tracks of V1500~Cyg, V1974~Cyg, and LV~Vul, respectively.
Each symbol is the same as that in Figure
\ref{hr_diagram_v574_pup_v679_car_v1369_cen_v5666_sgr_outburst_mv}.
\label{hr_diagram_v574_pup_v679_car_v1369_cen_v5666_sgr_outburst}}
\end{figure*}

\section{V574~Pup 2004}
\label{v574_pup_cmd}
\citet{hac10k} analyzed the optical and X-ray light curves of V574~Pup.
They assumed $A_V=2.2$ after \citet{burl08}, which will
be revised later.  \citet{hac15k} determined the absolute magnitude of 
theoretical $V$ light curve for various WD masses and chemical 
compositions.  Using their theoretical light curves,
we reexamine the light/color curves of V574~Pup,

\citet{nes07a} obtained $E(B-V)=0.50$ during the SSS phase
from the relation of $E(B-V)=N_{\rm H}/4.8\times 10^{21}$~cm$^{-2}$
\citep{boh78} together with their estimate of $N_{\rm H}= (2.5\pm0.6)
\times 10^{21}$~cm$^{-2}$ calculated with an X-ray model spectrum fitting.
Here, $N_{\rm H}$ is the neutral hydrogen column density.
\citet{nai10} derived the distance of $d=5.5$~kpc from the maximum
magnitude versus rate of decline (MMRD) relation of
\citet{del95} and obtained the reddening of 
$E(B-V)=A_V/3.1= 1.95/3.1= 0.63$ from the empirical relation of
$(B-V)_{0, \rm max}=+0.23\pm0.06$ at optical $V$ maximum \citep{van87}.

\subsection{Distance-reddening relations toward V574~Pup}
\label{distance_reddening_v574_pup}
We obtain the distance and reddening toward V574~Pup based on the
time-stretching method.   Figure
\ref{distance_reddening_v574_pup_v679_car_v1369_cen_v5666_sgr}(a)
shows four distance moduli in the $B$, $V$, $I_{\rm C}$,
and $K_{\rm s}$ bands by the thin solid magenta, blue, cyan, cyan-blue lines, 
that is, $(m-M)_B= 15.43$, $(m-M)_V= 15.0$, $(m-M)_I= 14.22$, and
$(m-M)_K= 13.68$, which are obtained in Appendix \ref{v574_pup},
together with Equations
(\ref{distance_modulus_rb}), (\ref{distance_modulus_rv}),
(\ref{distance_modulus_ri}), and (\ref{distance_modulus_rk}).
These four lines cross at $d=5.3$~kpc and $E(B-V)=0.45$.
Thus, we obtain the reddening toward V574~Pup, $E(B-V)=0.45\pm0.05$,
the distance of V574~Pup, $d=5.3\pm0.5$~kpc, and the timescaling factor,
$f_{\rm s}=1.26$ against LV~Vul.  The distance
modulus in the $V$ band is $(m-M)_V=15.0\pm0.1$.  These values
are listed in Table \ref{extinction_various_novae}.

We further check our estimates of the distance modulus in the $V$ band
$(m-M)_V=15.0\pm0.1$ and extinction $E(B-V)=0.45\pm0.05$.
Figure \ref{distance_reddening_v574_pup_v679_car_v1369_cen_v5666_sgr}(a)
shows the distance-reddening relations toward V574~Pup,
whose galactic coordinates are $(l,b)=(242\fdg5695, -1\fdg9933)$.
The thick solid black/orange lines denote
the distance-reddening relations given by \citet{gre15, gre18},
respectively.
The open cyan-blue diamonds with error bars are taken from \citet{ozd16}.
The thick solid cyan-blue line denotes the result of \citet{chen18}.
These four distance-reddening relations based on the dust map are
consistent with each other until $d \lesssim 5$~kpc.

Green et al.'s black/orange lines 
are roughly consistent with our results of $d=5.3\pm0.5$~kpc and 
$E(B-V)=0.45\pm0.05$.  The NASA/IPAC galactic 2D dust absorption map 
gives $E(B-V)=0.60\pm0.02$ toward V574~Pup, which is slightly
larger than our value of $E(B-V)=0.45\pm0.05$.
This is because the reddening increases further beyond 
the position of V574~Pup as shown in Figure 
\ref{distance_reddening_v574_pup_v679_car_v1369_cen_v5666_sgr}(a).

\subsection{Color-magnitude diagram}
\label{cmd_v574_pup}
Using $E(B-V)=0.45$ and $(m-M)_V=15.0$, we plot the color-magnitude 
diagram of V574~Pup in Figure
\ref{hr_diagram_v574_pup_v679_car_v1369_cen_v5666_sgr_outburst_mv}(a). 
We add tracks of V1500~Cyg (solid green line), LV~Vul (solid orange line),
and V1974~Cyg (solid magenta line) for comparison.
The data of V574~Pup are scattered in the early phase owing to 
its oscillatory behavior but close to that of V1974~Cyg (solid magenta line).
Therefore, V574~Pup belongs to the V1974~Cyg type in the 
$(B-V)_0$-$M_V$ color-magnitude diagram.
The track in the early oscillatory phase shows a circular movement
similar to that of V705~Cas as shown in Figure 37(b) of Paper II.

In Figure
\ref{hr_diagram_v574_pup_v679_car_v1369_cen_v5666_sgr_outburst}(a),
adopting $E(B-V)=0.45$ and $(m-M')_V=15.25$ in Equation
(\ref{absolute_mag_lv_vul_v574_pup_only}), we obtain the time-stretched
color-magnitude diagram of V574~Pup.
The text ``$(m-M')_V=15.25(-0.25)$'' in the figure means that
$(m-M')_V=15.25$ and $(m-M)_V=15.25 - 0.25 = 15.0$.

We note the starting epoch of the nebular phase in
Figures \ref{hr_diagram_v574_pup_v679_car_v1369_cen_v5666_sgr_outburst_mv}(a)
and \ref{hr_diagram_v574_pup_v679_car_v1369_cen_v5666_sgr_outburst}(a).   
A nebular phase could be identified by the first clear
appearance of the nebular emission lines [\ion{O}{3}] or [\ion{Ne}{3}]
which are stronger than permitted lines.   The nebular phase of V574~Pup
started at $m_V\sim12$ from the spectra of the 
Small Medium Aperture Telescope System (SMARTS) 
database\footnote{http://www.astro.sunysb.edu/fwalter/SMARTS/NovaAtlas/atlas.html} \citep{wal12}, which is denoted by the large open cyan square in Figures 
\ref{hr_diagram_v574_pup_v679_car_v1369_cen_v5666_sgr_outburst_mv}(a)
and \ref{hr_diagram_v574_pup_v679_car_v1369_cen_v5666_sgr_outburst}(a).  
After the onset of the nebular phase, the track turns to the right (red).
This is because the strong emission lines of [\ion{O}{3}] contribute
to the blue edge of the $V$ filter and its large contribution makes
the $B-V$ color redder.

\citet{nes07a} discussed that the SSS phase
started between May and July in 2005.  This epoch 
corresponds to $m_V \sim 13.2$ and denoted by ``SSS on'' in Figures
\ref{hr_diagram_v574_pup_v679_car_v1369_cen_v5666_sgr_outburst_mv}(a)
and \ref{hr_diagram_v574_pup_v679_car_v1369_cen_v5666_sgr_outburst}(a).  
The $(B-V)_0$ color seems to keep a constant value of $\sim 0.0$
during the SSS phase, as shown in Figure 
\ref{v574_pup_v1974_cyg_v_bv_logscale}(b).

In Figure
\ref{hr_diagram_v574_pup_v679_car_v1369_cen_v5666_sgr_outburst}(a),
the track of V574~Pup almost follows those of V1500~Cyg and V1974~Cyg
except for the early oscillation phase.
Thus, we conclude that V574~Pup belongs to the V1500~Cyg/V1974~Cyg type
in the time-stretched color-magnitude diagram.

\subsection{Model light curve fitting}
\label{model_light_curve}
\citet{kat94h} calculated evolutions of nova outbursts
based on the optically thick wind theory.  Their numerical models
provide various physical quantities such as the photospheric temperature,
radius, velocity, and wind mass-loss rate of a nova hydrogen-rich envelope
(mass of $M_{\rm env}$) for a specified WD mass ($M_{\rm WD}$) and
chemical composition of the envelope.  \citet{hac15k}
calculated the absolute $V$ magnitude light curves based on 
Kato \& Hachisu's envelope models.  Their model $V$
light curve is composed of photospheric emission and free-free emission.
The photospheric emission is approximated by blackbody emission at the
photosphere, while the free-free emission is calculated from 
the photospheric radius, velocity, and wind mass-loss rate \citep{hac15k}.
Fitting the absolute $V$ magnitudes ($M_V$) of model light curve
with the observed apparent $V$ magnitudes ($m_V$), we are able to specify
the $(m-M)_V$ (and even $M_{\rm WD}$) for the nova.
Such results are presented in \citet{hac15k, hac16k, hac18k, hac18kb}.

\citet{hac10k} estimated the WD mass of V574~Pup to be $1.05~M_\sun$. 
In Figure \ref{v574_pup_v1974_cyg_v_bv_logscale}(a),
we plot the $1.05~M_\sun$ WD model (solid black line for $V$ magnitudes
and solid red line for soft X-ray fluxes) with the chemical composition
of Ne nova 3 \citep{hac16k}.
We use the absolute $V$ magnitudes of the $1.05~M_\sun$ WD model
calculated in \citet{hac16k}.
We add a $0.98~M_\sun$ WD model (solid green lines both for
$V$ and soft X-ray) with the chemical
composition of CO nova 3 \citep{hac16k} for comparison, the timescale of
which is stretched by a factor of $f_{\rm s}= 1.17$.
The absolute $V$ light curve of the $1.05~M_\sun$ WD model reasonably follows 
the observed apparent $V$ light curves of V574~Pup for $(m-M)_V=15.0$.
In these models, the optically thick winds stop at the open black
circle (at the right edge of the solid black line). 

The model supersoft X-ray light curve also shows good fit with 
the X-ray observation, i.e., turn-on and turnoff times.
The X-ray data are the same as those of \citet{hac10k}.
Thus, the $1.05~M_\sun$ WD model is consistent with the X-ray data.
We list the WD mass of V574~Pup in Table \ref{wd_mass_novae}.

\section{V679~Car 2008}
\label{v679_car_cmd}
V679~Car was discovered by Katarzyna Malek at mag 7.9 on UT 2008
November 26.26 (JD~2454795.76).  It reached mag 7.55 on UT 2008
November 27.27 (JD~2454796.77) \citep{waa08}.
The spectrum showed that the object is a classical nova 
of the \ion{Fe}{2} type 
Recently, \citet{fra18} suggested that V679~Car is a candidate
for GeV gamma-ray detected nova with {\it Fermi}/Large Area Telescope (LAT).

\subsection{Distance-reddening relations toward V679~Car}
\label{distance_reddening_v679_car}
We plot three distance moduli in the $B$, $V$, and $I_{\rm C}$
bands in Figure 
\ref{distance_reddening_v574_pup_v679_car_v1369_cen_v5666_sgr}(b)
by the magenta, blue, and cyan lines, that is,
$(m-M)_B= 16.8$, $(m-M)_V= 16.1$, and $(m-M)_I= 15.0$
in Appendix \ref{v679_car}
together with Equations (\ref{distance_modulus_rb}), 
(\ref{distance_modulus_rv}), and (\ref{distance_modulus_ri}),
respectively.
These three lines cross at $d=6.2$~kpc and $E(B-V)=0.69$.
Thus, we obtain the reddening toward V679~Car, $E(B-V)=0.69\pm0.05$,
the distance of V574~Pup, $d=6.2\pm0.7$~kpc, and the timescaling factor,
$f_{\rm s}=1.0$ against LV~Vul.  The distance
modulus in the $V$ band is $(m-M)_V=16.1\pm0.1$.  These values
are listed in Table \ref{extinction_various_novae}.

We check the distance and reddening toward V679~Car, whose
galactic coordinates are $(l,b)=(291\fdg4697, -0\fdg5479)$,
on the basis of dust extinction maps calculated by \citet{mar06}
and \citet{chen18}.
Figure \ref{distance_reddening_v574_pup_v679_car_v1369_cen_v5666_sgr}(b)
shows the relations given by \citet{mar06}:
$(l, b)=(291\fdg25,  -0\fdg50)$ by open red squares,
$(291\fdg50,  -0\fdg50)$ by filled green squares,
$(291\fdg25,  -0\fdg75)$ by blue asterisks,
and $(291\fdg50,  -0\fdg75)$ by open magenta circles.
The filled green squares of \citet{mar06} are the closest direction
toward V679~Car among the four nearby directions.  
The solid cyan-blue line denotes the relation given by \citet{chen18}.
Neither Green et al.'s (2015, 2018) nor \"Ozd\"ormez et al.'s (2016)
data are available toward this direction.
The three relations of $(m-M)_B= 16.8$ (magenta line),
$(m-M)_V= 16.1$ (blue line), and $(m-M)_I= 15.0$ (cyan line)
cross at $d=6.2$~kpc and $E(B-V)=0.69$.
This crossing point is significantly different from the closest relation
(filled green squares) given by \citet{mar06}.
However, Chen et al.'s relation (solid cyan-blue line)
seems to show possible agreement
with our crossing point, although the data end at $d=4.5$~kpc.

This large discrepancy can be understood as follows: Marshall et al.'s
3D dust map gives an averaged value of a relatively broad region, 
and thus a pinpoint reddening toward the nova could be different
from the values of the 3D dust maps, because
the resolution of the dust map is considerably larger
than molecular cloud structures observed in the interstellar medium.
For example, Marshall et al.'s 3D dust map gave a significantly different
value for the reddening of LV~Vul as shown in Figure
\ref{distance_reddening_lv_vul_v1500_cyg_v1668_cyg_v1974_cyg}(a).

\subsection{Color-magnitude diagram}
\label{cmd_v679_car}
The $V$ light and $(B-V)_0$ color curve shapes of V679~Car are
similar to those of LV~Vul and V1668~Cyg as shown in Figures  
\ref{v679_car_lv_vul_v1668_cyg_b_v_logscale_2fig} and
\ref{v679_car_lv_vul_v1668_cyg_v_bv_ub_logscale}.
This suggests that the color-magnitude diagram of V679~Car is 
also similar to them.
Adopting $E(B-V)=0.69$ and $(m-M)_V=16.1$, we plot the color-magnitude
track in Figure
\ref{hr_diagram_v574_pup_v679_car_v1369_cen_v5666_sgr_outburst_mv}(b) 
together with V1500~Cyg (solid green line)
and LV~Vul (solid orange line).  The SMARTS spectra of V679~Car
show that the nebular phase started between
JD~2454845.71 (UT 2009 January 14.21) 
and JD~2454860.67 (UT 2009 January 29.17),
corresponding to $m_V=12.45$ and $B-V=0.40$, which is denoted by the 
large open red square in Figure
\ref{hr_diagram_v574_pup_v679_car_v1369_cen_v5666_sgr_outburst_mv}(b).
The start of the nebular phase usually corresponds to the turning point
from toward blue to toward red (from toward left to toward right)
as already discussed in Section \ref{v574_pup_cmd}.
The track of V679~Car broadly locates on the track of LV~Vul (orange line),
although the $(B-V)_0$ colors of the early phase are rather scattered
and slightly bluer than that of LV~Vul.  After this turning point, 
the track of V679~Car follows that of LV~Vul.
We conclude that V679~Car belongs
to the LV~Vul type in the color-magnitude diagram (Paper II).

Adopting $E(B-V)=0.69$ and $(m-M')_V=16.1$ from Equation
(\ref{absolute_mag_lv_vul_v679_car_only}), we plot the time-stretched
color-magnitude diagram of V679~Car in Figure
\ref{hr_diagram_v574_pup_v679_car_v1369_cen_v5666_sgr_outburst}(b). 
The $(B-V)_0$ intrinsic color is not affected by time-stretch
\citep[see][]{hac18k}.
Because $f_{\rm s}=1.0$, the time-stretched color-magnitude track
is the same as in Figure
\ref{hr_diagram_v574_pup_v679_car_v1369_cen_v5666_sgr_outburst_mv}(b),
but the track of V1500~Cyg is shifted down. 
The overlap of the track to the LV~Vul track supports our values
of $E(B-V)=0.69$ and $(m-M')_V=16.1$,
that is, $f_{\rm s}=1.0$, $E(B-V)=0.69\pm0.05$, $(m-M)_V=16.1\pm0.1$,
and $d=6.2\pm0.7$~kpc.
We list our results in Table \ref{extinction_various_novae}.

\subsection{Model light curve fitting with V679~Car}
\label{model_lc_v679_car}
Figure \ref{v679_car_lv_vul_v1668_cyg_v_bv_ub_logscale}(a) 
shows the model light curve of a $0.98~M_\sun$ WD (solid red lines)
with the envelope chemical composition of CO Nova 3 \citep{hac16k}.
Taking the distance modulus in the $V$ band of $(m-M)_V=16.1$,
we reasonably fit the model absolute $V$ light curve with the observed 
apparent $V$ magnitudes of V679~Car.  Thus, our obtained value of 
$(m-M)_V=16.1$ is consistent with the model light curve.
We also add the $0.98~M_\sun$ WD model (solid green lines) with the same
envelope chemical composition of CO nova 3 but a slightly larger initial
envelope mass, assuming $(m-M)_V=14.6$ for V1668~Cyg.
This model fits with the $V$ and UV~1455\AA\  light curves of V1668~Cyg
as already discussed in our previous paper \citep{hac16k}.

Our model light curve of the $0.98~M_\sun$ WD predicts the SSS phase
between day 250 and day 600 as shown in Figure
\ref{v679_car_lv_vul_v1668_cyg_v_bv_ub_logscale}(a).
V679~Car was observed with {\it Swift} eight times, but no SSS phase
was detected \citep{schw11}.  
The epochs of the {\it Swift} observations are indicated by the
downward magenta arrows.
The second and third arrows almost overlap. 
We expect detection of supersoft X-rays at the last one or two.
Non-detection may be owing to 
a large extinction toward V679~Car, as large as
$E(B-V)=0.69$, corresponding to $N_{\rm H}=
8.3\times 10^{21} E(B-V) = 0.6 \times 10^{22}$~cm$^{-2}$
\citep[][]{lis14}, or see also Figure 3 in 
\citet{schw11},

\section{V1369~Cen 2013}
\label{v1369_cen_cmd}
V1369~Cen was discovered by J. Seach at mag 5.5 on UT 2013 December
2.692 (JD 2456629.192) \citep{gui13} 
and reached $m_{V,\rm max}\approx3.0$ on JD 2456639.18
(VSOLJ data by S. Kiyota, however, we omit his $m_V=2.8$ on JD 2456637.21
because it is too bright compared with other data).
\citet{mas18} obtained the reddening to be $E(B-V)=0.15$ from 
the \ion{Na}{1}~D2 $\lambda 5890$ interstellar absorption line
and the best fit column density of $N_{\rm H}=7.2\times 10^{21}$~cm$^{-2}$
derived from the model of the interstellar Ly$\alpha$ absorption.
This value of $N_{\rm H}$ would be a typo 
of $N_{\rm H}=7.2\times 10^{20}$~cm$^{-2}$, 
because of $E(B-V)=N_{\rm H}/4.8\times 10^{21}$~cm$^{-2}$ \citep{boh78}.

This nova is characterized by pre-nova X-ray observation and GeV gamma-ray
detection.  \citet{kuu13} reported a pre-nova X-ray detection 
between UT 2002 February 20 16:57 and 21:08 with an absorbed 0.3-10 keV
flux of $(1.0\pm0.1)\times 10^{-12}$~erg~s$^{-1}$cm$^2$ and $N_{\rm H}=(2\pm1)
\times 10^{21}$~cm$^{-2}$ for a non-isothermal collisional plasma.
\citet{che13} reported a gamma-ray detection during
UT 2013 December 7-10 with an average flux of $F(E>100$~MeV$)\sim
(2.1\pm0.6) \times 10^{-7}$~ph~cm$^{-2}$~s$^{-1}$.

\subsection{Distance and reddening toward V1369~Cen}
\label{distance-reddening_relation_v1369_cen}
We obtain the distance and reddening toward the nova based on the
time-stretching method.
We plot four distance moduli in the $B$, $V$, $I_{\rm C}$, and
$K_{\rm s}$ bands in Figure 
\ref{distance_reddening_v574_pup_v679_car_v1369_cen_v5666_sgr}(c),
which are obtained in Appendix \ref{v1369_cen}.
These four (magenta, blue, cyan, cyan-blue) lines 
cross at the distance of $0.96$~kpc and the reddening of $E(B-V)=0.11$.
Thus, we obtain the distance $d= 0.96\pm0.1$~kpc, the reddening 
$E(B-V)= 0.11\pm0.03$, and the timescaling factor
$f_{\rm s}= 1.48$ against LV~Vul.

\citet{izz13} obtained an extinction of $E(B-V) = 0.11\pm0.08$ from the
relation given by \citet{poz12} or $E(B-V) = 0.14$ from the relation given
by \citet{mun97}, both between the widths of \ion{Na}{1} D doublet
and the extinction.   \citet{sho14} obtained the extinction of
$E(B-V)\sim0.1$, using UV interstellar spectra.
They also estimated the distance
of $d\sim 2.4$~kpc comparing the UV fluxes with those of V339~Del.
As already mentioned, \citet{mas18} obtained the reddening of 
$E(B-V)=0.15$ from the \ion{Na}{1}~D2 $\lambda 5890$ interstellar
absorption line and the best fit neutral hydrogen column density
derived from the model of the interstellar Ly$\alpha$ absorption.

We further examine the reddening and distance based on various
distance-reddening relations.
Figure \ref{distance_reddening_v574_pup_v679_car_v1369_cen_v5666_sgr}(c)
shows the distance-reddening relations toward V1369~Cen,
whose galactic coordinates are $(l,b)=(310\fdg9816, +2\fdg7274)$.
We plot Marshall et al.'s (2006) relations:
$(l, b)=(310\fdg75, +2\fdg50)$ by open red squares,
$(311\fdg00, +2\fdg50)$ by filled green squares,
$(310\fdg75, +2\fdg75)$ by blue asterisks,
and $(311\fdg00, +2\fdg75)$ by open magenta circles.
The open magenta circles are the closest direction
among the four nearby directions.  
Green et al.'s or \"Ozd\"ormez et al.'s data are not available.
We add the relation (cyan-blue line) given by \citet{chen18}, which
is consistent with our crossing point of $E(B-V)=0.11$ and $d=0.96$~kpc.
Although Marshall et al.'s relations do not reach $E(B-V)\lesssim 0.2$,
it seems that its trend of linear extension toward 
$E(B-V)= 0.11$ are roughly consistent with our crossing point
at $d=0.96\pm0.1$~kpc and $E(B-V)=0.11\pm0.03$.

\subsection{Color-magnitude diagram}
Adopting $E(B-V)=0.11$ and $(m-M)_V=10.25$, we plot the color-magnitude
diagram of V1369~Cen in Figure
\ref{hr_diagram_v574_pup_v679_car_v1369_cen_v5666_sgr_outburst_mv}(c). 
We also plot the tracks of V1500~Cyg (solid green line), 
LV~Vul (solid orange line), and V496~Sct (filled cyan triangles).
The data of V496~Sct are the same as those in Figures 72(b) and 73
of Paper II.
The track of V1369~Cen and V496~Sct almost follow that of LV~Vul
in the early phase until the dust-shell formation, which is denoted by the
large open blue square.  After that, the track of V1369~Cen underlies
the track of LV~Vul.

Figure \ref{hr_diagram_v574_pup_v679_car_v1369_cen_v5666_sgr_outburst}(c)
shows the time-stretched color-magnitude diagram of V1369~Cen.  We
adopt $E(B-V)=0.11$ and $(m-M')_V=10.65$ from Equation
(\ref{absolute_mag_v1369_cen_lv_vul}).
Now, the track of V1369~Cen and V496~Sct almost follow that of
LV~Vul.  Thus, we conclude that V1369~Cen belongs to the LV~Vul type
in the time-stretched color-magnitude diagram.  This overlapping of 
V1369 Cen track to that of LV~Vul suggests that our adopted values of
$E(B-V)=0.11$, $f_{\rm s}=1.48$, and $(m-M')_V=10.65$ are reasonable.

\subsection{Model light curve fitting}
Figure \ref{all_mass_v1369_cen_v1668_cyg_x45z02c15o20}(a) shows
the model absolute $V$ light curve of a $0.90~M_\sun$ WD 
(thick solid red line) with the envelope chemical
composition of CO Nova 3 \citep{hac16k}.
Here, we assume $(m-M)_V=10.25$.  The chemical composition of 
V1369~Cen ejecta is not known, so we assume 
CO Nova 3 because V1369~Cen showed a optically-thin dust blackout similar
to V1668~Cyg, of which the chemical composition is represented
roughly by CO Nova 3 \citep{hac06kb, hac16k}.

In free-free emission dominant spectra, optical and 
near-infrared (NIR), i.e., $B$, $V$, and $I_{\rm C}$
light curves should have the same shape \citep[see, e.g.,][]{hac06kb}.  
The V1369~Cen data show such a property in the middle phase of
light curves and are in good agreement with the model light curve.
The observed $B$ and $V$ magnitudes departed from
the theoretical light curve at $t\gtrsim 100$ day because of
strong contributions of emission lines to the $B$ and
$V$ bands in the nebular phase (see Paper I).
This model light curve again confirms our result of 
$(m-M)_V=10.25$.  The WD mass is around $0.9~M_\sun$.

Our $0.90~M_\sun$ WD model also predicts a supersoft X-ray source phase
from Day 340 until Day 960 (thin solid red line in Figure 
\ref{all_mass_v1369_cen_v1668_cyg_x45z02c15o20}(a)).
\citet{pag14} reported an X-ray spectrum on UT 2014 March 8
(JD 2456724.5).  This epoch (Day $\sim100$) corresponds to the
deepest dust blackout (see Figure \ref{v1369_cen_v_bv_ub_color_curve}),
and is much earlier than the X-ray turn-on time in our model.
\citet{mas18} presented the X-ray count rates of V1369~Cen obtained with
{\it Swift}.  We plot the X-ray data taken from the {\it Swift} Web 
site\footnote{http://www.swift.ac.uk/} \citep{eva09} in Figure
\ref{all_mass_v1369_cen_v1668_cyg_x45z02c15o20}(a).
The blue pluses represent the hard X-ray (1.5--10~kev) and
the magenta crosses denote the soft X-ray (0.3--1.5~keV) components.
No clear supersoft X-rays were detected until Day 370.
We may conclude that the X-rays detected on Day $\sim 100$ are
not from the WD surface but shock-origin.

\section{V5666~Sgr 2014}
\label{v5666_sgr_cmd}
V5666~Sgr was discovered by S. Furuyama at mag 8.7 on UT 2014 January
26.857 (JD~2456684.357) \citep{fur14}. 
It reached $m_V=10.0$ on JD~2456697.3 (estimated from
S. Kiyota's $m_V=9.8$ of VSOLJ).  The nova was identified by A. Arai
as an \ion{Fe}{2} type \citep{fur14}. 

We estimate the distance moduli of V5666~Sgr in the four bands, 
i.e., $(m-M)_B=15.87$, $(m-M)_V=15.38$, $(m-M)_I=14.59$, 
and $(m-M)_K=14.07$ in Appendix \ref{v5666_sgr}.
We plot these four relations by the magenta,
blue, cyan, and cyan-blue lines in 
Figure \ref{distance_reddening_v574_pup_v679_car_v1369_cen_v5666_sgr}(d).
These four lines cross at $d=5.8$~kpc and $E(B-V)=0.50$.
Thus, we adopt $d=5.8\pm0.6$, $E(B-V)=0.50\pm0.05$, 
and the timescaling factor $f_{\rm s}=1.78$ against LV~Vul.

Figure \ref{distance_reddening_v574_pup_v679_car_v1369_cen_v5666_sgr}(d)
also shows various distance-reddening relations toward V5666~Sgr, 
whose galactic coordinates are $(l,b)=(9\fdg8835, -4\fdg6567)$.
We add Marshall et al.'s (2006) relations: 
$(l, b)=(9\fdg75, -4\fdg50)$ by open red squares,
$(10\fdg00, -4\fdg50)$ by filled green squares,
$(9\fdg75, -4\fdg75)$ by blue asterisks,
and $(10\fdg00, -4\fdg75)$ by open magenta circles, each with error bars.
The closest direction among the four nearby directions
is that of open magenta circles.  
We also add the relations given by \citet{gre15, gre18} 
(solid black and orange lines, respectively) and the relation 
by \citet{chen18} (solid cyan-blue line).
\"Ozd\"ormez et al.'s data are not available.
The orange line is consistent with our results of
$d=5.8\pm0.6$~kpc and $E(B-V)=0.50\pm0.05$.

\citet{schu14} determined interstellar extinction
as a function of distance in the galactic bulge 
covering $-10\fdg0 < l < 10\fdg0$ and $-10\fdg0 <b< 5\fdg0$,
using data from the VISTA Variables in the Via Lactea (VVV) survey
together with the Besan\c{c}on stellar population synthesis model
of the Galaxy.  The resolution is $0\fdg1\times0\fdg1$
and the distance is extended up to 10 kpc in a 0.5~kpc step.
We plot four Schuletheis et al.'s 
distance-reddening relations toward near the direction of V5666~Sgr,
i.e., $(l,b)=(9\fdg8, -4\fdg6)$, $(9\fdg8, -4\fdg7)$, 
$(9\fdg9, -4\fdg6)$, and $(9\fdg9, -4\fdg7)$ by very thin solid cyan lines
in Figure \ref{distance_reddening_v574_pup_v679_car_v1369_cen_v5666_sgr}(d),
where we adopt the relations of $A_{K_s}=0.364~E(B-V)$ \citep{sai13} 
and $A_{K_s}=0.528~E(J-K_s)$ \citep{nis09shogo} in the conversion
of $E(B-V)$ from their $E(J-K_s)$ in their Table 1.  
The four lines show zigzag patterns, although the reddening should
increase monotonically with the distance.  In this sense, 
Schuletheis et al.'s relation may not be appropriate
in the middle distance.
Our crossing point, $d=5.8$~kpc and $E(B-V)=0.50$, however,
is consistent with theirs. 

Adopting $E(B-V)=0.50$ and $(m-M)_V=15.4$, we plot 
the color-magnitude diagram of V5666~Sgr in Figure
\ref{hr_diagram_v574_pup_v679_car_v1369_cen_v5666_sgr_outburst_mv}(d).
In the same figure, we add the tracks of V1500~Cyg (solid green line)
and LV~Vul (solid orange line).  Although the track of V5666~Sgr
almost follows that of LV~Vul,  its position is slightly lower than that
of LV~Vul.  

Figure \ref{hr_diagram_v574_pup_v679_car_v1369_cen_v5666_sgr_outburst}(d)
shows the time-stretched color-magnitude diagram of V5666~Sgr.
We adopt $E(B-V)=0.50$ and $(m-M')_V=16.05$ in Equation
(\ref{absolute_mag_v5666_sgr_lv_vul_stretch}).
Now, the track of V5666~Sgr follows well that of LV~Vul.
Thus, we conclude that V5666~Sgr belongs to the LV~Vul type 
in the time-stretched color-magnitude diagram.
This overlap supports our results of $E(B-V)=0.50$
and $(m-M')_V=16.05$, that is, $E(B-V)=0.50\pm0.05$, $(m-M)_V=15.4\pm0.1$, 
$f_{\rm s}=1.78$, and $d=5.8\pm0.6$~kpc.
We list our results in Table \ref{extinction_various_novae}.

We further confirm $(m-M)_V=15.4$ from our model light curve fitting.
Taking $(m-M)_V=15.4$, we plot a $0.85~M_\sun$ WD model
with the chemical composition of CO nova 3 \citep{hac16k}
in Figure \ref{all_mass_v5666_sgr_v1668_cyg_x45z02c15o20}(a).  
The $0.85~M_\sun$ WD (solid red lines) model does not match with 
the early phase of oscillatory flat peak,
but reasonably fit with the mid and later phases, where free-free
emission dominates in the $B$, $V$, and $I_{\rm C}$ magnitudes.
Thus, we again confirm $(m-M)_{V, \rm V5666~Sgr}=15.4\pm0.2$.
The estimated WD mass is about $M_{\rm WD}=0.85~M_\sun$ for the
chemical composition of CO Nova 3.

\section{Revisiting 12 Novae in the Time-stretched Color-Magnitude Diagram}
\label{twelve_novae}

In what follows, we reexamine 12 novae 
studied in Paper II with a new light of time-stretched
color-magnitude diagram.


\begin{figure*}
\plotone{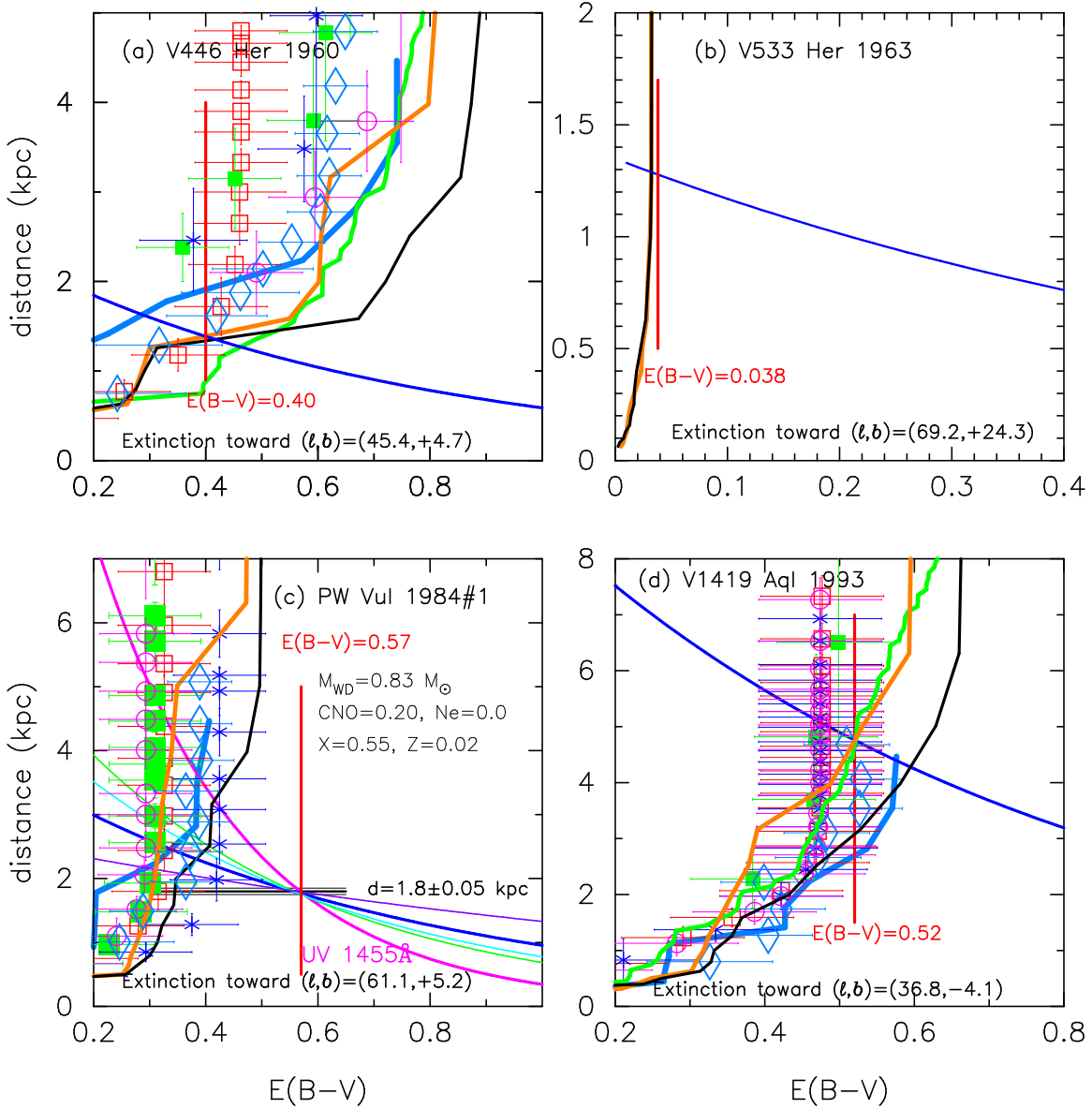}
\caption{
Same as Figure 
\ref{distance_reddening_lv_vul_v1500_cyg_v1668_cyg_v1974_cyg},
but for (a) V446~Her, (b) V533~Her, (c) PW~Vul, and (d) V1419~Aql.  
In panel (c), the solid magenta line
represents the UV~1455\AA\  flux fitting.  The thin solid
green, cyan, and blue-magenta lines denotes the relations of
$(m-M)_U=13.92$, $(m-M)_B=13.55$, $(m-M)_V=13.0$, and $(m-M)_I=12.12$,
respectively.
\label{distance_reddening_v446_her_v533_her_pw_vul_v1419_aql}}
\end{figure*}


\begin{figure*}
\plotone{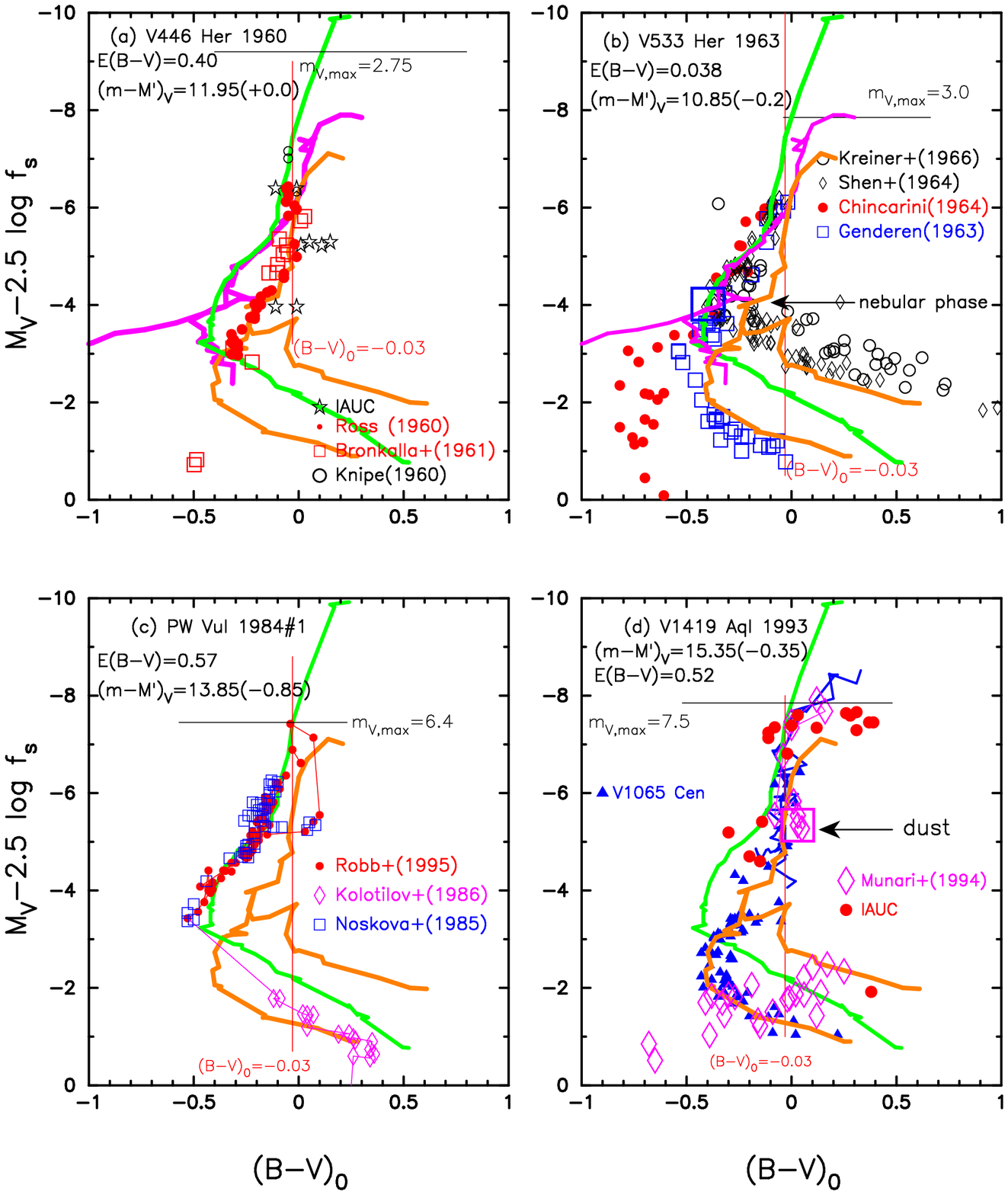}
\caption{
Same as Figure 
\ref{hr_diagram_v574_pup_v679_car_v1369_cen_v5666_sgr_outburst},
but for (a) V446~Her, (b) V533~Her, (c) PW~Vul, and (d) V1419~Aql.
The data are the same as those in Papers I \& II.
The solid orange lines show the template track of LV~Vul while  
the solid green lines show that of V1500~Cyg.  
In panels (a) and (b), we add the tracks of V1974~Cyg (solid magenta lines).
In panel (d), we add the track of V1668~Cyg (solid blue lines) and
V1065~Cen (filled blue triangles).
In panel (b), we depict the start of the nebular phase
by the large open blue square.
In panel (d), we denote the start of the dust blackout 
by the large open magenta square.
\label{hr_diagram_v446_her_v533_her_pw_vul_v1419_aql_outburst}}
\end{figure*}

\subsection{V446~Her 1960}
\label{v446_her_cmd}
For the reddening toward V446~Her, \citet{hac14k} obtained the arithmetic
average of the values in literature to be $E(B-V)=0.41\pm0.15$.  Recently,
\citet{ozd16} proposed another average of $E(B-V)=0.37\pm0.04$ including
the result of \citet{sel13}, $E(B-V)=0.38\pm0.04$, from the archival
2200\AA\  feature.  See Paper II and \citet{ozd16} for a summary
of other estimates on the distance and reddening.  In the present paper,
we adopt $E(B-V)=0.40\pm0.05$  after \citet{hac14k}.  We obtain the 
distance of $d=1.38\pm0.2$~kpc from Equation (\ref{distance_modulus_rv}),
$E(B-V)=0.40\pm0.05$, $f_{\rm s}= 1.0$ against LV~Vul,
and $(m-M)_V=11.95\pm0.1$ in Appendix \ref{v446_her}.

Figure \ref{distance_reddening_v446_her_v533_her_pw_vul_v1419_aql}(a) shows
various distance-reddening relations toward V446~Her, whose galactic
coordinates are $(l, b)= (45\fdg4092, +4\fdg7075)$.
The vertical solid red line denotes the reddening of $E(B-V)=0.40$. 
The solid blue line denotes the distance modulus in the $V$ band, i.e.,
$(m-M)_V=11.95$ and Equation (\ref{distance_modulus_rv}).
These two lines cross at $E(B-V)=0.40$ and $d=1.38$~kpc. 
We further plot Marshall et al.'s (2006) relations 
in four directions close to the direction of V446~Her;
toward $(l, b)=(45\fdg25, 4\fdg50)$ by open red squares,
$(45\fdg50, 4\fdg50)$ by filled green squares,
$(45\fdg25, 4\fdg75)$ by blue asterisks, and
$(45\fdg50, 4\fdg75)$ by open magenta circles.
The closest direction in the galactic coordinates
is that of open magenta circles among the four nearby directions.
The solid black/orange lines denote the distance-reddening relations
given by \citet{gre15, gre18}, respectively.
The solid green line represents the relation given by \citet{sal14}.
The open cyan diamonds depict the relation of \citet{ozd16}.  
The cyan-blue line represents the relation given by \citet{chen18}.
Our crossing point at $E(B-V)=0.40$ and $d=1.38$~kpc is consistent with
Marshall et al.'s (open magenta circles) and Green et al.'s 
(solid black/orange lines) relations.  
Our crossing point is also consistent with
the distance-reddening relation given by \citet{ozd16}.
\citet{coh85} obtained the distance of V446~Her by the expansion
parallax method to be $d=(v_{\rm exp} \times t)/r_{\rm shell} =
(1235~{\rm km~s}^{-1} \times 24~{\rm yr})/4\farcs5 =1.35$~kpc.
Our value of $d=1.38\pm0.2$~kpc is consistent with Cohen's value.
Thus, we confirm again that our obtained values of
$(m-M)_V=11.95\pm0.1$, $E(B-V)=0.40\pm0.05$, and $d=1.38\pm0.2$~kpc
are reasonable.

Adopting $E(B-V)=0.40$ and $(m-M')_V=11.95$ in Equation
(\ref{absolute_mag_lv_vul_v446_her}), we obtain the time-stretched
color-magnitude diagram of V446~Her in Figure
\ref{hr_diagram_v446_her_v533_her_pw_vul_v1419_aql_outburst}(a).
The track of V446~Her is just on the LV~Vul track (solid orange line).
Therefore, V446~Her belongs to the LV~Vul type.
This overlapping to the LV~Vul track supports 
our values of $E(B-V)=0.40$ and $(m-M')_V=11.95$, that is, $f_{\rm s}=1.0$,
$E(B-V)=0.40\pm0.05$, $(m-M)_V=11.95\pm0.1$, and $d=1.38\pm0.2$~kpc.
We list our results in Table \ref{extinction_various_novae}.

Taking $(m-M)_V=11.95$ for V446~Her, 
we add a $0.98~M_\sun$ WD model (solid red line) with
the chemical composition of CO nova 3 \citep{hac16k} in Figure 
\ref{v446_her_v1668_cyg_lv_vul_v_color_logscale}(a).
The model absolute $V$ light curve reasonably fits with the observed
apparent $V$ light curve of V446~Her.  This confirms that 
the distance modulus of $(m-M)_V=11.95$ is reasonable.
In the same figure, we depict another $0.98~M_\sun$ WD model 
with $(m-M)_V=14.6$ for V1668~Cyg (solid black lines).  
The difference between these two models represents the difference in
the initial hydrogen-rich envelope mass;
the initial envelope mass $M_{\rm env,0}$ of the solid red line
is slightly smaller than that of the solid black line.

\subsection{V533~Her 1963}
\label{v533_her_cmd}
V533~Her was studied in Paper II on the
$(B-V)_0$-$M_V$ diagram but we here examine it on the time-stretched
color-magnitude diagram.  
\citet{hac16kb} took $E(B-V)=0.038\pm0.002$ from the NASA/IPAC galactic
absorption map.  We adopt their value of $E(B-V)=0.038$.  
We determine the distance modulus in the $V$ band to be 
$(m-M)_V=10.65\pm0.1$ together with $f_{\rm s}= 1.20$ against LV~Vul
in Appendix \ref{v533_her}, and plot it by
the solid blue line in Figure 
\ref{distance_reddening_v446_her_v533_her_pw_vul_v1419_aql}(b).
The vertical solid red line is the color excess of $E(B-V)=0.038$.
These two lines cross at $d=1.28$~kpc (and $E(B-V)=0.038$).  
Thus, we obtain the distance of $d=1.28\pm0.2$~kpc.

Figure \ref{distance_reddening_v446_her_v533_her_pw_vul_v1419_aql}(b)
also shows a few distance-reddening relations toward V533~Her,
whose galactic coordinates are $(l, b)= (69\fdg1887, +24\fdg2733)$.
None of Marshall et al.'s, Sale et al.'s, \"Ozd\"ormez et al.'s,
and Chen et al.'s relations is available. 
The solid black/orange lines denote the distance-reddening relations
given by \citet{gre15, gre18}, respectively.
Our crossing point is broadly consistent with Green et al.'s 
distance-reddening relations. 

Adopting $E(B-V)=0.038$ and $(m-M')_V=10.85$ in Equation
(\ref{absolute_mag_v533_her_lv_vul_v}),
we obtain the time-stretched color-magnitude diagram of V533~Her in Figure 
\ref{hr_diagram_v446_her_v533_her_pw_vul_v1419_aql_outburst}(b). 
We also add the template tracks
of LV~Vul (thick solid orange lines), V1500~Cyg (thick solid green line),
and V1974~Cyg (solid magenta lines).
The V533~Her track broadly follows the mid part of 
V1974~Cyg tracks at least until the onset of the nebular phase (see also
Figure \ref{v533_her_lv_vul_v1974_cyg_v_bv_ub_logscale}(a) and (b)).
We regard that V533~Her belongs to the V1500~Cyg type in the time-stretched
color-magnitude diagram. 
This supports our values of $(m-M')_V=10.85$ and $E(B-V)=0.038$, that is,
$(m-M)_V=10.65\pm0.1$, $E(B-V)=0.038\pm0.01$, and $f_{\rm s}=1.20$.
Our results are listed in Table \ref{extinction_various_novae}.

We again discuss the effect of different responses
in the color filters.
The color-magnitude track of V533~Her bifurcates at the onset
of the nebular phase around UT 1963 April 19 \citep{chi64r}, i.e., 
at $m_V=7.2$ denoted by the large open blue square in Figure
\ref{hr_diagram_v446_her_v533_her_pw_vul_v1419_aql_outburst}(b).
We plot four color-magnitude data taken from 
\citet{kre66}, \citet{she64}, \citet{gen63}, and \citet{chi64}, 
from upper to lower (or from right to left), and one of the three major
branches immediately turns to the right.
After the onset of the nebular phase, strong emission lines 
[\ion{O}{3}] began to contribute to the blue edge of $V$ filter.
The response of each $V$ filter is different to each other 
at the blue edge and this difference makes a significant
difference in the $V$ magnitude.
This effect can be clearly seen in Figures 3, 4, and 5 of \citet{kre66},
in which the $V$ magnitude started to branch off after UT 1963 April 10 
while the $B$ magnitude did not among the various observers.

We check our obtained value of $(m-M)_V=10.65$ by fitting with our model $V$
light curve in Figure \ref{v533_her_lv_vul_v1974_cyg_v_bv_ub_logscale}(a).
We add a $1.03~M_\sun$ WD model (solid red line) with the chemical 
composition of Ne nova 2 \citep{hac10k}.  Taking $(m-M)_V=10.65$ for V533~Her,
the absolute model $V$ light curve reproduces the observed apparent 
$V$ light curve of V533~Her (filled red circles).
This confirms that the distance modulus of $(m-M)_V=10.65$ is reasonable.

\subsection{PW~Vul 1984\#1}
\label{pw_vul_cmd}
PW~Vul was examined in Paper II on the $(B-V)_0$-$M_V$ diagram.
Here, we reexamine it on the time-stretched color-magnitude diagram.
In Appendix \ref{pw_vul}, we obtain $(m-M)_V=13.0\pm0.2$ together with
$f_{\rm s}= 2.24$ against LV~Vul based on the 
time-stretching method.  Taking $(m-M)_V=13.0$ for PW~Vul, 
we plot a $0.83~M_\sun$ WD model (solid red line) with
the chemical composition of CO nova 4 \citep{hac15k} in Figure 
\ref{pw_vul_lv_vul_v1668_cyg_v1500_cyg_v_bv_ub_color_logscale}(a).
The model absolute $V$ light curve reasonably fits with the observed
apparent $V$ light curve of PW~Vul after $t > 60$~day. 
This confirms that the distance modulus of $(m-M)_V=13.0$ is reasonable.

We also fit our UV~1455\AA\  light curve model of the $0.83~M_\sun$ WD
with the observation as shown in Figure 
\ref{pw_vul_lv_vul_v1668_cyg_v1500_cyg_v_bv_ub_color_logscale}(a).
We obtain $F_{1455}^{\rm mod}= 15$ and
$F_{1455}^{\rm obs}= 6$ in units of 
$10^{-12}$~erg~cm$^{-2}$~s$^{-1}$~\AA$^{-1}$
at the upper bound of Figure 
\ref{pw_vul_lv_vul_v1668_cyg_v1500_cyg_v_bv_ub_color_logscale}(a).
With these values, we plot the distance-reddening relation of
Equation (\ref{qu_vul_uv1455_fit_eq2}) in Figure
\ref{distance_reddening_v446_her_v533_her_pw_vul_v1419_aql}(c)
by the solid magenta line.   The solid blue line of $(m-M)_V=13.0$
and this magenta line cross at $d=1.8$~kpc and $E(B-V)= 0.57$.
These light curve fittings are essentially the same as in the previous
work \citep{hac15k}.  Thus, we obtain the distance of $d=1.8\pm0.2$~kpc
from Equation (\ref{distance_modulus_rv}), $E(B-V)=0.57\pm0.05$, 
and $(m-M)_V=13.0\pm0.2$.  This distance estimate is consistent with
the distance estimated by \citet{dow00}, $d=1.8\pm0.05$~kpc,
with the nebular expansion parallax method.  We list the results
in Table \ref{extinction_various_novae}. 

For comarison, in Figure
\ref{pw_vul_lv_vul_v1668_cyg_v1500_cyg_v_bv_ub_color_logscale}(a),
we depict a $0.98~M_\sun$ WD model
(solid blue lines) with the chemical composition of CO nova 3 \citep{hac15k},
assuming that $(m-M)_V=14.6$ for V1668~Cyg.  With the same stretching factor
of $f_{\rm s}=2.24$, both the $V$ and UV~1455\AA\   light curves almost
overlap between PW~Vul and V1668~Cyg.  This confirms that
the optical light curves and UV~1455\AA\  light curves follow the 
same timescaling law as already explained in Section \ref{introduction}.

We check the distance and reddening based on various
distance-reddening relations.
\citet{hac15k} obtained $E(B-V)=0.55\pm0.05$ from the fitting
in the color-color diagram (Paper I), being consistent with
the crossing point mentioned above.  
Figure \ref{distance_reddening_v446_her_v533_her_pw_vul_v1419_aql}(c)
shows various distance-reddening relations toward PW~Vul, 
whose galactic coordinates are $(l, b)=(61\fdg0983, +5\fdg1967)$.  
We plot the constraint of $d=1.8\pm0.05$~kpc by \citet{dow00}
(solid horizontal black lines).
We further add other distance-reddening relations toward PW~Vul. 
The vertical solid red line is $E(B-V)=0.57$.
We plot the distance-reddening relations (solid black/orange lines)
given by \citet{gre15, gre18}, respectively.
We also plot four distance-reddening relations of \citet{mar06};
toward $(l, b)=(61\fdg00, +5\fdg00)$ denoted by open red squares, 
$(61\fdg25, +5\fdg00)$ by filled green squares,
$(61\fdg00, +5\fdg25)$ by blue asterisks,
and $(61\fdg25, +5\fdg25)$ by open magenta circles, each with error bars.
The closest direction in the galactic coordinates is that of blue asterisks.
The open cyan-blue diamonds represent the relation given by 
\citet{ozd16}.
Our crossing point at $d=1.8$~kpc and $E(B-V)=0.57$ 
is $\Delta E(B-V)=0.15$~mag
larger than the trends of Marshall et al.'s blue asterisks data.
To summarize, our set of $d=1.8$~kpc and $E(B-V)=0.57$ is not consistent
with all the 3D dust maps given by \citet{mar06}, \citet{ozd16},
\citet{gre15, gre18}, and \citet{chen18}.
This is the first and only the case that our crossing point is largely
different from the several 3D dust maps.  We discuss the reason of such 
a large discrepancy below.

The 3D dust maps basically give an averaged value of a relatively broad
region, so the pinpoint reddening could be different from the values of
the 3D dust maps.  The pinpoint estimate of the reddening toward PW~Vul
indicates $E(B-V)=0.55\pm0.05$ as already discussed in \citet{hac15k}.
For example, \citet{and91} obtained 
$E(B-V)=0.58 \pm 0.06$ from \ion{He}{2} $\lambda1640/\lambda4686$
ratio and $E(B-V)=0.55 \pm 0.1$ from the interstellar absorption
feature at 2200\AA\  for the reddening toward PW~Vul.  
\citet{sai91} reported $E(B-V)=0.60 \pm 0.06$ from \ion{He}{2}
$\lambda1640/\lambda4686$ ratio.  
\citet{hac14k} estimated the reddening to be $E(B-V)=0.55\pm0.05$
from the color-color diagram fit of PW~Vul.
As for the distance to PW~Vul, \citet{dow00} obtained $d=1.8\pm0.05$~kpc
with the nebular expansion parallax method.  These pinpoint constraints
are all consistent with our crossing point of
$E(B-V)=0.57\pm0.05$, $(m-M)_V = 13.0\pm0.2$, and $d=1.8\pm0.2$~kpc.

We further obtain the distance moduli of $UBVI$ bands and check
their crossing points in the distance-reddening plane.  We obtain
$(m-M)_U= 13.92$, $(m-M)_B= 13.55$, and $(m-M)_V= 13.0$ 
(and $(m-M)_I= 12.12$) in Appendix \ref{pw_vul}, and plot them in Figure 
\ref{distance_reddening_v446_her_v533_her_pw_vul_v1419_aql}(c)
by the thin solid green, cyan, thick solid blue (and thin
solid blue-magenta) lines, respectively.  These lines cross at
$E(B-V)=0.57$ and $d=1.8$~kpc, confirming again our results.

Adopting $E(B-V)=0.57$ and $(m-M')_V=13.85$ in Equation
(\ref{absolute_mag_pw_vul_lv_vul}), we obtain the time-stretched
color-magnitude diagram of PW~Vul in Figure
\ref{hr_diagram_v446_her_v533_her_pw_vul_v1419_aql_outburst}(c).
Here, we plot the data taken from \citet{nos85}, \citet{kol86},
and \citet{rob95}.  PW~Vul follows the V1500~Cyg track except for
the very early phase.  Therefore, PW~Vul belongs to the V1500~Cyg type
in the time-stretched color-magnitude diagram.
This overlapping to the V1500~Cyg track supports 
our estimates of $E(B-V)=0.57$ and $(m-M')_V=13.85$, that is,
$f_{\rm s}=2.24$, $E(B-V)=0.57\pm0.05$, $(m-M)_V=13.0\pm0.2$,
and $d=1.8\pm0.2$~kpc.
We list our results in Table \ref{extinction_various_novae}.

\subsection{V1419~Aql 1993}
\label{v1419_aql_cmd}
V1419~Aql was examined in Paper II on the $(B-V)_0$-$M_V$ diagram.
Here, we reexamine it on the time-stretched color-magnitude diagram.
We determine the distance modulus in the $V$ band to be
$(m-M)_V=15.0\pm0.2$ together with $f_{\rm s}= 1.41$ against LV~Vul
in Appendix \ref{v1419_aql}. 
For the reddening toward V1419~Aql, \citet{hac14k, hac16kb} obtained
$E(B-V)=0.50\pm0.05$ from the fitting in the color-color diagram
(Paper I).  On the other hand, the NASA/IPAC galactic 2D dust
absorption map gives $E(B-V)=0.55 \pm 0.01$ in the direction toward 
V1419~Aql.  See \citet{hac14k, hac16kb} and \citet{ozd16} for a summary
of other estimates on the extinction and distance of V1419~Aql.
We adopt the arithmetic mean of these two values, i.e.,
$E(B-V)=0.52\pm0.05$ in the present paper.
Then, we obtain the distance of $d=4.7\pm0.5$~kpc from Equation 
(\ref{distance_modulus_rv}).

We check the distance and reddening toward V1419~Aql, 
whose galactic coordinates are $(l, b)=(36\fdg8110, -4\fdg1000)$,
based on various distance-reddening relations in
Figure \ref{distance_reddening_v446_her_v533_her_pw_vul_v1419_aql}(d).
The solid blue line denotes the distance modulus of $(m-M)_V = 15.0$
and Equation (\ref{distance_modulus_rv}).
The vertical solid red line is $E(B-V)=0.52$.
These two lines cross at $d=4.7$~kpc (and $E(B-V)=0.52$).
We add the distance-reddening relations (solid black/orange lines)
given by \citet{gre15, gre18}, respectively.
We also plot four distance-reddening relations of \citet{mar06};
toward $(l, b)=(36\fdg75, -4\fdg00)$ denoted by open red squares, 
$(37\fdg00, -4\fdg00)$ by filled green squares,
$(36\fdg75, -4\fdg25)$ by blue asterisks,
and $(37\fdg00, -4\fdg25)$ by open magenta circles, each with error bars.
The closest direction is that of open red squares.
The solid green line is the relation given by \citet{sal14}.
The open cyan-blue diamonds with error bars are the relation given
by \citet{ozd16}.
The solid cyan-blue line is the relation given by \citet{chen18}.
Our set of $E(B-V)=0.52$ and $d=4.7$~kpc is broadly consistent with
the distance-reddening relations of \citet{mar06}, \citet{sal14},
\citet{gre18}, and \citet{ozd16}.
Thus, we confirm that our adopted values of $E(B-V)=0.52\pm0.05$, 
$(m-M)_V = 15.0\pm0.2$, and $d=4.7\pm0.5$~kpc are reasonable.
We list our results in Table \ref{extinction_various_novae}.

Adopting $E(B-V)=0.52$ and $(m-M')_V=15.35$ in Equation
(\ref{absolute_mag_v1419_aql_lv_vul}), we obtain the time-stretched
color-magnitude diagram of V1419~Aql in Figure
\ref{hr_diagram_v446_her_v533_her_pw_vul_v1419_aql_outburst}(d).
Here, we plot the data taken from \citet{mun94b} and
IAU Circular Nos. 5794, 5802, 5807, and 5829.
The track of V1419~Aql is considerably affected by dust formation,
the start of which is denoted by the large magenta square.  We regard that 
V1419~Aql belongs to the LV~Vul type because the V1419~Aql track follows
a part of the V1668~Cyg track (blue lines) until the dust blackout started.
For comparison, we also plot the V1065~Cen track (filled blue triangles)
that shows a similar shallow dust blackout \citep[see Figures 4 and 6 
of][]{hac18k}.
The track of V1065~Cen almost follows the lower branch of LV~Vul,
while the track of V1419~Aql is not clear owing to dust blackout.

Finally, we plot a $0.90~M_\sun$ WD model with the chemical composition
of CO nova 3 \citep{hac16k} by the solid red lines in Figure
\ref{v1419_aql_v1668_cyg_lv_vul_v_bv_ub_color_logscale}(a).
The $V$ model light curve reasonably fits with the $V$ light curve
of V1419~Aql during $\log t$~(day)$= 1.0-1.5$.  This confirms our
value of $(m-M)_V= 15.0$.


\begin{figure*}
\plotone{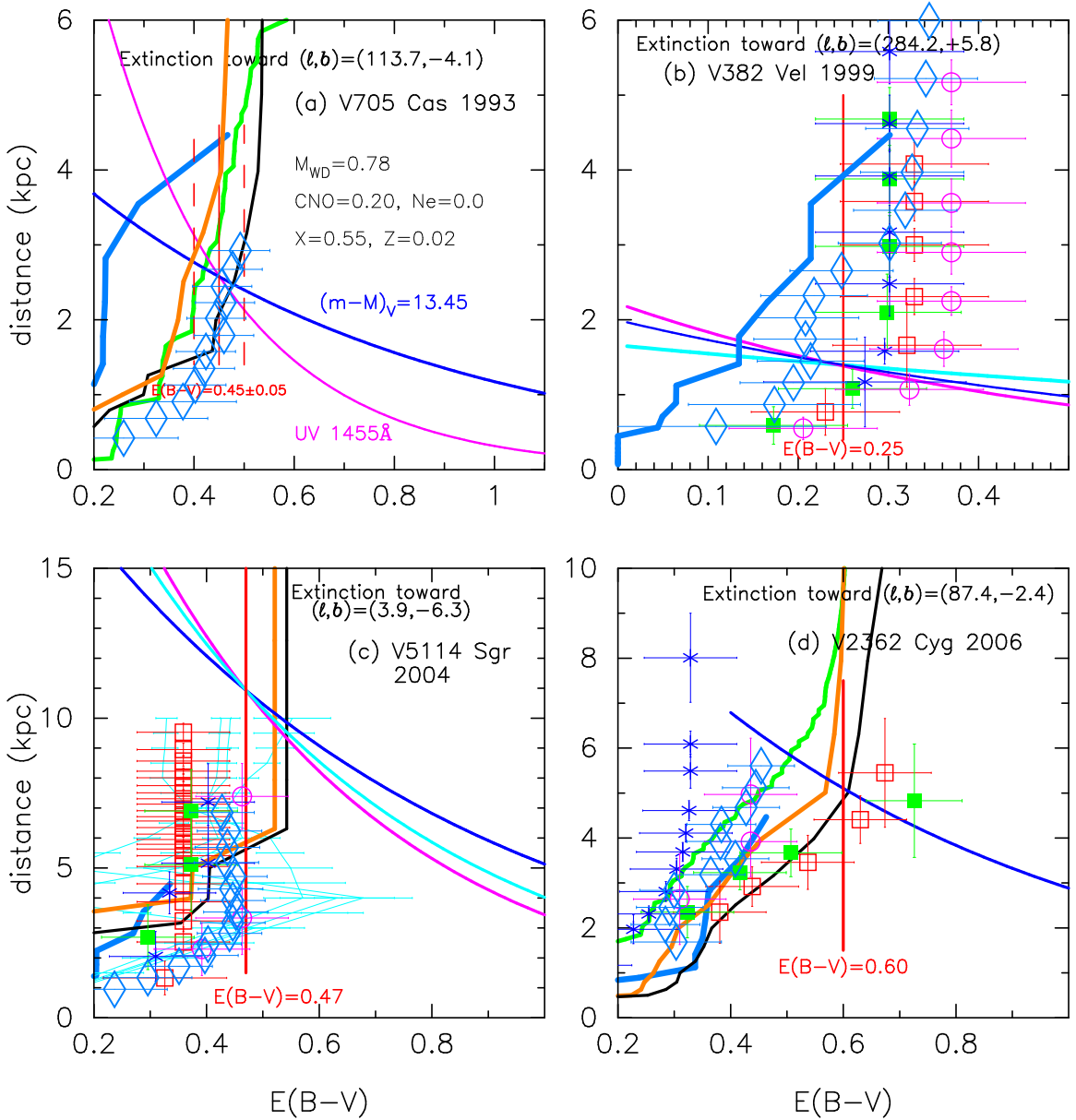}
\caption{
Various distance-reddening relations for (a) V705~Cas, (b) V382~Vel,
(c) V5114~Sgr, and (d) V2362~Cyg.
The vertical thick solid red lines represent the color excess of
each nova.  The solid black/orange lines represent the distance-reddening
relations given by \citet{gre15, gre18}, respectively.
Four sets (open red squares, filled green squares, blue asterisks,
and open magenta circles) of data, each with error bars, show 
distance-reddening relations given by \citet{mar06}
in four nearby directions of each nova.
The open cyan-blue diamonds with error bars are the relation given
by \citet{ozd16, ozd18}.
The thick solid green line represents the relation given by \citet{sal14}.
The thick solid cyan-blue line represents the relation given by \citet{chen18}.
In panel (c), the four very thin cyan lines with error bars denote 
the four nearby distance-reddening relations given by \citet{schu14}.
\label{distance_reddening_v705_cas_v382_vel_v5114_sgr_v2362_cyg}}
\end{figure*}


\begin{figure*}
\plotone{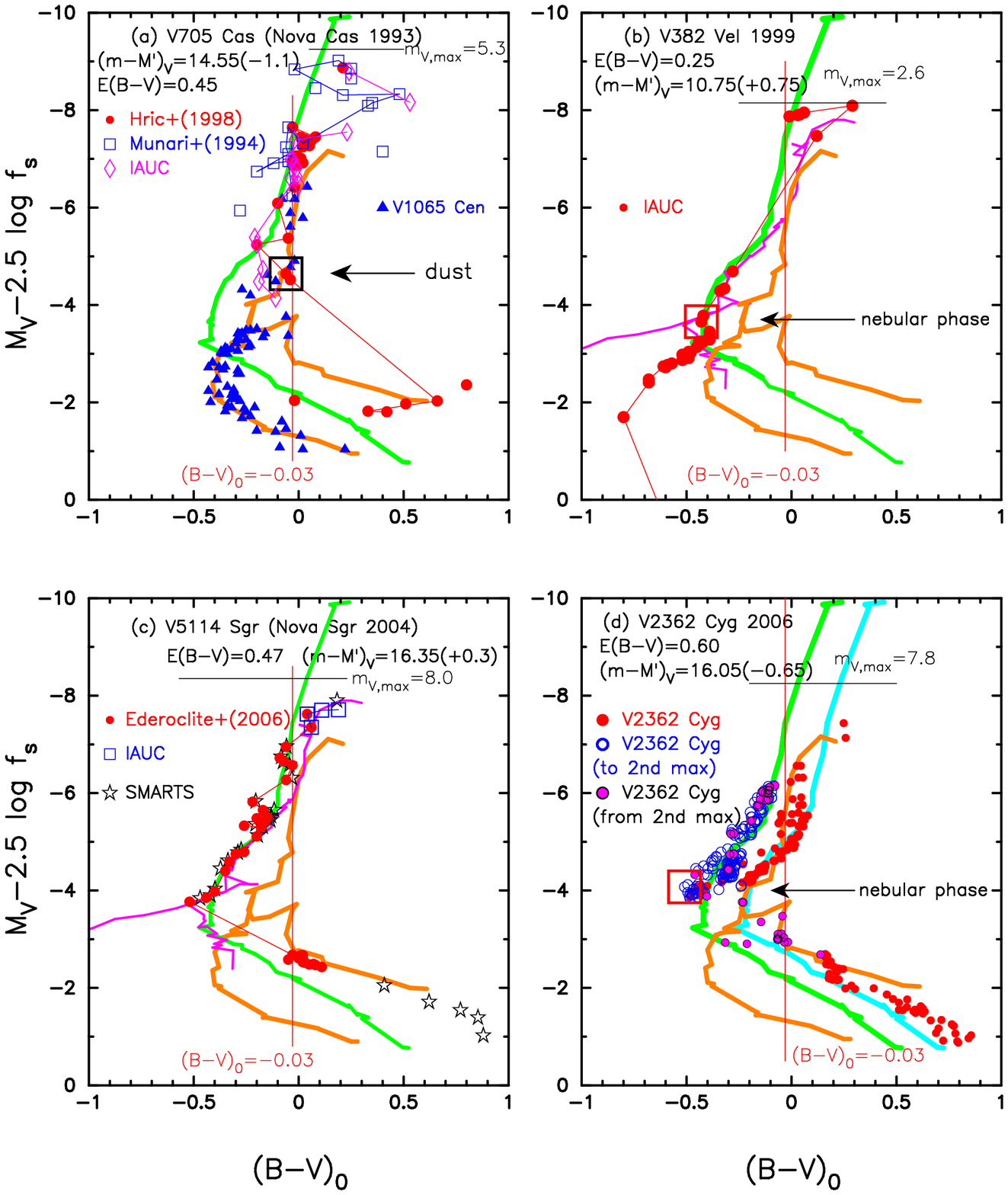}
\caption{
Same as Figure 
\ref{hr_diagram_v574_pup_v679_car_v1369_cen_v5666_sgr_outburst}, 
but for (a) V705~Cas, (b) V382~Vel, (c) V5114~Sgr, and (d) V2362~Cyg.
The solid orange, green, and magenta lines indicate the template tracks
of LV~Vul, V1500~Cyg, and V1974~Cyg, respectively.  
The onset of nebular phase is indicated by a large open red square.
The start of dust blackout is denoted by a large open black square.
In panel (d), we add the track of V1500~Cyg shifted redward
by $\Delta (B-V)=0.20$ (solid cyan lines).  
We indicate the rising/decay phase of the secondary maximum
by blue open circles/filled magenta circles with black outlines,
respectively.
\label{hr_diagram_v705_cas_v382_vel_v5114_sgr_v2362_cyg_outburst}}
\end{figure*}

\subsection{V705~Cas 1993}
\label{v705_cas_cmd}
V705~Cas was examined in Paper II,
but we reexamine it on the time-stretched color-magnitude diagram.
For the reddening toward V705~Cas, we adopt $E(B-V)=0.45\pm0.05$ 
after \citet{hac14k}.  We do not repeat the discussion. 
Papers I \& II give a summary on the distance
and reddening toward V705~Cas, $(l,b)=(113\fdg6595,-4\fdg0959)$.
We obtain the timescaling factor of $f_{\rm s}= 2.8$ against LV~Vul
and the distance of $d=2.6\pm0.3$~kpc from
Equation (\ref{distance_modulus_rv}), $E(B-V)=0.45\pm0.05$,
and $(m-M)_V=13.45\pm0.2$ in Appendix \ref{v705_cas}. 
Using the expansion parallax method, \citet{eyr96} and \citet{dia01}
obtained $d=2.5$~kpc and $d=2.9\pm0.4$~kpc, respectively.
These distances are all consistent with our value of $d=2.6\pm0.3$~kpc.
We list our results in Table \ref{extinction_various_novae}.

Next, we discuss the model light curve fitting.
We obtain $(m-M)_V=13.45$ based on the time-stretching method in Appendix
\ref{v705_cas}.  Assuming that $(m-M)_V=13.45$, 
we plot a $0.78~M_\sun$ WD model (solid magenta lines) with
the chemical composition of CO nova 4 \citep{hac15k} in Figure 
\ref{v705_cas_pw_vul_lv_vul_v_bv_ub_color_curve_logscale}(a).
The model absolute $V$ and UV~1455\AA\   light curves of 
the $0.78~M_\sun$ WD reasonably fit with the observed apparent $V$ 
and UV~1455\AA\  light curves of V705~Cas.  This model light curve
fitting is essentially the same as Figure 15 of \citet{hac15k} and
we again confirm that
the distance modulus of $(m-M)_V=13.45$ is reasonable for V705~Cas.
In the same figure, we add a $0.83~M_\sun$ WD model (solid blue lines) with
the same chemical composition of CO nova 4, assuming that $(m-M)_V=13.0$
for PW~Vul.  

We reanalyze the distance and reddening toward V705~Cas based on various
distance-reddening relations in
Figure \ref{distance_reddening_v705_cas_v382_vel_v5114_sgr_v2362_cyg}(a).
The solid blue line shows $(m-M)_V = 13.45$ and 
Equation (\ref{distance_modulus_rv}).  The solid magenta line
is the UV~1455\AA\  fit of our $0.78~M_\sun$ WD model. 
The UV~1455\AA\  light curve fit gives the relation of 
Equation (\ref{qu_vul_uv1455_fit_eq2}) together with
$F_{1455}^{\rm obs}=6.0$ and $F_{1455}^{\rm mod}=12.5$ in units of 
$10^{-12}$~erg~cm$^{-2}$~s$^{-1}$~\AA$^{-1}$
at the upper bound of Figure 
\ref{v705_cas_pw_vul_lv_vul_v_bv_ub_color_curve_logscale}(a).
We also plot the relations given by \citet{gre15, gre18} by the solid
black/orange lines, respectively. 
No data of \citet{mar06} are available in this direction.
The solid green line represent the relation given by \citet{sal14}.
The open cyan-blue diamonds with error bars denote the relation
given by \citet{ozd16}.  
The solid cyan-blue line denotes the relation given by \citet{chen18}.
The blue, magenta, and red lines cross at $E(B-V)=0.45$ and $d=2.6$~kpc.
This crossing point is also consistent with the distance-reddening relation
given by \citet[][black line]{gre15}, \citet[][green line]{sal14}, 
and \citet[][cyan-blue diamonds]{ozd16}.
 
Adopting $E(B-V)=0.45$ and $(m-M')_V=14.55$ in Equation
(\ref{absolute_mag_v705_cas_lv_vul}), we obtain the time-stretched
color-magnitude diagram of V705~Cas in Figure
\ref{hr_diagram_v705_cas_v382_vel_v5114_sgr_v2362_cyg_outburst}(a).
Here, we plot the data taken from \citet{mun94a}, \citet{hri98}, and
IAU Circular Nos. 5905, 5912, 5914, 5920, 5928, 5929, 5945, and 5957.
The track of V705~Cas is also affected by dust formation
as denoted by the large black square.  We also plot the V1065~Cen track 
\citep[see Figures 4 and 6 of][]{hac18k} in Figure 
\ref{hr_diagram_v705_cas_v382_vel_v5114_sgr_v2362_cyg_outburst}(a).  
V1065~Cen shows a similar dust blackout to V705~Cas
but its depth of dust blackout is much shallower 
(see Figure \ref{v496_sct_v1065_cen_lv_vul_v_bv_ub_color_logscale})
than that of V705~Cas.
The V1065~Cen track comes back soon after the dust blackout and follows
again the LV~Vul track, so that V1065~Cen belongs to
the LV~Vul type in the time-stretched color-magnitude diagram.
We regard that V705~Cas also belongs to
the LV~Vul type because the V705~Cas track almost follows that of 
LV~Vul and V1065~Cen until the dust blackout started.
The early pulse (oscillation) in the $V$ light curve makes a loop 
in the color-magnitude diagram.
This kind of loops were also seen in multiple-peak novae like
V723~Cas and V5558~Sgr as discussed by \citet{hac16kb}. 
We list our results in Table \ref{extinction_various_novae}.

\subsection{V382~Vel 1999}
\label{v382_vel_cmd}
V382~Vel was also studied in Paper II on the $(B-V)_0$-$M_V$ diagram.
In this subsection, we examine it on the 
$(B-V)_0$-$(M_V-2.5\log f_{\rm s}$) diagram.  
For the reddening toward V382~Vel, $(l,b)=(284\fdg1674, +5\fdg7715)$,
\citet{ste99} obtained $E(B-V)=A_V/3.1=0.8/3.1=0.26$ from an equivalent
width of 0.024 nm of the interstellar \ion{Ca}{2}~K line. 
\citet{ori02} obtained the hydrogen column density
toward V382~Vel to be $N_{\rm H}\sim 2 \times 10^{21}$~cm$^{-2}$
from X-ray spectrum fittings.  This value can be converted with
$E(B-V)= N_{\rm H}/ 4.8 \times 10^{21} \sim 0.4$ \citep{boh78},
$E(B-V)= N_{\rm H}/ 5.8 \times 10^{21} \sim 0.34$ \citep{gue09}, and
$E(B-V)= N_{\rm H}/ 6.8 \times 10^{21} \sim 0.3$ \citep{lis14}.
On the other hand, \citet{del02} obtained
$E(B-V)=0.05$ from various line ratios and $E(B-V)=0.09$ from 
\ion{Na}{1}~D interstellar absorption features.
\citet{sho03} obtained $E(B-V)= 0.20$ from the resemblance
of {\it IUE} spectra to that of V1974~Cyg.  
\citet{sho03} and \citet{nes05} obtained 
$N_{\rm H}=1.2\times 10^{21}$~cm$^{-2}$, 
corresponding to $E(B-V)= N_{\rm H}/ 4.8 \times 10^{21} = 0.25$,
$E(B-V)= N_{\rm H}/ 5.8 \times 10^{21} = 0.21$, and
$E(B-V)= N_{\rm H}/ 6.8 \times 10^{21} = 0.18$.  
\citet{hac14k} obtained $E(B-V)=0.15\pm0.05$ by fitting the color-color
track of V382~Vel with the general track of novae.
To summarize, there are two estimates,  $E(B-V)=0.1\pm0.05$ 
and $E(B-V)=0.25\pm0.05$.  

We obtain three distance moduli of $B$, $V$, and $I_{\rm C}$ bands 
in Appendix \ref{v382_vel} based on the time-stretching method.
We plot these distance moduli in Figure
\ref{distance_reddening_v705_cas_v382_vel_v5114_sgr_v2362_cyg}(b)
by the magenta, blue, and cyan lines, that is,
$(m-M)_B= 11.7$, $(m-M)_V= 11.48$, and $(m-M)_I= 11.09$
together with Equations (\ref{distance_modulus_rb}), 
(\ref{distance_modulus_rv}), and (\ref{distance_modulus_ri}).
These three lines broadly cross at $d=1.4$~kpc and $E(B-V)=0.25$.
Thus, we obtain the timescaling factor $f_{\rm s}= 0.51$ against LV~Vul,
the reddening toward V382~Vel, $E(B-V)=0.25\pm0.05$,
and the distance to V382~Vel, $d=1.4\pm0.2$~kpc.  The distance
modulus in the $V$ band is $(m-M)_V=11.5\pm0.2$.  These values
are listed in Table \ref{extinction_various_novae}.

Adopting $E(B-V)=0.25$ and $(m-M')_V=10.75$ in Equation 
(\ref{absolute_mag_v382_vel_lv_vul}),
we plot the time-stretched color-magnitude diagram of V382~Vel in Figure
\ref{hr_diagram_v705_cas_v382_vel_v5114_sgr_v2362_cyg_outburst}(b).
The nebular phase had started at least by the end of 1999 June, 
i.e., $\sim40$ days after the optical maximum \citep{del02}.
We plot this phase ($M_V=m_V-(m-M)_V\approx7.09-11.5=-4.41$)
by the large open red square in Figure
\ref{hr_diagram_v705_cas_v382_vel_v5114_sgr_v2362_cyg_outburst}(b).
The track of V382~Vel follows the V1500~Cyg/V1974~Cyg tracks
at least until the nebular phase started.
Thus, we regard that V382~Vel belongs to the V1500~Cyg type
in the time-stretched color-magnitude diagram.
This overlapping to the tracks of V1500~Cyg/V1974~Cyg 
supports our values of $E(B-V)=0.25$ and $(m-M')_V=10.75$,
that is, $E(B-V)=0.25\pm0.05$, $(m-M)_V=11.5\pm0.2$, $f_{\rm s}=0.51$,
and $d=1.4\pm0.2$~kpc.

Taking $(m-M)_V=11.5$ for V382~Vel, we plot the $V$ model
light curve (total flux of free-free plus photospheric emission:
solid red line) of a $1.23~M_\sun$ WD with the envelope chemical
composition of Ne nova 2 \citep{hac10k, hac16k} in Figure
\ref{v382_vel_lv_vul_v1668_cyg_v1974_cyg_v_bv_ub_color_logscale}(a).
The model $V$ light curve reasonably reproduces the observation 
(filled red circles).  This model light curve fitting gives the same 
result as in Figure 27 of \citet{hac16k}.  Thus, we again confirm
that the distance modulus of $(m-M)_V=11.5\pm0.2$ is reasonable.

We further examine the distance and reddening toward V382~Vel,
$(l, b)= (284\fdg1674, +5\fdg7715)$, based
on various distance-reddening relations in
Figure \ref{distance_reddening_v705_cas_v382_vel_v5114_sgr_v2362_cyg}(b).
We plot the four sets of distance-reddening relation 
calculated by \citet{mar06}; toward
$(l, b)= (284\fdg00, +5\fdg75)$ denoted by open red squares,
$(284\fdg25, +5\fdg75)$ by filled green squares,
$(284\fdg00, +6\fdg00)$ by blue asterisks, and
$(284\fdg25, +6\fdg00)$ by open magenta circles.
The filled green squares are the closest direction
among the four nearby directions.
The open cyan-blue diamonds with error bars denote the relation
given by \citet{ozd16}.
The solid cyan-blue line denotes the relation given by \citet{chen18}.
The vertical solid red line is the color excess of $E(B-V)=0.25$.
Our crossing point at $d=1.4$~kpc and $E(B-V)=0.25$
is broadly consistent with Marshall et al.'s 
and \"Ozd\"ormez et al.'s relations within error bars.

The distance toward V382~Vel was recently estimated by \citet{tom15} to be 
$d=(v_{\rm exp} \times t)/r_{\rm shell} = (1800\pm100~{\rm km~s}^{-1}
\times 12~{\rm yr})/(6\farcs0 \pm 0\farcs25) \approx 0.8\pm0.1$~kpc 
with the expansion parallax method.  This value is much smaller than 
our estimate of $d=1.4\pm0.2$~kpc.  
However, the distance obtained with the expansion
parallax method depends largely on the assumed velocity, $v_{\rm exp}$.
If we adopt other velocities estimated in the early outburst phase
$v_{\rm exp}\sim3500$~km~s$^{-1}$ \citep[e.g.,][]{del02} 
or $v_{\rm exp}\sim2900$~km~s$^{-1}$ \citep[e.g.,][]{nes05}, 
we have large distances such as $d\sim1.5$~kpc or $d\sim1.2$~kpc,
respectively.

\subsection{V5114~Sgr 2004}
\label{v5114_sgr_cmd}
V5114~Sgr was studied in Paper II on the $(B-V)_0$-$M_V$ diagram.
In this subsection, we reexamine it but on the 
$(B-V)_0$-$(M_V-2.5\log f_{\rm s}$) diagram.
We obtain three distance moduli in the $U$, $B$, and $V$ bands in
Appendix \ref{v5114_sgr}.
We plot these three distance moduli in
Figure \ref{distance_reddening_v705_cas_v382_vel_v5114_sgr_v2362_cyg}(c)
by the magenta, cyan, and blue lines, that is,
$(m-M)_U= 17.45$, $(m-M)_B= 17.15$, and $(m-M)_V= 16.65$ 
together with Equations (\ref{distance_modulus_ru}),
(\ref{distance_modulus_rb}), and
(\ref{distance_modulus_rv}), respectively.
These three lines do not exactly but broadly cross 
at $d=10.9$~kpc and $E(B-V)=0.47$.
Thus, we obtain the timescaling factor of $f_{\rm s}= 0.76$ against LV~Vul, 
the reddening of $E(B-V)=0.47\pm0.05$, and the distance of $d=10.9\pm1$~kpc.  
These values are listed in Table \ref{extinction_various_novae}.

Various reddening estimates were summarized in Paper II and
their arithmetic mean was calculated to be $E(B-V)=0.51\pm 0.09$.
The NASA/IPAC galactic dust absorption map gives $E(B-V)=0.49 \pm 0.02$
in the direction toward V5114~Sgr.  \citet{hac14k} derived the reddening
of $E(B-V)=0.45\pm 0.05$ on the basis of the general track of
color-color evolution of novae.  These are all consistent with our
new value of $E(B-V)=0.47\pm0.05$.

Next, we reanalyze the distance and reddening toward V5114~Sgr, 
$(l, b)=(3\fdg9429, -6\fdg3121)$, based on various distance-reddening
relations in Figure 
\ref{distance_reddening_v705_cas_v382_vel_v5114_sgr_v2362_cyg}(c).
The vertical solid red line is $E(B-V)=0.47$.
We plot four distance-reddening relations of \citet{mar06};
toward $(l, b)=(3\fdg75, -6\fdg50)$ denoted by open red squares, 
$(4\fdg00, -6\fdg50)$ by filled green squares,
$(3\fdg75, -6\fdg25)$ by blue asterisks,
and $(4\fdg00, -6\fdg25)$ by open magenta circles, each with error bars.
The closest direction in the galactic coordinates
is that of open magenta circles.
The black/orange lines are the relations given by \citet{gre15, gre18},
respectively. 
The cyan-blue line is the relation given by \citet{chen18}. 
We further add the 3D reddening map given by \citet{schu14}.
We plot four Schuletheis et al.'s distance-reddening relations toward
near the direction of V5114~Sgr: $(l,b)=(3\fdg9, -6\fdg4)$, 
$(3\fdg9, -6\fdg3)$, $(4\fdg0, -6\fdg4)$, and $(4\fdg0, -6\fdg3)$ 
by the very thin solid cyan lines.
The lines show zigzag patterns in this case, although the reddening
must increase monotonically with the distance.  In this sense,
Schuletheis et al.'s distance-reddening relation may not be
appropriate in the middle part of the lines.  
Our set of $E(B-V)=0.47$ and $d=10.9$~kpc
is consistent with the trend of Marshall et al.'s data
(open magenta circles), although $\Delta E(B-V)\sim0.05-0.1$ mag
smaller than Green et al.'s.  

Adopting $E(B-V)=0.47$ and $(m-M')_V=16.35$ in Equation
(\ref{absolute_mag_v5114_sgr_lv_vul}), we obtain the time-stretched
color-magnitude diagram of V5114~Sgr in Figure
\ref{hr_diagram_v705_cas_v382_vel_v5114_sgr_v2362_cyg_outburst}(c).
Here, we plot the data taken from \citet{ede06}, IAU Circular Nos. 8306
and 8310, and SMARTS.  V5114~Sgr follows those of V1500~Cyg/V1974~Cyg.
Therefore, we regard that V5114~Sgr belongs to the V1500~Cyg type
in the time-stretched color-magnitude diagram.
This overlapping supports our estimates of $E(B-V)=0.47$ and 
$(m-M')_V=16.35$, that is, $f_{\rm s}=0.76$, $E(B-V)=0.47\pm0.05$, 
$(m-M)_V=16.65\pm0.1$, and $d=10.9\pm1$~kpc.  We list our results in 
Table \ref{extinction_various_novae}.

Taking $(m-M)_V=16.65$ for V5114~Sgr, we plot the $V$ model
light curve (solid red line) of a $1.15~M_\sun$ WD with the envelope chemical
composition of Ne nova 2 \citep{hac10k, hac16k} in Figure
\ref{v5114_sgr_lv_vul_v1668_cyg_v_color_logscale}(a).
The model $V$ light curve reasonably reproduces the observation 
(filled red circles).  This again confirms that the distance modulus
of $(m-M)_V=16.65\pm0.1$ is reasonable.

\subsection{V2362~Cyg 2006}
\label{v2362_cyg_cmd}
V2362~Cyg was also studied in Paper II.
Here, we reexamine it on the $(B-V)_0$-$(M_V-2.5\log f_{\rm s}$) diagram.
V2362~Cyg shows a prominent secondary maximum \citep[e.g.,][see Figure 
\ref{v2362_cyg_lv_vul_v1500_cyg_v_bv_ub_color_logscale}]{kim08, mun08b}.  
The origin of the secondary maximum was discussed by \citet{hac09ka}.

The color excess of V2362~Cyg was estimated by various authors
(see Section 3.22 of Paper II).   We adopt $E(B-V)=0.60\pm0.05$
after Hachisu \& Kato (Paper II).  Then, the timescaling factor is
obtained to be $f_{\rm s}= 1.78$ against LV~Vul and the distance is 
$d=5.1\pm0.5$~kpc from Equation (\ref{distance_modulus_rv})
together with $(m-M)_V=15.4\pm0.2$ in Appendix \ref{v2362_cyg}. 

Adopting $E(B-V)=0.60$ and $(m-M')_V=16.05$ 
from Equation (\ref{absolute_mag_lv_vul_v2362_cyg_v}), 
we plot the time-stretched color-magnitude diagram of V2362~Cyg in Figure 
\ref{hr_diagram_v705_cas_v382_vel_v5114_sgr_v2362_cyg_outburst}(d). 
We also add the template tracks of 
LV~Vul (thick solid orange line) and V1500~Cyg (thick solid green line).
We further add another track of V1500~Cyg (thick solid cyan line) which is
shifted toward red by $\Delta (B-V)=0.20$~mag.  

V2362~Cyg shows an interesting evolution in the color-magnitude diagram.
V2362~Cyg goes down along the LV~Vul track (orange line) or the 
red-shifted V1500~Cyg track (cyan line) just after the optical maximum. 
In the secondary maximum phase, it goes up along the original
track of V1500~Cyg (solid green line) as denoted by open blue circles, 
and then goes down along the same V1500~Cyg track (solid green line)
as denoted by filled magenta circles with black outlines.
After the onset of the nebular phase, the track turns to the right
and follows again the upper LV~Vul track (orange line) or
the red-shifted V1500~Cyg track (cyan line).

These distinct two tracks are a hint to resolve the reason why 
the two types of V1500~Cyg and LV~Vul tracks are clearly separated
in the time-stretched color-magnitude diagram.
\citet{mun08b} wrote ``the emission spectrum 
at second maximum was quite different from that at first maximum,
reflecting the much higher temperature of the underlying continuum.
At second maximum, \ion{Fe}{2} emission lines were not seen and
were replaced by \ion{N}{2}, \ion{N}{3}, \ion{O}{2}, [\ion{O}{1}], 
and \ion{He}{1}.''   The strong emission lines (\ion{He}{2}/\ion{N}{3}
as well as Balmer lines such as H$\gamma$, H$\delta$, H$\epsilon$) 
contribute to the $B$-band and make $B-V$ blue in the secondary maximum
phase.
Thus, the difference in the $(B-V)_0$ color is real rather than
the difference in the $V$ filter response.
We can conclude that the two tracks reflect
the difference in the ionization state which originates from the   
temperature difference of the underlying continuum.
We call the redder location the LV~Vul type and the bluer location 
the V1500~Cyg type.
The separation is $\Delta (B-V)_0\sim 0.2$ mag.
Overlapping of these tracks of V2362~Cyg
with the two templates novae, LV~Vul and V1500~Cyg, supports 
our values of $(m-M')_V=16.05$ and $E(B-V)=0.60$, that is,
$f_{\rm s}=1.78$, $E(B-V)=0.60\pm0.05$, $(m-M)_V=15.4\pm0.2$,
and $d=5.1\pm0.5$~kpc.
We list our results in Table \ref{extinction_various_novae}.

We reexamine the distance-reddening relation toward V2362~Cyg,
$(l, b)= (87\fdg3724,-2\fdg3574)$, based on several distance-reddening
relations in 
Figure \ref{distance_reddening_v705_cas_v382_vel_v5114_sgr_v2362_cyg}(d).
The solid blue line denotes the distance-reddening relation of Equation 
(\ref{distance_modulus_rv}) together with $(m-M)_V=15.4$.
The vertical solid red line is the color excess of V2362~Cyg estimated
in Paper II.  These two lines cross at $d=5.1$~kpc (and $E(B-V)=0.60$).
We plot the four sets of distance-reddening relation 
calculated by \citet{mar06}; toward
$(l, b)= (87\fdg25,-2\fdg25)$ denoted by open red squares,
$(87\fdg50,-2\fdg25)$ by filled green squares,
$(87\fdg25,-2\fdg50)$ by blue asterisks, and
$(87\fdg50,-2\fdg50)$ by open magenta circles.
The open red squares are the closest direction
toward V2362~Cyg among the four nearby directions.
The solid black/orange lines denote the distance-reddening relations
given by \citet{gre15, gre18}, respectively.
The solid green line represent the relation given
by \citet{sal14} and the open cyan-blue diamonds with error bars 
indicate the relation given by \citet{ozd16}.  
The solid cyan-blue line is the relation given by \citet{chen18}.
Our crossing point at $d=5.1$~kpc and $E(B-V)=0.60$ is roughly 
consistent with Marshall et al.'s (open red squares) and 
Green et al.'s  (solid black/orange lines) relations,
although the reddenings
given by \citet{sal14} and \citet{ozd16} slightly deviates
from the crossing point.
Thus, we confirm again that our estimated values of 
$(m-M)_V=15.4\pm0.2$, $E(B-V)=0.60\pm0.05$, and $d=5.1\pm0.5$~kpc
are reasonable.


\begin{figure*}
\plotone{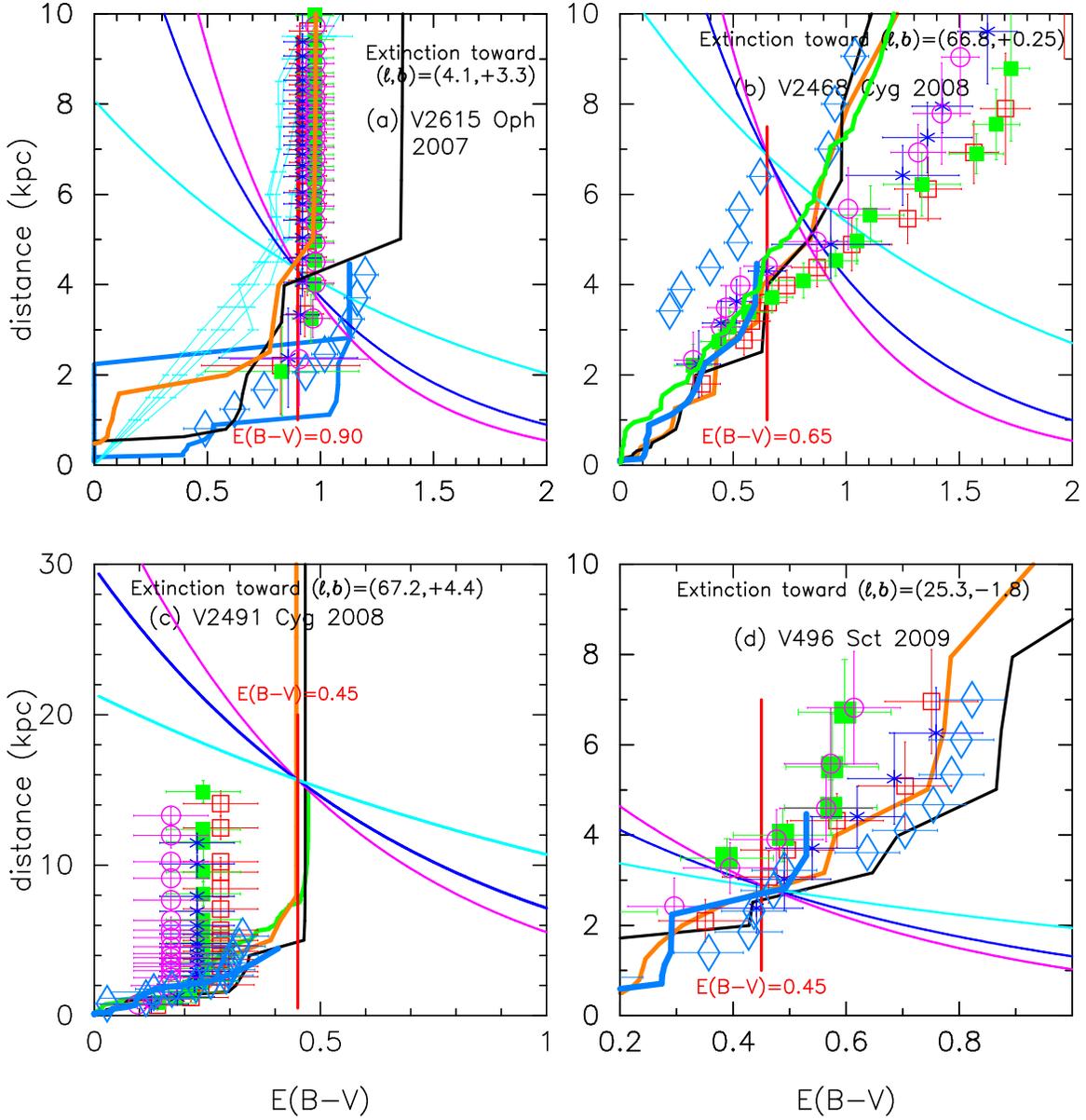}
\caption{
Same as Figures
\ref{distance_reddening_lv_vul_v1500_cyg_v1668_cyg_v1974_cyg} and
\ref{distance_reddening_v574_pup_v679_car_v1369_cen_v5666_sgr}, but for
(a) V2615~Oph, (b) V2468~Cyg, (c) V2491~Cyg, and (d) V496~Sct.
\label{distance_reddening_v2615_oph_v2468_cyg_v2491_cyg_v496_sct}}
\end{figure*}


\begin{figure*}
\plotone{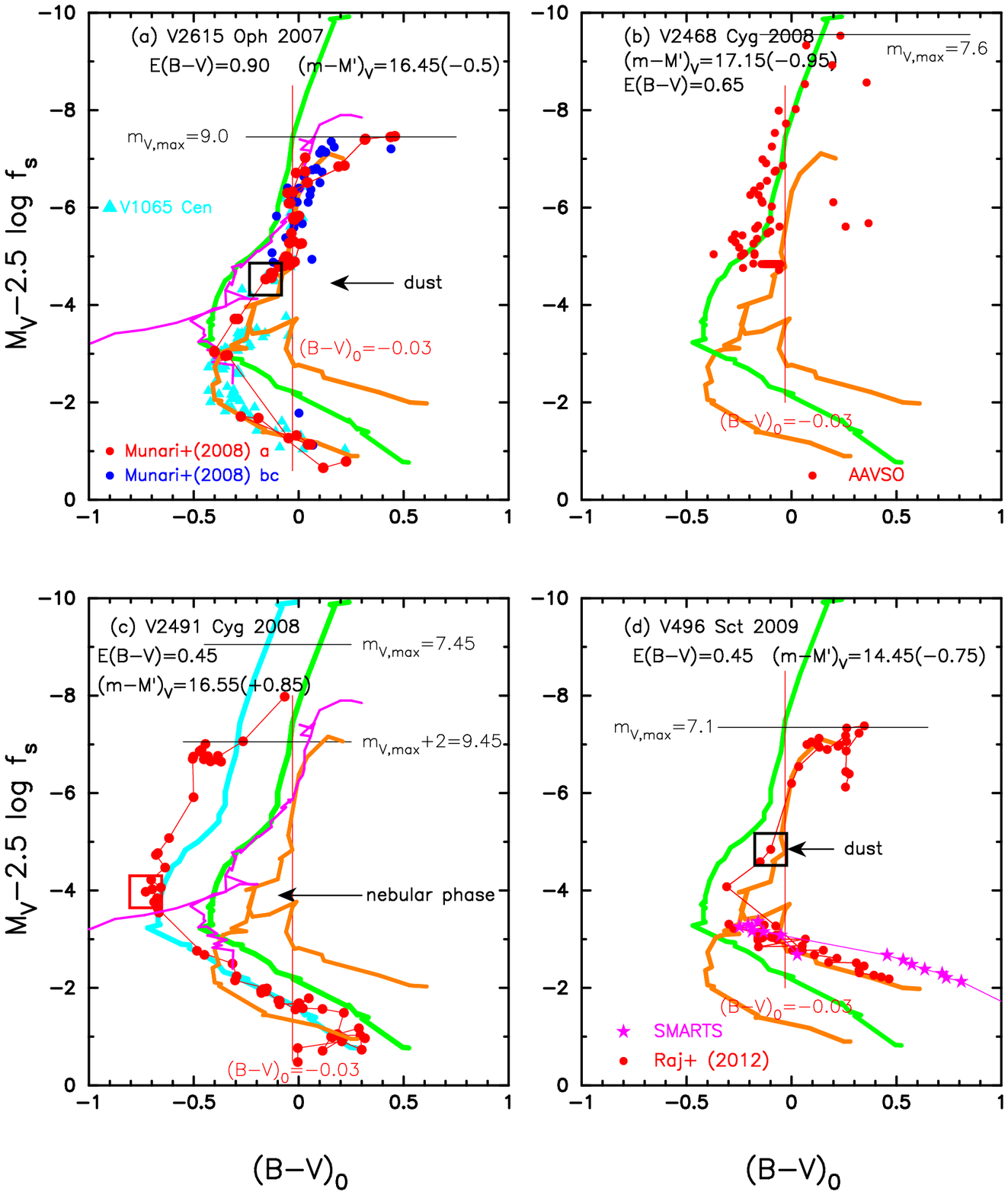}
\caption{
Same as Figure 
\ref{hr_diagram_v574_pup_v679_car_v1369_cen_v5666_sgr_outburst}, 
but for (a) V2615~Oph, (b) V2468~Cyg, (c) V2491~Cyg, and (d) V496~Sct.  
The solid orange, green, and magenta lines indicate the template tracks
of LV~Vul, V1500~Cyg, and V1974~Cyg, respectively.  
The onset of nebular phase is indicated by a large open red square while
the onset of dust blackout is indicated by a large open black square.
In panel (c), we add the track of V1500~Cyg shifted blueward
by $\Delta (B-V)=-0.25$ (solid cyan lines).  
\label{hr_diagram_v2615_oph_v2468_cyg_v2491_cyg_v496_sct_outburst}}
\end{figure*}

\subsection{V2615~Oph 2007}
\label{v2615_oph_cmd}
V2615~Oph was also studied in Paper II on the $(B-V)_0$-$M_V$ diagram.
In this subsection, we reexamine it on the 
$(B-V)_0$-$(M_V-2.5\log f_{\rm s}$) diagram.

We obtain three distance moduli in the $B$, $V$, and $I_{\rm C}$ bands
in Appendix \ref{v2615_oph}.
We plot these three distance moduli in Figure
\ref{distance_reddening_v2615_oph_v2468_cyg_v2491_cyg_v496_sct}(a) by
the thin solid magenta, blue, and cyan lines.  These three lines cross
at $d=4.3$~kpc and $E(B-V)=0.90$.
Therefore, we adopt $E(B-V)=0.90\pm0.05$.
See Paper II for other various estimates on the reddening and distance.
Thus, we obtain the timescaling factor of $f_{\rm s}= 1.58$ against LV~Vul
and the distance of $d=4.3\pm0.4$~kpc, 
$E(B-V)=0.90\pm0.05$, and $(m-M)_V=15.95\pm0.2$.

We reanalyze the distance and reddening toward V2615~Oph, 
$(l, b)=(4\fdg1475, +3\fdg3015)$,
based on various distance-reddening relations in 
Figure \ref{distance_reddening_v2615_oph_v2468_cyg_v2491_cyg_v496_sct}(a).
We plot four distance-reddening relations of \citet{mar06};
toward $(l, b)=(4\fdg00, +3\fdg25)$ denoted by open red squares, 
$(4\fdg25, +3\fdg25)$ by filled green squares,
$(4\fdg00, +3\fdg50)$ by blue asterisks,
and $(4\fdg25, +3\fdg50)$ by open magenta circles, each with error bars.
The direction toward V2615~Oph is between those of open red squares
and filled green squares.
We add the distance-reddening relations (solid black/orange lines)
given by \citet{gre15, gre18}, respectively.
We further add four distance-reddening relations 
(very thin solid cyan lines) given by \citet{schu14}.
The open cyan-blue diamonds are the relation given by \citet{ozd16}.
The cyan-blue line represents the relation given by \citet{chen18};
one is toward $(4\fdg15, +3\fdg25)$ and the other is toward
$(4\fdg15, +3\fdg35)$.
Our crossing point at $E(B-V)=0.90$ and $d=4.3$~kpc
is consistent with the both trends of Marshall et al.
and Green et al.  The NASA/IPAC galactic 2D dust absorption map
gives $E(B-V)=0.87\pm0.02$ toward V2615~Oph, being consistent with
our value of $E(B-V)=0.90\pm0.05$.
We list our results in Table \ref{extinction_various_novae}.

Adopting $E(B-V)=0.90$ and $(m-M')_V=16.45$ in Equation
(\ref{absolute_mag_v2615_oph_lv_vul}), we obtain the time-stretched
color-magnitude diagram of V2615~Oph in Figure
\ref{hr_diagram_v2615_oph_v2468_cyg_v2491_cyg_v496_sct_outburst}(a).
Here, we add the time-stretched track of V1065~Cen by the filled 
cyan triangles.
The track of V2615~Oph is similar to that of V1065~Cen until the dust
blackout.  It broadly follows the LV~Vul track. 
Therefore, V2615~Oph belongs to the LV~Vul type in the time-stretched
color-magnitude diagram.  
This overlapping supports our estimates of 
$E(B-V)=0.90$ and $(m-M')_V=16.45$, that is, $f_{\rm s}=1.58$, 
$E(B-V)=0.90\pm0.05$, $(m-M)_V=15.95\pm0.2$, and $d=4.3\pm0.4$~kpc.

Taking $(m-M)_V=15.95$ for V2615~Oph, we plot the $V$ model light curve 
(solid red line) of a $0.90~M_\sun$ WD with the envelope chemical
composition of CO nova 3 \citep{hac16k} in Figure
\ref{v2615_oph_lv_vul_v1419_aql_v_bv_ub_color_logscale}(a).
The model $V$ light curve reasonably reproduces the observation 
(filled red circles).
This confirms that the distance modulus of $(m-M)_V=15.95\pm0.2$
is reasonable.

\subsection{V2468~Cyg 2008}
\label{v2468_cyg_cmd}
V2468~Cyg was studied in Paper II.
Here, we reexamine it on the $(B-V)_0$-$(M_V-2.5\log f_{\rm s}$)
diagram.   

We obtain three distance moduli in the $B$, $V$, and $I_{\rm C}$ bands
in Appendix \ref{v2468_cyg}.
We plot these three distance moduli in Figure
\ref{distance_reddening_v2615_oph_v2468_cyg_v2491_cyg_v496_sct}(b) by
the thin solid magenta, blue, and cyan lines.  These three lines cross
at $d=6.9$~kpc and $E(B-V)=0.65$.
We adopt $f_{\rm s}= 2.4$ against LV~Vul, $E(B-V)=0.65\pm0.05$, and 
$(m-M)_V=16.2\pm0.2$.
The previous values in Paper II are $E(B-V)=0.75\pm0.05$, 
$(m-M)_V=15.6\pm0.2$, and $d=4.5\pm0.5$~kpc, being smaller than
our new values. 
The main difference is the improvement in the timescaling factor.  
See the discussion in Paper II for a summary
of other estimates on the reddening and distance toward V2468~Cyg.

Figure \ref{distance_reddening_v2615_oph_v2468_cyg_v2491_cyg_v496_sct}(b)
also shows several distance-reddening relations toward V2468~Cyg,
$(l, b)=(66\fdg8084, +0\fdg2455)$.
We plot four distance-reddening relations of 
\citet{mar06}; toward $(l, b)=(66\fdg75, +0\fdg00)$ denoted by open 
red squares, $(67\fdg00, +0\fdg00)$ by filled green squares,
$(66\fdg75, +0\fdg25)$ by blue asterisks,
and $(67\fdg00, +0\fdg25)$ by open magenta circles, each with error bars.
The closest direction in the galactic coordinates is that of blue asterisks.
We add the distance-reddening relations (solid black/orange lines)
given by \citet{gre15, gre18}, respectively.  
The solid green line represents the relation given by \citet{sal14}
and the open cyan-blue diamonds with error bars are given by \citet{ozd16}. 
The cyan-blue line represents the relation given by \citet{chen18}.
Our crossing point at $E(B-V)=0.65$ and $d=6.9$~kpc is consistent
with the distance-reddening relation given by \"Ozd\"ormez et al. 
Thus, we confirm that our set of $E(B-V)=0.65\pm0.05$ and $d=6.9\pm0.8$~kpc
are reasonable.
We list our results in Table \ref{extinction_various_novae}.

Adopting $E(B-V)=0.65$ and $(m-M')_V=17.15$ in Equation
(\ref{absolute_mag_v2468_cyg_lv_vul}), we obtain the time-stretched
color-magnitude diagram of V2468~Cyg in Figure
\ref{hr_diagram_v2615_oph_v2468_cyg_v2491_cyg_v496_sct_outburst}(b).
Here, we plot the data of V2468~Cyg taken from AAVSO.
Although the color data are rather scattered, V2468~Cyg broadly
follows the V1500~Cyg track.
Therefore, V2468~Cyg belongs to the V1500~Cyg type.
This overlapping supports our estimates of $E(B-V)=0.65$
and $(m-M')_V=17.15$, that is, $f_{\rm s}=2.4$, $E(B-V)=0.65\pm0.05$, 
$(m-M)_V=16.2\pm0.2$, and $d=6.9\pm0.8$~kpc.

Taking $(m-M)_V=16.2$ for V2468~Cyg, we plot the $V$ model light curve 
(solid red line) of a $0.85~M_\sun$ WD with the envelope chemical
composition of CO nova 4 \citep{hac15k} in Figure
\ref{v2468_cyg_lv_vul_v1500_cyg_v_color_logscale_no2}(a).
The model $V$ light curve reasonably reproduces the upper bound of
observation (filled red circles) during $\log t~({\rm day})\sim 1-2$.
This suggests that the distance modulus of $(m-M)_V=16.2\pm0.2$ is reasonable.


\begin{figure}
\plotone{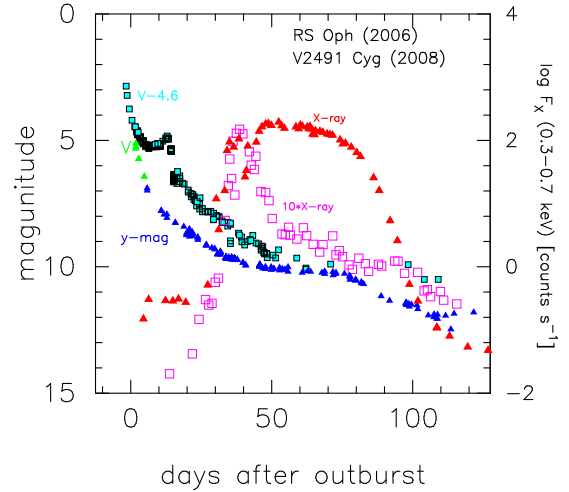}
\caption{
The $V$ (filled cyan squares with black outlines) 
and X-ray (open magenta squares)
light curves of V2491~Cyg are plotted on a linear
timescale together with the $V$ (filled green triangles) and $y$ 
(filled blue triangles), and supersoft X-ray (filled red triangles)
light curves of RS~Oph.  The $V$ 
magnitudes of V2491~Cyg are shifted upward by $4.6$ mag to match the 
absolute magnitudes of V2491~Cyg to those of RS~Oph.  The open magenta
triangles represent 10 times the supersoft X-ray fluxes of V2491~Cyg.  
The data of RS~Oph are the same as those in \citet{hac06b, hac08b}
and \citet{hac07kl}.
\label{light_rs_oph2006_v2491_cyg2008_xray}}
\end{figure}

\subsection{V2491~Cyg 2008\#2}
\label{v2491_cyg_cmd}
V2491~Cyg was studied in Paper II on the $(B-V)_0$-$M_V$ diagram.  In this
subsection, we reexamine it on the $(B-V)_0$-$(M_V-2.5\log f_{\rm s}$) diagram.
V2491~Cyg showed a small secondary maximum \citep[e.g.,][]{hac09ka}.
This nova is also characterized by the detection of pre-outburst X-ray
emission \citep{iba08a, iba08b}.  \citet{hac16kb} adopted $E(B-V)=0.23$
after \citet{mun11} while we here adopt a different
value of $E(B-V)=0.45$.  These two values are significantly different,
so we explained the reason below.

Figure \ref{distance_reddening_v2615_oph_v2468_cyg_v2491_cyg_v496_sct}(c)
plots three distance moduli in the $B$, $V$, and $I_{\rm C}$ bands
obtained from the time-stretching method in Appendix \ref{v2491_cyg}
by the magenta, blue, and cyan lines, that is,
$(m-M)_B= 17.83$, $(m-M)_V= 17.41$, and $(m-M)_I= 16.62$
together with Equations (\ref{distance_modulus_rb}), 
(\ref{distance_modulus_rv}), and (\ref{distance_modulus_ri}).
These three lines broadly cross at $d=15.9$~kpc and $E(B-V)=0.45$.

Figure \ref{distance_reddening_v2615_oph_v2468_cyg_v2491_cyg_v496_sct}(c)
also shows several distance-reddening relations toward V2491~Cyg,
$(l, b)= (67\fdg2287,+4\fdg3532)$.
We plot the distance-reddening relations given by \citet{mar06}; toward
$(l, b)= (67\fdg00,4\fdg25)$ denoted by open red squares,
$(67\fdg25,4\fdg25)$ by filled green squares,
$(67\fdg00,4\fdg50)$ by blue asterisks, and
$(67\fdg25,4\fdg50)$ by open magenta circles.
The closest direction in the galactic coordinates
is that denoted by filled green squares.
The solid green line represents the relation given by \citet{sal14},
black/orange lines by \citet{gre15, gre18}.
The open cyan-blue diamonds with error bars denote the relation given
by \citet{ozd16} and the solid cyan-blue line represents the
relation given by \citet{chen18}.   

Marshall et al.'s distance-reddening relations (green squares and 
blue asterisks) match the smaller value of $E(B-V)=0.23\pm0.01$.  
On the other hand, Green et al.'s and Sale et al.'s relations 
are consistent with the larger value of $E(B-V)=0.45\pm0.05$,
which is consistent with our new estimate of $E(B-V)=0.45\pm0.05$.

The reddening toward V2491~Cyg was obtained by \citet{lyn08b} to be 
$E(B-V)=0.3$ from \ion{O}{1} lines, which was revised by \citet{rud08b}
to be $E(B-V)=0.43$ from the \ion{O}{1} lines at $0.84~\mu$m and $1.13~\mu$m.
\citet{mun11} obtained $E(B-V)=0.23\pm0.01$ from an average of
$E(B-V)=0.24$ from \ion{Na}{1} 5889.953 line profiles,
$E(B-V)= (B-V)_{\rm max} - (B-V)_{0, \rm max} =
0.46 - (0.23 \pm 0.06) = 0.23\pm 0.06$,
and $E(B-V)= (B-V)_{t2} - (B-V)_{0, t2} = 0.20 - (-0.02\pm 0.04)
= 0.22\pm 0.04$, from the intrinsic colors at maximum and $t_2$ time
\citep{van87}.
The distance modulus and distance to V2491~Cyg were estimated
by \citet{mun11} as
$(m-M)_V= m_{V,\rm max} - M_{V,\rm max}= 7.45 - (-9.06)=16.51$
from the MMRD relation \citep{coh88} together with $t_2=4.8$~days, and
then derived the distance of $d=14$~kpc. 
On the other hand, the NASA/IPAC Galactic 2D dust absorption map gives 
$E(B-V)=0.48 \pm 0.03$ in the direction toward V2491~Cyg.
To summarize, there are two estimates on the reddening, i.e.,
$E(B-V)=0.23\pm0.01$ and $E(B-V)=0.43\pm0.05$.
Our value is consistent with the larger one. 
We here adopt $E(B-V)=0.45\pm0.05$ and $d=15.9\pm2$~kpc.  The distance
modulus in the $V$ band is $(m-M)_V=17.4\pm0.2$.  
These values are listed in Table \ref{extinction_various_novae}.

Using $E(B-V)=0.45$ and $(m-M')_V=16.55$ from 
Equation (\ref{absolute_mag_lv_vul_v2491_cyg}), 
we plot the time-stretched color-magnitude diagram of V2491~Cyg in Figure 
\ref{hr_diagram_v2615_oph_v2468_cyg_v2491_cyg_v496_sct_outburst}(c). 
We add the template tracks of LV~Vul
(thick solid orange line), V1500~Cyg (thick solid green line), and 
V1974~Cyg (solid magenta line).  We also add the V1500~Cyg track
shifted toward blue by $\Delta (B-V)=-0.25$
(thick solid cyan line).  V2491~Cyg approximately 
follows the blue-shifted V1500~Cyg track (cyan line) except 
for during the tiny secondary maximum.  The track turns to the right
after the onset of the nebular phase (Paper II).
This redward excursion is due to large contribution of the strong
emission lines [\ion{O}{3}] to the $V$ magnitude
as already discussed in the previous subsections.  

\citet{mun11} suggested the low metallicity of [Fe/H]$=-0.25$ for V2491~Cyg.
This sub-solar metallicity is roughly consistent with
$-0.5<$ [Fe/H] $<-0.3$ at the galacto-centric distance ($\sim 14$~kpc)
of V2491~Cyg.  \citet{hac18kb} clarified that the color-magnitude 
tracks of LMC novae are located at the bluer side by $\Delta (B-V)=-0.3$
than those of galactic novae.  
For example, the color-magnitude track of YY~Dor overlaps with
the V1668~Cyg track shifted by $\Delta (B-V)=-0.3$.
They supposed that this blue-shift is caused by the lower metallicity
of LMC stars, [Fe/H]$=-0.55$ \citep[see, e.g.,][]{pia13}.
The LMC~N~2009a track is in good agreement with
the 0.2 mag blue-shifted V1500~Cyg and 0.15 mag blue-shifted
U~Sco tracks \citep{hac18kb}.  This bluer feature of the LMC~N~2009a track
is also due to the lower metallicity.
The blue-shifted nature of the V2491~Cyg track is consistent with
the nature of LMC novae \citep{hac18kb}.  

It should be noted that the empirical relations proposed by
\citet{van87}, i.e., $(B-V)_0= 0.23\pm0.06$ at maximum
and $(B-V)_0= -0.02\pm0.04$ at $t_2$, are not applicable to
novae of sub-solar metallicity as clearly shown in Figure
\ref{hr_diagram_v2615_oph_v2468_cyg_v2491_cyg_v496_sct_outburst}(c).
This is one of the reason why \citet{mun11} obtain the smaller
reddening of $E(B-V)=0.23\pm0.01$.  

\citet{mun11} obtained the abundance of ejecta to be
$X = 0.573$, $Y = 0.287$, $Z = 0.140$ by mass weight, 
with those of individual elements being $X_{\rm N} = 0.074$, 
$X_{\rm O} = 0.049$, and $X_{\rm Ne} = 0.015$
\citep[see also][for another abundance estimate]{tar14b}. 
This abundance is close to that of Ne nova 2 \citep{hac10k}.
Taking $(m-M)_V=17.4$ for V2491~Cyg, we plot the $V$ model
light curve of $1.35~M_\sun$ (solid red lines) and $1.30~M_\sun$
(solid green lines) WDs with the envelope chemical composition
of Ne nova 2 in Figure
\ref{v2491_cyg_lv_vul_v1500cyg_cyg_v_color_logscale}(a).
The model $V$ light curve reasonably reproduces the observation 
(filled red circles) except during the secondary maximum.
This suggests that the distance modulus of $(m-M)_V=17.4\pm0.2$ is reasonable.

Our model light curve fitting suggests that the WD mass is approximately
$1.35~M_\sun$.  The WD mass of RS~Oph has been estimated to be
$\sim 1.35~M_\sun$ by \citet{hac06b, hac07kl} and \citet{hac18kb}.
It is interesting to compare the $V$ (or $y$) and supersoft X-ray
light curves of these two novae 
(Figure \ref{light_rs_oph2006_v2491_cyg2008_xray}).
The starting time of the SSS phases are almost the same between
the two novae while the duration of the SSS phase is rather different
and much longer in the recurrent nova RS~Oph than in the classical
nova V2491~Cyg.

\subsection{V496~Sct 2009}
\label{v496_sct_cmd}
V496~Sct was studied in Paper II on the $(B-V)_0$-$M_V$ diagram.  In this
subsection, we reexamine it on the $(B-V)_0$-$(M_V-2.5\log f_{\rm s}$) diagram.

We obtain the distance moduli of $BVI_{\rm C}$  
bands in Appendix \ref{v496_sct}
and plot them by the magenta, blue, and cyan lines, 
i.e., $(m-M)_B=14.15$, $(m-M)_V = 13.71$, and $(m-M)_I=13.02$, 
in Figure 
\ref{distance_reddening_v2615_oph_v2468_cyg_v2491_cyg_v496_sct}(d).
These three lines broadly cross each other at $d=2.9$~kpc and $E(B-V)=0.45$.
Thus, we determine the distance and reddening to be
$d=2.9\pm0.3$~kpc and $E(B-V)=0.45\pm0.05$.  The distance modulus
in the $V$ band is $(m-M)_V = 13.7\pm0.2$.

\citet{hac16kb} obtained $E(B-V)=0.50\pm0.05$ by assuming 
that the intrinsic $B-V$ color curve of
V496~Sct is the same as those of FH~Ser and NQ~Vul.
See Paper II for a summary for the other estimates on the reddening. 
We reanalyze the distance and reddening toward V496~Sct, 
$(l, b)=(25\fdg2838, -1\fdg7678)$,
based on various distance-reddening relations in
Figure \ref{distance_reddening_v2615_oph_v2468_cyg_v2491_cyg_v496_sct}(d).
We plot four distance-reddening relations of \citet{mar06};
toward $(l, b)=(25\fdg25, -1\fdg75)$ denoted by open red squares, 
$(25\fdg50, -1\fdg75)$ by filled green squares,
$(25\fdg25, -2\fdg00)$ by blue asterisks,
and $(25\fdg50, -2\fdg00)$ by open magenta circles, each with error bars.
The closest direction in the galactic coordinates
is that of open red squares.
We add the distance-reddening relations (solid black/orange lines)
given by \citet{gre15, gre18}, respectively.
The open cyan-blue diamonds with error bars represent the relation
given by \citet{ozd16}.
The solid cyan-blue line denotes the relation given by \citet{chen18}. 
Our set of $d=2.9$~kpc and $E(B-V)=0.45$
is consistent with the trends of Marshall et al., Green et al.,
\"Ozd\"ormez et al, and Chen et al.   
Our results are listed in Table \ref{extinction_various_novae}.

Adopting $E(B-V)=0.45$ and $(m-M')_V=14.45$ in Equation
(\ref{absolute_mag_v496_sct_lv_vul}), we obtain the time-stretched
color-magnitude diagram of V496~Sct in Figure
\ref{hr_diagram_v2615_oph_v2468_cyg_v2491_cyg_v496_sct_outburst}(d).
Here, we plot the data taken from \citet{raj12} and SMARTS \citep{wal12}.
V496~Sct follows the LV~Vul track until the dust blackout.
Note that, even after the dust blackout, V496~Sct comes back 
and follows the LV~Vul track again 
(upper branch after the onset of nebular phase).
Therefore, V496~Sct belongs to the LV~Vul type in the time-stretched
color-magnitude diagram.  This overlapping supports 
our estimates of $E(B-V)=0.45$ and $(m-M')_V=14.45$, that is,
$f_{\rm s}=2.0$, $E(B-V)=0.45\pm0.05$, $(m-M)_V=13.7\pm0.2$,
and $d=2.9\pm0.3$~kpc.

Taking $(m-M)_V=13.7$ for V496~Sct, we plot the $V$ model light curve 
(solid red line) of a $0.85~M_\sun$ WD with the envelope chemical
composition of CO nova 3 \citep{hac16k} in Figure
\ref{v496_sct_lv_vul_v1668_cyg_v_bv_ub_color_logscale_no2}(a).
The model $V$ light curve reasonably reproduces the observation 
(filled red circles).
This confirms that the distance modulus of $(m-M)_V=13.7\pm0.2$ is reasonable.

\section{Discussion}
\label{discussion}
\subsection{Comparison with Gaia DR2 distances}
\label{gaia_distance}

Recently, \citet{schaefer18} listed distances of 64 novae from
Gaia data release 2 (DR2) in his Table 1.
Among them, seven novae dubbed
the ``very well observed light curve ($< 30$\% error)'' coincide with
our analyzed novae in Table \ref{extinction_various_novae}.
These distances are compared with the present results (Gaia DR2)
as follows: CI Aql, 3.3 kpc (3.19 kpc);   V705 Cas, 2.6 kpc (2.16 kpc);
V1974 Cyg, 1.8 kpc (1.63 kpc):  V446 Her, 1.38 kpc (1.36 kpc);
V533 Her, 1.28  kpc (1.20 kpc); V382 Vel 1.4 kpc (1.80 kpc),
PW Vul 1.8 kpc (2.42 kpc).
These values are in good agreement within each error box.



\section{Conclusions}
\label{conclusions}
We have obtained the following results:\\

\noindent
{\bf 1.} We improved the timescaling factor
$f_{\rm s}$ of each nova by overlapping not only $V$
light curves but also $(B-V)_0$ and $(U-B)_0$ color curves. 
Applying the improved $f_{\rm s}$  to the time-stretching method, we 
revised the distance modulus in the $V$ band $\mu_V=(m-M)_V$. 
The results are summarized in Table \ref{extinction_various_novae}.\\

\noindent
{\bf 2.} We also reexamine the color excess $E(B-V)$ from 
the multi-band time-stretching method.  We obtain the distance moduli
in the $UBVI_{\rm C}K_{\rm s}$ bands, $(m-M)_U$,
$(m-M)_B$, $(m-M)_V$, $(m-M)_I$, and $(m-M)_K$,
of each nova.  The crossing point of these distance-reddening
relations gives a reasonable reddening and distance toward the nova.
The results are also summarized in Table \ref{extinction_various_novae}.\\

\noindent
{\bf 3.} With the improved values of $f_{\rm s}$, $(m-M)_V$, and $E(B-V)$,
we obtain the $(B-V)_0$-$(M_V-2.5\log f_{\rm s})$ color-magnitude diagram.
We call it the time-stretched color-magnitude diagram.
In general, each nova evolves 
from the upper right (red) to the lower left (blue) and then turns back
toward the right (red) at the onset of the nebular phase. 
We found two representative tracks in this diagram,
the tracks of LV~Vul and V1500~Cyg.  The template LV~Vul goes down
along the line of $(B-V)_0=-0.03$, the color of optically thick free-free
emission, in the middle part of the track.  The template V1500~Cyg track
is almost parallel to, but located at the bluer side of
$\Delta (B-V)_0\sim -0.2$ mag than 
the LV~Vul track.  This difference is caused by
the difference in the ionization state which originates from the   
temperature difference of the underlying continuum. \\

\noindent
{\bf 4.} Among the eight novae studied in Sections 
\ref{method_example}--\ref{v5666_sgr_cmd}, V1668~Cyg (see Figure 
\ref{hr_diagram_v446_her_v533_her_pw_vul_v1419_aql_outburst}(d)),
V679~Car, V1369~Cen, and 
V5666~Sgr follow the template track of LV~Vul while
V1974~Cyg (see Figure 
\ref{hr_diagram_v574_pup_v679_car_v1369_cen_v5666_sgr_outburst}(a))
and V574~Pup follow the template track of V1500~Cyg.\\

\noindent
{\bf 5.} We reanalyzed additional 12 novae in Section \ref{twelve_novae}
on the time-stretched color-magnitude diagram.  
Among the 12 novae, V446~Her, V1419~Aql, V705~Cas, V2615~Oph, and V496~Sct
broadly follow the LV~Vul (or V1668~Cyg) template track, while  
V533~Her, PW~Vul, V382~Vel, V5114~Sgr, V2468~Cyg, 
and V2491~Cyg follow the template track of V1500~Cyg (or V1974~Cyg). 
Only V2362~Cyg follows the track of LV~Vul in the first
decline and then follows the V1500~Cyg track during
the secondary maximum (a large rebrightening).   
Thus, we establish the two representative tracks
in the $(B-V)_0$-$(M_V-2.5\log f_{\rm s})$ diagram.\\

\noindent
{\bf 6.} 
The location of V2491~Cyg track is about 0.25 mag bluer than that of
the original V1500~Cyg track (Figure
\ref{hr_diagram_v2615_oph_v2468_cyg_v2491_cyg_v496_sct_outburst}(c)).
This bluer location is due to a lower metallicity of the nova ejecta
(e.g., subsolar by [Fe$/$H]$= -0.25$) as suggested
by \citet{mun11}.  This kind of low metallicity effect is already
discussed in \citet{hac18kb} for LMC novae.\\

\noindent
{\bf 7.} We estimated the white dwarf masses and $(m-M)_V$ of the novae
by directly fitting the absolute $V$ model light curves ($M_V$)
with observational apparent $V$ magnitudes ($m_V$).  
The obtained results of $(m-M)_V$ are in good agreement with
the estimates by the time-stretched color-magnitude diagram method.\\

\noindent
{\bf 8.} The white dwarf masses are estimated from the $V$, UV~1455\AA,
and supersoft X-ray light curve fittings, assuming an appropriate
chemical composition of ejecta.  They are 
$0.98~M_\sun$ (LV~Vul, CO3),
$1.2~M_\sun$ (V1500~Cyg, Ne2),
$0.98~M_\sun$ (V1668~Cyg, CO3),
$0.98~M_\sun$ (V1974~Cyg, CO3),
$1.05~M_\sun$ (V574~Pup, Ne3),
$0.98~M_\sun$ (V679~Car, CO3),
$0.90~M_\sun$ (V1369~Cen, CO3),
$0.85~M_\sun$ (V5666~Sgr, CO3),
$0.98~M_\sun$ (V446~Her, CO3),
$1.03~M_\sun$ (V533~Her, Ne2),
$0.83~M_\sun$ (PW~Vul, CO4),
$0.90~M_\sun$ (V1419~Aql, CO3),
$0.78~M_\sun$ (V705~Cas, CO4),
$1.23~M_\sun$ (V382~Vel, Ne2),
$1.15~M_\sun$ (V5114~Sgr, Ne2),
$0.85~M_\sun$ (V2362~Cyg, interpolation),
$0.90~M_\sun$ (V2615~Oph, CO3),
$0.85~M_\sun$ (V2468~Cyg, CO3),
$1.35~M_\sun$ (V2491~Cyg, Ne2), and
$0.85~M_\sun$ (V496~Sct, CO3).  
These results are summarized in Table \ref{wd_mass_novae}.\\

\noindent
{\bf 9.}
Our distance estimates are in good agreement with the results 
of Gaia Data Release 2.

\acknowledgments
     We express our gratitude to T. Iijima 
and the Astronomical Observatory of Padova (Asiago)
for the warm hospitality during which we initiated the present work.
     We are grateful to 
the late A. Cassatella for providing us with  
UV 1455 \AA~data for {\it IUE} novae.
     We thank
the American Association of Variable Star Observers
(AAVSO) and the Variable Star Observers League of Japan (VSOLJ)
for the archival data of various novae.
We are also grateful to the anonymous referee for useful comments
regarding how to improve the manuscript.
This research has been supported in part by Grants-in-Aid for
Scientific Research (15K05026, 16K05289) 
from the Japan Society for the Promotion of Science.



\appendix
\section{Time-stretched Light Curves of V574~Pup, V679~Car, V1369~Cen,
and V5666~Sgr}
\label{v574_pup_v679_car_v1369_cen_v5666_sgr}


\begin{figure}
\plotone{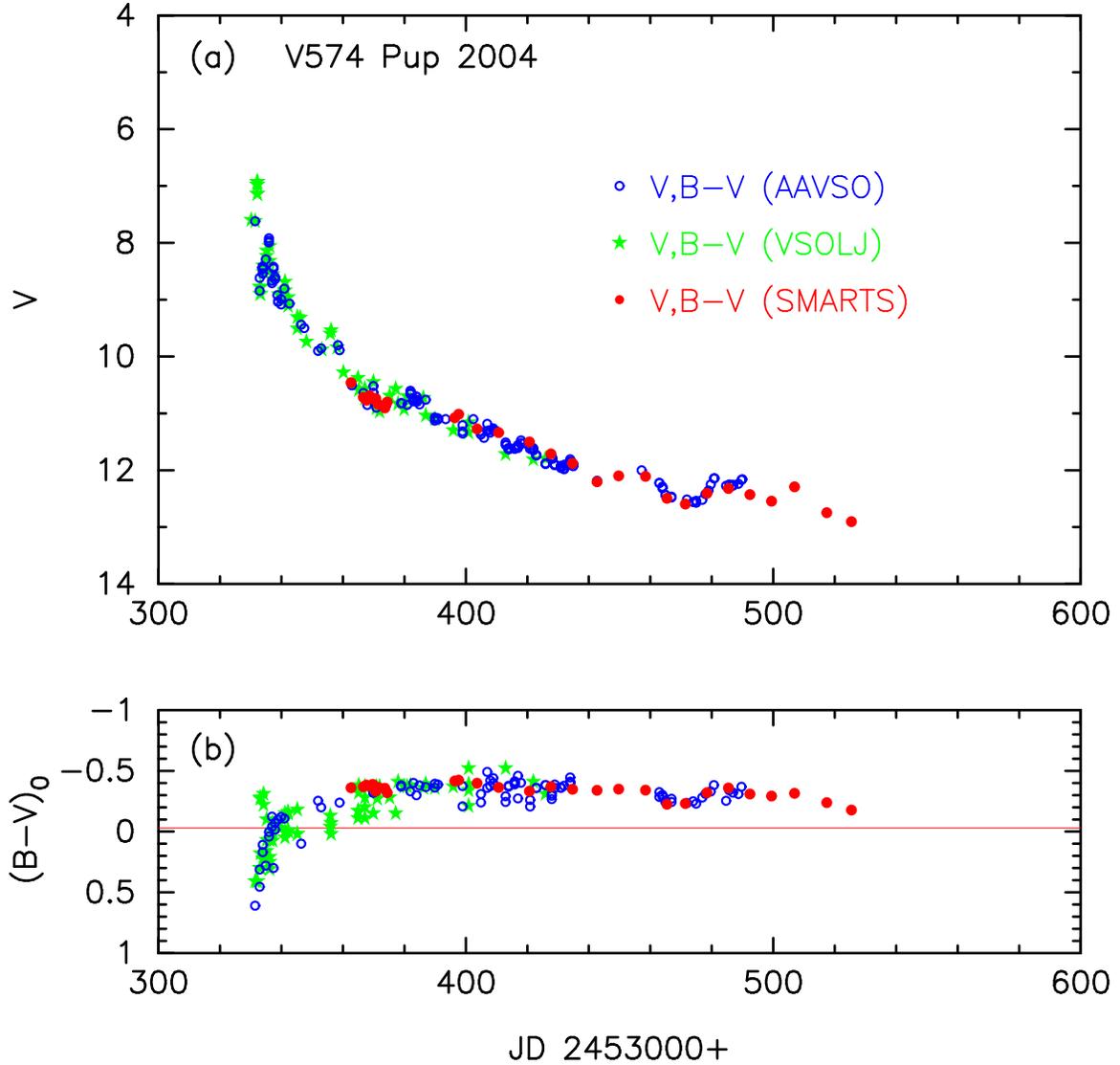}
\caption{
(a) The $V$ light and (b) $(B-V)_0$ color curves of V574~Pup.
The $BV$ data are taken from AAVSO (open blue circles), 
VSOLJ (filled green stars), and SMARTS (filled red circles).
In panel (b), the $(B-V)_0$ are dereddened using Equation 
(\ref{dereddening_eq_bv}) with $E(B-V)=0.45$.  
The $B-V$ data of AAVSO are systematically shifted
toward red by 0.2 mag.  The horizontal solid red 
line denotes $(B-V)_0=-0.03$, which is the intrinsic $B-V$ color
of optically thick free-free emission.  
\label{v574_pup_v_bv_ub_color_curve}}
\end{figure}


\begin{figure}
\plottwo{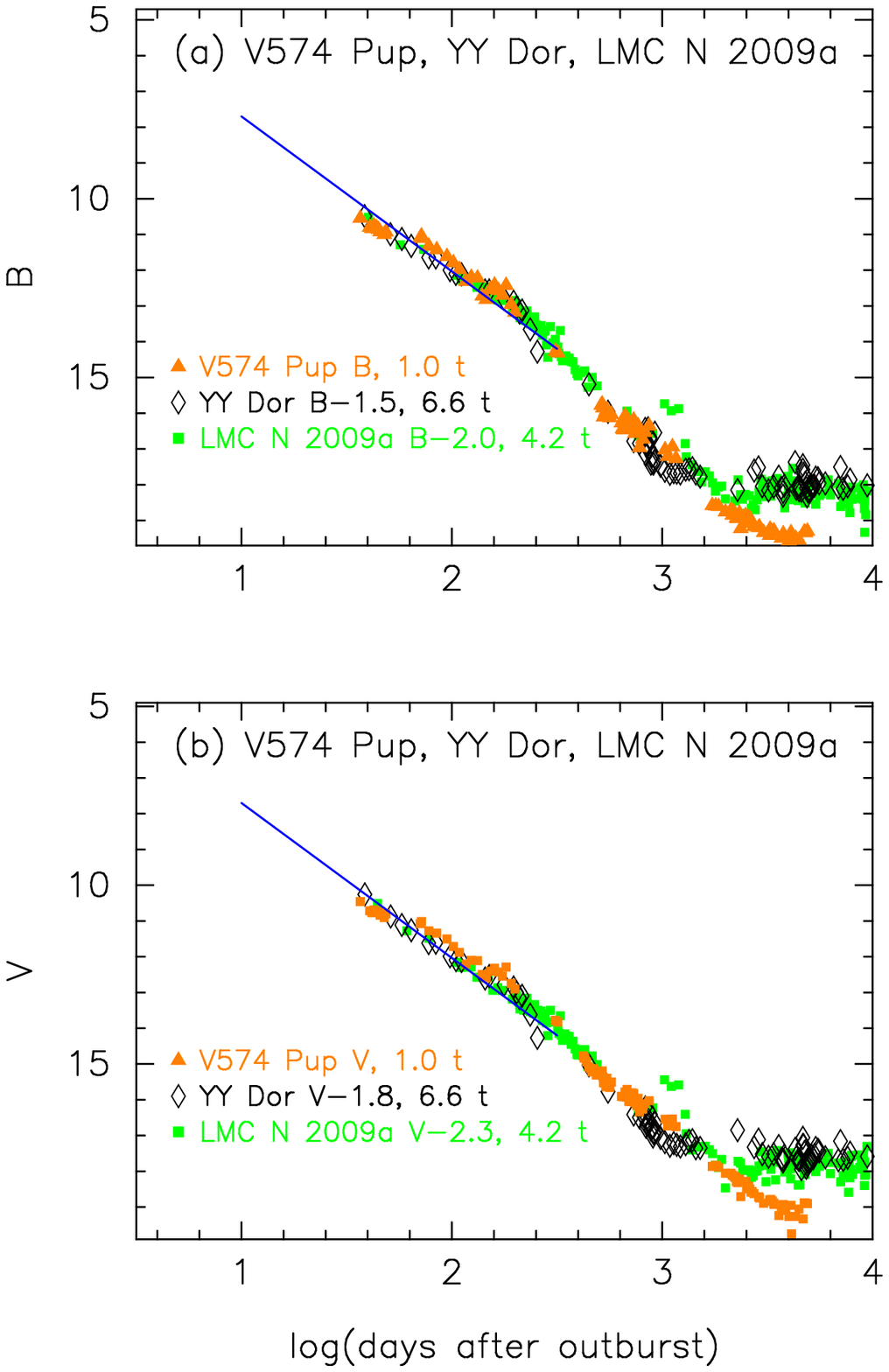}{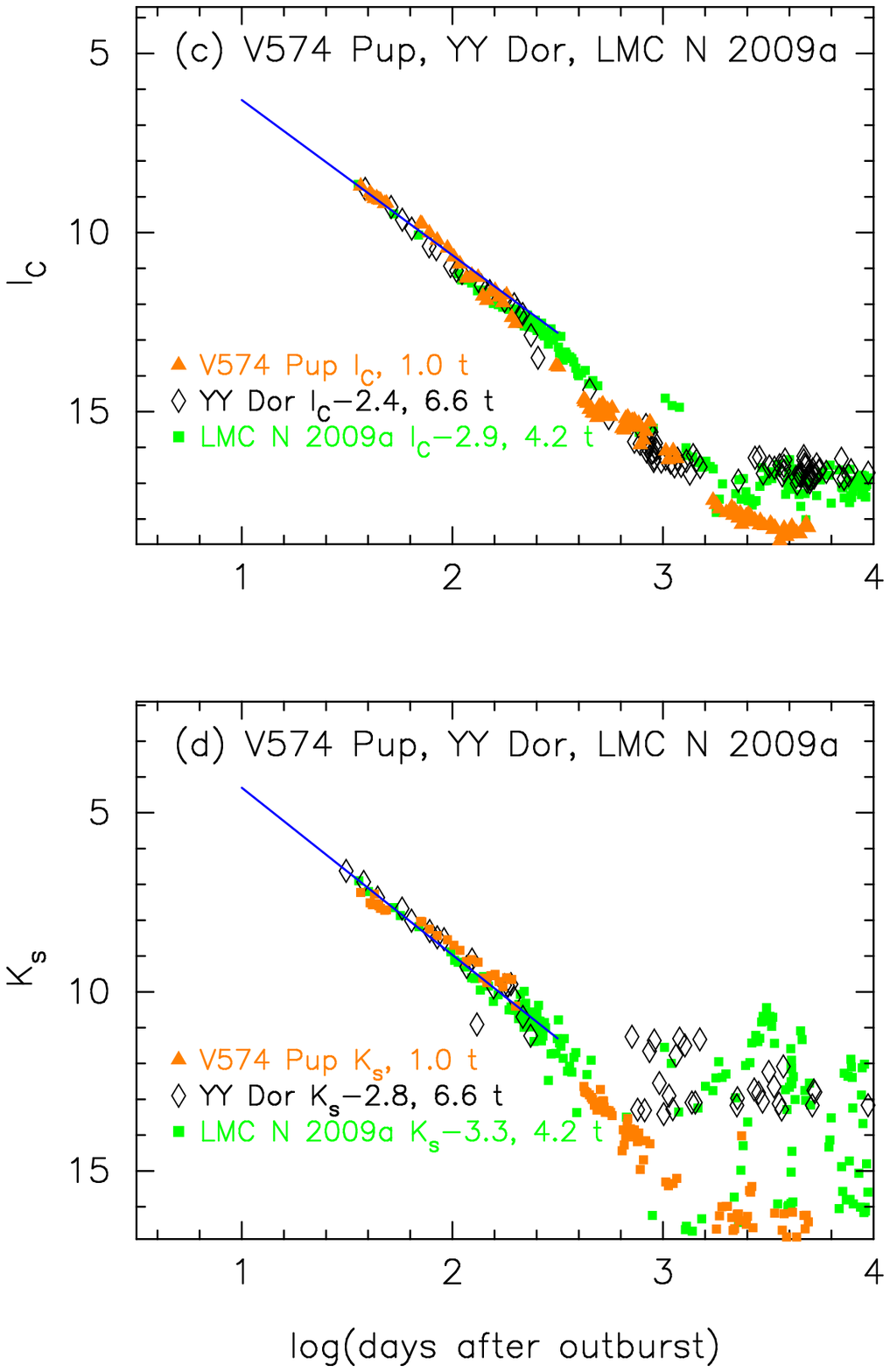}
\caption{
Same as Figure \ref{lv_vul_yy_dor_lmcn_2009a_v_b_logscale_2fig}, but
for the (a) $B$, (b) $V$, (c) $I_{\rm C}$, and (d) $K_{\rm s}$
light curves of V574~Pup, YY~Dor, and LMC~N~2009a.  
The solid blue lines denote the slope of $F_\nu\propto t^{-1.75}$,
which represents well the optically thick wind phase \citep{hac06kb}. 
The $BV$ data of YY~Dor and LMC~N~2009a are taken
from SMARTS.  The $BV$ data of V574~Pup are the same as those in
Figure \ref{v574_pup_v_bv_ub_color_curve}.
The $I_{\rm C} K_{\rm s}$ data 
of V574~Pup, YY~Dor,  and LMC~N~2009a are taken from SMARTS.
\label{v574_pup_yy_dor_lmcn_2009a_b_v_i_k_logscale_4fig}}
\end{figure}


\begin{figure}
\plotone{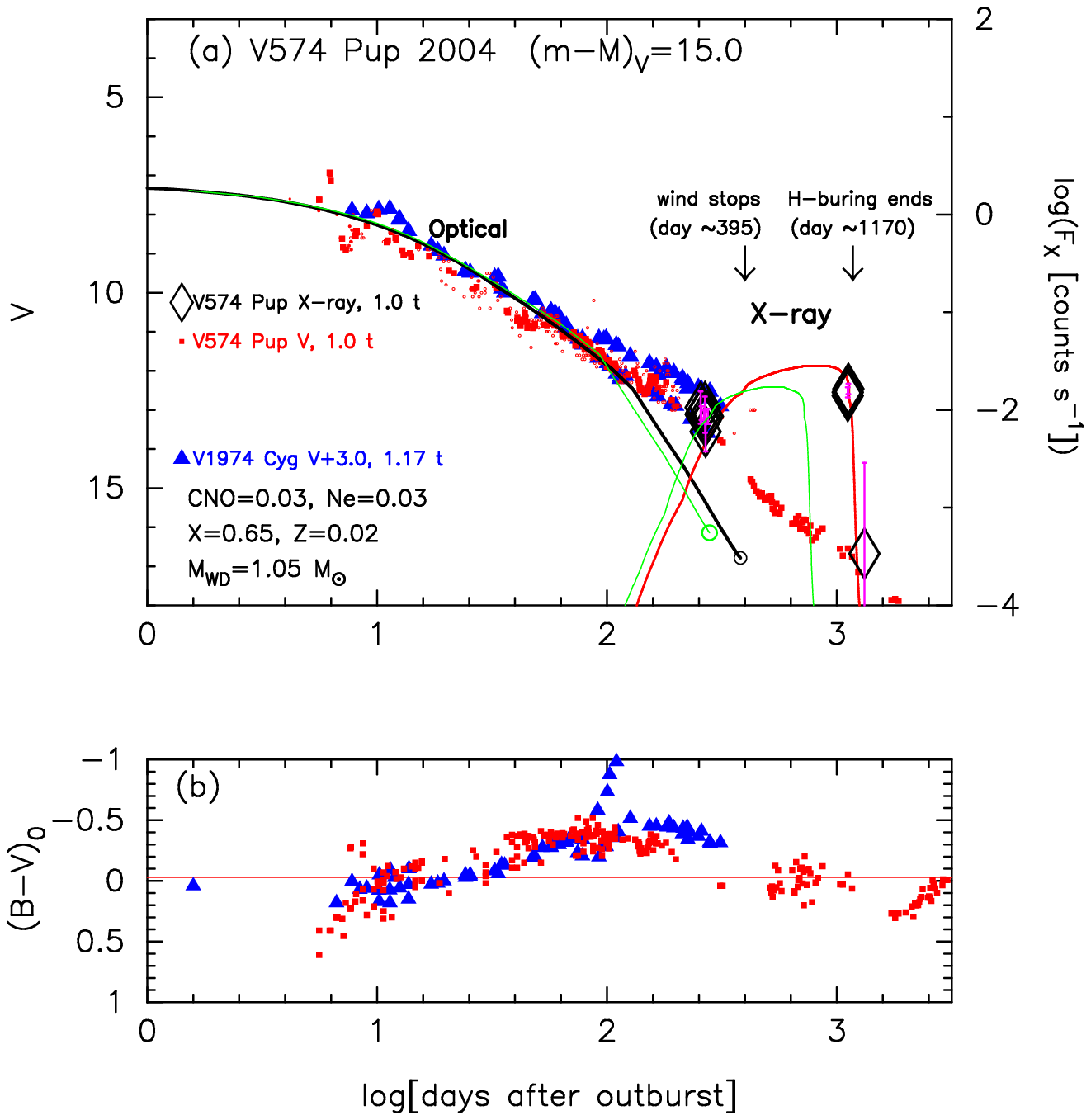}
\caption{
The (a) $V$ and X-ray light curves and (b) $(B-V)_0$ 
color curves of V574~Pup (filled red squares) as well as V1974~Cyg
(filled blue triangles) on a logarithmic timescale.
The large open black diamonds with error bars (vertical magenta lines)
denote the soft X-ray flux obtained by \citet{nes07a}. 
The $BV$ data of V574~Pup are the same as those in Figure
\ref{v574_pup_v_bv_ub_color_curve}.  The other data are the same as
those in \citet{hac16k}.  In panel (a), assuming that 
$(m-M)_V=15.0$ for V574~Pup, we add a model $V$ light curve
(solid black line) of a $1.05~M_\sun$ WD with the envelope chemical
composition of Ne nova 3 \citep{hac16k}.
We also add a $0.98~M_\sun$ WD (left solid green line) 
with the chemical composition of CO nova 3 \citep{hac16k}, assuming
that $(m-M)_V=12.2$ for V1974~Cyg (with a stretching factor of 
$f_{\rm s}= 1.17$ against V574~Pup).
We also add model X-ray light curves of $1.05~M_\sun$ (solid red line) and
$0.98~M_\sun$ (right solid green line) WDs.
Optically thick winds stop at the right edge of
each model light curve (denoted by an open circle). 
\label{v574_pup_v1974_cyg_v_bv_logscale}}
\end{figure}

\subsection{V574~Pup 2004}
\label{v574_pup}
Figure \ref{v574_pup_v_bv_ub_color_curve} shows (a) the $V$ light curve and
(b) $(B-V)_0$ color curve of V574~Pup on a linear timescale.  We deredden
the color with $E(B-V)=0.45$ as obtained in Section 
\ref{distance_reddening_v574_pup}.
The $BV$ data are taken from the Small Medium Aperture Telescope System
(SMARTS) data base \citep{wal12}, 
American Association of Variable Star Observers (AAVSO),
and the Variable Star Observers League of Japan (VSOLJ).
The $B-V$ data of AAVSO (open blue circles) are systematically bluer
by 0.2 mag compared with the other two colors of SMARTS and VSOLJ,
so we shift them toward red by 0.2 mag in 
Figure \ref{v574_pup_v_bv_ub_color_curve}(b).
Such differences come from slightly different responses
of $V$ filters at different observatories \citep[see, e.g., 
discussion in][]{hac06kb}.
The horizontal solid red line indicates the $B-V$ color of
optically thick free-free emission, i.e., $(B-V)_0=-0.03$
(see Paper I).
The typical error of SMARTS $BVRI$ data is about 0.001 mag, being
much smaller than the size of each symbol.  The error of AAVSO and
VSOLJ data are not reported.

Figure \ref{v574_pup_yy_dor_lmcn_2009a_b_v_i_k_logscale_4fig} shows 
the (a) $B$, (b) $V$, (c) $I_{\rm C}$, and (d) $K_{\rm s}$ light curves
of V574~Pup together with YY~Dor and LMC~N~2009a.
Here, we assume that V574~Pup outbursted on JD~2453326.0 (day 0).
\citet{hac18kb} have already analyzed YY~Dor and LMC~N~2009a 
and their various properties are summarized in their Tables 1, 2, and 3.
We increase/decrease the horizontal shift by a step of 
$\delta (\Delta \log t) = \delta \log f_{\rm s} = 0.01$ 
and the vertical shift by a step of 
$\delta (\Delta V)= 0.1$~mag, and determine the best overlapping one by eye.
As mentioned in Section \ref{introduction},
the model light curves obey the universal decline law.  
This phase corresponds to the optically thick wind phase
before the nebular phase started.  Therefore, we should use only
the optically thick phase.  However, we usually use all the phases
if they overlap each other even after the nebular phase
in order to determine the timescaling factor accurately as much as
possible.  If not, we use only the part
of optically thick wind phase before the nebular (or dust blackout)
phase started.    It should be noted that the timescaling factor of 
$f_{\rm s}$ is common among these four band light curves.  
In the case of Figure
\ref{v574_pup_yy_dor_lmcn_2009a_b_v_i_k_logscale_4fig}, the four
band light curves well overlap each other even after the nebular phase
started.
For the $B$ band, we apply Equation (\ref{distance_modulus_general_temp_b})
to Figure \ref{v574_pup_yy_dor_lmcn_2009a_b_v_i_k_logscale_4fig}(a)
and obtain
\begin{eqnarray}
(m&-&M)_{B, \rm V574~Pup} \cr
&=& ((m - M)_B + \Delta B)_{\rm YY~Dor} - 2.5 \log 6.6 \cr
&=& 18.98 - 1.5\pm0.2 - 2.05 = 15.43\pm0.2 \cr
&=& ((m - M)_B + \Delta B)_{\rm LMC~N~2009a} - 2.5 \log 4.2 \cr
&=& 18.98 - 2.0\pm0.2 - 1.55 = 15.43\pm0.2, 
\label{distance_modulus_b_yy_dor_u_sco_lmcn2009a}
\end{eqnarray}
where we adopt 
$(m-M)_B = 18.49 + 4.1\times 0.12=18.98$ for the LMC novae.
Thus, we obtain $(m-M)_{B, \rm V574~Pup}= 15.43\pm0.1$.

For the $V$ light curves in Figure 
\ref{v574_pup_yy_dor_lmcn_2009a_b_v_i_k_logscale_4fig}(b), we similarly obtain
\begin{eqnarray}
(m&-&M)_{V, \rm V574~Pup} \cr
&=& ((m - M)_V + \Delta V)_{\rm YY~Dor} - 2.5 \log 6.6 \cr
&=& 18.86 - 1.8\pm0.2 - 2.05 = 15.01\pm0.2 \cr
&=& ((m - M)_V + \Delta V)_{\rm LMC~N~2009a} - 2.5 \log 4.2 \cr
&=& 18.86 - 2.3\pm0.2 -1.55 = 15.01\pm0.2,
\label{distance_modulus_v_yy_dor_u_sco_lmcn2009a}
\end{eqnarray}
where we adopt 
$(m-M)_V = 18.49 + 3.1\times 0.12= 18.86$ for the LMC novae.
Thus, we obtain $(m-M)_{V, \rm V574~Pup}= 15.0\pm0.1$.

We apply Equation (\ref{distance_modulus_general_temp_i}) for
the $I_{\rm C}$ band to Figure 
\ref{v574_pup_yy_dor_lmcn_2009a_b_v_i_k_logscale_4fig}(c)
and obtain
\begin{eqnarray}
(m&-&M)_{I, \rm V574~Pup} \cr
&=& ((m - M)_I + \Delta I_C)_{\rm YY~Dor} - 2.5 \log 6.6 \cr
&=& 18.67 - 2.4\pm0.3 - 2.05 = 14.22\pm0.3 \cr
&=& ((m - M)_I + \Delta I_C)_{\rm LMC~N~2009a} - 2.5 \log 4.2 \cr
&=& 18.67 - 2.9\pm0.3 -1.55 = 14.22\pm0.3, 
\label{distance_modulus_i_yy_dor_u_sco_lmcn2009a}
\end{eqnarray}
$(m-M)_I = 18.49 + 1.5\times 0.12=18.67$ for the LMC novae.
Thus, we obtain $(m-M)_{I, \rm V574~Pup}= 14.22\pm0.2$.

For the $K_s$ light curves in Figure
\ref{v574_pup_yy_dor_lmcn_2009a_b_v_i_k_logscale_4fig}(d), 
we apply Equation (\ref{distance_modulus_general_temp_k}) and obtain
\begin{eqnarray}
(m&-&M)_{K, \rm V574~Pup} \cr
&=& ((m - M)_K + \Delta K_s)_{\rm YY~Dor} - 2.5 \log 6.6 \cr
&=& 18.53 - 2.8\pm0.4 - 2.05 = 13.68\pm0.4 \cr
&=& ((m - M)_K + \Delta K_s)_{\rm LMC~N~2009a} - 2.5 \log 4.2 \cr
&=& 18.53 - 3.3\pm0.4 -1.55 = 13.68\pm0.4,
\label{distance_modulus_k_yy_dor_u_sco_lmcn2009a}
\end{eqnarray}
$(m-M)_{K, \rm YY~Dor}=18.49 + 0.35\times 0.12= 18.53$ for the LMC novae.
Thus, we obtain $(m-M)_{K, \rm V574~Pup}= 13.68\pm0.2$.

Figure \ref{v574_pup_v1974_cyg_v_bv_logscale} shows
the $V$ light curve and $(B-V)_0$ color curve of V574~Pup
on a logarithmic timescale in comparison with V1974~Cyg.
The V1974~Cyg data are taken from Figure 38 of Paper II.
We add visual magnitudes (small red dots) of V574~Pup,
which are taken from AAVSO.
The $V$ light curves of these two novae do not exactly but broadly
overlap each other.  Note that there are two different values of
V1974~Cyg $V$ magnitudes in the nebular phase.
This is because the response of each $V$ filter is different between
these two observations.
We set the $V$ light curve of V574~Pup 
to overlap with the lower (fainter) branch of the V1974~Cyg light curves.
We also take into account the $(B-V)_0$ color evolution in Figure
\ref{v574_pup_v1974_cyg_v_bv_logscale}(b) to determine the 
horizontal shift of $\Delta \log t= \log f_{\rm s}$.
We set $\log f_{\rm s}$ to overlap the two $(B-V)_0$ color evolution 
as much as possible.

Applying the obtained $f_{\rm s}$ and $\Delta V$ to Equation
(\ref{distance_modulus_general_temp}), we have the relation of
\begin{eqnarray}
(m&-&M)_{V, \rm V574~Pup} \cr
&=& \left( (m-M)_V + \Delta V\right)_{\rm V1974~Cyg} - 2.5 \log 1.17 \cr 
&=& 12.2 + 3.0\pm0.2 - 0.175 = 15.025\pm0.3,
\label{distance_modulus_v574_pup_v1500_cyg_v1668_cyg_v1974_cyg}
\end{eqnarray}
where we adopt $(m-M)_{V, \rm V1974~Cyg}=12.2$ in Section \ref{v1974_cyg}.
Thus, we obtain $f_{\rm s}=1.07\times 1.17= 1.26$ against LV~Vul 
(see Table \ref{extinction_various_novae})
and $(m-M)_V=15.0\pm0.3$ for V574~Pup.

Using $(m-M)_V=15.0\pm0.3$ and $f_{\rm s}= 1.26$, we have the relation of
\begin{eqnarray}
(m-M')_{V, \rm V574~Pup} 
&\equiv& (m_V - (M_V - 2.5\log f_{\rm s}))_{\rm V574~Pup} \cr
&=& ((m - M)_V + 2.5\log f_{\rm s})_{\rm V574~Pup} \cr
&=& 15.0\pm0.3 + 2.5\times 0.10 = 15.25\pm0.3. 
\label{absolute_mag_lv_vul_v574_pup_only}
\end{eqnarray}


\begin{figure}
\plotone{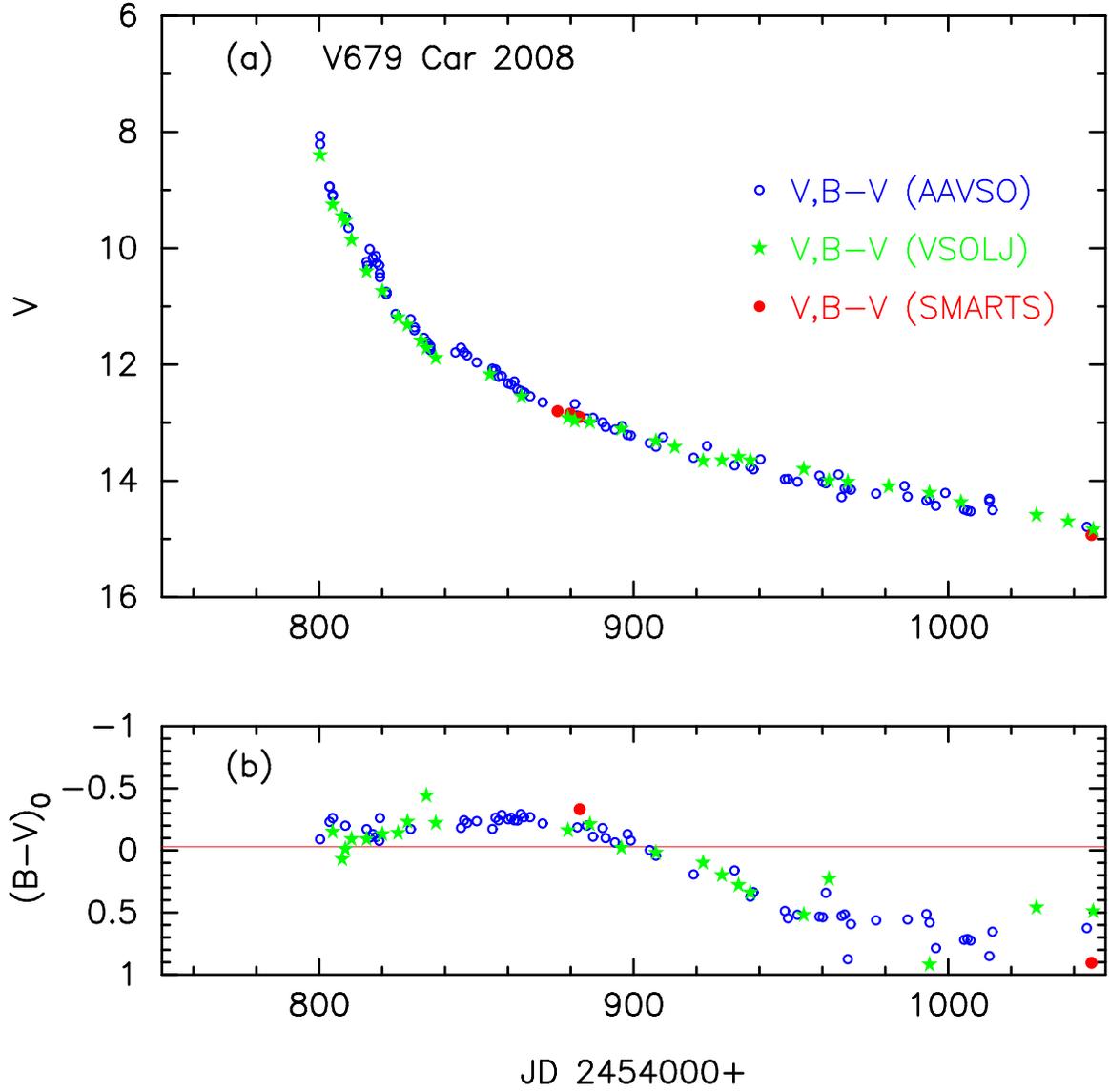}
\caption{
Same as Figure \ref{v574_pup_v_bv_ub_color_curve}, but for V679~Car.
In panel (b), the $(B-V)_0$ data are dereddened with $E(B-V)=0.69$ but
the $B-V$ data of VSOLJ (filled green stars) are systematically shifted
toward red by 0.25 mag to overlap them to the other color data.  
\label{v679_car_v_bv_ub_color_curve}}
\end{figure}


\begin{figure}
\epsscale{0.5}
\plotone{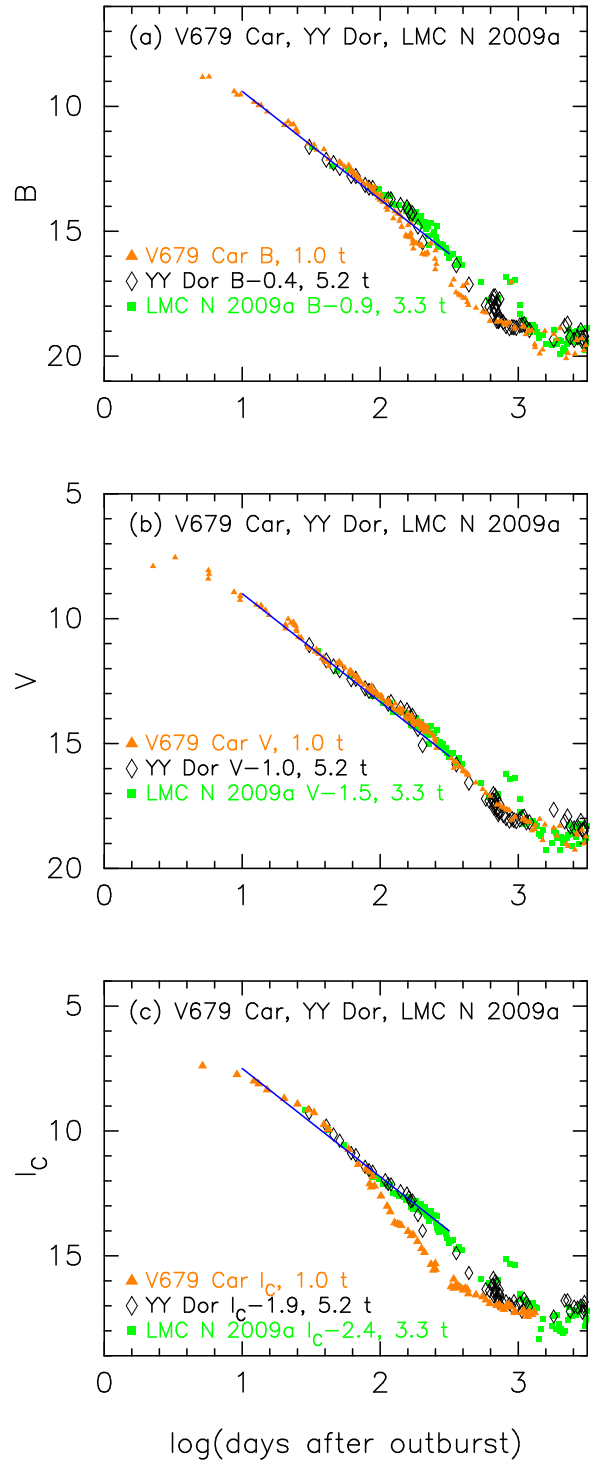}
\caption{
The (a) $B$, (b) $V$, and (c) $I_{\rm C}$ light curves of V679~Car
as well as YY~Dor and LMC~N~2009a.  
The solid blue lines denote the slope of $F_\nu\propto t^{-1.75}$. 
The $BV$ data of V679~Car are the same as those in Figure
\ref{v679_car_v_bv_ub_color_curve}.
The $I_{\rm C}$ data of V679~Car, YY~Dor, and LMC~N~2009a 
are taken from AAVSO, VSOLJ, and SMARTS.
\label{v679_car_yy_dor_lmcn_2009a_b_v_i_logscale_3fig}}
\end{figure}


\begin{figure}
\epsscale{0.5}
\plotone{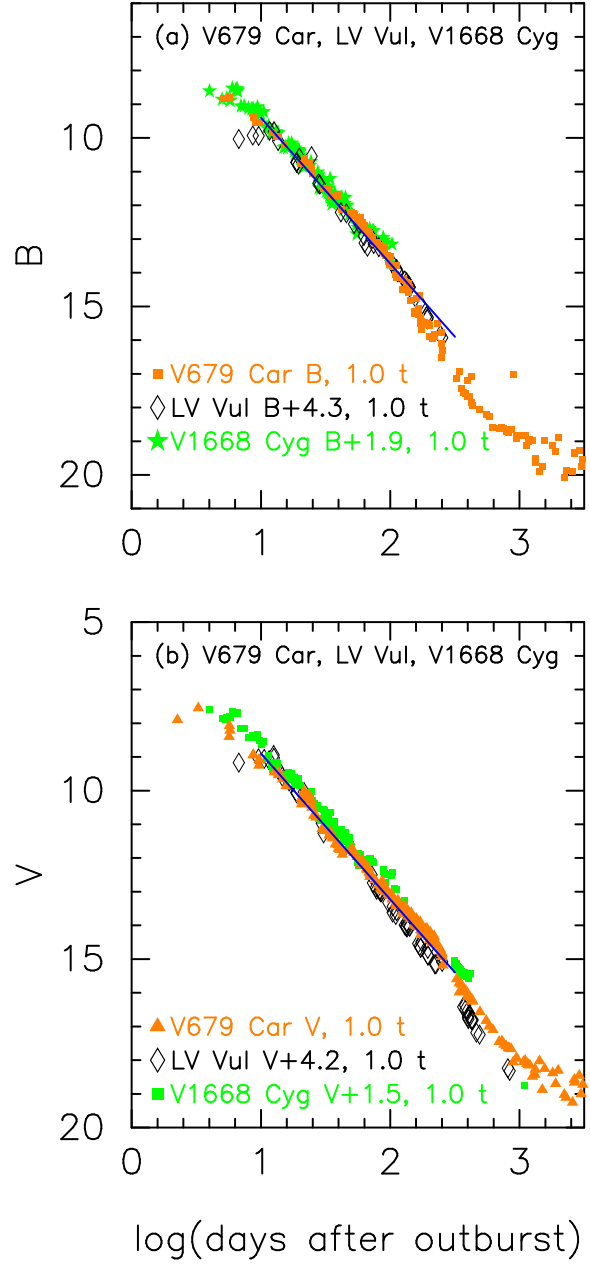}
\caption{
The (a) $B$ and (b) $V$ light curves of V679~Car as well as LV~Vul 
and V1668~Cyg.  
The $BV$ data of V679~Car are the same as those in Figure
\ref{v679_car_v_bv_ub_color_curve}.
The $BV$ data of LV~Vul and V1668~Cyg are taken from various
literature and the same as those in Paper I.
\label{v679_car_lv_vul_v1668_cyg_b_v_logscale_2fig}}
\end{figure}


\begin{figure}
\plotone{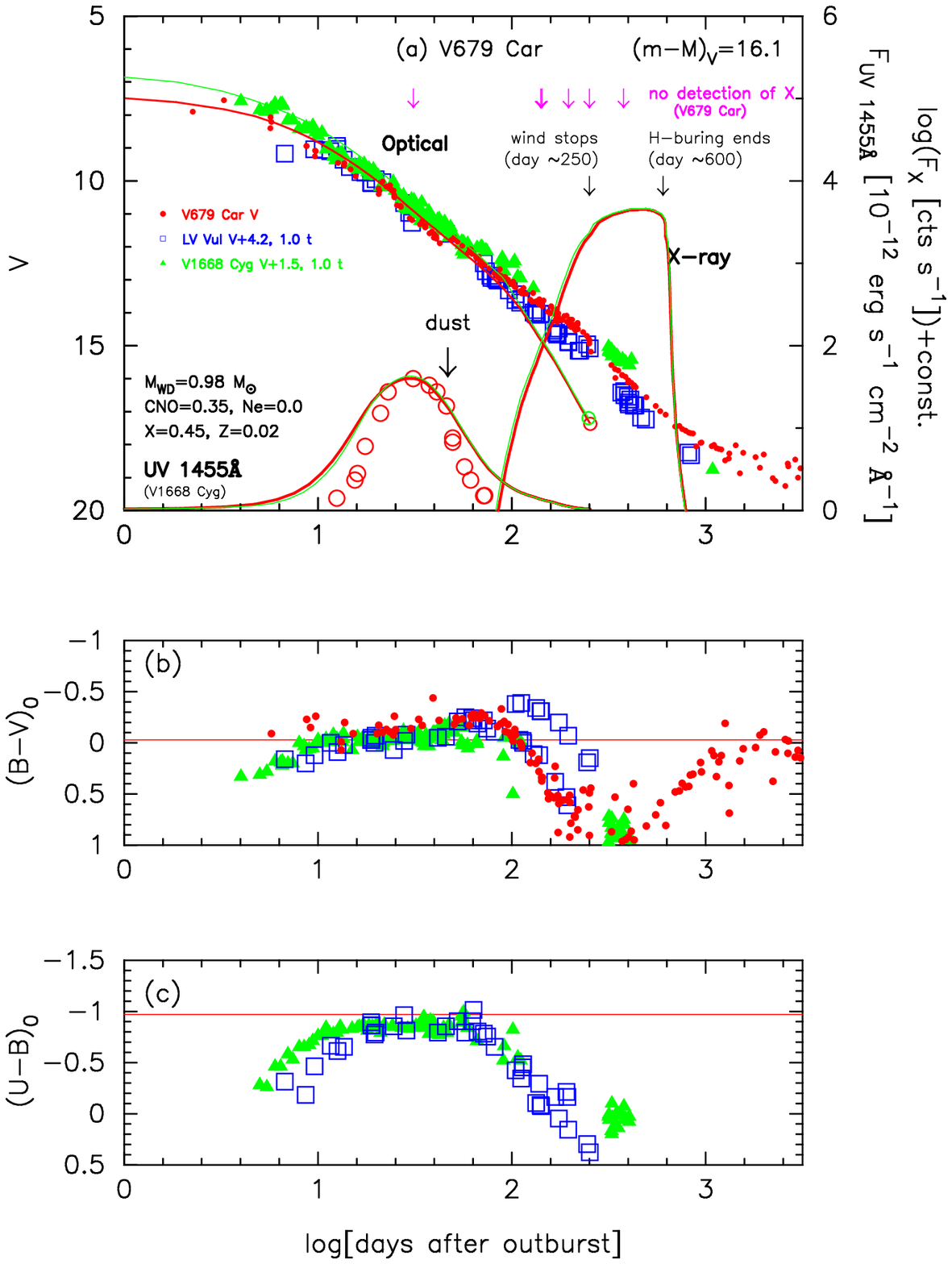}
\caption{
Same as Figure 
\ref{v574_pup_v1974_cyg_v_bv_logscale}, bur for
V679~Car.  We add the light/color curves of LV~Vul and V1668~Cyg.
We also add the observed UV~1455\AA\  flux data of V1668~Cyg. 
The timescales of LV~Vul and V1668~Cyg are the same as that of V679~Car.
The UV~1455\AA\   flux of V1668~Cyg sharply
drops at Day 60 owing to a shallow dust blackout
(denoted by the arrow labeled ``dust'').  
No supersoft X-ray phase of V679~Car was detected with {\it Swift} 
as denoted by downward magenta arrows \citep{schw11}.
We add two $0.98~M_\sun$ WD model (solid red/green lines)
with the chemical composition of 
CO nova 3 \citep{hac16k}, which reproduces the $V$ light curves of V679~Car
(red) and V1668~Cyg (green), respectively.  
These two models also reproduce the UV~1455\AA\  light curve of V1668~Cyg.
The difference of these two models (red and green) exists
only in their initial envelope masses.  See text for more details.
\label{v679_car_lv_vul_v1668_cyg_v_bv_ub_logscale}}
\end{figure}

\subsection{V679~Car 2008}
\label{v679_car}
Figure \ref{v679_car_v_bv_ub_color_curve} shows the (a) $V$ and 
(b) $(B-V)_0$ evolutions of V679~Car on a linear timescale.
The $BV$ data are taken from AAVSO (open blue circles), 
VSOLJ (filled green stars), and SMARTS (filled red circles).  
Here, $(B-V)_0$ are dereddened with $E(B-V)=0.69$ 
as obtained in Section \ref{distance_reddening_v679_car},
but the $B-V$ data of VSOLJ (filled green stars)
are systematically shifted toward red by 0.25 mag to overlap 
them to the other color data. 
The typical error of SMARTS $BVRI$ data is about 0.003 mag, being
much smaller than the size of each symbol.  The error of AAVSO and
VSOLJ data are not reported.

Figure \ref{v679_car_yy_dor_lmcn_2009a_b_v_i_logscale_3fig} shows
the $B$, $V$, and $I_{\rm C}$ light curves of V679~Car
together with those of the LMC novae YY~Dor and LMC~N~2009a.
Here, we assume that the V679~Car outburst started on JD~2454794.5
(day 0).  
As in Figure \ref{v679_car_yy_dor_lmcn_2009a_b_v_i_logscale_3fig}(c),
the $I_{\rm C}$ light curve of V679~Car overlaps with
those of YY~Dor and LMC~N~2009a only in the first 60 days
because of the different contributions of line emission in the
nebular phase.  Thus, we use only the $V$ light curves that overlap
longer period to determine the timescaling factor $f_{\rm s}$.

We apply Equation (\ref{distance_modulus_general_temp_b}) 
for the $B$ band to Figure 
\ref{v679_car_yy_dor_lmcn_2009a_b_v_i_logscale_3fig}(a)
and obtain
\begin{eqnarray}
(m&-&M)_{B, \rm V679~Car} \cr
&=& ((m - M)_B + \Delta B)_{\rm YY~Dor} - 2.5 \log 5.2 \cr
&=& 18.98 - 0.4\pm0.2 - 1.8 = 16.78\pm0.2 \cr
&=& ((m - M)_B + \Delta B)_{\rm LMC~N~2009a} - 2.5 \log 3.3 \cr
&=& 18.98 - 0.9\pm0.2 - 1.3 = 16.78\pm0.2. 
\label{distance_modulus_b_v679_car_yy_dor_lmcn2009a}
\end{eqnarray}
Thus, we obtain $(m-M)_{B, \rm V679~Car}= 16.78\pm0.1$.

For the $V$ light curves in Figure 
\ref{v679_car_yy_dor_lmcn_2009a_b_v_i_logscale_3fig}(b),
we similarly obtain
\begin{eqnarray}
(m&-&M)_{V, \rm V679~Car} \cr
&=& ((m - M)_V + \Delta V)_{\rm YY~Dor} - 2.5 \log 5.2 \cr
&=& 18.86 - 1.0\pm0.2 - 1.8 = 16.06\pm0.2 \cr
&=& ((m - M)_V + \Delta V)_{\rm LMC~N~2009a} - 2.5 \log 3.3 \cr
&=& 18.86 - 1.5\pm0.2 - 1.3 = 16.06\pm0.2.
\label{distance_modulus_v_v679_car_yy_dor_lmcn2009a}
\end{eqnarray}
Thus, we obtain $(m-M)_{V, \rm V679~Car}= 16.06\pm0.1$.

We apply Equation (\ref{distance_modulus_general_temp_i}) to Figure
\ref{v679_car_yy_dor_lmcn_2009a_b_v_i_logscale_3fig}(c) and obtain
\begin{eqnarray}
(m&-&M)_{I, \rm V679~Car} \cr
&=& ((m - M)_I + \Delta I_C)_{\rm YY~Dor} - 2.5 \log 5.2 \cr
&=& 18.67 - 1.9\pm0.3 - 1.8 = 14.97\pm 0.3 \cr
&=& ((m - M)_I + \Delta I_C)_{\rm LMC~N~2009a} - 2.5 \log 3.3 \cr
&=& 18.67 - 2.4\pm0.3 -1.3 = 14.97\pm 0.3. 
\label{distance_modulus_i_v679_car_yy_dor_lmcn2009a}
\end{eqnarray}
Thus, we obtain $(m-M)_{I, \rm V679~Car}= 14.97\pm0.2$.

Figure \ref{v679_car_lv_vul_v1668_cyg_b_v_logscale_2fig} show
the comparison with the galactic novae LV~Vul and V1668~Cyg in
the (a) $B$ and (b) $V$ light curves.
Note that we do not time-stretch these two novae because their
timescales are almost the same as that of V679~Car.

Applying Equation (\ref{distance_modulus_general_temp_b}) for the $B$ band
to Figure \ref{v679_car_lv_vul_v1668_cyg_b_v_logscale_2fig}(a),
we have the relation of
\begin{eqnarray}
(m&-&M)_{B, \rm V679~Car} \cr
&=& ((m - M)_B + \Delta B)_{\rm LV~Vul} - 2.5 \log 1.0 \cr
&=& 12.45 + 4.3\pm0.2 + 0.0 = 16.75\pm 0.2 \cr
&=& ((m - M)_B + \Delta B)_{\rm V1668~Cyg} - 2.5 \log 1.0 \cr
&=& 14.9 + 1.9\pm0.2 + 0.0 = 16.8\pm 0.2. 
\label{distance_modulus_b_v679_car_lv_vul_v1668_cyg}
\end{eqnarray}
Thus, we obtain $(m-M)_{B, \rm V679~Car}= 16.78\pm0.2$, being consistent
with Equation (\ref{distance_modulus_b_v679_car_yy_dor_lmcn2009a}).

Applying Equation (\ref{distance_modulus_general_temp}) to Figure 
\ref{v679_car_lv_vul_v1668_cyg_b_v_logscale_2fig}(b),
we have the relation of
\begin{eqnarray}
(m &-& M)_{V, \rm V679~Car} \cr 
&=& (m-M)_{V, \rm LV~Vul} + \Delta V - 2.5 \log 1.0 \cr
&=& 11.85 + 4.2\pm 0.2 + 0.0 = 16.05\pm 0.2 \cr
&=& (m-M)_{V, \rm V1668~Cyg} + \Delta V - 2.5 \log 1.0 \cr
&=& 14.6 + 1.5\pm 0.2 + 0.0 = 16.1\pm 0.2.
\label{distance_modulus_v679_car_lv_vul_v1668_cyg_v}
\end{eqnarray}
Thus, we obtain $f_{\rm s}=1.0$ against LV~Vul 
and $(m-M)_V=16.08\pm0.1$ for V679~Car, again being consistent
with Equation (\ref{distance_modulus_v_v679_car_yy_dor_lmcn2009a}).
Thus, we confirm that both the time-stretching methods for the galactic
novae and LMC novae give the same results for the $B$ and $V$ bands
of V679~Car.

Figure \ref{v679_car_lv_vul_v1668_cyg_v_bv_ub_logscale}
shows the same $V$ light curve as in Figure
\ref{v679_car_lv_vul_v1668_cyg_b_v_logscale_2fig}(b), but also shows
the color evolution.  We confirm that we do not need to stretch
the light/color curves because its timescale
is almost the same as those of LV~Vul and V1668~Cyg.
We shift the $V$ light curves of LV~Vul by 4.2 mag 
and V1668~Cyg by 1.5 mag downward
in order to overlap these $V$ light curves.

The $(B-V)_0$ color evolution in Figure 
\ref{v679_car_lv_vul_v1668_cyg_v_bv_ub_logscale}(b)
is also used to constrain the 
horizontal shift of $\Delta \log t= \log f_{\rm s}= 0.0$.
The $(B-V)_0$
color evolution of LV~Vul splits into two branches just after the
nebular phase started as shown in Figure
\ref{v679_car_lv_vul_v1668_cyg_v_bv_ub_logscale}(b).
This is because each response function of $V$ filters is slightly
different from each other between these two observations.
We set the $(B-V)_0$ color evolution
of V679~Car to overlap with the lower branch of LV~Vul in Figure
\ref{v679_car_lv_vul_v1668_cyg_v_bv_ub_logscale}(b).  The timescaling
factor of $f_{\rm s}=1.0$ gives a good match between the two
$(B-V)_0$ color curves.

From Equations (\ref{time-stretching_general}),
(\ref{distance_modulus_general_temp}), and
(\ref{distance_modulus_v679_car_lv_vul_v1668_cyg_v}),
we have the relation of
\begin{eqnarray}
(m- M')_{V, \rm V679~Car} 
&\equiv& (m_V - (M_V - 2.5\log f_{\rm s}))_{\rm V679~Car} \cr
&=& \left( (m-M)_V + \Delta V \right)_{\rm LV~Vul} \cr
&=& 11.85 + 4.2\pm0.2 = 16.05\pm0.2.
\label{absolute_mag_lv_vul_v679_car_only}
\end{eqnarray}


\begin{figure}
\plotone{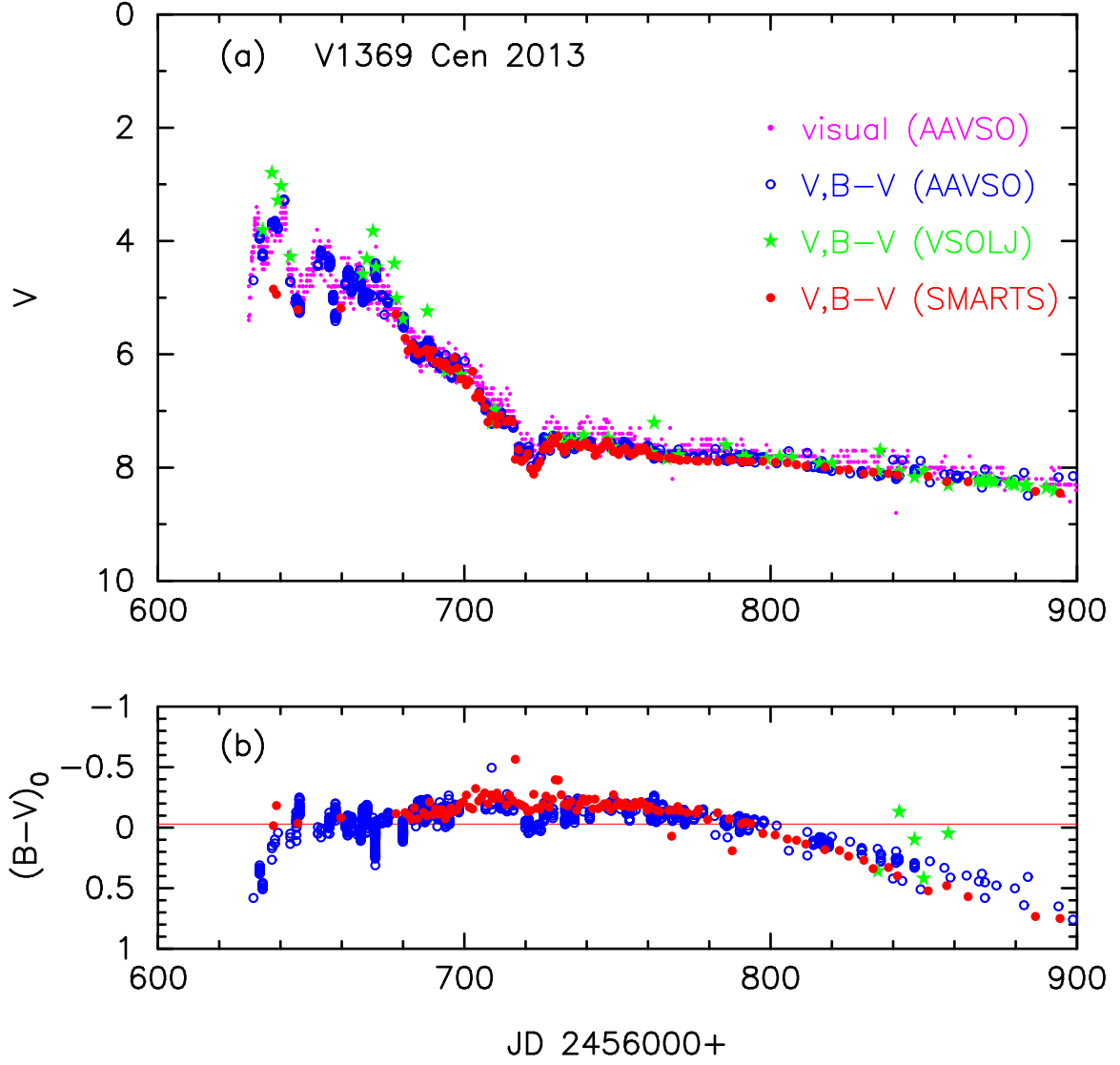}
\caption{
Same as Figure \ref{v574_pup_v_bv_ub_color_curve}, but for V1369~Cen.
(a) The visual (magenta dots), $V$ (blue open circles) data of AAVSO,
and $V$ data of VSOLJ (filled green stars) and
SMARTS (filled red circles) well overlap each other.
(b) The $(B-V)_0$ color is dereddened with $E(B-V)=0.11$.
\label{v1369_cen_v_bv_ub_color_curve}}
\end{figure}


\begin{figure}
\plotone{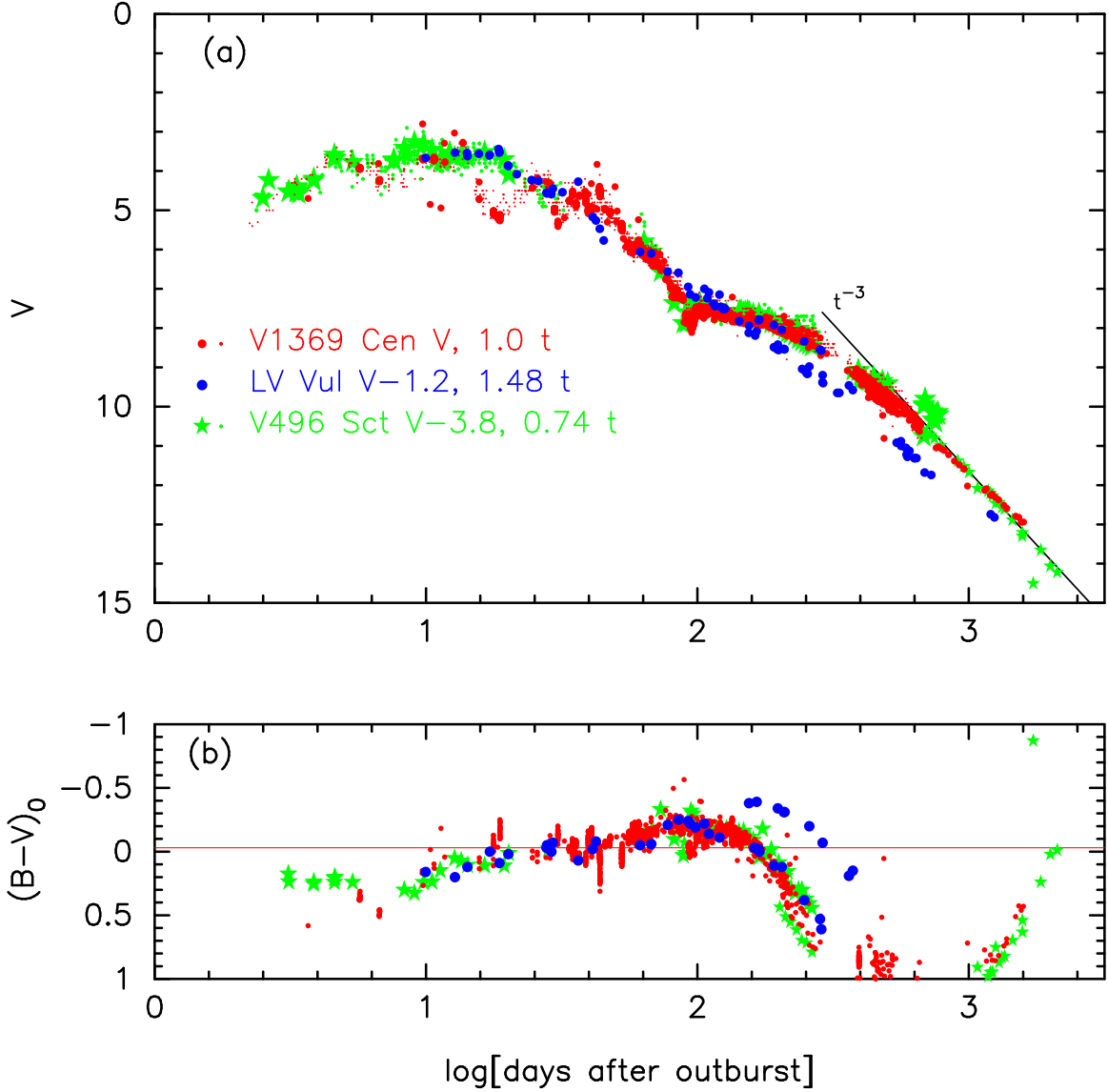}
\caption{
Same as Figure 
\ref{v574_pup_v1974_cyg_v_bv_logscale},
but for V1369~Cen.
The straight solid black line labeled ``$t^{-3}$''
indicates the flux from homologously expanding ejecta, i.e.,
free expansion after the optically thick winds stop
\citep[see, e.g.,][]{woo97, hac06kb}.
The $BV$ data of V1369~Cen are the same as those in Figure
\ref{v1369_cen_v_bv_ub_color_curve}.
We also plot the $BV$ data of V496~Sct and LV~Vul.  
The data of V496~Sct are the same as those in Figure 47
of Paper II.  
\label{v1369_cen_v496_sct_lv_vul_v_bv_ub_color_logscale}}
\end{figure}


\begin{figure}
\plottwo{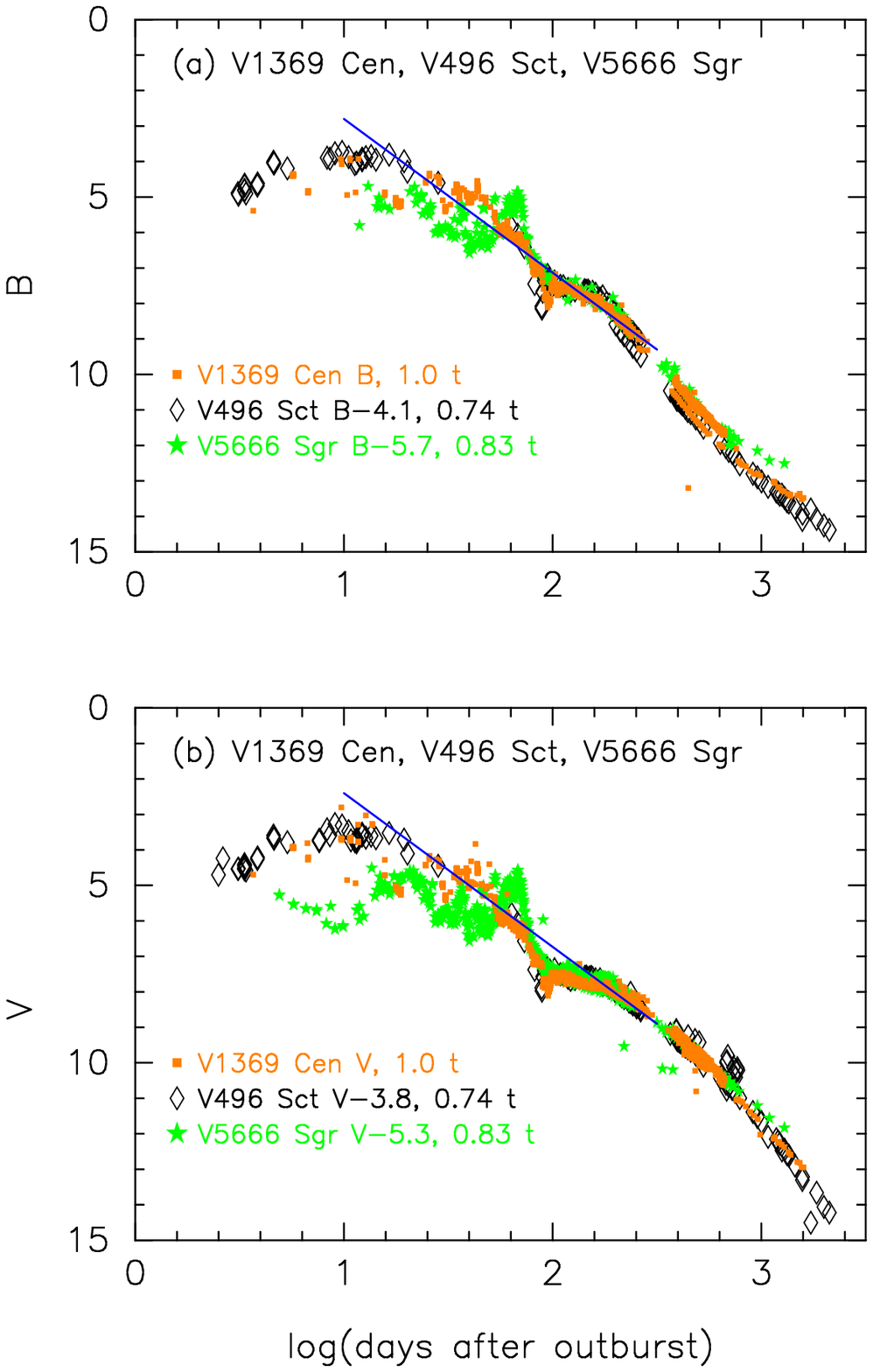}{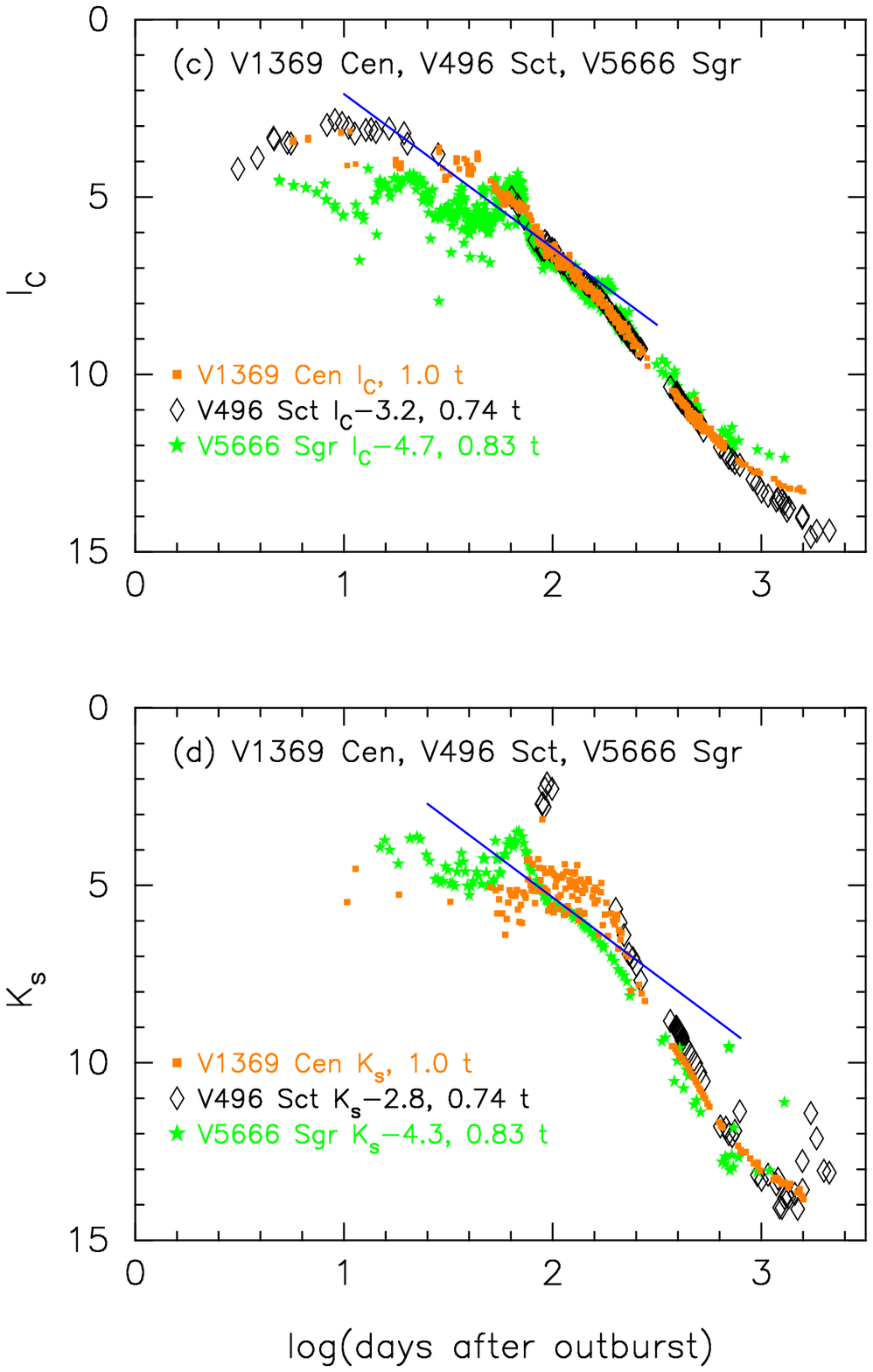}
\caption{
Same as Figure 
\ref{v574_pup_yy_dor_lmcn_2009a_b_v_i_k_logscale_4fig},
but for a different set of V1369~Cen, V496~Sct, and V5666~Sgr.
The $BV$ data of V1369~Cen are the same
as those in Figure \ref{v1369_cen_v_bv_ub_color_curve}.  
The $BV$ data of V496~Sct are the same as those in Figure
\ref{v496_sct_v1065_cen_lv_vul_v_bv_ub_color_logscale}.
The $BV$ data of V5666~Sgr are the same as those in Figure 
\ref{v5666_sgr_v_bv_ub_color_curve}.
The $I_{\rm C}$ data of V1369~Cen are taken from AAVSO, VSOLJ,
and SMARTS.  
The $I_{\rm C}$ data of V496~Sct are from \citet{raj12}
and SMARTS.  
The $I_{\rm C}$ data of V5666~Sgr are from AAVSO, VSOLJ, 
and SMARTS.  
The $K_{\rm s}$ data are all taken from SMARTS.
\label{v1369_cen_v496_sct_v5666_sgr_b_v_i_k_logscale_4fig}}
\end{figure}


\begin{figure}
\plotone{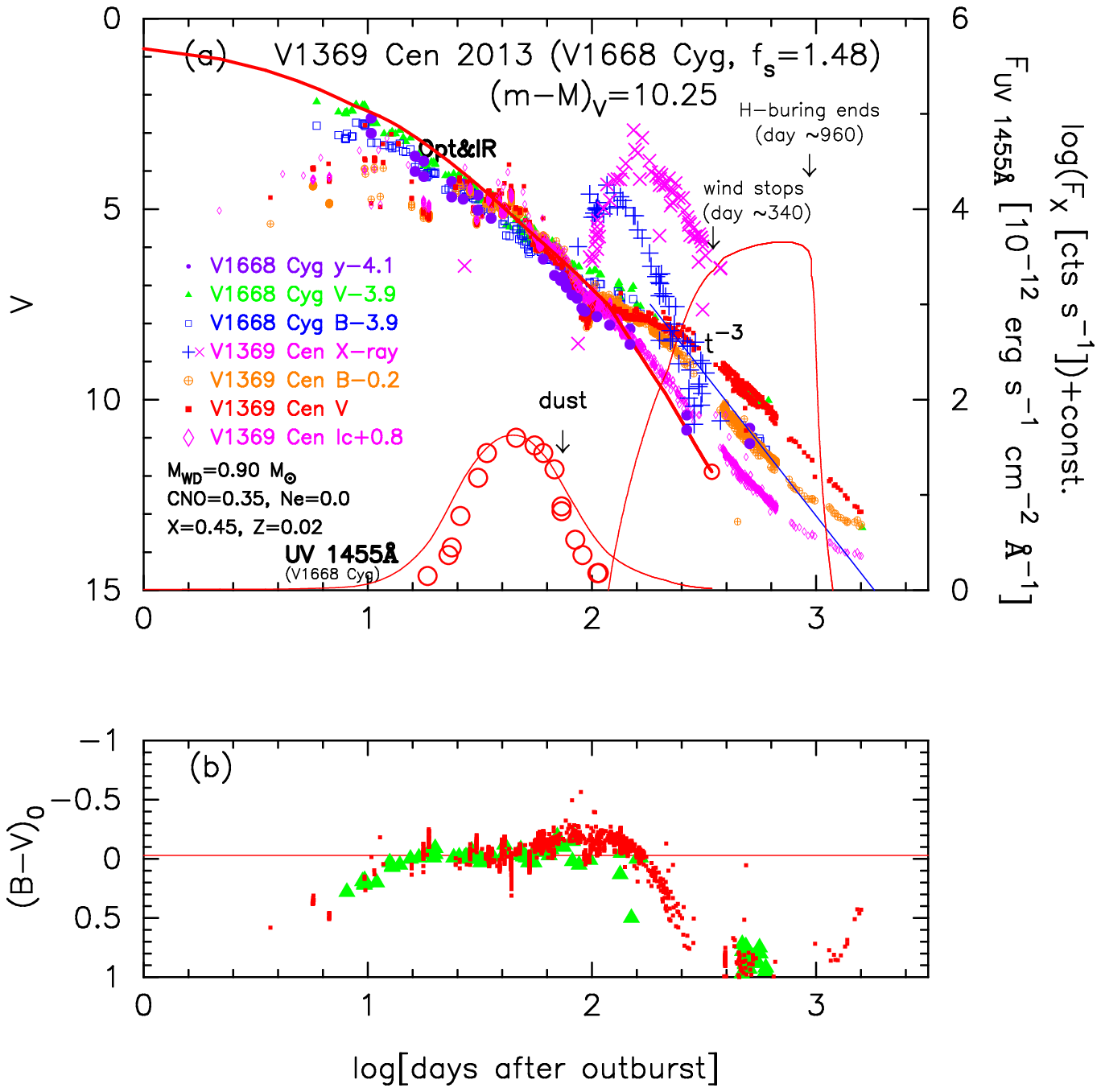}
\caption{
(a) The $V$ and X-ray light and (b) $(B-V)_0$ color curves of V1369~Cen. 
We also plot the  $V$ light curve and UV~1455\AA\  data (large open 
red circles)  
of V1668~Cyg, the light curves of which are stretched by a factor
of 1.48 in the time-direction against V1369~Cen.  
We add the $B$ and $I_C$ magnitudes of V1369~Cen taken from AAVSO,
VSOLJ, and SMARTS.  We add a $0.90 ~M_\sun$ WD model (solid red lines)
with the chemical composition of CO nova 3 \citep{hac16k},
which reproduces the optical and NIR light curves of V1369~Cen.
The $V$ model light curve followed the shape of $I_{\rm C}$ magnitudes 
after the nebular phase started while the observed $B$ and $V$ magnitudes
departed from it because of strong emission line effects.
We add a straight solid black line labeled ``$t^{-3}$''
that indicates the homologously expanding ejecta, i.e.,
free expansion after the optically thick winds stop.
\label{all_mass_v1369_cen_v1668_cyg_x45z02c15o20}}
\end{figure}

\subsection{V1369~Cen 2013}
\label{v1369_cen}
Figure \ref{v1369_cen_v_bv_ub_color_curve} shows the (a) $V$ light and 
(b) $(B-V)_0$ color curves of V1369~Cen, where $(B-V)_0$ are
dereddened with $E(B-V)=0.11$ as obtained in Section \ref{v1369_cen_cmd}.
The $V$ data of AAVSO (blue open circles), VSOLJ (filled green stars),
and SMARTS (filled red circles) are very similar to each other.

Figure \ref{v1369_cen_v496_sct_lv_vul_v_bv_ub_color_logscale} shows the
light/color curves of V1369~Cen as well as LV~Vul and V496~Sct on a 
logarithmic timescale.  Here we assume that the nova outbursted
on JD~2456627.5 (Day 0).  
We add a straight solid black line labeled ``$t^{-3}$''
that indicates the homologously expanding ejecta, i.e.,
free expansion, after the optically thick winds stop
\citep[see, e.g.,][]{woo97, hac06kb}.
Both the V1369~Cen and V496~Sct light curves have 
wavy structures in the early phase, but we overlap these three novae 
light/color curves as much as possible.  In the middle phase, both
V1369~Cen and V496~Sct show a sharp and shallow dip due to dust blackout.
In the later nebular phase, the $V$ light curves of V1369~Cen and
V496~Sct well overlap the upper branch of LV~Vul.
The $V$ light curve of LV~Vul splits into two branches in the
nebular phase due to the different responses of their $V$ filters
as discussed in Paper II.
This is because strong [\ion{O}{3}] lines contributes to the blue edge
of $V$ filter in the nebular phase and small differences among 
the response functions of $V$ filters make large differences
in the $V$ magnitudes. 
The $(B-V)_0$ color curves of LV~Vul also splits into two branches
for $\log t\gtrsim 2.0$, that is, in the nebular phase.  
The $V$ light curve of V1369~Cen follows V496~Sct and 
the upper branch of LV~Vul, while the $(B-V)_0$ color curve of V1369~Cen
follows V496~Sct and the lower branch of LV~Vul.  

Applying Equation (\ref{distance_modulus_general_temp})
to Figure \ref{v1369_cen_v496_sct_lv_vul_v_bv_ub_color_logscale},
we have the relation of
\begin{eqnarray}
(m&-&M)_{V, \rm V1369~Cen} \cr
&=& (m-M + \Delta V)_{V, \rm LV~Vul} - 2.5 \log 1.48 \cr
&=& 11.85 - 1.2\pm 0.2 - 0.425 = 10.23\pm 0.2 \cr
&=& (m-M + \Delta V)_{V, \rm V496~Sct} - 2.5 \log 0.74 \cr
&=& 13.7 - 3.8\pm 0.2 + 0.325 = 10.23\pm 0.2.
\label{distance_modulus_v1369_cen_v496_sct}
\end{eqnarray}
Thus, we obtain $f_{\rm s}=1.48$ against LV~Vul and $(m-M)_V=10.23\pm0.2$
for V1369~Cen.  These values are summarized in
Table \ref{extinction_various_novae}

From Equations (\ref{time-stretching_general}),
(\ref{distance_modulus_general_temp}), and
(\ref{distance_modulus_v1369_cen_v496_sct}),
we have the relation of
\begin{eqnarray}
(m&-& M')_{V, \rm V1369~Cen} \cr 
&\equiv& (m_V - (M_V - 2.5\log f_{\rm s}))_{\rm V1369~Cen} \cr
&=& \left( (m-M)_V + \Delta V \right)_{\rm LV~Vul} \cr
&=& 11.85 - 1.2\pm0.2 = 10.65\pm0.2.
\label{absolute_mag_v1369_cen_lv_vul}
\end{eqnarray}

We further compare with other novae having similar light curves to
V1369~Cen, i.e., V496~Sct and V5666~Sgr.   Figure 
\ref{v1369_cen_v496_sct_v5666_sgr_b_v_i_k_logscale_4fig} shows
the time-stretched light curves of $B$, $V$, $I_{\rm C}$, and $K_{\rm s}$
bands.  The light curves well overlap except for the $K_{\rm s}$ band.
Applying Equation (\ref{distance_modulus_general_temp_b})
for the $B$ band to Figure 
\ref{v1369_cen_v496_sct_v5666_sgr_b_v_i_k_logscale_4fig}(a), 
we have the relation of
\begin{eqnarray}
(m&-&M)_{B, \rm V1369~Cen} \cr 
&=& \left( (m-M)_B + \Delta B\right)_{\rm V496~Sct} - 2.5 \log 0.74 \cr 
&=& 14.15 - 4.1\pm0.2 + 0.325 = 10.38\pm0.2,
\label{distance_modulus_v1369_cen_v496_sct_v5666_sgr_b}
\end{eqnarray}
where we adopt $(m-M)_{B, \rm V496~Sct}=13.7 + 1.0\times 0.45= 14.15$ 
from Section \ref{v496_sct_cmd}.
Thus, we obtain $(m-M)_B=10.38\pm0.2$ for V1369~Cen.
We obtain $d=0.97\pm0.1$~kpc from Equation (\ref{distance_modulus_rb})
together with $E(B-V)=0.11$ and $(m-M)_B=10.38\pm0.2$.
We plot this relation of $(m-M)_B=10.38$ by the solid magenta line 
in Figure \ref{distance_reddening_v574_pup_v679_car_v1369_cen_v5666_sgr}(c).

Applying Equation (\ref{distance_modulus_general_temp}) to
Figure \ref{v1369_cen_v496_sct_v5666_sgr_b_v_i_k_logscale_4fig}(b), 
we have the relation of
\begin{eqnarray}
(m&-&M)_{V, \rm V1369~Cen} \cr 
&=& \left( (m-M)_V + \Delta V\right)_{\rm V496~Sct} - 2.5 \log 0.74 \cr 
&=& 13.7 - 3.8\pm0.2 + 0.325 = 10.23\pm0.2,
\label{distance_modulus_v1369_cen_v496_sct_v5666_sgr_v}
\end{eqnarray}
where we adopt $(m-M)_{V, \rm V496~Sct}=13.7$ from Section \ref{v496_sct_cmd}.
Thus, we obtain $(m-M)_V=10.23\pm0.2$ for V1369~Cen.
We obtain $d=0.95\pm0.1$~kpc from
Equation (\ref{distance_modulus_rv}) together with $E(B-V)=0.11$.
We plot this relation of $(m-M)_V=10.23$ by the solid blue line 
in Figure \ref{distance_reddening_v574_pup_v679_car_v1369_cen_v5666_sgr}(c).

From the $I_{\rm C}$ band data in Figure
\ref{v1369_cen_v496_sct_v5666_sgr_b_v_i_k_logscale_4fig}(a),
we obtain 
\begin{eqnarray}
(m&-&M)_{I, \rm V1369~Cen} \cr
&=& ((m - M)_I + \Delta I_C)_{\rm V496~Sct} - 2.5 \log 0.74 \cr
&=& 12.98 - 3.2\pm0.2 + 0.325 = 10.11\pm0.2,
\label{distance_modulus_i_v1369_cen_v496_sct_v5666_sgr}
\end{eqnarray}
where we adopt $(m-M)_{I, \rm V496~Sct}=13.7 - 1.6\times 0.45= 12.98$.
Thus, we obtain $(m-M)_{I, \rm V574~Pup}= 10.11\pm0.2$.
We obtain $d=0.97\pm0.1$~kpc from
Equation (\ref{distance_modulus_ri}) together with $E(B-V)=0.11$.
We plot this relation of $(m-M)_I=10.11$ by the solid cyan line 
in Figure \ref{distance_reddening_v574_pup_v679_car_v1369_cen_v5666_sgr}(c).

V1369~Cen and V496~Sct showed a shallow dust blackout around 100 days
($\log t \sim 2.0$) after the outbursts (Figure 
\ref{v1369_cen_v496_sct_v5666_sgr_b_v_i_k_logscale_4fig}).
Correspondingly, the $K_{\rm s}$ light curves have an enhancement
above the line of universal decline law ($F_\nu\propto t^{-1.75}$),
as shown in Figure
\ref{v1369_cen_v496_sct_v5666_sgr_b_v_i_k_logscale_4fig}(d).
In such a case, we do not use this part for overlapping.
In the later phase, however, the decline trend of V1369~Cen
becomes similar to those of V496~Sct and V5666~Sgr.
We, thus, try to overlap as much as possible in the later phase.
Applying Equation (\ref{distance_modulus_general_temp_k}) 
for the $K_{\rm s}$ band, 
we obtain
\begin{eqnarray}
(m&-&M)_{K, \rm V1369~Cen} \cr
&=& ((m - M)_K + \Delta K_{\rm s})_{\rm V496~Sct} - 2.5 \log 0.74 \cr
&=& 12.46 - 2.8\pm0.3 + 0.325 = 9.99\pm 0.3,
\label{distance_modulus_k_v1369_cen_v496_sct_v5666_sgr}
\end{eqnarray}
where we adopt $(m-M)_{K, \rm V496~Sct}=13.7 - 2.75\times 0.45= 12.46$.
Thus, we obtain $(m-M)_{K, \rm V1369~Cen}= 9.99\pm0.3$.
The distance is determined to be $d=0.98\pm0.3$~kpc from
Equation (\ref{distance_modulus_rk}) together with $E(B-V)=0.11$.
This distance is consistent with those determined from
$B$, $V$, and $I_{\rm C}$ bands.
We plot this relation of $(m-M)_K=9.99$ by the solid cyan-blue line 
in Figure \ref{distance_reddening_v574_pup_v679_car_v1369_cen_v5666_sgr}(c).

Figure \ref{all_mass_v1369_cen_v1668_cyg_x45z02c15o20} shows
comparison with V1668~Cyg, one of well studied novae.
The timescale of V1668~Cyg is stretched with $f_{\rm s}=1.48$
and the $V$ magnitude difference is $\Delta V=-3.9$.  
From the time-stretching method, we have the relation of
\begin{eqnarray}
(m&-&M)_{V, \rm V1369~Cen} \cr 
&=& (m-M + \Delta V)_{V, \rm V1668~Cyg} - 2.5 \log 1.48 \cr
&=& 14.6 - 3.9\pm0.2 - 0.43 = 10.27\pm0.2,
\label{distance_modulus_v1369_cen_v1668_cyg}
\end{eqnarray}
where we adopt $(m-M)_{V, \rm V1668~Cyg}=14.6$ from Section \ref{v1668_cyg}.
Thus, we again confirm that $f_{\rm s}=1.48$ against LV~Vul
and $(m-M)_V=10.27\pm0.2$.  To summarize, we adopt
$(m-M)_V=10.25\pm0.1$ from various estimates based on the time-stretching
method.


\begin{figure}
\plotone{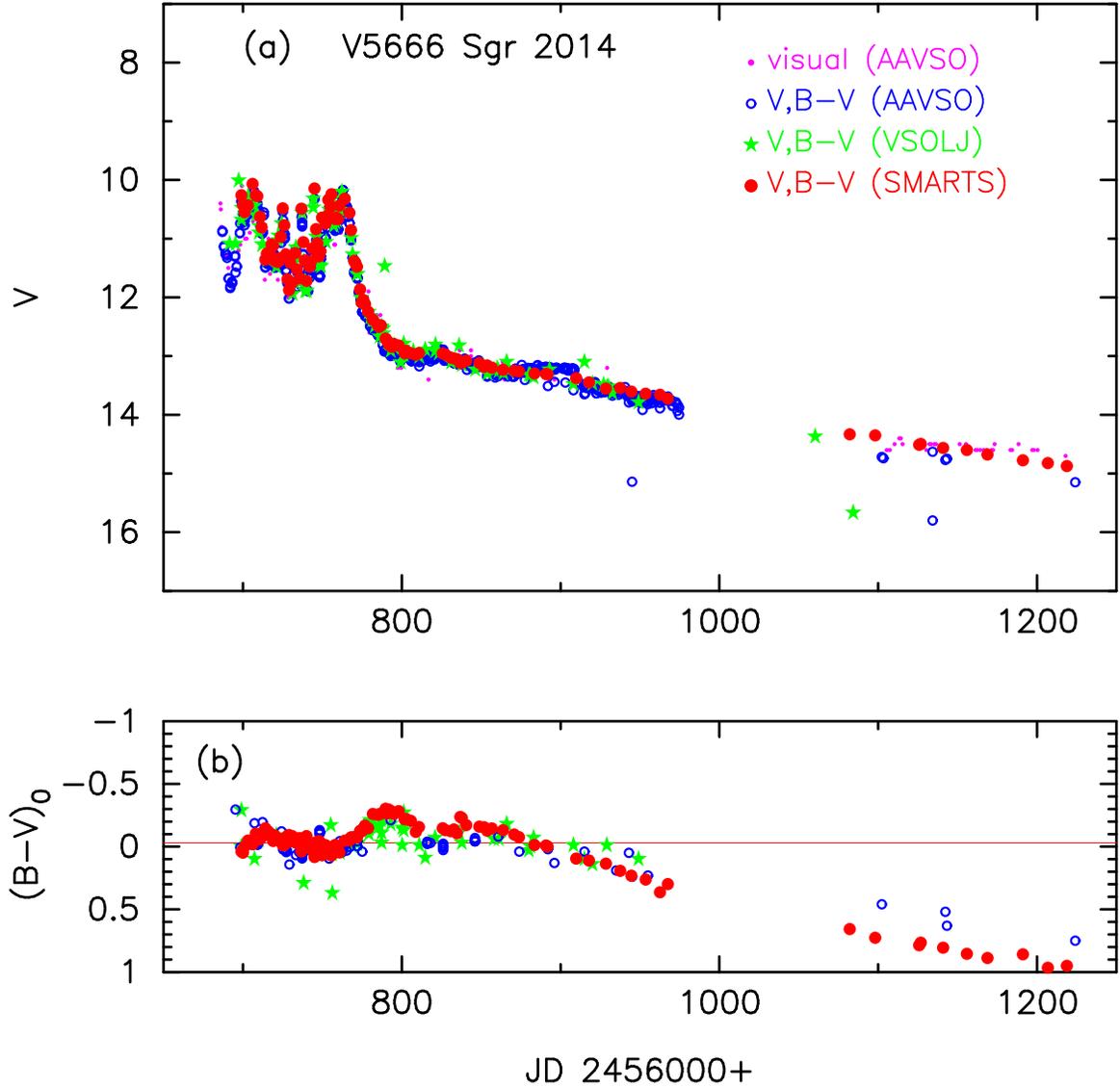}
\caption{
Same as Figure \ref{v574_pup_v_bv_ub_color_curve}, but for V5666~Sgr.
(a) The $V$ and visual data of AAVSO (blue open circles and magenta dots)
and $V$ data of VSOLJ (filled green stars) are systematically
shifted downward by 0.3 mag and 0.2 mag, respectively, to overlap them
to the SMARTS data (filled red circles).  (b) The $(B-V)_0$ 
are dereddened with $E(B-V)=0.50$.  The $B-V$ data of AAVSO
(blue open circles) and VSOLJ (filled green stars) are systematically
shifted up by 0.03 mag and 0.02 mag, respectively, to overlap them
to the SMARTS data (filled red circles).  
\label{v5666_sgr_v_bv_ub_color_curve}}
\end{figure}


\begin{figure}
\plotone{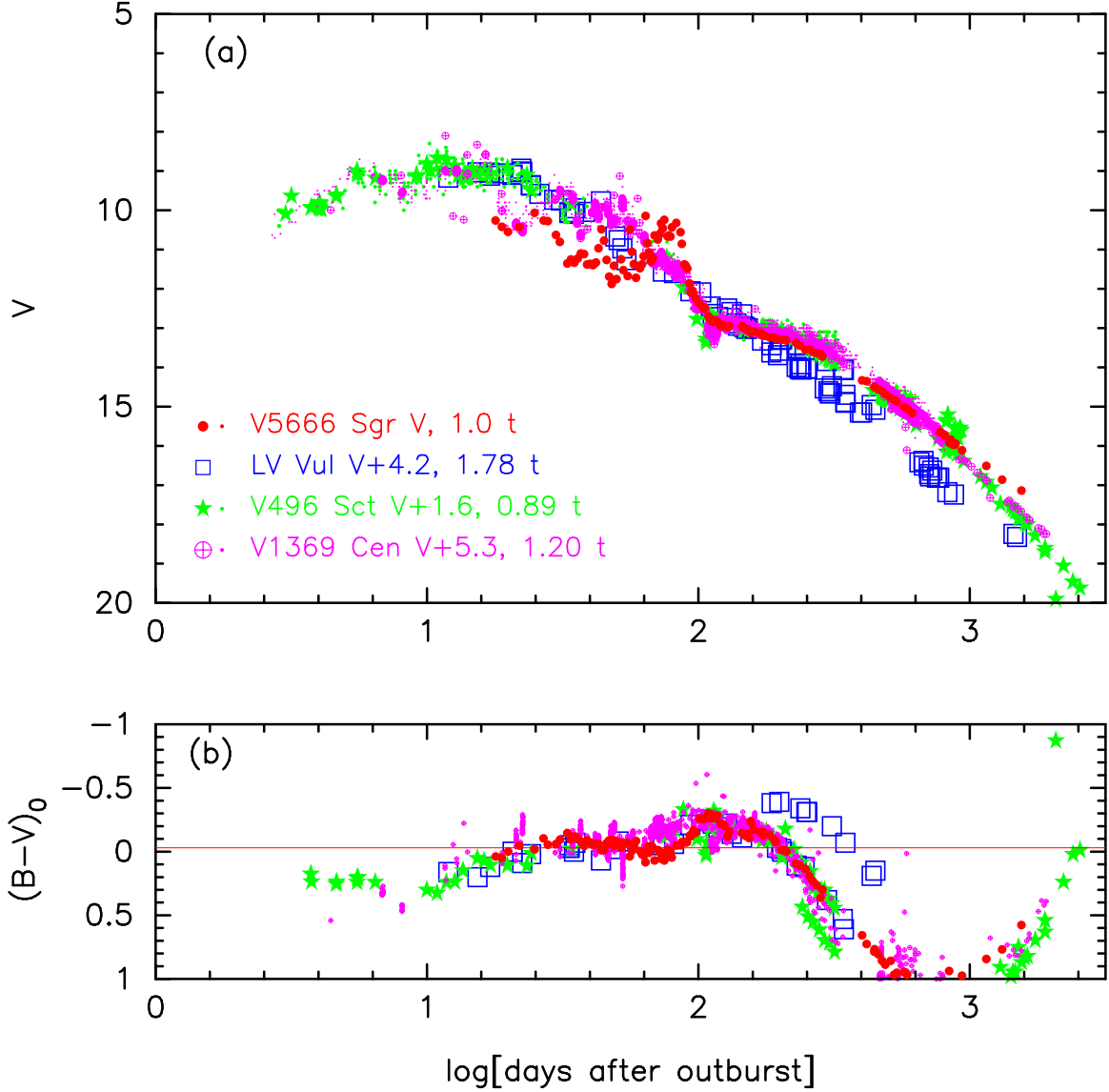}
\caption{
Same as Figure \ref{v1369_cen_v496_sct_lv_vul_v_bv_ub_color_logscale},
but for V5666~Sgr.  The $BV$ data of V5666~Sgr are the same as those
in Figure \ref{v5666_sgr_v_bv_ub_color_curve}.  The $B-V$ colors 
of V5666~Sgr are dereddened with $E(B-V)=0.50$.  
We also plot the $BV$ data of LV~Vul, V496~Sct, and V1369~Cen.
The timescales of LV~Vul, V496~Sct, and V1369~Cen are stretched by
1.78, 0.89, and 1.20, respectively, against V5666~Sgr.
\label{v5666_sgr_v1369_cen_v496_sct_v_bv_ub_color_logscale}}
\end{figure}


\begin{figure}
\plottwo{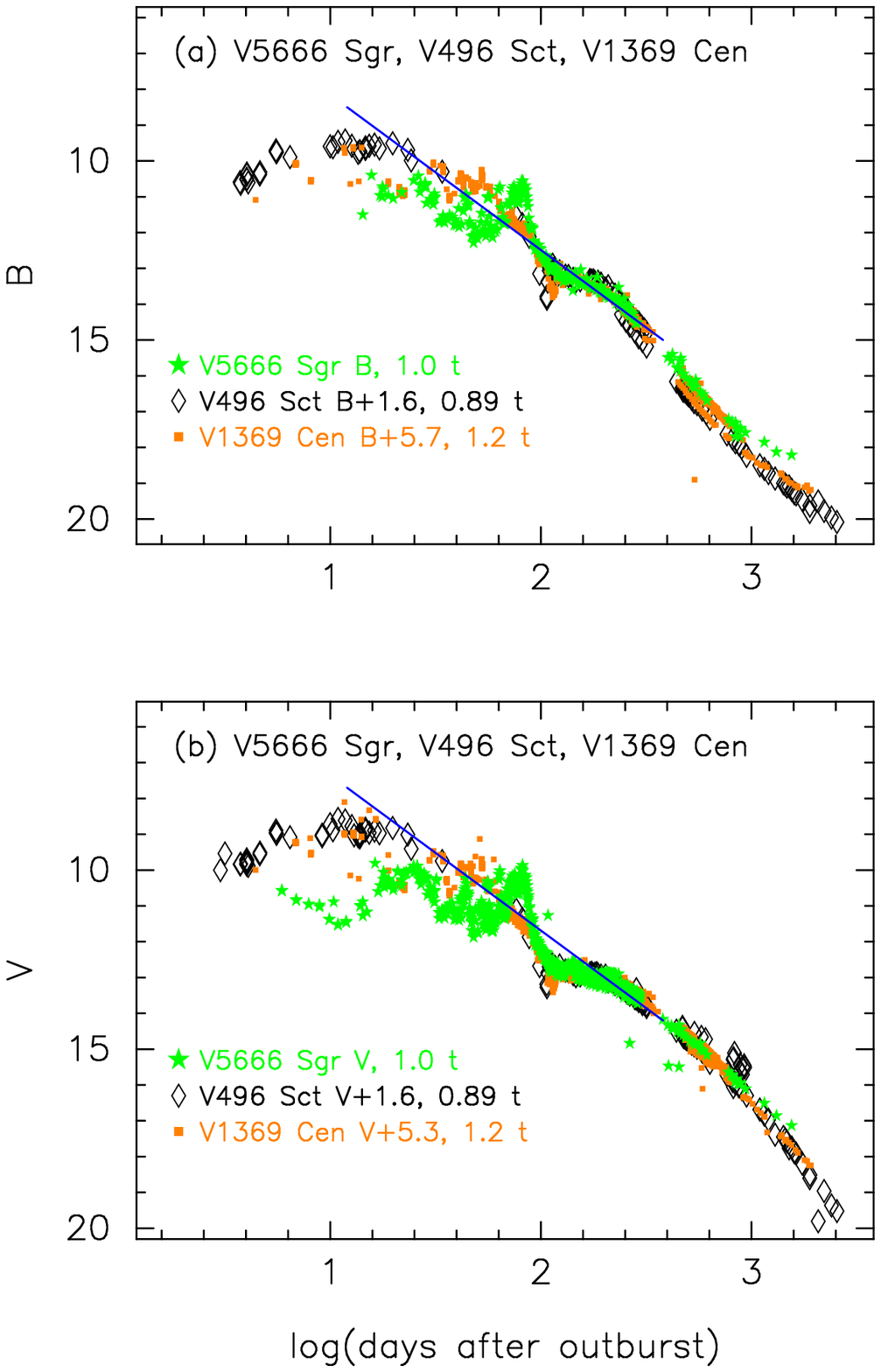}{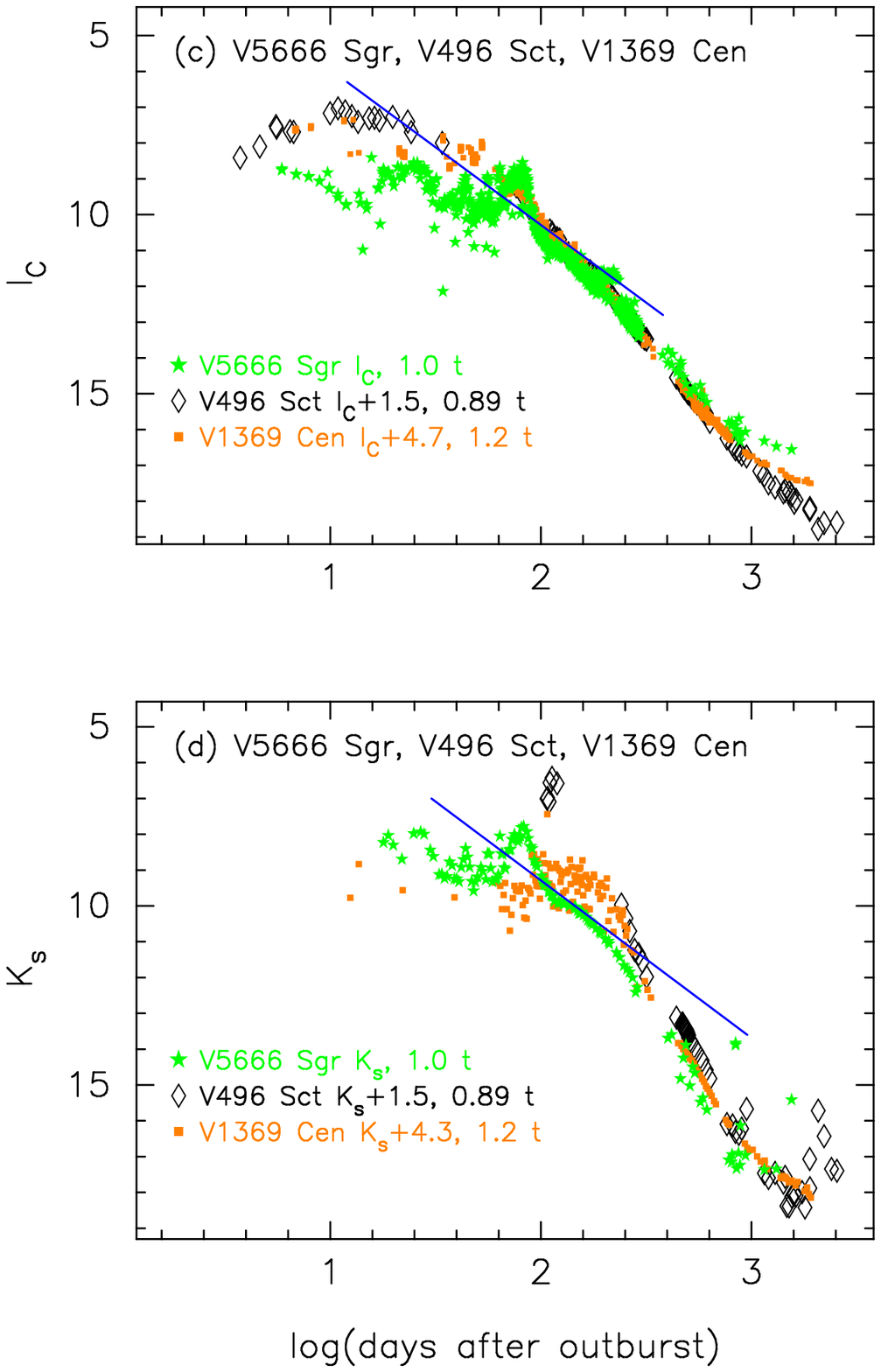}
\caption{
Same as Figure 
\ref{v1369_cen_v496_sct_v5666_sgr_b_v_i_k_logscale_4fig},
but for V5666~Sgr.
The data of V1369~Cen and V496~Sct are the same
as those in Figure 
\ref{v1369_cen_v496_sct_v5666_sgr_b_v_i_k_logscale_4fig}.
\label{v5666_sgr_v496_sct_v1369_cen_b_v_i_k_logscale_4fig}}
\end{figure}


\begin{figure}
\plotone{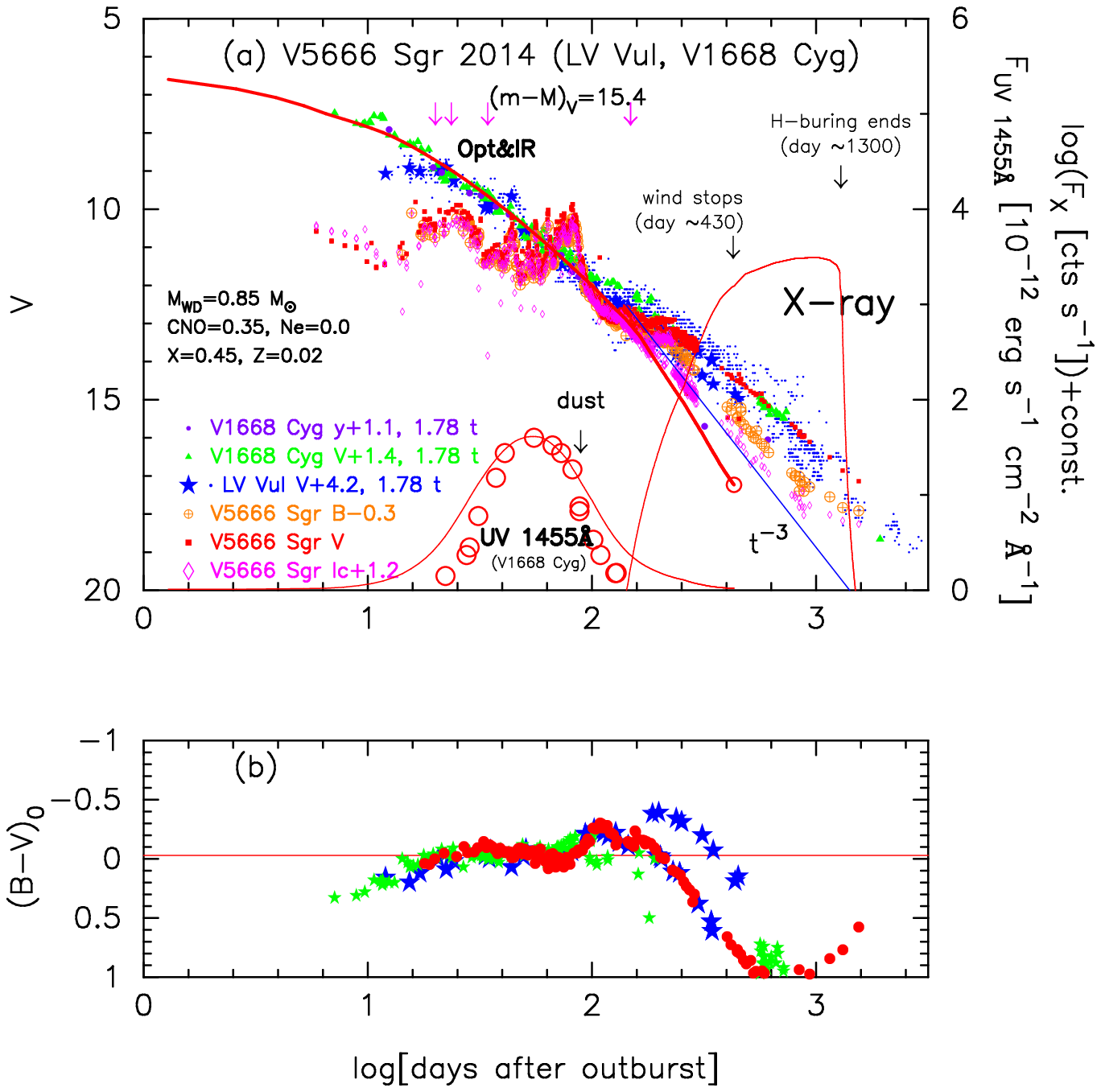}
\caption{
Same as Figure \ref{all_mass_v1369_cen_v1668_cyg_x45z02c15o20},
but for V5666~Sgr.  We also add (a) the $V$ light and (b)
$(B-V)_0$ color curves of LV~Vul and V1668~Cyg.
Their light/color curves and the UV~1455\AA\   flux (large open 
red circles) of V1668~Cyg are
stretched by a factor of 1.78 against V5666~Sgr in the time-direction.
We add the $B$ and $I_C$ magnitudes of V5666~Sgr taken 
from AAVSO, VSOLJ, and SMARTS.
We add a $0.85 ~M_\sun$ WD (solid red lines) models with the chemical
composition of CO nova 3 \citep{hac16k}, which approximately 
reproduce the optical and NIR light curves of V5666~Sgr.
The {\it Swift} XRT observed V5666~Sgr five times as denoted by
the downward magenta arrows, but did not detect X-rays.  
\label{all_mass_v5666_sgr_v1668_cyg_x45z02c15o20}}
\end{figure}

\subsection{V5666~Sgr 2014}
\label{v5666_sgr}
 
Figure \ref{v5666_sgr_v_bv_ub_color_curve} shows (a) the $V$ and 
visual light and (b) $(B-V)_0$ color curves of V5666~Sgr.  
The $V$ and visual data of AAVSO (blue open circles and magenta dots)
and $V$ data of VSOLJ (filled green stars) are systematically
shifted downward by 0.3 mag and 0.2 mag, respectively, to overlap them
to the SMARTS data (filled red circles).  Here, $(B-V)_0$ 
are dereddened with $E(B-V)=0.50$ as obtained in Section \ref{v5666_sgr_cmd}.
The $B-V$ data of AAVSO (blue open circles) and VSOLJ (filled green stars)
are systematically shifted upward by 0.03 mag and 0.02 mag, respectively, 
to overlap them to the SMARTS data (filled red circles).  

Figure \ref{v5666_sgr_v1369_cen_v496_sct_v_bv_ub_color_logscale}
shows (a) the $V$ light curves of V5666~Sgr, LV~Vul, V496~Sct, and V1369~Cen,
and (b) their $(B-V)_0$ color curves on logarithmic timescales.
Here we assume that the V5666~Sgr outbursted on JD~2456681.0 (Day 0).
Figure \ref{v5666_sgr_v1369_cen_v496_sct_v_bv_ub_color_logscale}
compares V5666~Sgr with the similar $V$ light curve novae, V496~Sct 
and V1369~Cen, in addition to LV~Vul.
We stretch the timescales of LV~Vul, V496~Sct, and V1369~Cen by 
1.78, 0.89,  and 1.20, respectively, against V5666~Sgr,
to overlap them with the V5666~Sgr $V$ light and $(B-V)_0$ color curves.
From the time-stretching method, we have the relation of
\begin{eqnarray}
(m&-&M)_{V, \rm V5666~Sgr} \cr 
&=& (m-M + \Delta V)_{V, \rm LV~Vul} - 2.5 \log 1.78 \cr
&=& 11.85 + 4.2\pm0.2 - 0.63 = 15.42\pm0.2 \cr
&=& (m-M + \Delta V)_{V, \rm V496~Sct} - 2.5 \log 0.89 \cr
&=& 13.7 + 1.6\pm0.2 + 0.13 = 15.43\pm0.2 \cr
&=& (m-M + \Delta V)_{V, \rm V1369~Cen} - 2.5 \log 1.20 \cr
&=& 10.25 + 5.3\pm0.2 - 0.20  = 15.35\pm0.2. 
\label{distance_modulus_v5666_sgr_v}
\end{eqnarray}
Thus, we obtained $(m-M)_{V, \rm V5666~Sgr}=15.4\pm0.1$.  

From Equations (\ref{time-stretching_general}),
(\ref{distance_modulus_general_temp}), and
(\ref{distance_modulus_v5666_sgr_v}),
we have the relation of
\begin{eqnarray}
(m&-& M')_{V, \rm V5666~Sgr} \cr 
&\equiv& (m_V - (M_V - 2.5\log f_{\rm s}))_{\rm V5666~Sgr} \cr
&=& \left( (m-M)_V + \Delta V \right)_{\rm LV~Vul} \cr
&=& 11.85 + 4.2\pm0.2 = 16.05\pm0.2.
\label{absolute_mag_v5666_sgr_lv_vul_stretch}
\end{eqnarray}

We further compare the V5666~Sgr light curves with other novae 
having similar light curves, V1369~Cen and V496~Sct.   Figure 
\ref{v5666_sgr_v496_sct_v1369_cen_b_v_i_k_logscale_4fig} shows
the same time-stretched light curves of $B$, $V$, $I_{\rm C}$,
and $K_{\rm s}$ bands as Figure
\ref{v1369_cen_v496_sct_v5666_sgr_b_v_i_k_logscale_4fig}.
Applying Equation (\ref{distance_modulus_general_temp_b})
for the $B$ band to Figure 
\ref{v5666_sgr_v496_sct_v1369_cen_b_v_i_k_logscale_4fig}(a), 
we have the relation of
\begin{eqnarray}
(m&-&M)_{B, \rm V5666~Sgr} \cr 
&=& \left( (m-M)_B + \Delta B\right)_{\rm V496~Sct} - 2.5 \log 0.89 \cr 
&=& 14.15 + 1.6\pm0.2 + 0.125 = 15.88\pm0.2 \cr
&=& \left( (m-M)_B + \Delta B\right)_{\rm V1369~Cen} - 2.5 \log 1.2 \cr 
&=& 10.36 + 5.7\pm0.2 - 0.20 = 15.86\pm0.2,
\label{distance_modulus_v5666_sgr_v496_sct_v1369_cen_b}
\end{eqnarray}
where we adopt $(m-M)_{B, \rm V496~Sct}=13.7 + 1.0\times 0.45= 14.15$ 
from Section \ref{v496_sct_cmd} and
$(m-M)_{B, \rm V1369~Cen}=10.25 + 1.0\times 0.11=10.36$
from Section \ref{v1369_cen_cmd}.
Thus, we obtain $(m-M)_B=15.87\pm0.1$ for V5666~Sgr.
We obtain $d=5.8\pm0.6$~kpc from Equation (\ref{distance_modulus_rb})
together with $E(B-V)=0.50$ and $(m-M)_B=15.87\pm0.1$.
We plot this relation of $(m-M)_B=15.87$ by the solid magenta line 
in Figure \ref{distance_reddening_v574_pup_v679_car_v1369_cen_v5666_sgr}(d).

Applying Equation (\ref{distance_modulus_general_temp}) to
Figure \ref{v5666_sgr_v496_sct_v1369_cen_b_v_i_k_logscale_4fig}(b), 
we have the relation of
\begin{eqnarray}
(m&-&M)_{V, \rm V5666~Sgr} \cr 
&=& \left( (m-M)_V + \Delta V\right)_{\rm V496~Sct} - 2.5 \log 0.89 \cr 
&=& 13.7 + 1.6\pm0.2 + 0.125 = 15.4\pm0.2 \cr
&=& \left( (m-M)_V + \Delta V\right)_{\rm V1369~Cen} - 2.5 \log 1.2 \cr 
&=& 10.25 + 5.3\pm0.2 - 0.20 = 15.35\pm0.2,
\label{distance_modulus_v5666_sgr_v496_sct_v1369_cen_v}
\end{eqnarray}
where we adopt $(m-M)_{V, \rm V496~Sct}=13.7$ from Section \ref{v496_sct_cmd}
and $(m-M)_{V, \rm V1369~Cen}=10.25$ from Section \ref{v1369_cen_cmd}.
Thus, we obtain $(m-M)_V=15.38\pm0.1$ for V5666~Sgr.
We obtain $d=5.8\pm0.6$~kpc from
Equation (\ref{distance_modulus_rv}) together with $E(B-V)=0.50\pm0.05$
and $(m-M)_V=15.38\pm0.1$.
We plot this relation of $(m-M)_V=15.38$ by the solid blue line 
in Figure \ref{distance_reddening_v574_pup_v679_car_v1369_cen_v5666_sgr}(d).

From the $I_{\rm C}$ band data in Figure
\ref{v5666_sgr_v496_sct_v1369_cen_b_v_i_k_logscale_4fig}(c),
we obtain 
\begin{eqnarray}
(m&-&M)_{I, \rm V5666~Sgr} \cr
&=& ((m - M)_I + \Delta I_C)_{\rm V496~Sct} - 2.5 \log 0.89 \cr
&=& 12.98 + 1.5\pm0.2 + 0.125 = 14.61\pm0.2 \cr
&=& ((m - M)_I + \Delta I_C)_{\rm V1369~Cen} - 2.5 \log 1.2 \cr
&=& 10.07 + 4.7\pm0.2 - 0.20 = 14.57\pm0.2, 
\label{distance_modulus_i_v5666_sgr_v496_sct_v1369_cen}
\end{eqnarray}
where we adopt $(m-M)_{I, \rm V496~Sct}=13.7 - 1.6\times 0.45= 12.98$
and $(m-M)_{I, \rm V1369~Cen}=10.25 - 1.6\times 0.11= 10.07$.
Thus, we obtain $(m-M)_{I, \rm V5666~Sgr}= 14.59\pm0.1$.
We obtain $d=5.9\pm0.6$~kpc from
Equation (\ref{distance_modulus_ri}) together with $E(B-V)=0.50\pm0.05$
and $(m-M)_I=14.59\pm0.1$.
We plot this relation of $(m-M)_I=14.59$ by the solid cyan line 
in Figure \ref{distance_reddening_v574_pup_v679_car_v1369_cen_v5666_sgr}(d).

We try to overlap the $K_{\rm s}$ light curves
as much as possible in the later phase of Figure
\ref{v5666_sgr_v496_sct_v1369_cen_b_v_i_k_logscale_4fig}(d).
Applying Equation (\ref{distance_modulus_general_temp_k}) 
for the $K_{\rm s}$ band, 
we obtain
\begin{eqnarray}
(m&-&M)_{K, \rm V5666~Sgr} \cr
&=& ((m - M)_K + \Delta K_{\rm s})_{\rm V496~Sct} - 2.5 \log 0.89 \cr
&=& 12.46 + 1.5\pm0.3 + 0.125 = 14.09\pm 0.3 \cr
&=& ((m - M)_K + \Delta K_{\rm s})_{\rm V1369~Cen} - 2.5 \log 1.2 \cr
&=& 9.95 + 4.3\pm0.3 - 0.20 = 14.05\pm 0.3,
\label{distance_modulus_k_v5666_sgr_v496_sct_v1369_cen}
\end{eqnarray}
where we adopt $(m-M)_{K, \rm V496~Sct}=13.7 - 2.75\times 0.45= 12.46$,
$(m-M)_{K, \rm V1369~Cen}=10.25 - 2.75\times 0.11= 9.95$.
Thus, we obtain $(m-M)_{K, \rm V5666~Sgr}= 14.07\pm0.2$.
We obtain $d=6.0\pm0.6$~kpc from
Equation (\ref{distance_modulus_rk}) together with $E(B-V)=0.50\pm0.05$
and $(m-M)_K=14.07\pm0.2$.
We plot this relation of $(m-M)_K=14.07$ by the solid cyan-blue line 
in Figure \ref{distance_reddening_v574_pup_v679_car_v1369_cen_v5666_sgr}(d).

Figure \ref{all_mass_v5666_sgr_v1668_cyg_x45z02c15o20} also shows
another comparison with LV~Vul and V1668~Cyg.
The light/color curves of LV~Vul and V1668~Cyg are stretched by a factor of 
1.78, so we obtain
\begin{eqnarray}
(m&-&M)_{V, \rm V5666~Sgr} \cr 
&=& (m-M + \Delta V)_{V, \rm LV~Vul} - 2.5 \log 1.78 \cr
&=& 11.85 + 4.2\pm0.2 - 0.63 = 15.42\pm0.2 \cr 
&=& (m-M + \Delta V)_{V, \rm V1668~Cyg} - 2.5 \log 1.78 \cr
&=& 14.6 + 1.4\pm0.2 - 0.63 = 15.37\pm0.2, 
\label{distance_modulus_v5666_sgr_v1668_cyg}
\end{eqnarray}
from our time-stretching method. 
Thus, we again confirm $(m-M)_{V, \rm V5666~Sgr}=15.4\pm0.2$.

\section{Reanalyzed Light Curves of 12 Novae}
\label{reanalyse_twelve_novae}


\begin{figure}
\plotone{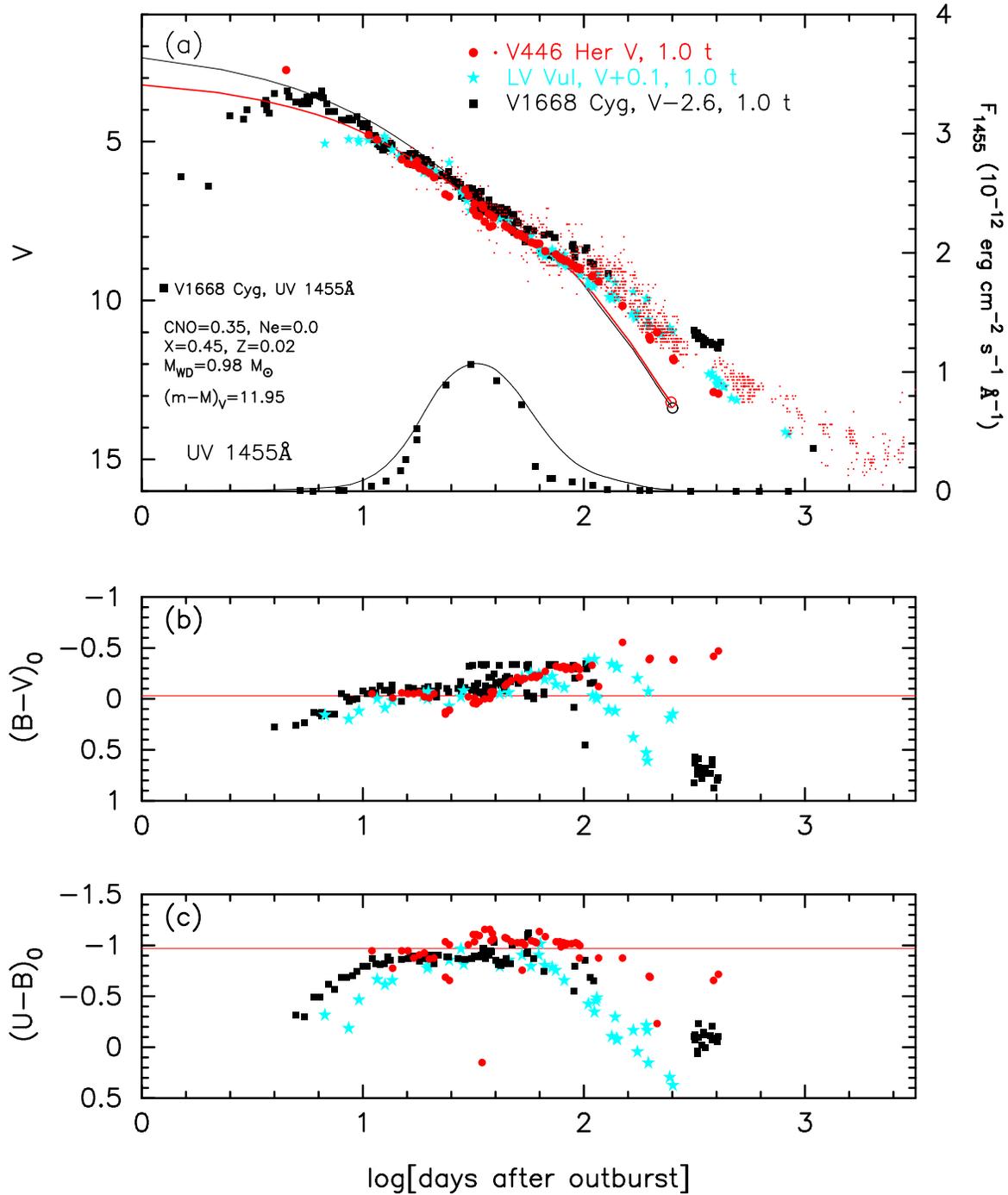}
\caption{
Same as Figure 
\ref{v574_pup_v1974_cyg_v_bv_logscale},
but for V446~Her (filled red circles).
The data of V446~Her are the same as those in Figure 42 of Paper I,
but we assume the outburst day of $t_{\rm OB}=$JD~2436993.0 for V446~Her.
The $(B-V)_0$ and $(U-B)_0$ of V446~Her are dereddened with $E(B-V)=0.40$.
We also plot (a) the visual magnitudes (red dots) of V446~Her.
We add (a) the $V$ light, (b) $(B-V)_0$, and (c) $(U-B)_0$ color
curves of LV~Vul and V1668~Cyg.  We also add (a)
the UV~1455\AA\   fluxes (filled black squares) of V1668~Cyg.
Assuming $(m-M)_V=11.95$ for V446~Her, we plot a model $V$ light
curve (solid red line) of a $0.98~M_\sun$ WD with the envelope chemical
composition of CO nova 3 \citep{hac16k}.  
We also add another $0.98~M_\sun$ WD model (solid black lines)
with the chemical composition of 
CO nova 3, assuming $(m-M)_V=14.6$ for V1668~Cyg.
This model (black lines) has a slightly larger initial envelope mass
than the model (red line) for V446~Her.
\label{v446_her_v1668_cyg_lv_vul_v_color_logscale}}
\end{figure}

\subsection{V446~Her 1960}
\label{v446_her}
Figure \ref{v446_her_v1668_cyg_lv_vul_v_color_logscale} shows
the light/color curves of V446~Her on a logarithmic timescale 
as well as LV~Vul and V1668~Cyg.  We overlap the light/color curves
of these three novae as much as possible.
Based on the time-stretching method, we have the relation of
\begin{eqnarray}
(m&-&M)_{V, \rm V446~Her} \cr 
&=& (m - M + \Delta V)_{V, \rm LV Vul} - 2.5 \log 1.0 \cr
&=& 11.85 + 0.1\pm0.2 + 0.0 = 11.95\pm0.2 \cr
&=& (m - M + \Delta V)_{V, \rm V1668~Cyg} - 2.5 \log 1.0 \cr
&=& 14.6 - 2.6\pm0.2 + 0.0 = 12.0\pm0.2.
\label{distance_modulus_v466_her_lv_vul_v1668_cyg}
\end{eqnarray}
Thus, we obtain $f_{\rm s}=1.0$ and $(m-M)_V=11.95\pm0.1$ for V446~Her.
The new distance modulus is slightly larger than the previous
value of $(m-M)_V=11.7\pm0.1$ (Paper II).  This difference
comes from the improved values of $f_{\rm s}$ and $\Delta V$. 
From Equations (\ref{distance_modulus_general_dot}) and
(\ref{distance_modulus_v466_her_lv_vul_v1668_cyg}),
we have the relation of
\begin{eqnarray}
(m&-& M')_{V, \rm V446~Her} \cr 
&\equiv& (m_V - (M_V - 2.5\log f_{\rm s}))_{\rm V446~Her} \cr
&=& (m-M+ \Delta V)_{V, \rm LV~Vul} \cr
&=& 11.85 + 0.1\pm0.2 = 11.95\pm0.2.
\label{absolute_mag_lv_vul_v446_her}
\end{eqnarray}


\begin{figure}
\plotone{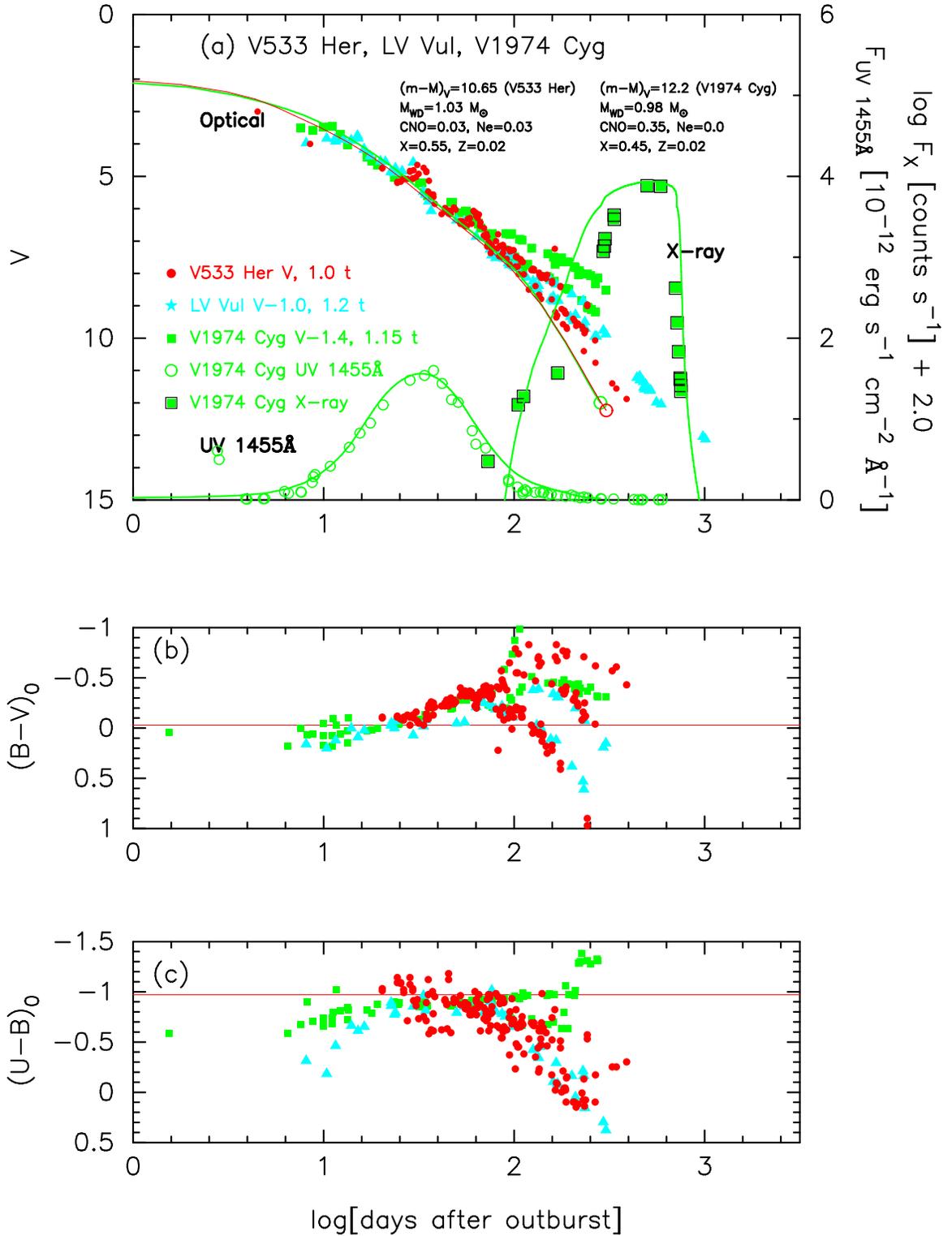}
\caption{
Same as Figure \ref{v574_pup_v1974_cyg_v_bv_logscale},
but for V533~Her (filled red circles).  
We add the data of LV~Vul and V1974~Cyg.
The $(B-V)_0$ and $(U-B)_0$ data of V533~Her are dereddened 
by Equations (\ref{dereddening_eq_bv}) and (\ref{dereddening_eq_ub})
with $E(B-V)=0.038$.
We also add the UV~1455\AA\  (open green circles) and supersoft X-ray
(filled green squares with black outlines) fluxes of V1974~Cyg.
The UV and X-ray data of V1974~Cyg are the same as those in Figure 13
of Paper I.
Taking $(m-M)_V=10.65$ for V533~Her, we plot a model $V$ light
curve (solid red line) of a $1.03~M_\sun$ WD with the envelope chemical
composition of Ne nova 2 \citep{hac10k}.  
We also add a $0.98~M_\sun$ WD model (solid green lines)
with the chemical composition of 
CO nova 3 \citep{hac16k}, taking $(m-M)_V=12.2$ for V1974~Cyg.
\label{v533_her_lv_vul_v1974_cyg_v_bv_ub_logscale}}
\end{figure}

\subsection{V533~Her 1963}
\label{v533_her}
We obtain $f_{\rm s}$ and $(m-M)_V$ 
by the time-stretching method as in Figure 
\ref{v533_her_lv_vul_v1974_cyg_v_bv_ub_logscale}, which shows
the $V$ light curve as well as the dereddened $(B-V)_0$ 
and $(U-B)_0$ color curves.  
We add the light/color curves of LV~Vul and V1974~Cyg
and overlap these light/color curves as much as possible.  
Applying the time-stretching method of Equation
(\ref{distance_modulus_general_temp}) to Figure 
\ref{v533_her_lv_vul_v1974_cyg_v_bv_ub_logscale}(a), 
we have the relation of
\begin{eqnarray}
(m&-&M)_{V, \rm V533~Her} \cr 
&=& ((m - M)_V + \Delta V)_{\rm LV~Vul} - 2.5 \log 1.20 \cr 
&=& 11.85 - 1.0\pm0.3 - 0.20 = 10.65\pm0.3 \cr
&=& ((m - M)_V + \Delta V)_{\rm V1974~Cyg} - 2.5 \log 1.15 \cr 
&=& 12.2 - 1.4\pm0.3 - 0.15 = 10.65\pm0.3.
\label{distance_modulus_v533_her_v1974_cyg_v1668_cyg_v}
\end{eqnarray}
The new result of $(m-M)_V=10.65\pm0.2$ is almost the same as
the previous result of $(m-M)_V=10.8\pm0.2$ (Paper II),
but slightly improve the timescaling factors of $f_{\rm s}$ and vertical
fit of $\Delta V$.  We obtain $f_{\rm s}=1.20$ against the template LV~Vul.
The distance is calculated to be $d= 1.28$~kpc from Equation
(\ref{distance_modulus_rv}) together with $(m-M)_V=10.65$ and
$E(B-V)= 0.038$. 
This value is almost the same as what \citet{hac16kb} concluded
from various results in the literature.

From Equations (\ref{distance_modulus_general_dot}) and
(\ref{distance_modulus_v533_her_v1974_cyg_v1668_cyg_v}), 
we have the relation of
\begin{eqnarray}
(m&-& M')_{V, \rm V533~Her} \cr 
&\equiv& (m_V - (M_V - 2.5\log f_{\rm s}))_{\rm V533~Her} \cr
&=& \left( (m-M)_V + \Delta V \right)_{\rm LV~Vul} \cr
&=& 11.85 - 1.0\pm0.3 = 10.85\pm0.3.
\label{absolute_mag_v533_her_lv_vul_v}
\end{eqnarray}


\begin{figure}
\plotone{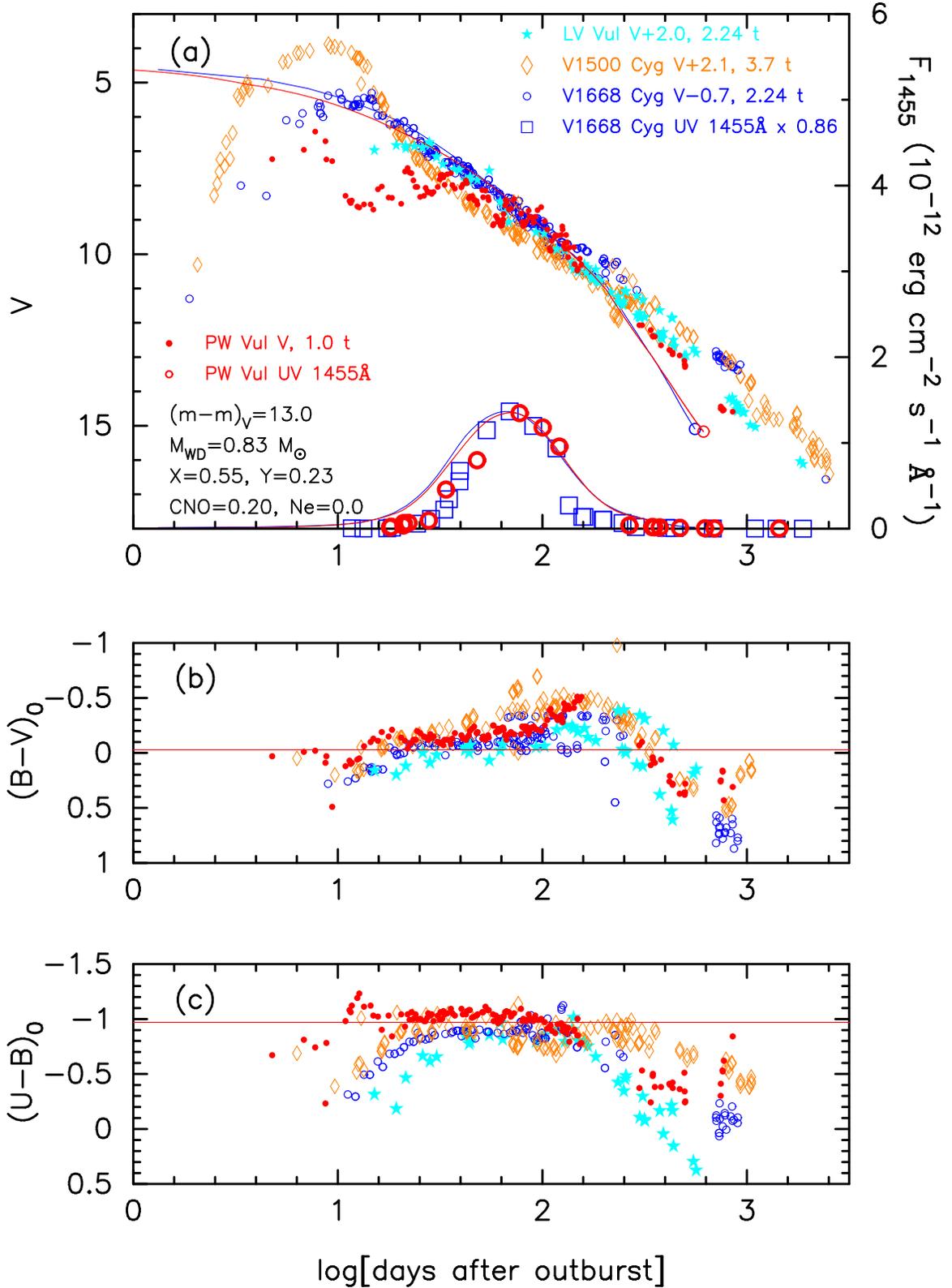}
\caption{
Same as Figure \ref{v574_pup_v1974_cyg_v_bv_logscale},
but for PW~Vul (filled red circles).
We add the UV~1455\AA\  flux of PW~Vul (large open red circles)
and V1668~Cyg (open blue squares). 
The data of PW~Vul are the same as those in Figure 6 of Paper I.
In panel (a), taking $(m-M)_V=13.0$ for PW~Vul, 
we add model light curves (solid red lines) of a $0.83~M_\sun$ WD
with the envelope chemical composition of CO nova 4 \citep{hac15k}.
We also add a $0.98~M_\sun$ WD model (solid blue lines) 
with the chemical composition of CO nova 3 \citep{hac16k}, taking
$(m-M)_V=14.6$ for V1668~Cyg.
\label{pw_vul_lv_vul_v1668_cyg_v1500_cyg_v_bv_ub_color_logscale}}
\end{figure}


\begin{figure}
\epsscale{0.55}
\plotone{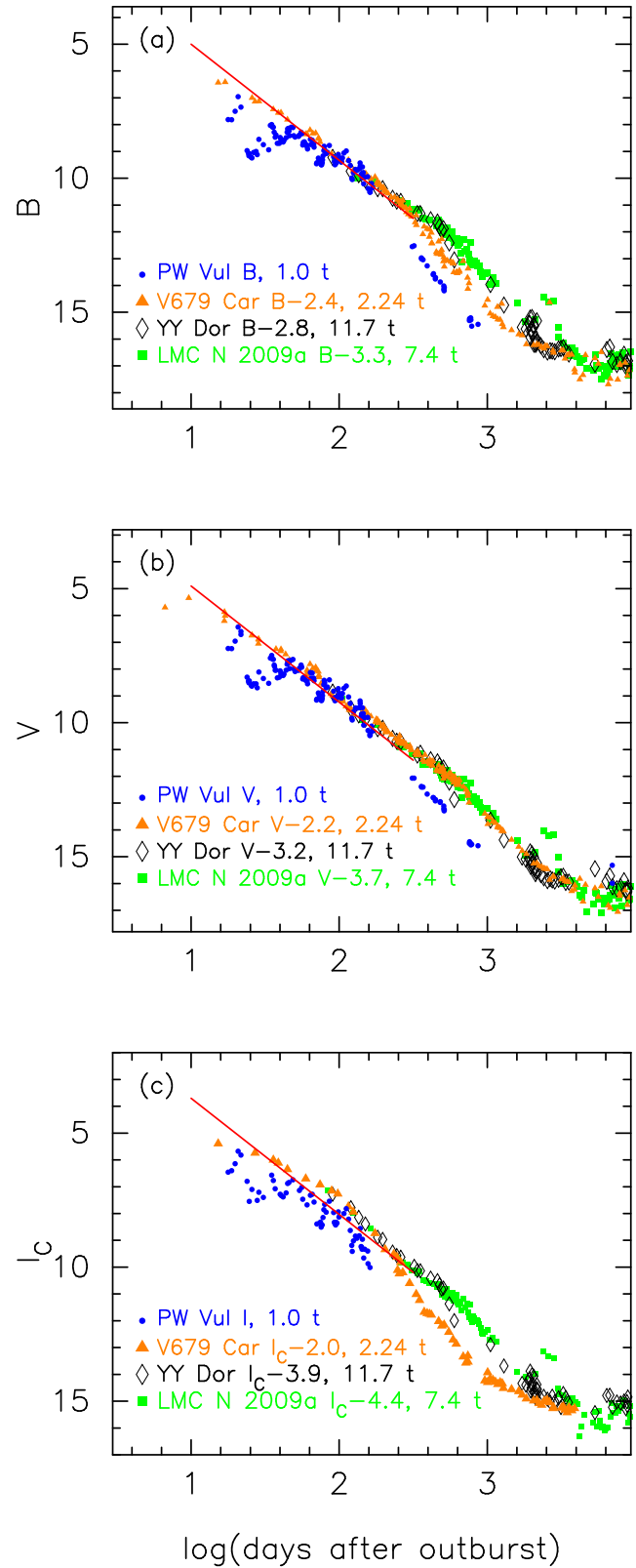}
\caption{
The $BVI$ light curves of PW~Vul are plotted together with
those of V679~Car, YY~Dor, and LMC~N~2009a
The $BV$ data of PW~Vul are teken from \citet{nos85}, \citet{kol86}, 
and \citet{rob95}. 
The $I$ data of PW~Vul are taken from \citet{rob95}. 
\label{pw_vul_v679_car_yy_dor_lmcn_2009a_b_v_i_logscale_3fig}}
\end{figure}


\begin{figure}
\epsscale{0.55}
\plotone{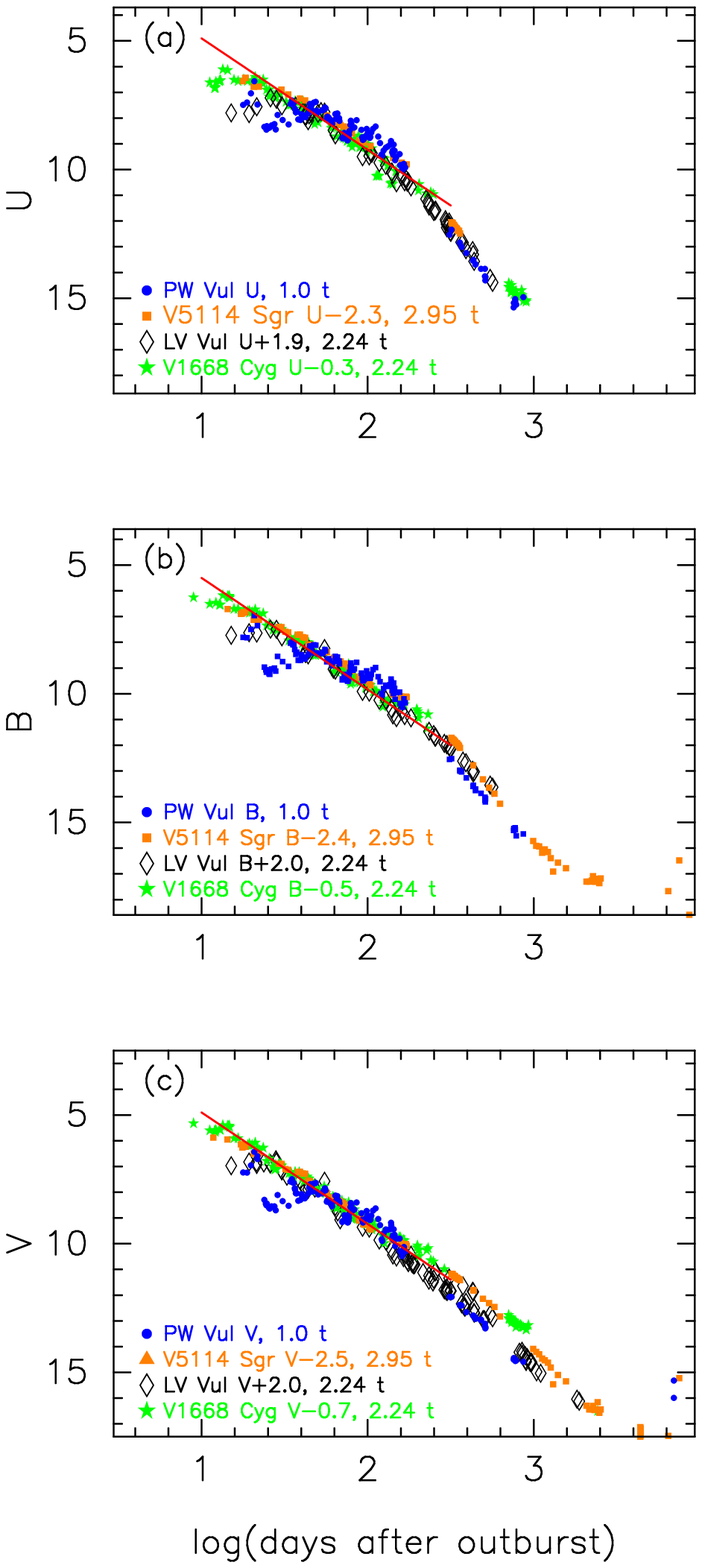}
\caption{
The $UBV$ light curves of PW~Vul are plotted together with
those of V5114~Sgr, LV~Vul, and V1668~Cyg.  
The $UBV$ data of PW~Vul are taken from \citet{nos85}, \citet{kol86}, and 
\citet{rob95}. 
\label{pw_vul_v5114_sgr_lv_vul_v1668_cyg_u_b_v_logscale_3fig}}
\end{figure}

\subsection{PW~Vul 1984\#1}
\label{pw_vul}
Figure \ref{pw_vul_lv_vul_v1668_cyg_v1500_cyg_v_bv_ub_color_logscale} 
shows the light/color curves of PW~Vul on a logarithmic timescale
as well as LV~Vul, V1500~Cyg, and V1668~Cyg.  We deredden the colors
of PW~Vul with $E(B-V)= 0.57$ as obtained in Section \ref{pw_vul_cmd}.
In this figure, we regard
that the $V$ light curve of PW~Vul oscillates between the $V$
light curves of V1668~Cyg and V1500~Cyg 
during $\log t~({\rm day}) \sim1.4$--$2.1$.
Based on the time-stretching method, we have the relation of
\begin{eqnarray}
(m&-&M)_{V, \rm PW~Vul} \cr 
&=& (m - M + \Delta V)_{V, \rm LV~Vul} - 2.5 \log 2.24 \cr
&=& 11.85 + 2.0\pm0.3 - 0.88 = 12.97\pm0.3 \cr
&=& (m - M + \Delta V)_{V, \rm V1668~Cyg} - 2.5 \log 2.24 \cr
&=& 14.6 - 0.7\pm0.3 - 0.88 = 13.02\pm0.3 \cr
&=& (m - M + \Delta V)_{V, \rm V1500~Cyg} - 2.5 \log 3.7 \cr
&=& 12.3 + 2.1\pm0.3 - 1.43 = 12.97\pm0.3.
\label{distance_modulus_pw_vul_t_pyx_lv_vul_v1668_cyg_v1500_cyg}
\end{eqnarray}
Thus, we obtain $f_{\rm s}=2.24$ against the template nova LV~Vul
and confirm the previous result
of $(m-M)_V=13.0\pm0.2$ for PW~Vul \citep{hac15k, hac16kb}.  
From Equations (\ref{distance_modulus_general_dot}) and
(\ref{distance_modulus_pw_vul_t_pyx_lv_vul_v1668_cyg_v1500_cyg}),
we have the relation of
\begin{eqnarray}
(m&-& M')_{V, \rm PW~Vul} \cr 
&\equiv& (m_V - (M_V - 2.5\log f_{\rm s}))_{\rm PW~Vul} \cr
&=& (m-M + \Delta V)_{V,\rm LV~Vul} \cr
&=& 11.85 + 2.0\pm0.3 =13.85\pm0.3.
\label{absolute_mag_pw_vul_lv_vul}
\end{eqnarray}

We further obtain the distance moduli of $BVI_{\rm C}$ bands and examine
the distance and reddening toward PW~Vul.
Figure \ref{pw_vul_v679_car_yy_dor_lmcn_2009a_b_v_i_logscale_3fig} shows
the three band light curves of PW~Vul as well as V679~Car, YY~Dor, and
LMC~N~2009a.
For the $B$ band, we apply Equation (\ref{distance_modulus_general_temp_b})
to Figure \ref{pw_vul_v679_car_yy_dor_lmcn_2009a_b_v_i_logscale_3fig}(a)
and obtain
\begin{eqnarray}
(m&-&M)_{B, \rm PW~Vul} \cr
&=& ((m - M)_B + \Delta B)_{\rm YY~Dor} - 2.5 \log 11.7 \cr
&=& 18.98 - 2.8\pm0.2 - 2.67 = 13.51\pm0.2 \cr
&=& ((m - M)_B + \Delta B)_{\rm LMC~N~2009a} - 2.5 \log 7.4 \cr
&=& 18.98 - 3.3\pm0.2 - 2.17 = 13.51\pm0.2 \cr 
&=& ((m - M)_B + \Delta B)_{\rm V679~Car} - 2.5 \log 2.24 \cr
&=& 16.79 - 2.4\pm0.2 - 0.87 = 13.52\pm0.2, 
\label{distance_modulus_b_pw_vul_v679_car_yy_dor_lmcn2009a}
\end{eqnarray}
where we adopt 
$(m-M)_{B, \rm V679~Car}=16.1 + 1.0\times 0.69= 16.79$.
Thus, we obtain $(m-M)_{B, \rm PW~Vul}= 13.52\pm0.1$.

For the $V$ light curves in Figure 
\ref{pw_vul_v679_car_yy_dor_lmcn_2009a_b_v_i_logscale_3fig}(b),
we similarly obtain
\begin{eqnarray}
(m&-&M)_{V, \rm PW~Vul} \cr
&=& ((m - M)_V + \Delta V)_{\rm YY~Dor} - 2.5 \log 11.7 \cr
&=& 18.86 - 3.2\pm0.2 - 2.67 = 12.99\pm0.2 \cr
&=& ((m - M)_V + \Delta V)_{\rm LMC~N~2009a} - 2.5 \log 7.4 \cr
&=& 18.86 - 3.7\pm0.2 - 2.17 = 12.99\pm0.2 \cr
&=& ((m - M)_V + \Delta V)_{\rm V679~Car} - 2.5 \log 2.24 \cr
&=& 16.1 - 2.2\pm0.2 - 0.87 = 13.03\pm0.2,
\label{distance_modulus_v_pw_vul_v679_car_yy_dor_lmcn2009a}
\end{eqnarray}
where we adopt 
$(m-M)_{V, \rm V679~Car}=16.1$ in Section \ref{v679_car_cmd}.
Thus, we obtain $(m-M)_{V, \rm PW~Vul}= 13.0\pm0.1$.

Applying Equation (\ref{distance_modulus_general_temp_i}) for
the $I_{\rm C}$ band to Figure 
\ref{pw_vul_v679_car_yy_dor_lmcn_2009a_b_v_i_logscale_3fig}(c),
we obtain
\begin{eqnarray}
(m&-&M)_{I, \rm PW~Vul} \cr
&=& ((m - M)_I + \Delta I_C)_{\rm YY~Dor} - 2.5 \log 11.7 \cr
&=& 18.67 - 3.9\pm0.3 - 2.67 = 12.1\pm0.3 \cr
&=& ((m - M)_I + \Delta I_C)_{\rm LMC~N~2009a} - 2.5 \log 7.4 \cr
&=& 18.67 - 4.4\pm0.3 - 2.17 = 12.1\pm0.3 \cr
&=& ((m - M)_I + \Delta I_C)_{\rm V679~Car} - 2.5 \log 2.24 \cr
&=& 15.0 - 2.0\pm0.3 - 0.87 = 12.13\pm0.3, 
\label{distance_modulus_i_pw_vul_v679_car_yy_dor_lmcn2009a}
\end{eqnarray}
where we adopt
$(m-M)_{I, \rm V679~Car}=16.1 - 1.6\times 0.69= 15.0$.
However, we should not use this result because
no $I_{\rm C}$ but only $I$ data of PW~Vul are available \citep{rob95}
and the $I$ light curve of PW~Vul (filled blue circles)
does not accurately follow the other $I_{\rm C}$ data which follow the
universal decline law as shown in Figure
\ref{pw_vul_v679_car_yy_dor_lmcn_2009a_b_v_i_logscale_3fig}(c). 

We also plot the $U$, $B$, and $V$ light curves of PW~Vul together
with those of LV~Vul, V1668~Cyg, and V5114~Sgr in Figure 
\ref{pw_vul_v5114_sgr_lv_vul_v1668_cyg_u_b_v_logscale_3fig}.
We apply Equation (\ref{distance_modulus_general_temp_u}) for the $U$ band to
Figure \ref{pw_vul_v5114_sgr_lv_vul_v1668_cyg_u_b_v_logscale_3fig}(a)
and obtain
\begin{eqnarray}
(m&-&M)_{U, \rm PW~Vul} \cr
&=& ((m - M)_U + \Delta U)_{\rm LV~Vul} - 2.5 \log 2.24 \cr
&=& 12.85 + 1.9\pm0.2 - 0.87 = 13.88\pm0.2 \cr
&=& ((m - M)_U + \Delta U)_{\rm V1668~Cyg} - 2.5 \log 2.24 \cr
&=& 15.1 - 0.3\pm0.2 - 0.87 = 13.93\pm0.2 \cr 
&=& ((m - M)_U + \Delta U)_{\rm V5114~Sgr} - 2.5 \log 2.95 \cr
&=& 17.43 - 2.3\pm0.2 - 1.17 = 13.96\pm0.2, 
\label{distance_modulus_u_pw_vul_v5114_sgr_lv_vul_v1668_cyg}
\end{eqnarray}
where we adopt 
$(m-M)_{U, \rm LV~Vul}=11.85 + (4.75-3.1) \times 0.60 =12.85$, and 
$(m-M)_{U, \rm V1668~Cyg}=14.6 + (4.75-3.1) \times 0.30 =15.10$, and
$(m-M)_{U, \rm V5114~Sgr}=16.65 + (4.75-3.1) \times 0.47 =17.43$.
Thus, we obtain $(m-M)_{U, \rm PW~Vul}= 13.92\pm0.2$.
For the $B$ light curves in Figure 
\ref{pw_vul_v5114_sgr_lv_vul_v1668_cyg_u_b_v_logscale_3fig}(b),
we similarly obtain
\begin{eqnarray}
(m&-&M)_{B, \rm PW~Vul} \cr
&=& ((m - M)_B + \Delta B)_{\rm LV~Vul} - 2.5 \log 2.24 \cr
&=& 12.45 + 2.0\pm0.2 - 0.87 = 13.58\pm0.2 \cr
&=& ((m - M)_B + \Delta B)_{\rm V1668~Cyg} - 2.5 \log 2.24 \cr
&=& 14.9 - 0.5\pm0.2 - 0.87 = 13.53\pm0.2 \cr 
&=& ((m - M)_B + \Delta B)_{\rm V5114~Sgr} - 2.5 \log 2.95 \cr
&=& 17.12 - 2.4\pm0.2 - 1.17 = 13.55\pm0.2, 
\label{distance_modulus_b_pw_vul_v5114_sgr_lv_vul_v1668_cyg}
\end{eqnarray}
where we adopt $(m-M)_{B, \rm LV~Vul}= 11.85 + 1.0\times 0.6 =12.45$, 
$(m-M)_{B, \rm V1668~Cyg}= 14.6 + 1.0\times 0.3 =14.9$, and
$(m-M)_{B, \rm V5114~Sgr}=16.65 + 1.0\times 0.47=17.12$.
Thus, we obtain $(m-M)_{B, \rm PW~Vul}= 13.55\pm0.2$.
We apply Equation (\ref{distance_modulus_general_temp}) to Figure
\ref{pw_vul_v5114_sgr_lv_vul_v1668_cyg_u_b_v_logscale_3fig}(c) and obtain
\begin{eqnarray}
(m&-&M)_{V, \rm PW~Vul} \cr
&=& ((m - M)_V + \Delta V)_{\rm LV~Vul} - 2.5 \log 2.24 \cr
&=& 11.85 + 2.0\pm0.2 - 0.87 = 12.98\pm0.2 \cr
&=& ((m - M)_V + \Delta V)_{\rm V1668~Cyg} - 2.5 \log 2.24 \cr
&=& 14.6 - 0.7\pm0.2 - 0.87 = 13.03\pm0.2 \cr 
&=& ((m - M)_V + \Delta V)_{\rm V5114~Sgr} - 2.5 \log 2.95 \cr
&=& 16.65 - 2.5\pm0.2 - 1.17 = 12.98\pm0.2.
\label{distance_modulus_v_pw_vul_v5114_sgr_lv_vul_v1668_cyg}
\end{eqnarray}
Thus, we obtain $(m-M)_{V, \rm PW~Vul}= 13.0\pm0.1$.
This result is essentially the same as those in Equations
(\ref{distance_modulus_pw_vul_t_pyx_lv_vul_v1668_cyg_v1500_cyg})
and (\ref{distance_modulus_v_pw_vul_v679_car_yy_dor_lmcn2009a}).

We plot these four distance moduli of $U$, $B$, $V$, and $I_{\rm C}$
bands in Figure 
\ref{distance_reddening_v446_her_v533_her_pw_vul_v1419_aql}(c)
by the thin solid green, cyan, and thick solid blue, and thin
solid blue-magenta lines, that is,
$(m-M)_U= 13.92$, $(m-M)_B= 13.55$, and $(m-M)_V= 13.0$ 
and $(m-M)_I= 12.12$, respectively. 
These four lines cross at $d=1.8$~kpc and $E(B-V)=0.57$.


\begin{figure}
\plotone{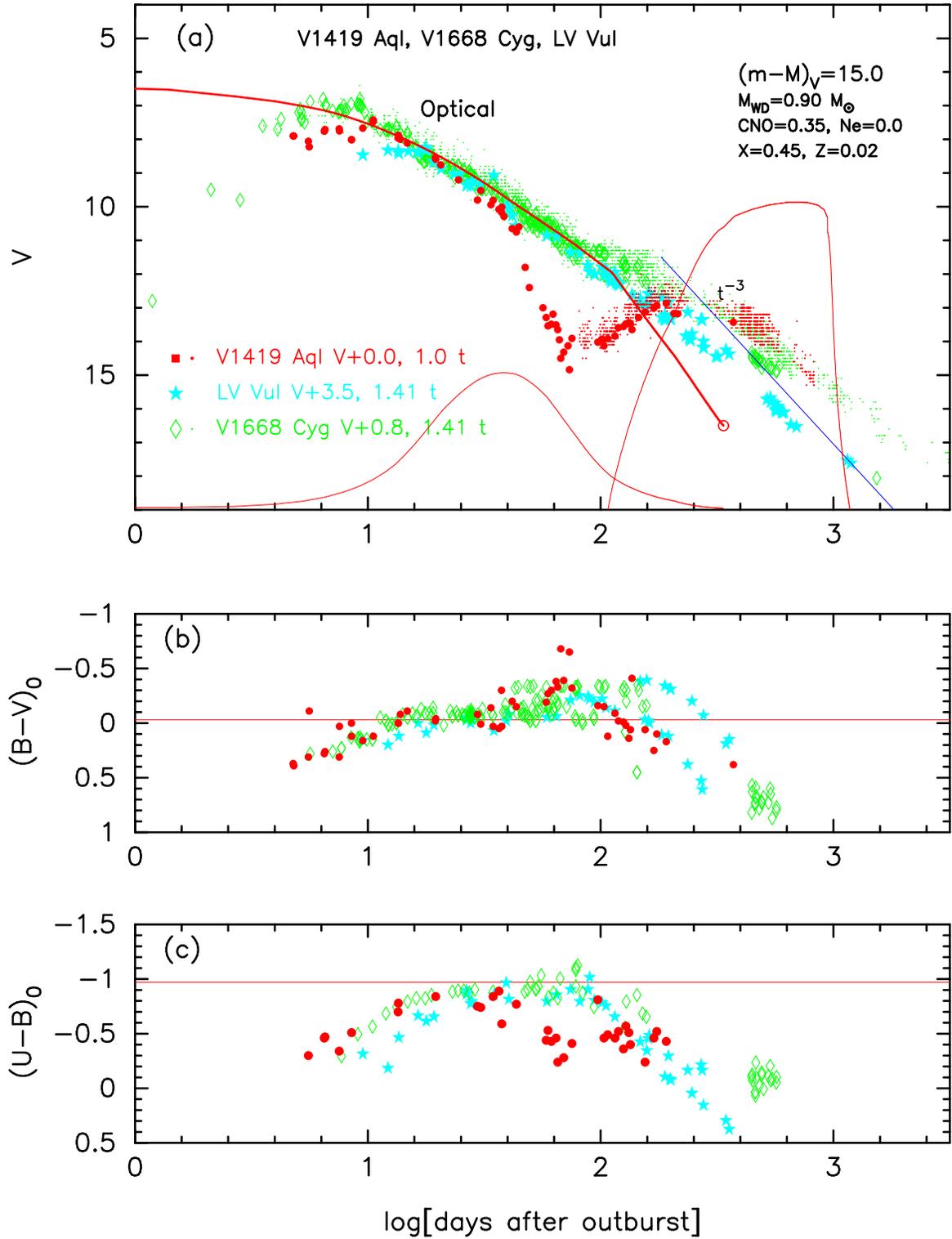}
\caption{
Same as Figure \ref{v574_pup_v1974_cyg_v_bv_logscale},
but for V1419~Aql (filled red circles).  The colors of V1419~Aql
are dereddened with $E(B-V)=0.52$.
The data of V1419~Aql are the same as those in Figure 35 
of Paper II.  
Adopting $(m-M)_V=15.0$ for V1419~Aql,
we plot a $0.90~M_\sun$ WD model (solid red lines) 
with the chemical composition of CO nova 3 \citep{hac16k}.
We add the UV~1455\AA\  flux (left thin solid red line)
and supersoft X-ray flux (right thin solid red line)
of the $0.90~M_\sun$ WD model.
\label{v1419_aql_v1668_cyg_lv_vul_v_bv_ub_color_logscale}}
\end{figure}

\subsection{V1419~Aql 1993}
\label{v1419_aql}
Figure \ref{v1419_aql_v1668_cyg_lv_vul_v_bv_ub_color_logscale}
shows the light/color curves of V1419~Aql
as well as LV~Vul and V1668~Cyg.  We regard that the $V$ light
curve of V1419~Aql follows V1668~Cyg and the upper branch of LV~Vul during
$\log t~({\rm day})\sim1.0$--$1.4$.
Based on the time-stretching method, we have the relation of
\begin{eqnarray}
(m&-&M)_{V, \rm V1419~Aql} \cr 
&=& (m - M + \Delta V)_{V, \rm LV~Vul} - 2.5 \log 1.41 \cr
&=& 11.85 + 3.5\pm0.3 - 0.38 = 14.97\pm0.3 \cr
&=& (m - M + \Delta V)_{V, \rm V1668~Cyg} - 2.5 \log 1.41 \cr
&=& 14.6 + 0.8\pm0.3 - 0.38 = 15.02\pm0.3.
\label{distance_modulus_v1419_aql_lv_vul_v1668_cyg}
\end{eqnarray}
Thus, we obtain $f_{\rm s}=1.41$ against the template nova LV~Vul
and $(m-M)_V=15.0\pm0.2$.
The new value is slightly larger than the previous estimate of 
$(m-M)_V=14.6\pm0.1$ by \citet{hac16kb}, because the timescaling factor
of $f_{\rm s}$ and $\Delta V$ are improved.
From Equations (\ref{distance_modulus_general_dot}) and
(\ref{distance_modulus_v1419_aql_lv_vul_v1668_cyg}),
we have the relation of
\begin{eqnarray}
(m&-& M')_{V, \rm V1419~Aql} \cr 
&\equiv& (m_V - (M_V - 2.5\log f_{\rm s}))_{\rm V1419~Aql} \cr
&=& (m-M + \Delta V)_{V,\rm LV~Vul} \cr
&=& 11.85 + 3.5\pm0.3 =15.35\pm0.3.
\label{absolute_mag_v1419_aql_lv_vul}
\end{eqnarray}


\begin{figure}
\plotone{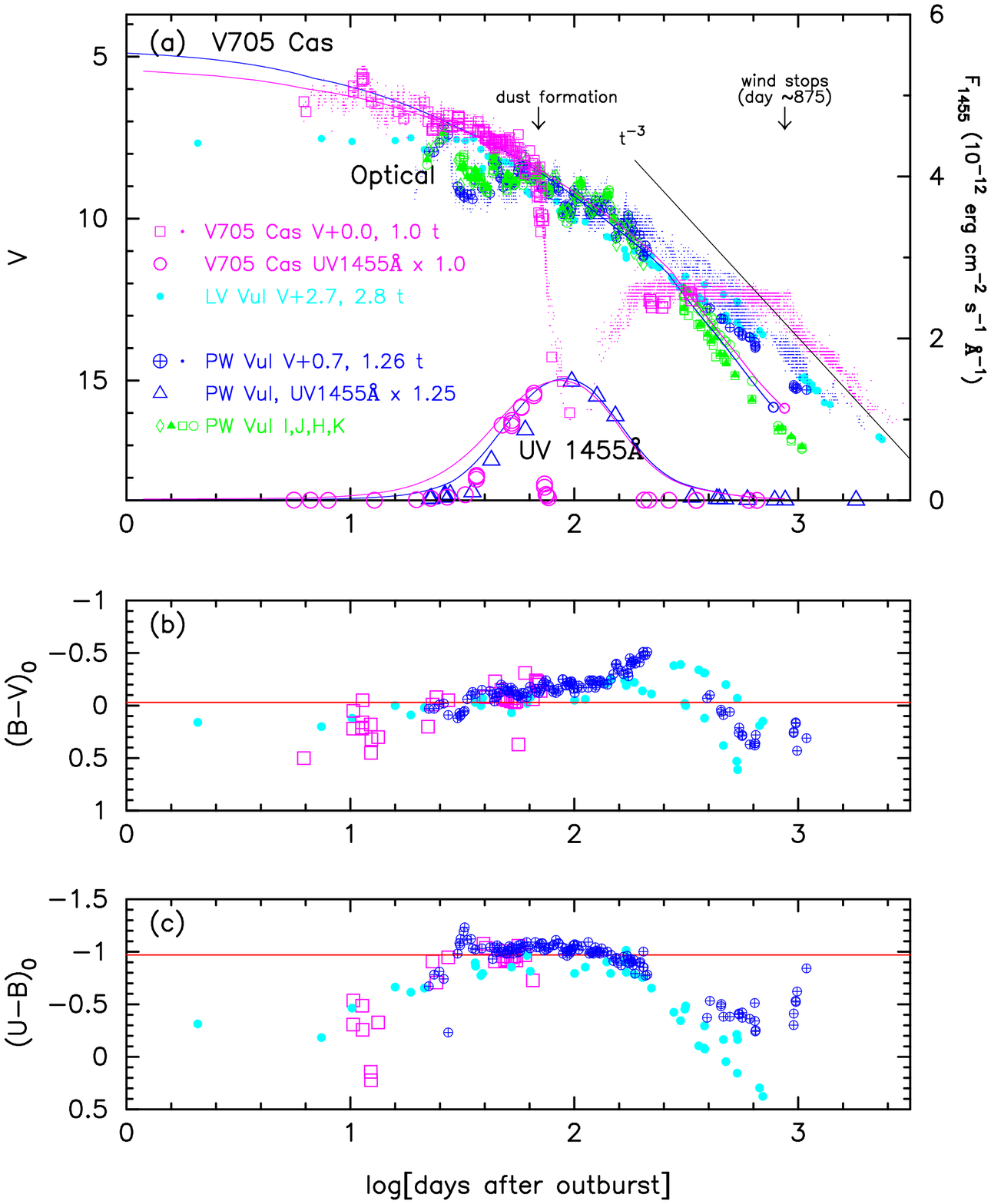}
\caption{
Same as Figure \ref{v574_pup_v1974_cyg_v_bv_logscale},
but for V705~Cas (open magenta squares for optical $V$ and circles for UV).
We also plot (a) the visual magnitudes (magenta dots for V705~Cas and
blue dots for PW~Vul).
The data of V705~Cas and PW~Vul are the same as those in Figure 13 
of \citet{hac15k}.  In panel (a), taking $(m-M)_V=13.45$ for V705~Cas, 
we plot model light curves (solid magenta lines) of a $0.78~M_\sun$ WD
with the envelope chemical composition of CO nova 4 \citep{hac15k}.
We also add a $0.83~M_\sun$ WD (solid blue lines) 
with the same chemical composition of CO nova 4, taking
$(m-M)_V=13.0$ for PW~Vul.
\label{v705_cas_pw_vul_lv_vul_v_bv_ub_color_curve_logscale}}
\end{figure}

\subsection{V705~Cas 1993}
\label{v705_cas}
Figure \ref{v705_cas_pw_vul_lv_vul_v_bv_ub_color_curve_logscale} shows
the light/color curves of V705~Cas 
as well as LV~Vul and PW~Vul.  We deredden the colors of V705~Cas 
with $E(B-V)=0.45$ as obtained in Section \ref{v705_cas_cmd}.
Based on the time-stretching method, we have the relation of
\begin{eqnarray}
(m&-&M)_{V, \rm V705~Cas} \cr 
&=& (m - M + \Delta V)_{V, \rm LV~Vul} - 2.5 \log 2.8 \cr
&=& 11.85 + 2.7\pm0.3 - 1.13 = 13.42\pm0.3 \cr
&=& (m - M + \Delta V)_{V, \rm PW~Vul} - 2.5 \log 1.26 \cr
&=& 13.0 + 0.7\pm0.3 - 0.25 = 13.45\pm0.3.
\label{distance_modulus_v705_cas_lv_vul_pw_vul}
\end{eqnarray}
Thus, we obtain $f_{\rm s}=2.8$ against the template nova LV~Vul
and $(m-M)_V=13.45\pm0.2$.
This value is consistent with the previous value of 
$(m-M)_V=13.4\pm0.1$ estimated by \citet{hac15k}.
From Equations (\ref{distance_modulus_general_dot}) and
(\ref{distance_modulus_v705_cas_lv_vul_pw_vul}), we have the relation of
\begin{eqnarray}
(m&-& M')_{V, \rm V705~Cas} \cr 
&\equiv& (m_V - (M_V - 2.5\log f_{\rm s}))_{\rm V705~Cas} \cr
&=& (m-M + \Delta V)_{V,\rm LV~Vul} \cr
&=& 11.85 + 2.7\pm0.3 =14.55\pm0.3.
\label{absolute_mag_v705_cas_lv_vul}
\end{eqnarray}


\begin{figure}
\plotone{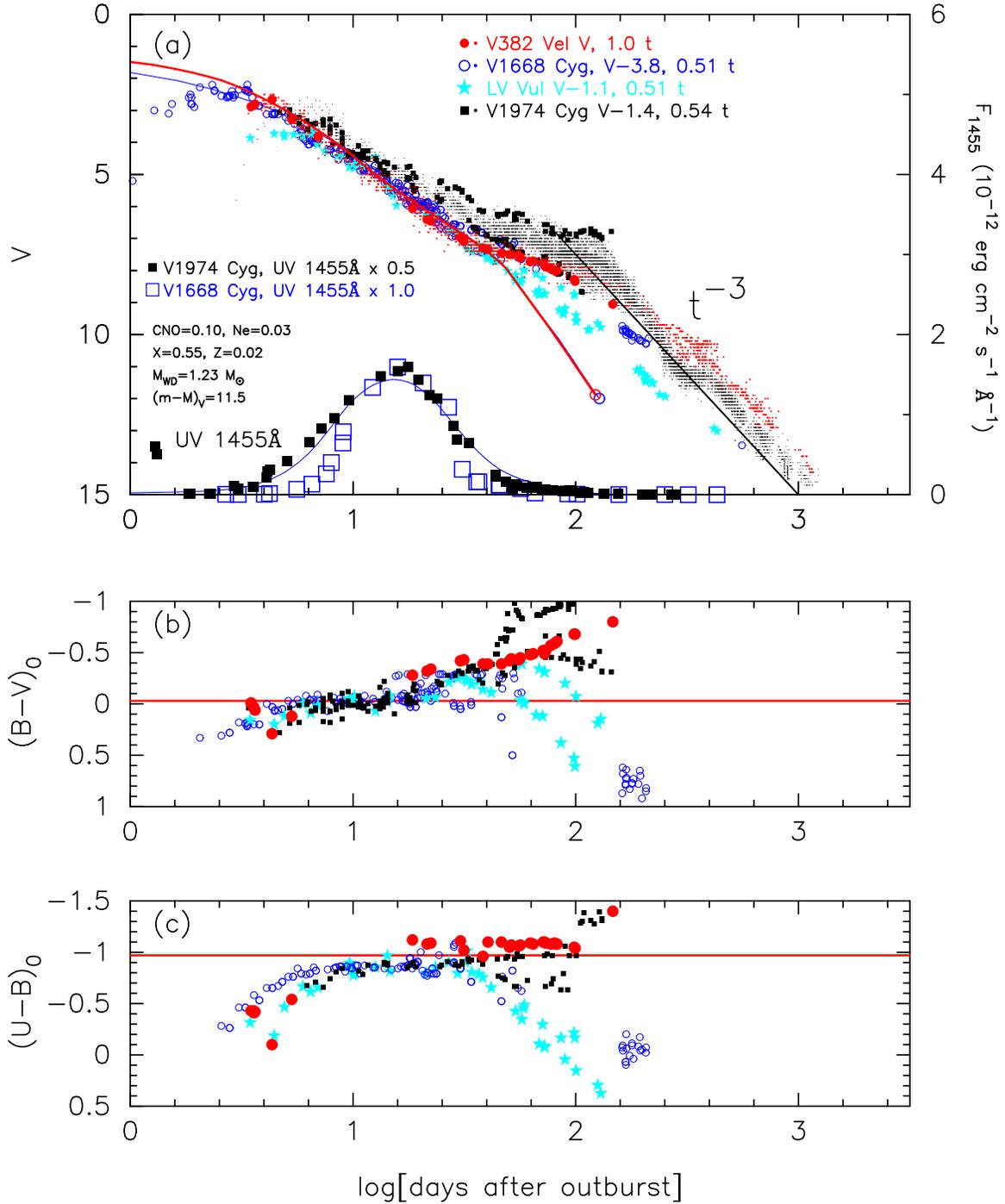}
\caption{
Same as Figure \ref{v574_pup_v1974_cyg_v_bv_logscale}, but for
V382~Vel.  We also plot (a) the visual magnitudes (red dots for V382~Vel,
blue dots for V1668~Cyg, and black dots for V1974~Cyg).   The sources
of V382~Vel data are the same as those in Figure 39 of Paper II.
The other data are the same as those in Figures 
\ref{v574_pup_v1974_cyg_v_bv_logscale} and
\ref{v679_car_lv_vul_v1668_cyg_v_bv_ub_logscale}. 
Taking $(m-M)_V=11.5$ for V382~Vel, we plot a $1.23~M_\sun$ WD model 
(solid red lines) with the chemical composition of Ne nova 2 \citep{hac10k}.
We also add a $0.98~M_\sun$ WD model (solid blue lines)
with the chemical composition of CO nova 3 \citep{hac16k},
taking $(m-M)_V=14.6$ for V1668~Cyg.  
We add a straight solid black line labeled ``$t^{-3}$''
that indicates the homologously expanding ejecta, i.e.,
free expansion, after the optically thick winds stop.
\label{v382_vel_lv_vul_v1668_cyg_v1974_cyg_v_bv_ub_color_logscale}}
\end{figure}


\begin{figure}
\epsscale{0.5}
\plotone{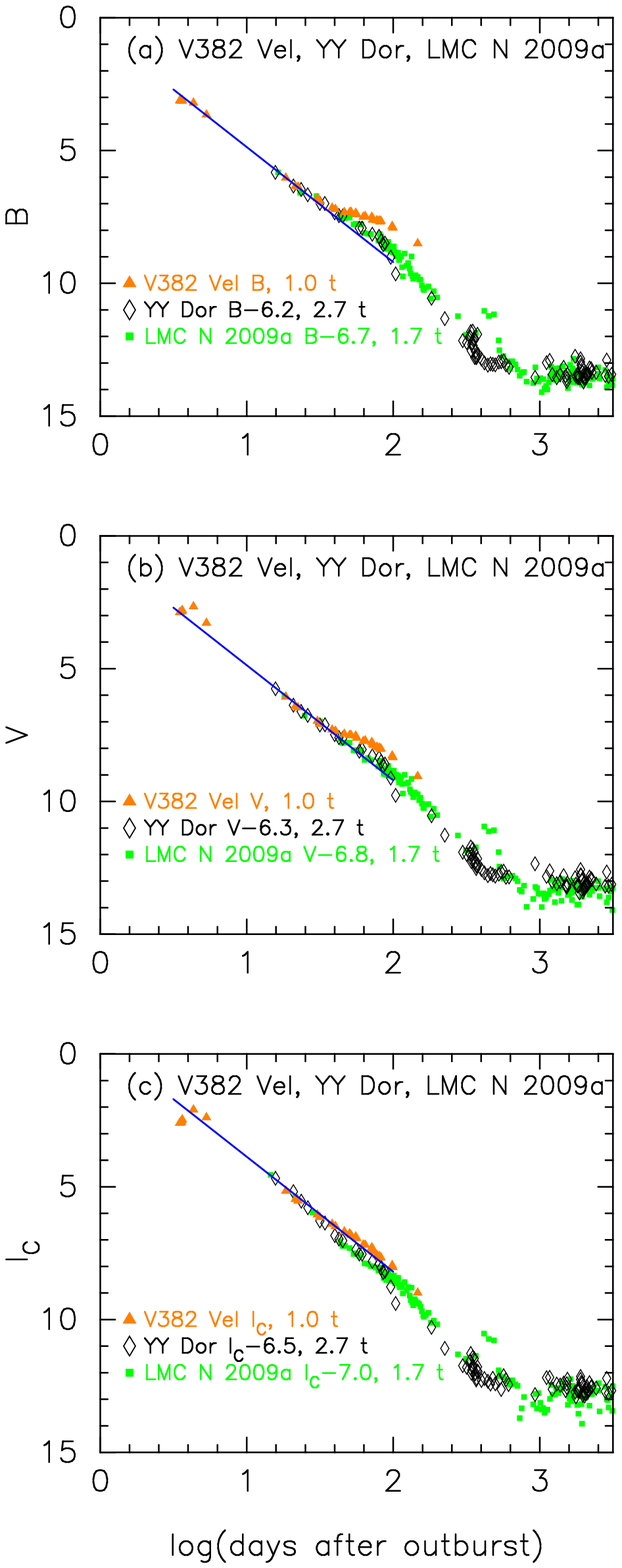}
\caption{
Same as Figure \ref{v679_car_yy_dor_lmcn_2009a_b_v_i_logscale_3fig},
but for V382~Vel.  
The $BVI_{\rm C}$ data of V382~Vel are taken from IAU Circular
Nos. 7176, 7179, 7196, 7209, 7216, 7226, 7232, 7238, and 7277.
\label{v382_vel_yy_dor_lmcn_2009a_b_v_i_logscale_3fig}}
\end{figure}

\subsection{V382~Vel 1999}
\label{v382_vel}
Figure
\ref{v382_vel_lv_vul_v1668_cyg_v1974_cyg_v_bv_ub_color_logscale}
shows the light/color curves of V382~Vel 
as well as LV~Vul, V1668~Cyg, and V1974~Cyg.  The data of V382~Vel are
the same as those in Figures 1 and 27 of Paper II.
We add a straight solid black line labeled ``$t^{-3}$'' 
that indicates the homologously expanding ejecta, i.e., 
free expansion after the optically thick winds stop 
\citep[see, e.g.,][]{woo97, hac06kb}. 
The $V$ light curves of these four novae roughly
overlap each other except for the very early phase of LV~Vul.
Based on the time-stretching method, we have the relation of
\begin{eqnarray}
(m&-&M)_{V, \rm V382~Vel} \cr
&=& (m - M + \Delta V)_{V, \rm LV~Vul} - 2.5 \log 0.51 \cr
&=& 11.85 - 1.1\pm0.3 + 0.73 = 11.48\pm0.3 \cr
&=& (m - M + \Delta V)_{V, \rm V1668~Cyg} - 2.5 \log 0.51 \cr
&=& 14.6 - 3.8\pm0.3 + 0.73 = 11.53\pm0.3 \cr 
&=& (m - M + \Delta V)_{V, \rm V1974~Cyg} - 2.5 \log 0.54 \cr
&=& 12.2 - 1.4\pm0.3 + 0.68 = 11.48\pm0.3.
\label{distance_modulus_v382_vel_lv_vul}
\end{eqnarray}
Thus, we obtain $f_{\rm s}=0.51$ against the template nova LV~Vul
and $(m-M)_V=11.5\pm0.2$.
The result of $(m-M)_V=11.5\pm0.2$ is the same as 
$(m-M)_V=11.5\pm0.2$ in the previous work \citep{hac16k}. 
From Equations (\ref{distance_modulus_general_dot}) and
(\ref{distance_modulus_v382_vel_lv_vul}), 
we have the relation of
\begin{eqnarray}
(m&-& M')_{V, \rm V382~Vel} \cr 
&\equiv& (m_V - (M_V - 2.5\log f_{\rm s}))_{\rm V382~Vel} \cr
&=& \left( (m-M)_V + \Delta V \right)_{\rm LV~Vul} \cr
&=& 11.85 - 1.1\pm0.3 = 10.75\pm0.3.
\label{absolute_mag_v382_vel_lv_vul}
\end{eqnarray}

We obtain the reddening and distance from the time-stretching method.
We plot the $B$, $V$, and $I_{\rm C}$ light curves of V382~Vel together
with those of the LMC novae YY~Dor and LMC~N~2009a in Figure 
\ref{v382_vel_yy_dor_lmcn_2009a_b_v_i_logscale_3fig}.
We apply Equation (\ref{distance_modulus_general_temp_b}) for the $B$ band
to Figure \ref{v382_vel_yy_dor_lmcn_2009a_b_v_i_logscale_3fig}(a) 
and obtain
\begin{eqnarray}
(m&-&M)_{B, \rm V382~Vel} \cr
&=& ((m - M)_B + \Delta B)_{\rm YY~Dor} - 2.5 \log 2.7 \cr
&=& 18.98 - 6.2\pm0.3 - 1.08 = 11.7\pm0.3 \cr
&=& ((m - M)_B + \Delta B)_{\rm LMC~N~2009a} - 2.5 \log 1.7 \cr
&=& 18.98 - 6.7\pm0.3 - 0.58 = 11.7\pm0.3. 
\label{distance_modulus_b_v382_vel_yy_dor_lmcn2009a}
\end{eqnarray}
Thus, we obtain $(m-M)_{B, \rm V382~Vel}= 11.7\pm0.2$.
For the $V$ light curves in Figure 
\ref{v382_vel_yy_dor_lmcn_2009a_b_v_i_logscale_3fig}(b),
we similarly obtain
\begin{eqnarray}
(m&-&M)_{V, \rm V382~Vel} \cr
&=& ((m - M)_V + \Delta V)_{\rm YY~Dor} - 2.5 \log 2.7 \cr
&=& 18.86 - 6.3\pm0.3 - 1.08 = 11.48\pm0.3 \cr
&=& ((m - M)_V + \Delta V)_{\rm LMC~N~2009a} - 2.5 \log 1.7 \cr
&=& 18.86 - 6.8\pm0.3 -0.58 = 11.48\pm0.3.
\label{distance_modulus_v_v382_vel_yy_dor_lmcn2009a}
\end{eqnarray}
Thus, we obtain $(m-M)_{V, \rm V382~Vel}= 11.48\pm0.2$.  We apply 
Equation (\ref{distance_modulus_general_temp_i})  for the $I_{\rm C}$ band
to Figure \ref{v382_vel_yy_dor_lmcn_2009a_b_v_i_logscale_3fig}(c)
and obtain
\begin{eqnarray}
(m&-&M)_{I, \rm V382~Vel} \cr
&=& ((m - M)_I + \Delta I_C)_{\rm YY~Dor} - 2.5 \log 2.7 \cr
&=& 18.67 - 6.5\pm0.3 - 1.08 = 11.09\pm 0.3 \cr
&=& ((m - M)_I + \Delta I_C)_{\rm LMC~N~2009a} - 2.5 \log 1.7 \cr
&=& 18.67 - 7.0\pm0.3 -0.58 = 11.09\pm 0.3. 
\label{distance_modulus_i_v382_vel_yy_dor_lmcn2009a}
\end{eqnarray}
Thus, we obtain $(m-M)_{I, \rm V382~Vel}= 11.09\pm0.2$.


\begin{figure}
\plotone{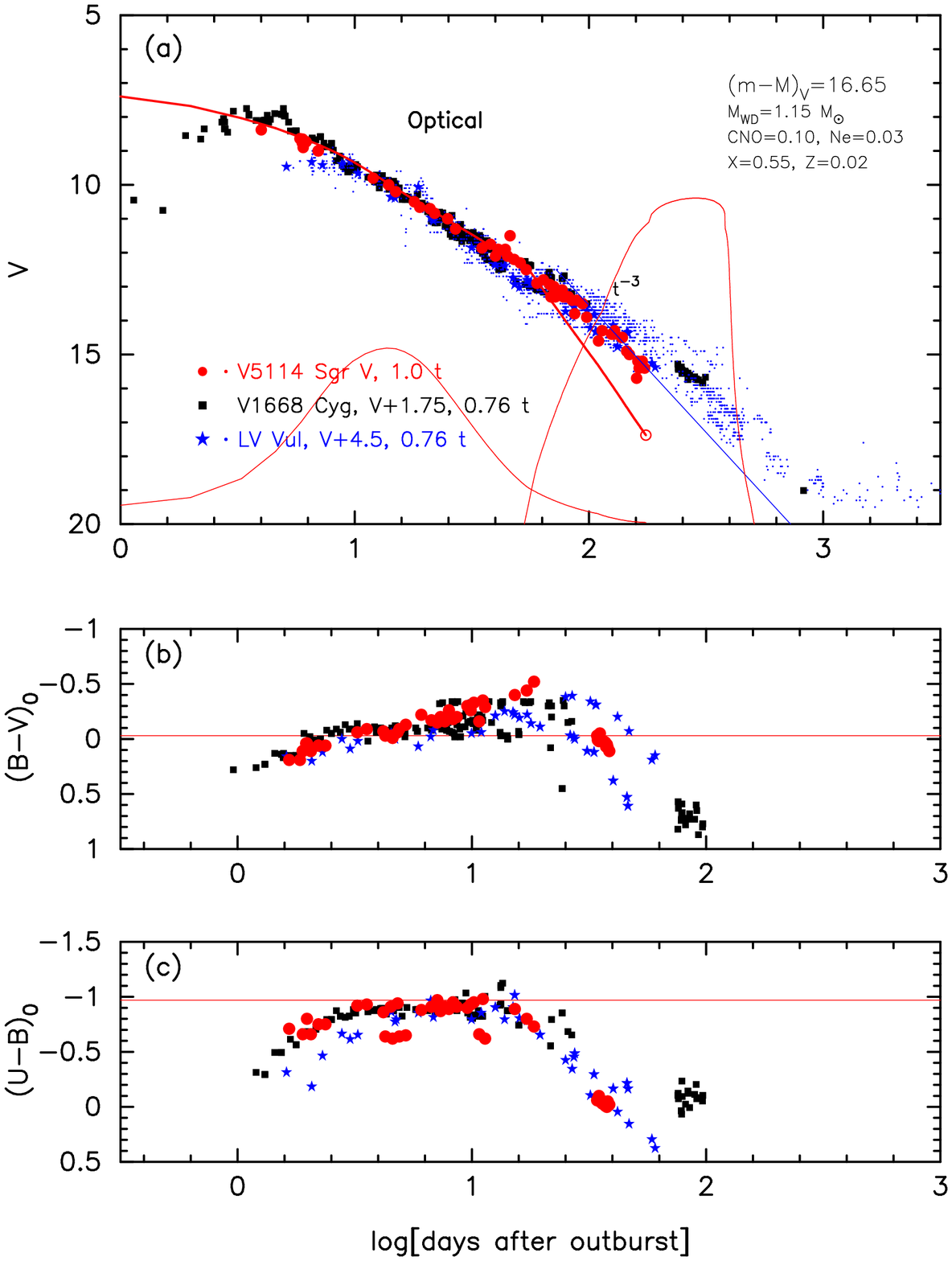}
\caption{
Same as Figure \ref{v574_pup_v1974_cyg_v_bv_logscale},
but for V5114~Sgr (filled red circles).
The data of V5114~Sgr are the same as in Figure 52 of Paper II,
which were taken from \citet{ede06}, SMARTS \citep{wal12},
and IAU Circular Nos. 8306 and 8310.  
Adopting $(m-M)_V=16.65$ for V5114~Sgr, we plot a $1.15~M_\sun$ WD model 
(solid red lines) with the chemical composition of Ne nova 2 \citep{hac10k}.
We add the UV~1455\AA\  flux (left thin solid red line)
and supersoft X-ray flux (right thin solid red line) of 
the $1.15~M_\sun$ WD model.
\label{v5114_sgr_lv_vul_v1668_cyg_v_color_logscale}}
\end{figure}


\begin{figure}
\epsscale{0.55}
\plotone{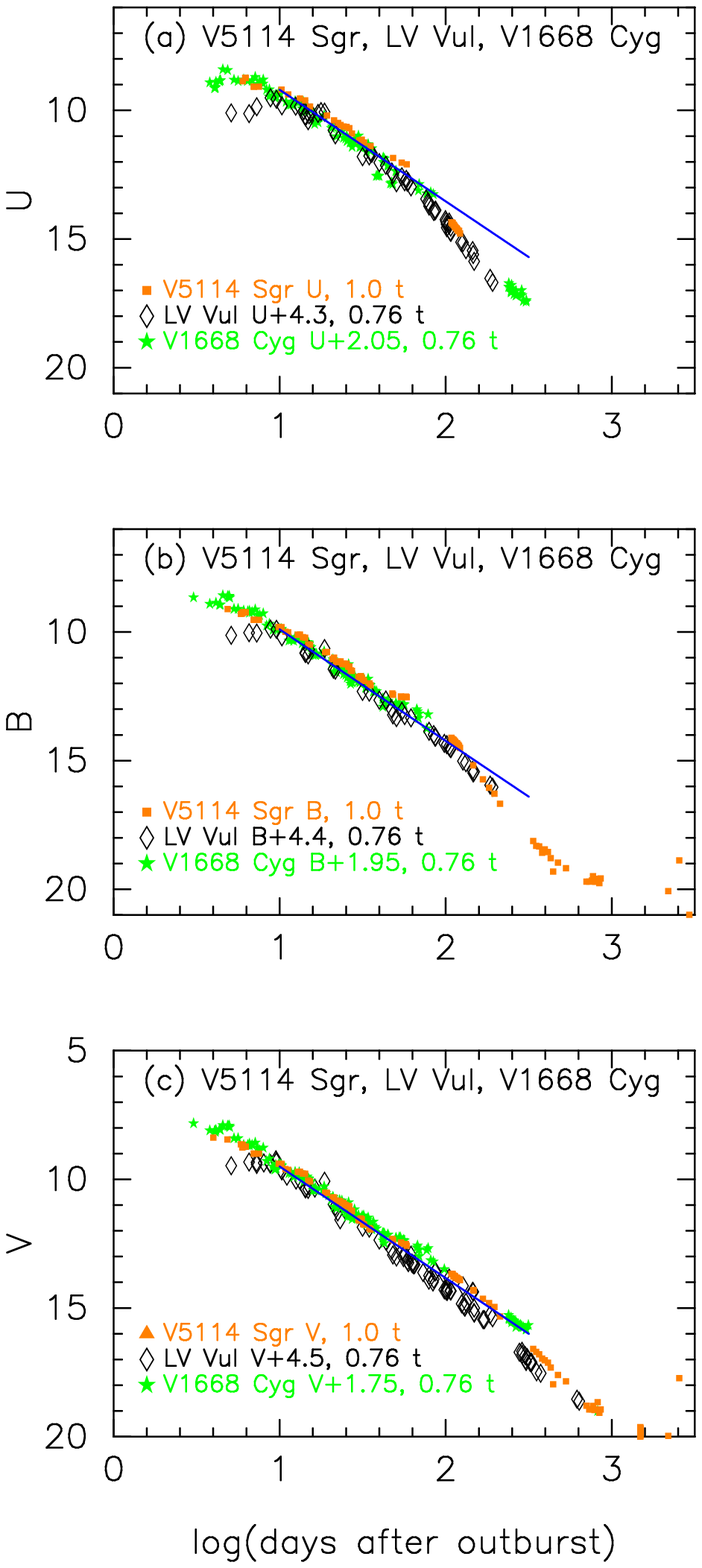}
\caption{
The $UBV$ light curves of V5114~Sgr are plotted together with
those of LV~Vul and V1668~Cyg.  
The $UBV$ data of V5114~Sgr are taken from \citet{ede06} and
IAU Circular No. 8310.  The $BV$ data of V5114~Sgr are taken from 
SMARTS \citep{wal12} and IAU Circular No. 8306.
\label{v5114_sgr_lv_vul_v1668_cyg_u_b_v_logscale_3fig}}
\end{figure}

\subsection{V5114~Sgr 2004}
\label{v5114_sgr}

Figure \ref{v5114_sgr_lv_vul_v1668_cyg_v_color_logscale} shows
the light/color curves of V5114~Sgr as well as LV~Vul and V1668~Cyg.
We regard that the $V$ light curve of V5114~Sgr follows V1668~Cyg and LV~Vul.
Based on the time-stretching method, we have the relation of
\begin{eqnarray}
(m&-&M)_{V, \rm V5114~Sgr} \cr 
&=& (m - M + \Delta V)_{V, \rm LV~Vul} - 2.5 \log 0.76 \cr
&=& 11.85 + 4.5\pm0.2 + 0.30 = 16.65\pm0.2 \cr
&=& (m - M + \Delta V)_{V, \rm V1668~Cyg} - 2.5 \log 0.76 \cr
&=& 14.6 + 1.75\pm0.2 + 0.30 = 16.65\pm0.2.
\label{distance_modulus_v5114_sgr_lv_vul_iv_cep_v1668_cyg}
\end{eqnarray}
Thus, we obtain $f_{\rm s}=0.76$ against the template nova LV~Vul
and $(m-M)_V=16.65\pm0.1$.
This value is slightly larger than the previous value of 
$(m-M)_V=16.5\pm0.1$ estimated by \citet{hac16kb}.
From Equations (\ref{distance_modulus_general_dot}) and
(\ref{distance_modulus_v5114_sgr_lv_vul_iv_cep_v1668_cyg}),
we have the relation of
\begin{eqnarray}
(m&-& M')_{V, \rm V5114~Sgr} \cr 
&\equiv& (m_V - (M_V - 2.5\log f_{\rm s}))_{\rm V5114~Sgr} \cr
&=& (m-M + \Delta V)_{V,\rm LV~Vul} \cr
&=& 11.85 + 4.5\pm0.2 =16.35\pm0.2.
\label{absolute_mag_v5114_sgr_lv_vul}
\end{eqnarray}

We obtain the reddening and distance from the time-stretching method.
We plot the $U$, $B$, and $V$ light curves of V5114~Sgr together
with those of LV~Vul and V1668~Cyg in Figure 
\ref{v5114_sgr_lv_vul_v1668_cyg_u_b_v_logscale_3fig}.
We apply Equation (\ref{distance_modulus_general_temp_u}) for the $U$ band to
Figure \ref{v5114_sgr_lv_vul_v1668_cyg_u_b_v_logscale_3fig}(a) and obtain
\begin{eqnarray}
(m&-&M)_{U, \rm V5114~Sgr} \cr
&=& ((m - M)_U + \Delta U)_{\rm LV~Vul} - 2.5 \log 0.76 \cr
&=& 12.85 + 4.3\pm0.2 + 0.30 = 17.45\pm0.2 \cr
&=& ((m - M)_U + \Delta U)_{\rm V1668~Cyg} - 2.5 \log 0.76 \cr
&=& 15.1 + 2.05\pm0.2 + 0.30 = 17.45\pm0.2, 
\label{distance_modulus_u_v5114_sgr_lv_vul_v1668_cyg}
\end{eqnarray}
where we adopt $(m-M)_{U, \rm LV~Vul}=11.85 + (4.75-3.1) \times 0.60
=12.85$, and $(m-M)_{U, \rm V1668~Cyg}=14.6 + (4.75-3.1) \times 0.30
=15.10$.
Thus, we obtain $(m-M)_{U, \rm V5114~Sgr}= 17.45\pm0.2$.
For the $B$ light curves in Figure 
\ref{v5114_sgr_lv_vul_v1668_cyg_u_b_v_logscale_3fig}(b),
we similarly obtain
\begin{eqnarray}
(m&-&M)_{B, \rm V5114~Sgr} \cr
&=& ((m - M)_B + \Delta B)_{\rm LV~Vul} - 2.5 \log 0.76 \cr
&=& 12.45 + 4.4\pm0.2 + 0.30 = 17.15\pm0.2 \cr
&=& ((m - M)_B + \Delta B)_{\rm LMC~N~2009a} - 2.5 \log 0.76 \cr
&=& 14.9 + 1.95\pm0.2 + 0.30 = 17.15\pm0.2, 
\label{distance_modulus_b_v5114_sgr_yy_dor_lmcn2009a}
\end{eqnarray}
where we adopt $(m-M)_{B, \rm LV~Vul}= 11.85 + 1.0\times 0.6 =12.45$ 
and $(m-M)_{B, \rm V1668~Cyg}= 14.6 + 1.0\times 0.3 =14.9$.
Thus, we obtain $(m-M)_{B, \rm V5114~Sgr}= 17.15\pm0.2$.
We apply Equation (\ref{distance_modulus_general_temp}) to Figure
\ref{v5114_sgr_lv_vul_v1668_cyg_u_b_v_logscale_3fig}(c) and obtain
\begin{eqnarray}
(m&-&M)_{V, \rm V5114~Sgr} \cr
&=& ((m - M)_V + \Delta V)_{\rm LV~Vul} - 2.5 \log 0.76 \cr
&=& 11.85 + 4.5\pm0.2 + 0.30 = 16.65\pm0.2 \cr
&=& ((m - M)_V + \Delta V)_{\rm V1668~Cyg} - 2.5 \log 0.76 \cr
&=& 14.6 + 1.75\pm0.2 + 0.30 = 16.65\pm0.2.
\label{distance_modulus_v_v5114_sgr_yy_dor_lmcn2009a}
\end{eqnarray}
Thus, we obtain $(m-M)_{V, \rm V5114~Sgr}= 16.65\pm0.2$.
This result is essentially the same as that in Equation
(\ref{distance_modulus_v5114_sgr_lv_vul_iv_cep_v1668_cyg}).
We plot these three distance moduli of $U$, $B$, and $V$ bands in
Figure \ref{distance_reddening_v705_cas_v382_vel_v5114_sgr_v2362_cyg}(c)
by the magenta, cyan, and blue lines, that is,
$(m-M)_U= 17.45$, $(m-M)_B= 17.15$, and $(m-M)_V= 16.65$ 
together with Equations (\ref{distance_modulus_ru}),
(\ref{distance_modulus_rb}), and
(\ref{distance_modulus_rv}), respectively.
These three lines cross at $d=10.9$~kpc and $E(B-V)=0.47$.


\begin{figure}
\plotone{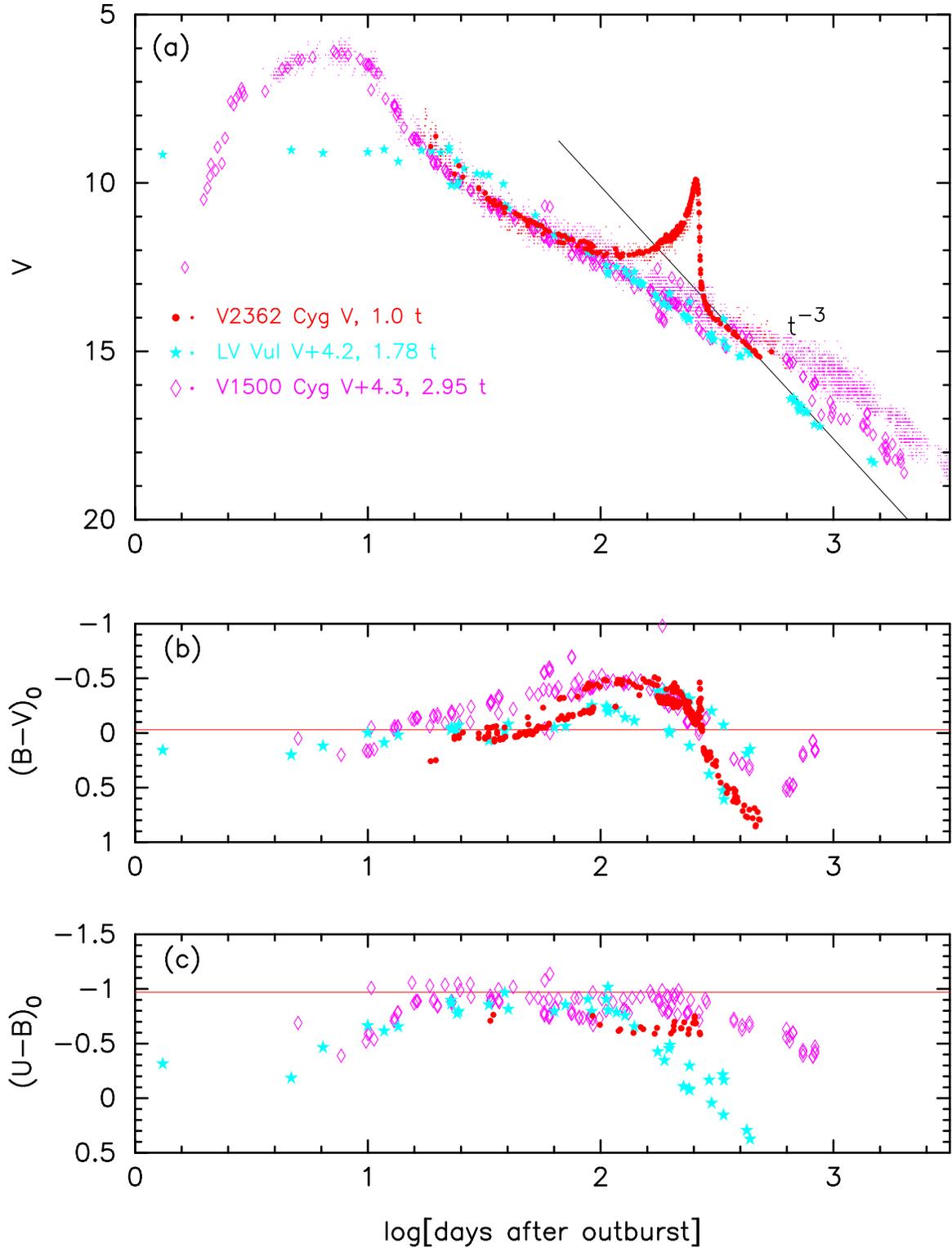}
\caption{
Same as Figure \ref{v574_pup_v1974_cyg_v_bv_logscale}, but for V2362~Cyg
(filled red circles) as well as LV~Vul and V1500~Cyg.  The $(B-V)_0$ 
and $(U-B)_0$ data of V2362~Cyg are dereddened with $E(B-V)=0.60$.
The data of V1500~Cyg and V2362~Cyg are the same as those in Figure 6
of Paper I and Figure 54 of Paper II, respectively.  
The free-free flux decays along the $F_\nu \propto t^{-3}$ line 
after optically thick winds stop.  See text for detail.
\label{v2362_cyg_lv_vul_v1500_cyg_v_bv_ub_color_logscale}}
\end{figure}

\subsection{V2362~Cyg 2006}
\label{v2362_cyg}

Figure \ref{v2362_cyg_lv_vul_v1500_cyg_v_bv_ub_color_logscale}
shows the light/color curves of V2362~Cyg on a logarithmic timescale
as well as LV~Vul and V1500~Cyg.
We regard that the $V$ light curve of V2362~Cyg follows that of V1500~Cyg.
Based on the time-stretching method, we have the relation of
\begin{eqnarray}
(m&-&M)_{V, \rm V2362~Cyg} \cr 
&=& (m - M + \Delta V)_{V, \rm LV~Vul} - 2.5 \log 1.78 \cr
&=& 11.85 + 4.2\pm0.3 - 0.63 = 15.42\pm0.3 \cr
&=& (m - M + \Delta V)_{V, \rm V1500~Cyg} - 2.5 \log 2.95 \cr
&=& 12.3 + 4.3\pm0.2 - 1.18 = 15.42\pm0.2.
\label{distance_modulus_v2362_cyg_lv_vul_v1500_cyg_v1668_cyg_v}
\end{eqnarray}
Thus, we obtain $f_{\rm s}=1.78$ against the template nova LV~Vul
and $(m-M)_V=15.4\pm0.2$.

This new distance modulus in the $V$ band is smaller than the previous value
of $(m-M)_V=15.9\pm0.1$ in Paper II, because 
we improved both the timescaling factor of $f_{\rm s}$
and the vertical fit of $\Delta V$.
The LV~Vul upper branch in the time-stretched color-magnitude diagram
corresponds to the redder (lower) branch after 
the onset of the nebular phase in the $(B-V)_0$ color curves of Figure 
\ref{v2362_cyg_lv_vul_v1500_cyg_v_bv_ub_color_logscale}(b).
We regard that the V2362~Cyg color curves overlap
with this redder branch of LV~Vul in the $(B-V)_0$ color curve.
Thus, we have more carefully determined the timescaling factor of 
$f_{\rm s}$ than the previous work (Paper II).

From Equations (\ref{distance_modulus_general_dot}) and
(\ref{distance_modulus_v2362_cyg_lv_vul_v1500_cyg_v1668_cyg_v}),
we obtain the relation of
\begin{eqnarray}
(m&-& M')_{V, \rm V2362~Cyg} \cr 
&\equiv& (m_V - (M_V - 2.5\log f_{\rm s}))_{\rm V2362~Cyg} \cr
&=& (m-M+ \Delta V)_{V, \rm LV~Vul} \cr
&=& 11.85 +4.2\pm0.3 = 16.05\pm0.3.
\label{absolute_mag_lv_vul_v2362_cyg_v}
\end{eqnarray}


\begin{figure}
\plotone{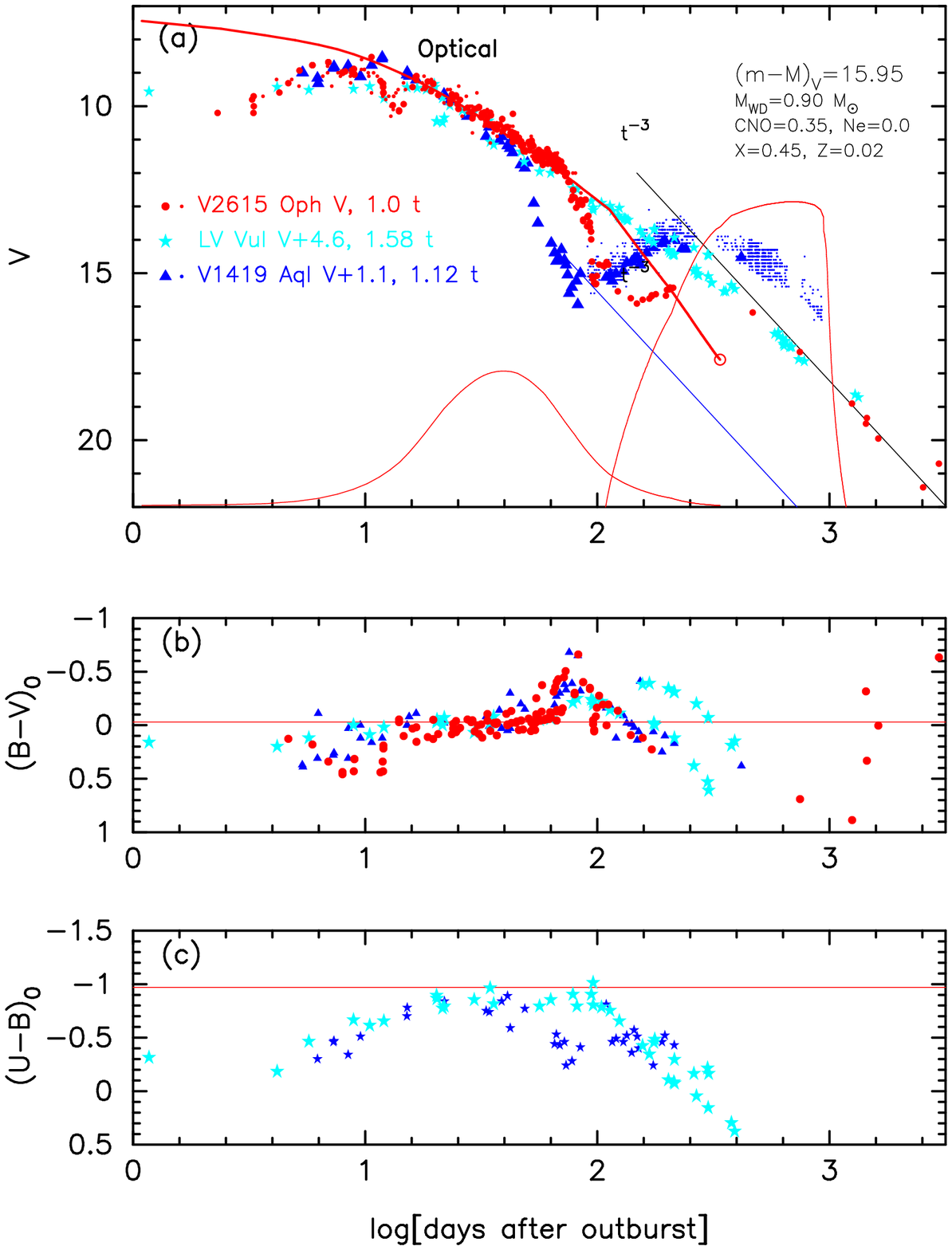}
\caption{
Same as Figure \ref{v574_pup_v1974_cyg_v_bv_logscale}, 
but for V2615~Oph (filled red circles for $V$ and small red dots for visual).
The data of V2615~Oph are the same as those in Figure 63 of Paper II,
but only the $B-V$ data (filled red circles in Figure
\ref{hr_diagram_v2615_oph_v2468_cyg_v2491_cyg_v496_sct_outburst}(a))
of group ``a'' of \citet{mun08a} are shifted by 0.15 mag bluer to overlap
them to the other $B-V$ data of group ``b'' and ``c''
(filled blue circles in Figure
\ref{hr_diagram_v2615_oph_v2468_cyg_v2491_cyg_v496_sct_outburst}(a)).
The other nova data are the same as those in Figure 
\ref{v1419_aql_v1668_cyg_lv_vul_v_bv_ub_color_logscale}.
We plot a $0.90~M_\sun$ WD model (solid red lines)
with the chemical composition of CO nova 3 \citep{hac16k},
taking $(m-M)_V=15.95$ for V2615~Oph.  
We add the UV~1455\AA\  flux (left thin solid red line)
and supersoft X-ray flux (right thin solid red line) of 
the $0.90~M_\sun$ WD model.
See the text for more details.  
\label{v2615_oph_lv_vul_v1419_aql_v_bv_ub_color_logscale}}
\end{figure}


\begin{figure}
\epsscale{0.55}
\plotone{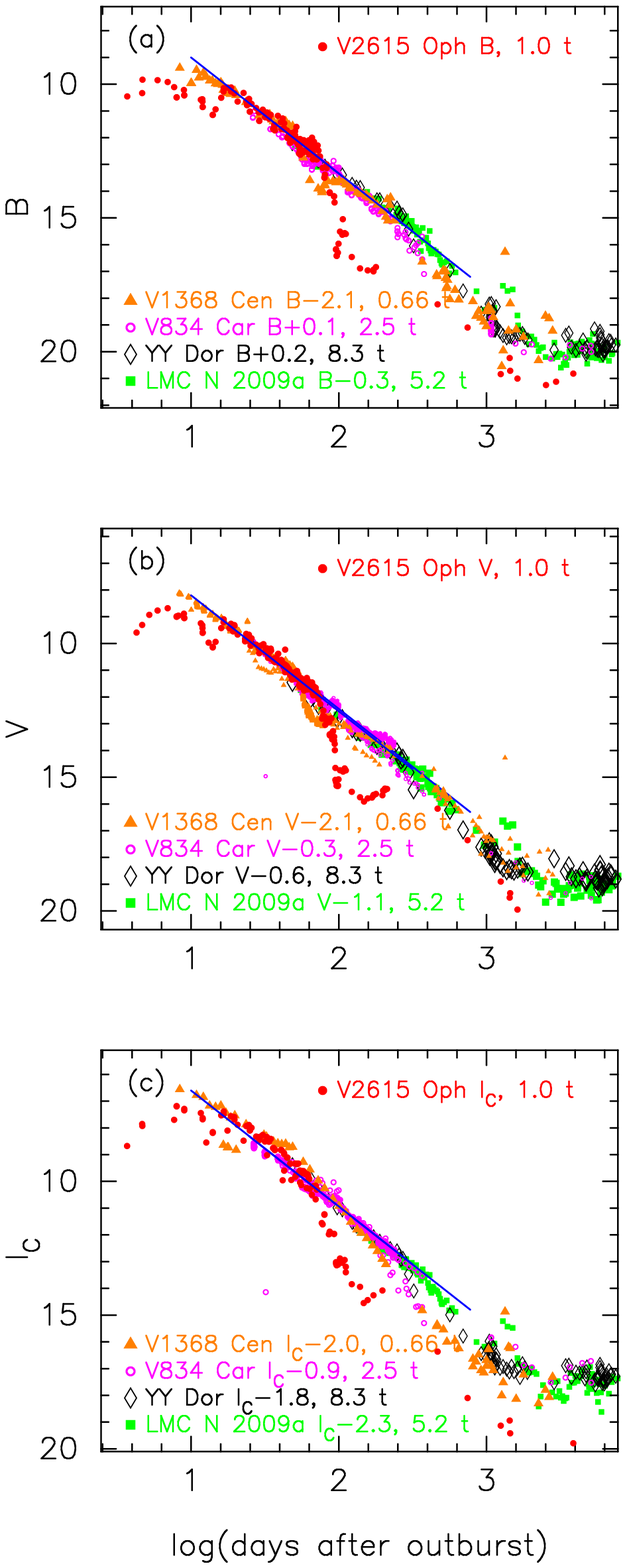}
\caption{
The $BVI_{\rm C}$ light curves of V2615~Oph are plotted together with
those of V1368~Cen, V834~Car, YY~Dor, and LMC~N~2009a.
The $BVI_{\rm C}$ data of V2615~Oph are taken from \citet{mun08a}.
The $BVI_{\rm C}$ data of V2615~Oph are also taken from SMARTS, AAVSO,
and VSOLJ.  The $BVI_{\rm C}$ data of V834~Car and V1368~Cen are taken
from SMARTS and AAVSO.    
\label{v2615_oph_v1368_cen_v834_car_yy_dor_lmcn_2009a_b_v_i_logscale_3fig}}
\end{figure}

\subsection{V2615~Oph 2007}
\label{v2615_oph}

Figure \ref{v2615_oph_lv_vul_v1419_aql_v_bv_ub_color_logscale} shows 
the light/color curves of V2615~Oph as well as LV~Vul and V1419~Aql.
The colors of V2615~Oph are dereddened with $E(B-V)=0.90$ as obtained
in Section \ref{v2615_oph_cmd}.
The $BV$ light curve data of this nova were obtained by \citet{mun08a}. 
However, their $B-V$ data show systematic differences among
three observatories, a, b, and c (see their Figure 1).  
So, we shift the $B-V$ data of group ``a'' (filled red circles in Figure 
\ref{hr_diagram_v2615_oph_v2468_cyg_v2491_cyg_v496_sct_outburst}(a))  
by 0.15 mag bluer.  The $B-V$ data of group ``b'' and ``c'' 
(filled blue circles in Figure 
\ref{hr_diagram_v2615_oph_v2468_cyg_v2491_cyg_v496_sct_outburst}(a))
are not shifted. 
As a result, the $B-V$ data of these three groups broadly overlap each other
in Figures \ref{v2615_oph_lv_vul_v1419_aql_v_bv_ub_color_logscale}(b)
and \ref{hr_diagram_v2615_oph_v2468_cyg_v2491_cyg_v496_sct_outburst}(a).
We also add the $V$ data taken from SMARTS \citep{wal12}.
Based on the time-stretching method, we have the relation of
\begin{eqnarray}
(m&-&M)_{V, \rm V2615~Oph} \cr 
&=& (m - M + \Delta V)_{V, \rm LV~Vul} - 2.5 \log 1.58 \cr
&=& 11.85 + 4.6\pm0.2 - 0.50 = 15.95\pm0.2 \cr
&=& (m - M + \Delta V)_{V, \rm V1419~Aql} - 2.5 \log 1.12 \cr
&=& 15.0 + 1.1\pm0.2 - 0.13 = 15.97\pm0.2.
\label{distance_modulus_v2615_oph_v1419_aql_lv_vul}
\end{eqnarray}
Thus, we obtain $f_{\rm s}=1.58$ against the template nova LV~Vul
and $(m-M)_V=15.95\pm0.2$.
This value is smaller than the previous estimate
of $(m-M)_V=16.5\pm0.1$ (Paper II).
This is due not only to our careful $(B-V)_0$ fitting that improves 
the timescaling factor of $f_{\rm s}$ but also to
the revised vertical $V$ fit to the LV~Vul light curve.
From Equations (\ref{distance_modulus_general_dot}) and
(\ref{distance_modulus_v2615_oph_v1419_aql_lv_vul}), 
we have the relation of
\begin{eqnarray}
(m&-& M')_{V, \rm V2615~Oph} \cr 
&\equiv& (m_V - (M_V + 2.5\log f_s))_{\rm V2615~Oph} \cr
&=& (m-M + \Delta V)_{V,\rm LV~Vul} \cr
&=& 11.85 + 4.6\pm0.2 =16.45\pm0.2.
\label{absolute_mag_v2615_oph_lv_vul}
\end{eqnarray}

We obtain the reddening and distance from the time-stretching method.
We plot the $B$, $V$, and $I_{\rm C}$ light curves of V2615~Oph together
with those of V1368~Cen, V834~Car, and the LMC novae, 
YY~Dor and LMC~N~2009a, in Figure 
\ref{v2615_oph_v1368_cen_v834_car_yy_dor_lmcn_2009a_b_v_i_logscale_3fig}
on a logarithmic timescale.  We apply Equation 
(\ref{distance_modulus_general_temp_b}) 
for the $B$ band  to Figure 
\ref{v2615_oph_v1368_cen_v834_car_yy_dor_lmcn_2009a_b_v_i_logscale_3fig}(a)
and obtain
\begin{eqnarray}
(m&-&M)_{B, \rm V2615~Oph} \cr
&=& ((m - M)_B + \Delta B)_{\rm V1368~Cen} - 2.5 \log 0.66 \cr
&=& 18.53 - 2.1\pm0.4 + 0.45 = 16.88\pm0.3 \cr
&=& ((m - M)_B + \Delta B)_{\rm V834~Car} - 2.5 \log 2.5 \cr
&=& 17.75 + 0.1\pm0.4 - 0.97 = 16.88\pm0.3 \cr
&=& ((m - M)_B + \Delta B)_{\rm YY~Dor} - 2.5 \log 8.3 \cr
&=& 18.98 + 0.2\pm0.4 - 2.3 = 16.88\pm0.3 \cr
&=& ((m - M)_B + \Delta B)_{\rm LMC~N~2009a} - 2.5 \log 5.2 \cr
&=& 18.98 - 0.3\pm0.4 - 1.8 = 16.88\pm0.3, 
\label{distance_modulus_b_v2615_oph_v1535_sco_v834_car_yy_dor_lmcn2009a}
\end{eqnarray}
where we adopt $(m-M)_{B, \rm V1368~Cen}=17.6 + 0.93 = 18.53$ and
$(m-M)_{B, \rm V834~Car}= 17.25 + 0.50 = 17.75$ from \citet{hac19k}.
Thus, we obtain $(m-M)_{B, \rm V2615~Oph}= 16.88\pm0.2$.
For the $V$ light curves in Figure 
\ref{v2615_oph_v1368_cen_v834_car_yy_dor_lmcn_2009a_b_v_i_logscale_3fig}(b),
we similarly obtain
\begin{eqnarray}
(m&-&M)_{V, \rm V2615~Oph} \cr
&=& ((m - M)_V + \Delta V)_{\rm V1368~Cen} - 2.5 \log 0.66 \cr
&=& 17.6 - 2.1\pm0.3 + 0.45 = 15.95\pm0.3 \cr
&=& ((m - M)_V + \Delta V)_{\rm V834~Car} - 2.5 \log 2.5 \cr
&=& 17.25 - 0.3\pm0.3 - 0.97 = 15.98\pm0.3 \cr
&=& ((m - M)_V + \Delta V)_{\rm YY~Dor} - 2.5 \log 8.3 \cr
&=& 18.86 - 0.6\pm0.3 - 2.3 = 15.96\pm0.3 \cr
&=& ((m - M)_V + \Delta V)_{\rm LMC~N~2009a} - 2.5 \log 5.2 \cr
&=& 18.86 - 1.1\pm0.3 - 1.8 = 15.96\pm0.3,
\label{distance_modulus_v_v2615_oph_v1535_sco_v834_car_yy_dor_lmcn2009a}
\end{eqnarray}
where we adopt $(m-M)_{V, \rm V1368~Cen}= 17.6$ 
and $(m-M)_{V, \rm V834~Car}= 17.25$ from \citet{hac19k}.
Thus, we obtain $(m-M)_{V, \rm V2615~Oph}= 15.96\pm0.2$, being consistent
with Equation (\ref{distance_modulus_v2615_oph_v1419_aql_lv_vul}).
We apply Equation (\ref{distance_modulus_general_temp_i}) for the $I_{\rm C}$
band to Figure
\ref{v2615_oph_v1368_cen_v834_car_yy_dor_lmcn_2009a_b_v_i_logscale_3fig}(c)
and obtain
\begin{eqnarray}
(m&-&M)_{I, \rm V2615~Oph} \cr
&=& ((m - M)_I + \Delta I_C)_{\rm V1368~Cen} - 2.5 \log 0.66 \cr
&=& 16.11 - 2.0\pm0.5 + 0.45 = 14.56\pm 0.3 \cr
&=& ((m - M)_I + \Delta I_C)_{\rm V834~Car} - 2.5 \log 2.5 \cr
&=& 16.45 - 0.9\pm0.5 - 0.97 = 14.58\pm 0.3 \cr
&=& ((m - M)_I + \Delta I_C)_{\rm YY~Dor} - 2.5 \log 8.3 \cr
&=& 18.67 - 1.8\pm0.5 - 2.3 = 14.57\pm 0.3 \cr
&=& ((m - M)_I + \Delta I_C)_{\rm LMC~N~2009a} - 2.5 \log 5.2 \cr
&=& 18.67 - 2.3\pm0.5 - 1.8 = 14.57\pm 0.3, 
\label{distance_modulus_i_v2615_oph_v1535_sco_v834_car_yy_dor_lmcn2009a}
\end{eqnarray}
where we adopt $(m-M)_{I, \rm V1368~Cen}=17.6 - 1.6\times 0.93 = 16.11$, and
$(m-M)_{I, \rm V834~Car}=17.25 - 1.6\times 0.50 = 16.45$ from \citet{hac19k}.
Thus, we obtain $(m-M)_{I, \rm V2615~Oph}= 14.57\pm0.2$.
We plot these three distance moduli in Figure
\ref{distance_reddening_v2615_oph_v2468_cyg_v2491_cyg_v496_sct}(a) by
the thin solid magenta, blue, and cyan lines.  These three lines cross
at $d=4.3$~kpc and $E(B-V)=0.90$.


\begin{figure}
\plotone{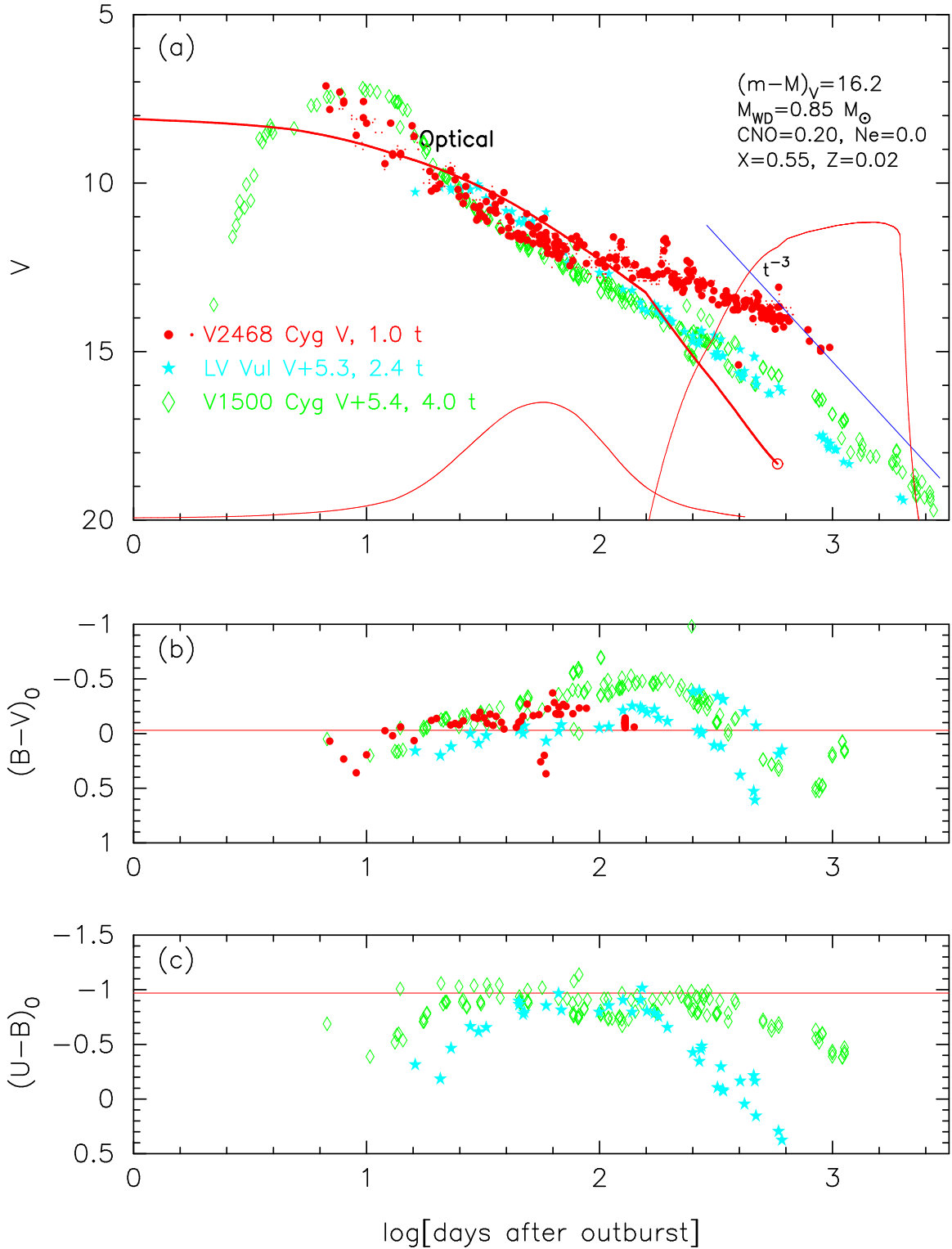}
\caption{
Same as Figure \ref{v574_pup_v1974_cyg_v_bv_logscale}, 
but for V2468~Cyg (filled red circles for $V$ and small red dots for visual).
The data of V2468~Cyg are the same as those in Figure 68 of Paper II.
We plot a $0.85~M_\sun$ WD model (solid red lines)
with the chemical composition of CO nova 4 \citep{hac15k},
taking $(m-M)_V=16.2$ for V2468~Cyg.  
We add the UV~1455\AA\  flux (left thin solid red line)
and supersoft X-ray flux (right thin solid red line) of 
the $0.85~M_\sun$ WD model.
See text for more details.  
\label{v2468_cyg_lv_vul_v1500_cyg_v_color_logscale_no2}}
\end{figure}


\begin{figure}
\epsscale{0.55}
\plotone{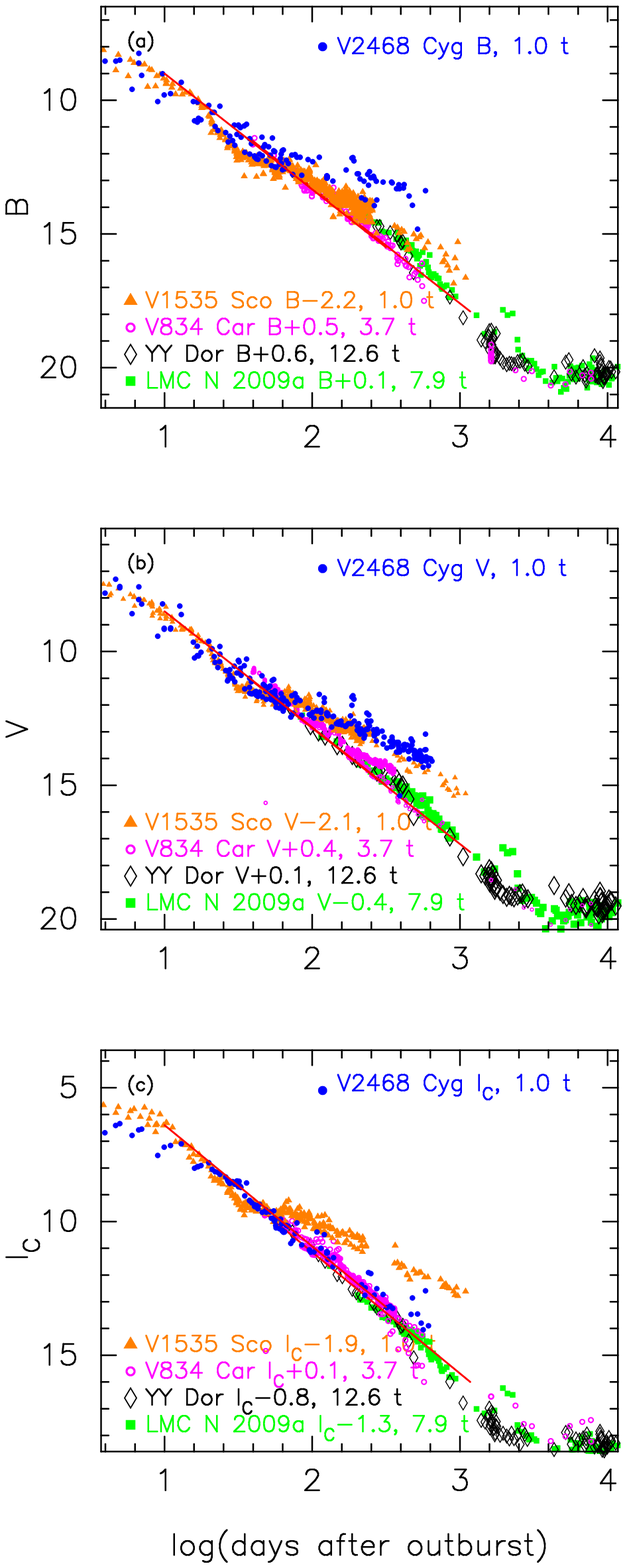}
\caption{
The $BVI_{\rm C}$ light curves of V2468~Cyg are plotted together with
those of V1535~Sco, V834~Car, YY~Dor, and LMC~N~2009a
The $BVI_{\rm C}$ data of V2468~Cyg are taken from AAVSO and VSOLJ.  
The $BVI_{\rm C}$ data of V834~Car are taken from SMARTS and AAVSO.
The $BVI_{\rm C}$ data of V1535~Sco are taken from SMARTS, AAVSO, and
VSOLJ. 
\label{v2468_cyg_v1535_sco_v834_car_yy_dor_lmcn_2009a_b_v_i_logscale_3fig}}
\end{figure}

\subsection{V2468~Cyg 2008\#1}
\label{v2468_cyg}
Figure \ref{v2468_cyg_lv_vul_v1500_cyg_v_color_logscale_no2}
shows the light/color curves of V2468~Cyg 
as well as LV~Vul and V1500~Cyg.  The $(B-V)_0$ color is dereddened
with $E(B-V)=0.65$ as obtained in Section \ref{v2468_cyg_cmd}.
We regard that the early phase
of the V2468~Cyg $V$ light curve is superbright like V1500~Cyg and
then oscillates between LV~Vul and V1500~Cyg
in the phase of $\log t~({\rm day})=1.0$--$2.0$.
The spectra of V1500~Cyg during the superbright
phase (until day $\sim 5$ and above the model $V$ light curve of red line)
are approximated with blackbody \citep{gal76, enn77}.  
After the superbright phase ended,
the spectrum of V1500~Cyg changed to be of free-free emission,
so that $V$ magnitudes are fitted with our model light curve
(solid red line).  Our time-stretching method is applicable to
the part of free-free emission dominated phase of optical light curves.

Based on the time-stretching method, we have the relation of
\begin{eqnarray}
(m&-&M)_{V, \rm V2468~Cyg} \cr 
&=& (m - M + \Delta V)_{V, \rm LV~Vul} - 2.5 \log 2.4 \cr
&=& 11.85 + 5.3\pm0.3 - 0.95 = 16.2\pm0.3 \cr
&=& (m - M + \Delta V)_{V, \rm V1500~Cyg} - 2.5 \log 4.0 \cr
&=& 12.3 + 5.4\pm0.3 - 1.5 = 16.2\pm0.3.
\label{distance_modulus_v2468_cyg_v1668_cyg_lv_vul_iv_cep}
\end{eqnarray}
Thus, we obtain $f_{\rm s}=2.4$ against the template nova LV~Vul
and $(m-M)_V=16.2\pm0.2$.
The newly obtained value is larger than the previous value
of $(m-M)_V=15.6\pm0.1$ estimated in Paper II because we improved
the timescaling factor $f_{\rm s}$ and the vertical $\Delta V$ fit.
From Equations (\ref{distance_modulus_general_dot}) and
(\ref{distance_modulus_v2468_cyg_v1668_cyg_lv_vul_iv_cep}),
we have the relation of
\begin{eqnarray}
(m&-& M')_{V, \rm V2468~Cyg} \cr 
&\equiv& (m_V - (M_V - 2.5\log f_s))_{\rm V2468~Cyg} \cr
&=& (m-M + \Delta V)_{V,\rm LV~Vul} \cr
&=& 11.85 + 5.3\pm0.3 = 17.15\pm0.3.
\label{absolute_mag_v2468_cyg_lv_vul}
\end{eqnarray}

We obtain the reddening and distance from the time-stretching method.
We plot the $B$, $V$, and $I_{\rm C}$ light curves of V2468~Cyg together
with those of V1535~Sco, V834~Car, and the LMC novae, 
YY~Dor and LMC~N~2009a, in Figure 
\ref{v2468_cyg_v1535_sco_v834_car_yy_dor_lmcn_2009a_b_v_i_logscale_3fig}
on a logarithmic timescale.  We apply Equation 
(\ref{distance_modulus_general_temp_b}) 
for the $B$ band  to Figure 
\ref{v2468_cyg_v1535_sco_v834_car_yy_dor_lmcn_2009a_b_v_i_logscale_3fig}(a)
and obtain
\begin{eqnarray}
(m&-&M)_{B, \rm V2468~Cyg} \cr
&=& ((m - M)_B + \Delta B)_{\rm V1535~Sco} - 2.5 \log 1.0 \cr
&=& 19.08 - 2.2\pm0.4 - 0.0 = 16.88\pm0.3 \cr
&=& ((m - M)_B + \Delta B)_{\rm V834~Car} - 2.5 \log 3.7 \cr
&=& 17.75 + 0.5\pm0.4 - 1.42 = 16.83\pm0.3 \cr
&=& ((m - M)_B + \Delta B)_{\rm YY~Dor} - 2.5 \log 12.6 \cr
&=& 18.98 + 0.6\pm0.4 - 2.75 = 16.83\pm0.3 \cr
&=& ((m - M)_B + \Delta B)_{\rm LMC~N~2009a} - 2.5 \log 7.9 \cr
&=& 18.98 + 0.1\pm0.4 - 2.25 = 16.83\pm0.3, 
\label{distance_modulus_b_v2468_cyg_v1535_sco_v834_car_yy_dor_lmcn2009a}
\end{eqnarray}
where we adopt $(m-M)_{B, \rm V1535~Sco}=18.3 + 0.78 = 19.08$ and
$(m-M)_{B, \rm V834~Car}= 17.25 + 0.50 = 17.75$ from \citet{hac19k}.
Thus, we obtain $(m-M)_{B, \rm V2468~Cyg}= 16.85\pm0.2$.
For the $V$ light curves in Figure 
\ref{v2468_cyg_v1535_sco_v834_car_yy_dor_lmcn_2009a_b_v_i_logscale_3fig}(b),
we similarly obtain
\begin{eqnarray}
(m&-&M)_{V, \rm V2468~Cyg} \cr
&=& ((m - M)_V + \Delta V)_{\rm V1535~Sco} - 2.5 \log 1.0 \cr
&=& 18.3 - 2.1\pm0.3 - 0.0 = 16.2\pm0.3 \cr
&=& ((m - M)_V + \Delta V)_{\rm V834~Car} - 2.5 \log 3.7 \cr
&=& 17.25 + 0.4\pm0.3 - 1.42 = 16.23\pm0.3 \cr
&=& ((m - M)_V + \Delta V)_{\rm YY~Dor} - 2.5 \log 12.6 \cr
&=& 18.86 + 0.1\pm0.3 - 2.75 = 16.21\pm0.3 \cr
&=& ((m - M)_V + \Delta V)_{\rm LMC~N~2009a} - 2.5 \log 7.9 \cr
&=& 18.86 - 0.4\pm0.3 - 2.25 = 16.21\pm0.3,
\label{distance_modulus_v_v2468_cyg_v1535_sco_v834_car_yy_dor_lmcn2009a}
\end{eqnarray}
where we adopt $(m-M)_{V, \rm V1535~Sco}= 18.3$ 
and $(m-M)_{V, \rm V834~Car}= 17.25$ from \citet{hac19k}.
Thus, we obtain $(m-M)_{V, \rm V2468~Cyg}= 16.21\pm0.2$, being consistent
with Equation (\ref{distance_modulus_v2468_cyg_v1668_cyg_lv_vul_iv_cep}).
We apply Equation (\ref{distance_modulus_general_temp_i}) for the $I_{\rm C}$
band to Figure
\ref{v2468_cyg_v1535_sco_v834_car_yy_dor_lmcn_2009a_b_v_i_logscale_3fig}(c)
and obtain
\begin{eqnarray}
(m&-&M)_{I, \rm V2468~Cyg} \cr
&=& ((m - M)_I + \Delta I_C)_{\rm V1535~Sco} - 2.5 \log 1.0 \cr
&=& 17.05 - 1.9\pm0.5 - 0.0 = 15.15\pm 0.3 \cr
&=& ((m - M)_I + \Delta I_C)_{\rm V834~Car} - 2.5 \log 3.7 \cr
&=& 16.45 + 0.1\pm0.5 - 1.42 = 15.13\pm 0.3 \cr
&=& ((m - M)_I + \Delta I_C)_{\rm YY~Dor} - 2.5 \log 12.6 \cr
&=& 18.67 - 0.8\pm0.5 - 2.75 = 15.12\pm 0.3 \cr
&=& ((m - M)_I + \Delta I_C)_{\rm LMC~N~2009a} - 2.5 \log 7.9 \cr
&=& 18.67 - 1.3\pm0.5 - 2.25 = 15.12\pm 0.3, 
\label{distance_modulus_i_v2468_cyg_v1535_sco_v834_car_yy_dor_lmcn2009a}
\end{eqnarray}
where we adopt $(m-M)_{I, \rm V1535~Sco}=18.3 - 1.6\times 0.78 = 17.05$, and
$(m-M)_{I, \rm V834~Car}=17.25 - 1.6\times 0.50 = 16.45$ from \citet{hac19k}.
Thus, we obtain $(m-M)_{I, \rm V2468~Cyg}= 15.13\pm0.2$.


\begin{figure}
\plotone{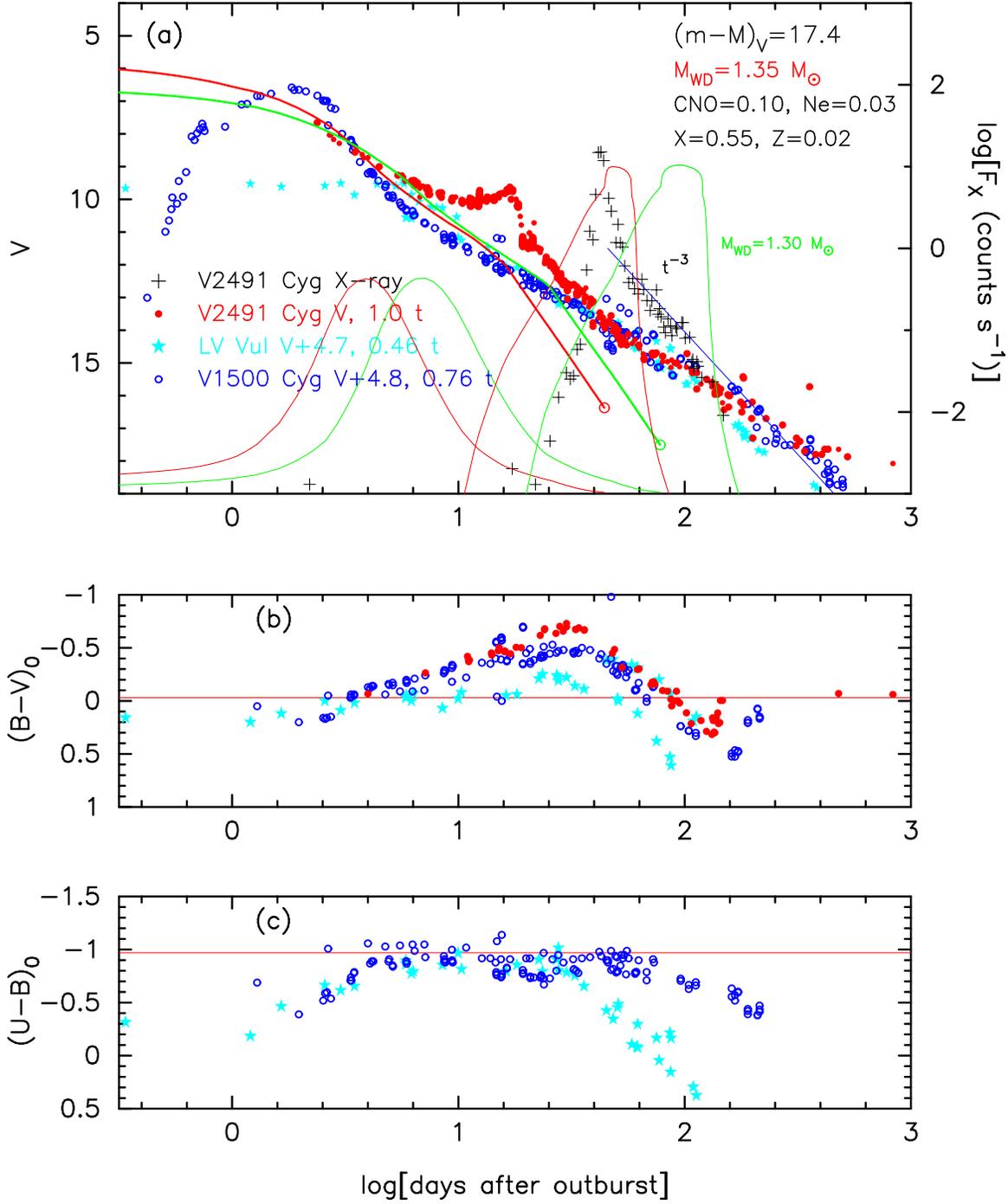}
\caption{
Same as Figure \ref{v574_pup_v1974_cyg_v_bv_logscale},
but for V2491~Cyg (filled red circles).
The $(B-V)_0$ color of V2491~Cyg is dereddened with $E(B-V)=0.45$.
The data of V2491~Cyg are the same as those in Figure 70 of Paper II.
We plot $1.35~M_\sun$ and $1.30~M_\sun$ WD models (solid red/green lines,
respectively) with the chemical composition of Ne nova 2 \citep{hac10k},
taking $(m-M)_V=17.4$ for V2491~Cyg.  
We add the UV~1455\AA\  flux (left thin solid red/green line)
and supersoft X-ray flux (right thin solid red/green line) of 
the $1.35/1.30~M_\sun$ WD models.
See text for more details.  
\label{v2491_cyg_lv_vul_v1500cyg_cyg_v_color_logscale}}
\end{figure}


\begin{figure}
\epsscale{0.5}
\plotone{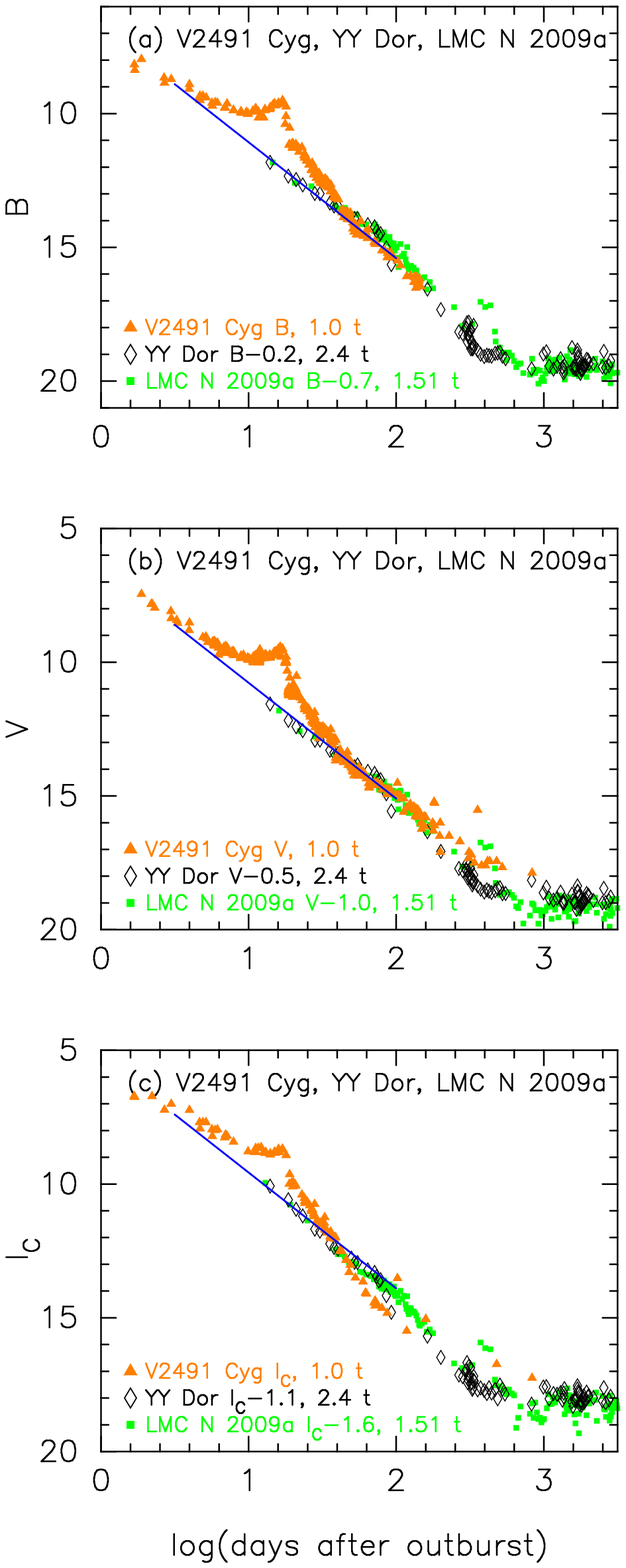}
\caption{
Same as Figure \ref{v679_car_yy_dor_lmcn_2009a_b_v_i_logscale_3fig},
but for V2491~Cyg.  
The $BVI_{\rm C}$ data of V2491~Cyg are taken from AAVSO, VSOLJ,
and \citet{mun11}.
\label{v2491_cyg_yy_dor_lmcn_2009a_b_v_i_logscale_3fig}}
\end{figure}

\subsection{V2491~Cyg 2008\#2}
\label{v2491_cyg}

Figure \ref{v2491_cyg_lv_vul_v1500cyg_cyg_v_color_logscale} shows
the light/color curves of V2491~Cyg as well as LV~Vul and V1500~Cyg.
Based on the time-stretching method, we have the relation of
\begin{eqnarray}
(m&-&M)_{V, \rm V2491~Cyg} \cr 
&=& (m - M + \Delta V)_{V, \rm LV~Vul} - 2.5 \log 0.46 \cr
&=& 11.85 + 4.7\pm0.4 + 0.85 = 17.4\pm0.4 \cr
&=& (m - M + \Delta V)_{V, \rm V1500~Cyg} - 2.5 \log 0.76 \cr
&=& 12.3 + 4.8\pm0.4 + 0.30 = 17.4\pm0.4.
\label{distance_modulus_v2491_cyg_lv_vul_iv_cep_v1668_cyg}
\end{eqnarray}
Thus, we obtain $f_{\rm s}=0.46$ against the template nova LV~Vul
and $(m-M)_V=17.4\pm0.3$.  The new distance modulus 
in the $V$ band is larger than the previous value
of $(m-M)_V=16.5\pm0.1$ (Paper II).  This is because the reddening
is revised to be $E(B-V)=0.45$ and the timescaling factor is improved
to be $f_{\rm s}=0.46$.  We carefully
redetermine the timescaling factor by fitting the $(B-V)_0$ color
curve of V2491~Cyg (filled red circles) with that of LV~Vul (filled
cyan stars) and V1500~Cyg (open blue circles) as shown in Figure
\ref{v2491_cyg_lv_vul_v1500cyg_cyg_v_color_logscale}(b).
From Equations (\ref{distance_modulus_general_dot}) and
(\ref{distance_modulus_v2491_cyg_lv_vul_iv_cep_v1668_cyg}),
we obtain
\begin{eqnarray}
(m&-& M')_{V, \rm V2491~Cyg} \cr 
&\equiv& (m_V - (M_V - 2.5\log f_{\rm s}))_{\rm V2491~Cyg} \cr
&=& (m-M+ \Delta V)_{V, \rm LV~Vul} \cr
&=& 11.85 +4.7\pm0.4 = 16.55\pm0.4.
\label{absolute_mag_lv_vul_v2491_cyg}
\end{eqnarray}

We obtain the reddening and distance from the time-stretching method.
We plot the $B$, $V$, and $I_{\rm C}$ light curves of V2491~Cyg together
with those of the LMC novae YY~Dor and LMC~N~2009a in Figure 
\ref{v2491_cyg_yy_dor_lmcn_2009a_b_v_i_logscale_3fig} on a logarithmic
timescale.  We apply Equation (\ref{distance_modulus_general_temp_b}) 
for the $B$ band  to Figure 
\ref{v2491_cyg_yy_dor_lmcn_2009a_b_v_i_logscale_3fig}(a) and obtain
\begin{eqnarray}
(m&-&M)_{B, \rm V2491~Cyg} \cr
&=& ((m - M)_B + \Delta B)_{\rm YY~Dor} - 2.5 \log 2.4 \cr
&=& 18.98 - 0.2\pm0.4 - 0.95 = 17.83\pm0.4 \cr
&=& ((m - M)_B + \Delta B)_{\rm LMC~N~2009a} - 2.5 \log 1.51 \cr
&=& 18.98 - 0.7\pm0.4 - 0.45 = 17.83\pm0.4. 
\label{distance_modulus_b_v2491_cyg_yy_dor_lmcn2009a}
\end{eqnarray}
Thus, we obtain $(m-M)_{B, \rm V2491~Cyg}= 17.83\pm0.3$.
For the $V$ light curves in Figure 
\ref{v2491_cyg_yy_dor_lmcn_2009a_b_v_i_logscale_3fig}(b),
we similarly obtain
\begin{eqnarray}
(m&-&M)_{V, \rm V2491~Cyg} \cr
&=& ((m - M)_V + \Delta V)_{\rm YY~Dor} - 2.5 \log 2.4 \cr
&=& 18.86 - 0.5\pm0.3 - 0.95 = 17.41\pm0.3 \cr
&=& ((m - M)_V + \Delta V)_{\rm LMC~N~2009a} - 2.5 \log 1.51 \cr
&=& 18.86 - 1.0\pm0.3 -0.45 = 17.41\pm0.3.
\label{distance_modulus_v_v2491_cyg_yy_dor_lmcn2009a}
\end{eqnarray}
Thus, we obtain $(m-M)_{V, \rm V2491~Cyg}= 17.41\pm0.2$.
We apply Equation (\ref{distance_modulus_general_temp_i}) for the $I_{\rm C}$
band to Figure
\ref{v2491_cyg_yy_dor_lmcn_2009a_b_v_i_logscale_3fig}(c) and obtain
\begin{eqnarray}
(m&-&M)_{I, \rm V2491~Cyg} \cr
&=& ((m - M)_I + \Delta I_C)_{\rm YY~Dor} - 2.5 \log 2.4 \cr
&=& 18.67 - 1.1\pm0.5 - 0.95 = 16.62\pm 0.5 \cr
&=& ((m - M)_I + \Delta I_C)_{\rm LMC~N~2009a} - 2.5 \log 1.51 \cr
&=& 18.67 - 1.6\pm0.5 - 0.45 = 16.62\pm 0.5. 
\label{distance_modulus_i_v2491_cyg_yy_dor_lmcn2009a}
\end{eqnarray}
Thus, we obtain $(m-M)_{I, \rm V2491~Cyg}= 16.62\pm0.3$.

We plot these three distance moduli of $B$, $V$, and $I_{\rm C}$ bands in
Figure \ref{distance_reddening_v2615_oph_v2468_cyg_v2491_cyg_v496_sct}(c)
by the magenta, blue, and cyan lines, that is,
$(m-M)_B= 17.83$, $(m-M)_V= 17.41$, and $(m-M)_I= 16.62$
together with Equations (\ref{distance_modulus_rb}), 
(\ref{distance_modulus_rv}), and (\ref{distance_modulus_ri}).
These three lines broadly cross at $d=15.9$~kpc and $E(B-V)=0.45$.


\begin{figure}
\plotone{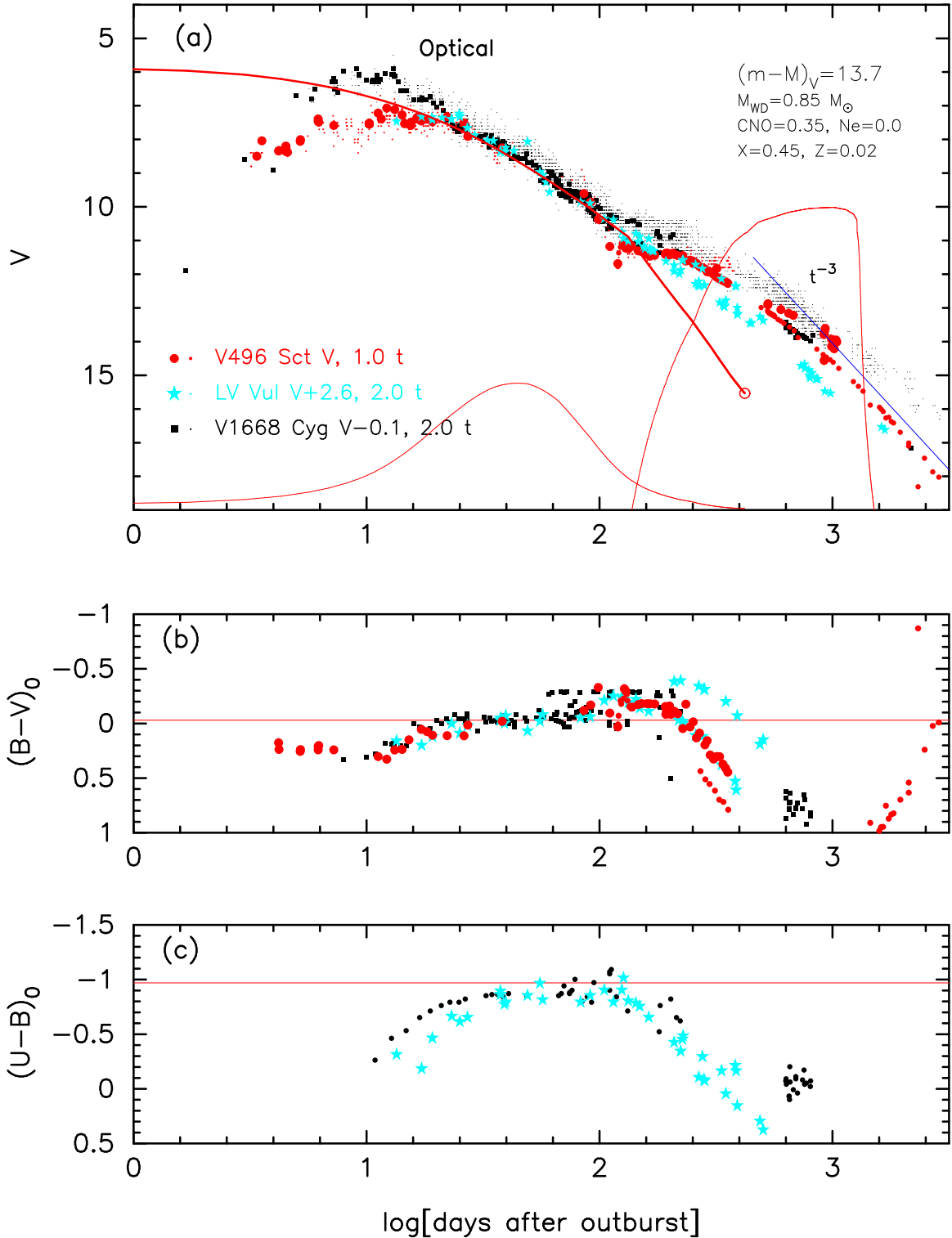}
\caption{
Same as Figure \ref{v574_pup_v1974_cyg_v_bv_logscale},
but for V496~Sct (filled red circles for $V$ and small red dots for visual).
The data of V496~Sct are the same as those in Figure 73 of Paper II.
We plot a $0.85~M_\sun$ WD model (solid red lines)
with the chemical composition of CO nova 3 \citep{hac16k},
taking $(m-M)_V=13.7$ for V496~Sct.  
We add the UV~1455\AA\  flux (left thin solid red line)
and supersoft X-ray flux (right thin solid red line) of 
the $0.85~M_\sun$ WD model.
See the text for more details.  
\label{v496_sct_lv_vul_v1668_cyg_v_bv_ub_color_logscale_no2}}
\end{figure}


\begin{figure}
\plotone{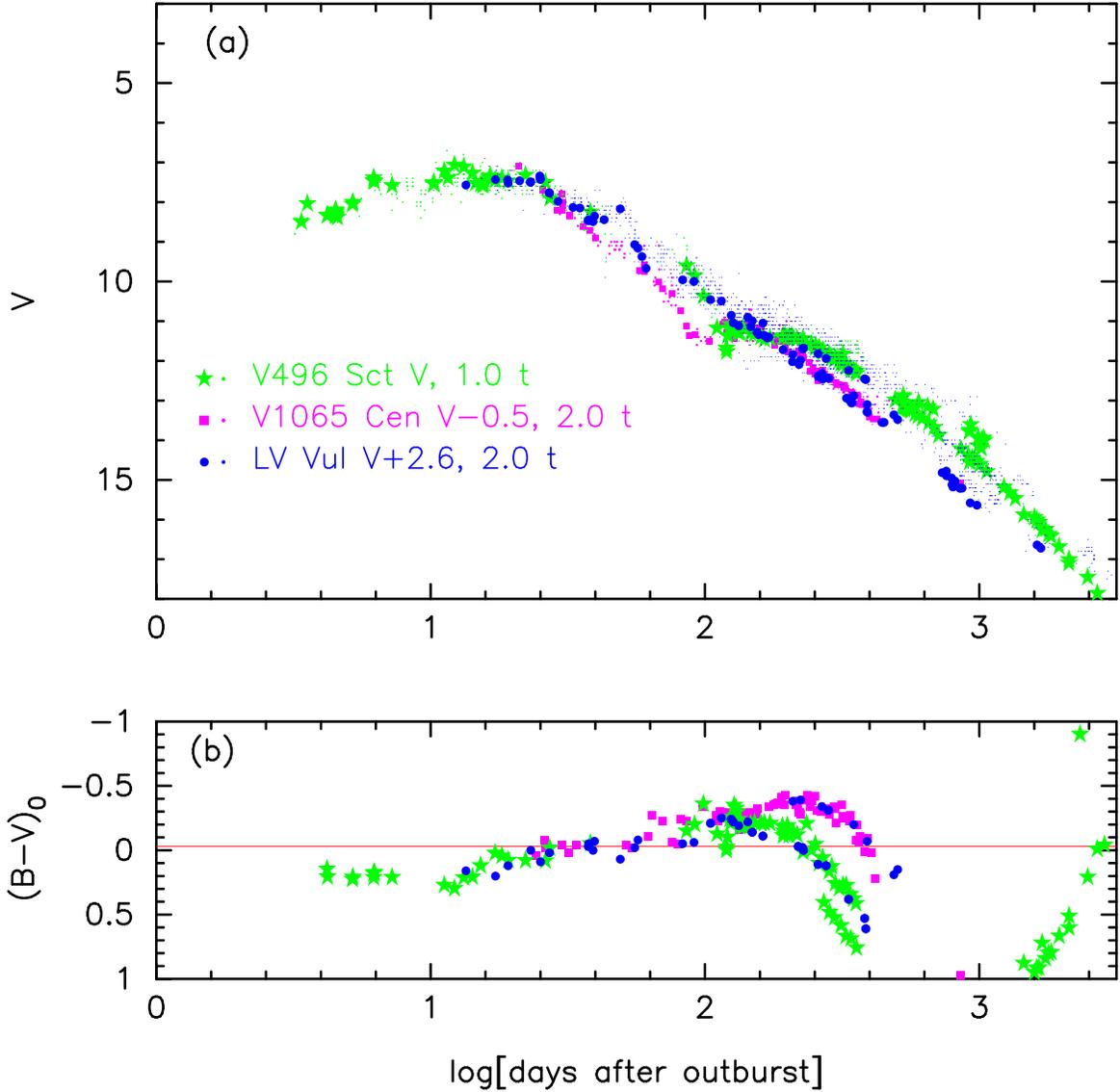}
\caption{
Same as Figure 
\ref{v496_sct_lv_vul_v1668_cyg_v_bv_ub_color_logscale_no2},
but for a different set of V496~Sct, V1065~Cen, and LV~Vul.
The data of V496~Sct are the same as those in Figure
\ref{v496_sct_lv_vul_v1668_cyg_v_bv_ub_color_logscale_no2}.
We also plot the $BV$ data of V1065~Cen and LV~Vul.
The data of V1065~Cen are the same as those in Figure 4 of \citet{hac18k}.  
\label{v496_sct_v1065_cen_lv_vul_v_bv_ub_color_logscale}}
\end{figure}


\begin{figure}
\epsscale{0.5}
\plotone{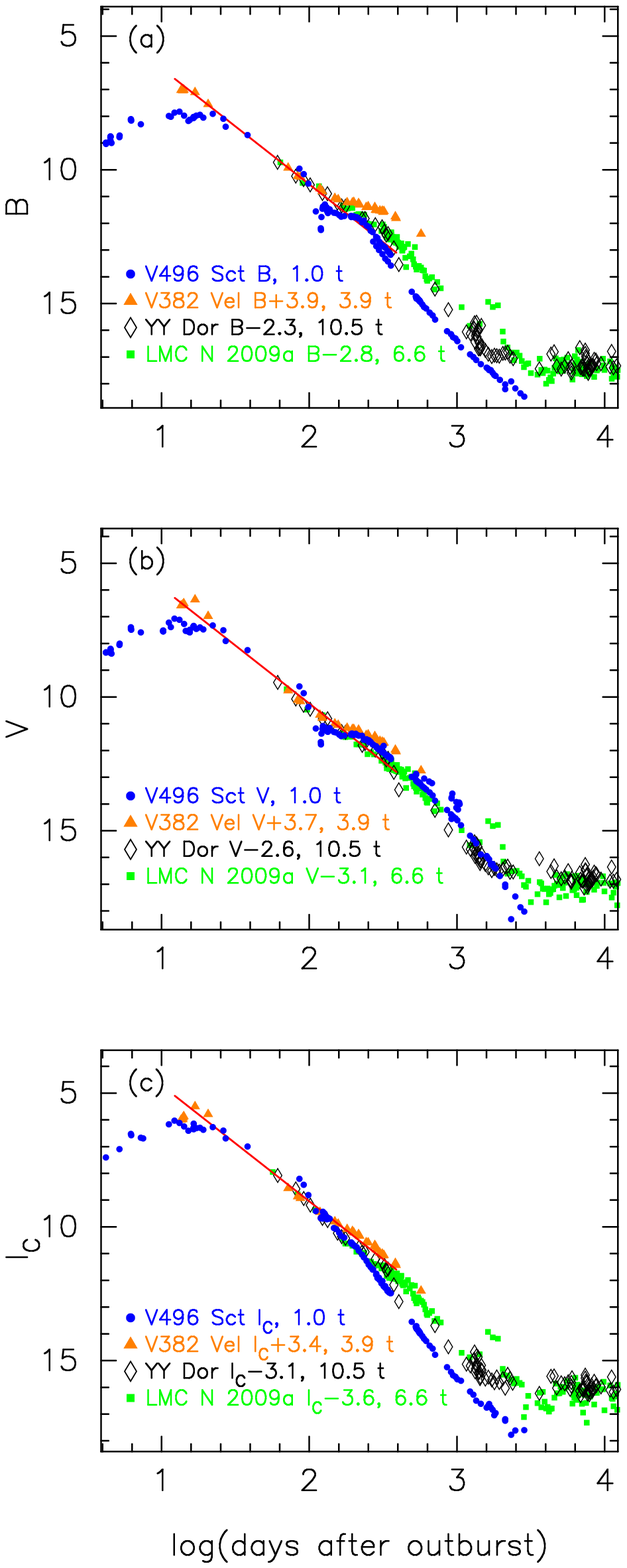}
\caption{
Same as Figure \ref{v679_car_yy_dor_lmcn_2009a_b_v_i_logscale_3fig},
but for V496~Sct.  
The $BVI_{\rm C}$ data of V496~Sct are taken from AAVSO, SMARTS,
and \citet{raj12}.
\label{v496_sct_v382_vel_yy_dor_lmcn_2009a_b_v_i_logscale_3fig}}
\end{figure}

\subsection{V496~Sct 2009}
\label{v496_sct}
Figure \ref{v496_sct_lv_vul_v1668_cyg_v_bv_ub_color_logscale_no2}
shows the light/color curves of V496~Sct 
as well as LV~Vul and V1668~Cyg. 
In the early phase, the $V$ light curve of V496~Sct overlaps that of
LV~Vul.   In the middle phase, V496~Sct show a sharp and shallow dip
due to dust blackout.  In the later nebular phase, V496~Sct 
overlaps well with the upper branch of LV~Vul, where the $V$ light curve
of LV~Vul splits into two branches in the nebular phase due to
the different responses of their $V$ filters as discussed in Paper II.
This is because strong [\ion{O}{3}] lines contributes to the blue edge
of $V$ band filter in the nebular phase and small differences among 
the response functions of $V$ filters make large differences
in the $V$ magnitudes. 
The $(B-V)_0$ color curves of LV~Vul also splits into two branches
for $\log t\gtrsim 2.0$, that is, in the nebular phase.  
The lower branch of LV~Vul in the $(B-V)_0$ color curve
is close to that of V496~Sct.
We regard that the $V$ light curve of V496~Sct follows V1668~Cyg and 
the upper branch of LV~Vul and the $(B-V)_0$ color curve of V496~Sct
follows the lower branch of LV~Vul.  

Based on the time-stretching method, we have the relation of
\begin{eqnarray}
(m&-&M)_{V, \rm V496~Sct} \cr 
&=& (m - M + \Delta V)_{V, \rm LV~Vul} - 2.5 \log 2.0 \cr
&=& 11.85 + 2.6\pm0.3 - 0.75 = 13.7\pm0.3 \cr
&=& (m - M + \Delta V)_{V, \rm V1668~Cyg} - 2.5 \log 2.0 \cr
&=& 14.6 - 0.1\pm0.3 - 0.75 = 13.75 \pm0.3 .
\label{distance_modulus_v496_sct_lv_vul_v1668_cyg}
\end{eqnarray}
Thus, we obtain $f_{\rm s}=2.0$ against the template nova LV~Vul
and $(m-M)_V=13.7\pm0.2$.
The new value is much smaller than the previous estimate of 
$(m-M)_V=14.4\pm0.1$ (Paper II).  This is because
we improved the timescaling factor of $f_{\rm s}$ and the
vertical $V$ light curve fitting.
From Equations (\ref{distance_modulus_general_dot}) and
(\ref{distance_modulus_v496_sct_lv_vul_v1668_cyg}),
we have the relation of
\begin{eqnarray}
(m&-& M')_{V, \rm V496~Sct} \cr 
&\equiv& (m_V - (M_V - 2.5\log f_s))_{\rm V496~Sct} \cr
&=& (m-M + \Delta V)_{V,\rm LV~Vul} \cr
&=& 11.85 + 2.6\pm0.3 =14.45\pm0.3.
\label{absolute_mag_v496_sct_lv_vul}
\end{eqnarray}

Figure \ref{v496_sct_v1065_cen_lv_vul_v_bv_ub_color_logscale} shows the
light/color curves of V496~Sct as well as LV~Vul and V1065~Cen.
The light/color curves of V1065~Cen show different paths of V496~Sct.
We regard that the $V$ light curve of V1065~Cen follows the lower
branch of LV~Vul and the $(B-V)_0$ color curve evolves along
the upper branch of LV~Vul in the nebular phase ($t > 100$~day).
For V1065~Cen, \citet{hac18k} derived $(m-M)_V=15.0\pm0.2$ and 
$E(B-V)=0.45\pm0.05$ (and $f_{\rm s}=1.0$ against LV~Vul) as summarized in
Table \ref{extinction_various_novae}.  
Applying Equation (\ref{distance_modulus_general_temp})
to Figure \ref{v496_sct_v1065_cen_lv_vul_v_bv_ub_color_logscale},
we have the relation of
\begin{eqnarray}
(m&-&M)_{V, \rm V496~Sct} \cr 
&=& (m-M + \Delta V)_{V, \rm LV~Vul} - 2.5 \log 2.0 \cr
&=& 11.85 + 2.6\pm 0.2 - 0.75 = 13.7\pm 0.2 \cr
&=& (m-M + \Delta V)_{V, \rm V1065~Cen} - 2.5 \log 2.0 \cr
&=& 15.0  - 0.5\pm 0.2 - 0.75 = 13.75\pm 0.2.
\label{distance_modulus_v496_sct_lv_vul_v1065_cen_v}
\end{eqnarray}
This value of $(m-M)_V=13.7\pm0.2$ is consistent with
Equation (\ref{distance_modulus_v496_sct_lv_vul_v1668_cyg}).

Figure 
\ref{v496_sct_v382_vel_yy_dor_lmcn_2009a_b_v_i_logscale_3fig}
plots the $B$, $V$, and $I_{\rm C}$ light curves of V496~Sct together
with those of V382~Vel, and the LMC novae YY~Dor and LMC~N~2009a.
We apply Equation (\ref{distance_modulus_general_temp_b}) for the $B$ band
to Figure \ref{v496_sct_v382_vel_yy_dor_lmcn_2009a_b_v_i_logscale_3fig}(a) 
and obtain
\begin{eqnarray}
(m&-&M)_{B, \rm V496~Sct} \cr
&=& ((m - M)_B + \Delta B)_{\rm YY~Dor} - 2.5 \log 10.5 \cr
&=& 18.98 - 2.3\pm0.2 - 2.55 = 14.13\pm0.2 \cr
&=& ((m - M)_B + \Delta B)_{\rm LMC~N~2009a} - 2.5 \log 6.6 \cr
&=& 18.98 - 2.8\pm0.2 - 2.05 = 14.13\pm0.2 \cr 
&=& ((m - M)_B + \Delta B)_{\rm V382~Vel} - 2.5 \log 3.9 \cr
&=& 11.75 + 3.9\pm0.2 - 1.47 = 14.18\pm0.2. 
\label{distance_modulus_b_v496_sct_v382_vel_yy_dor_lmcn2009a}
\end{eqnarray}
Thus, we obtain $(m-M)_{B, \rm V496~Sct}= 14.15\pm0.1$,
For the $V$ light curves in Figure 
\ref{v496_sct_v382_vel_yy_dor_lmcn_2009a_b_v_i_logscale_3fig}(b),
we similarly obtain
\begin{eqnarray}
(m&-&M)_{V, \rm V496~Sct} \cr
&=& ((m - M)_V + \Delta V)_{\rm YY~Dor} - 2.5 \log 10.5 \cr
&=& 18.86 - 2.6\pm0.2 - 2.55 = 13.71\pm0.2 \cr
&=& ((m - M)_V + \Delta V)_{\rm LMC~N~2009a} - 2.5 \log 6.6 \cr
&=& 18.86 - 3.1\pm0.2 - 2.05 = 13.71\pm0.2 \cr
&=& ((m - M)_V + \Delta V)_{\rm V382~Vel} - 2.5 \log 3.9 \cr
&=& 11.5 + 3.7\pm0.2 - 1.47 = 13.73\pm0.2.
\label{distance_modulus_v_v496_sct_v382_vel_yy_dor_lmcn2009a}
\end{eqnarray}
Thus, we obtain $(m-M)_{V, \rm V496~Sct}= 13.71\pm0.1$.  We apply 
Equation (\ref{distance_modulus_general_temp_i})  for the $I_{\rm C}$ band
to Figure \ref{v496_sct_v382_vel_yy_dor_lmcn_2009a_b_v_i_logscale_3fig}(c)
and obtain
\begin{eqnarray}
(m&-&M)_{I, \rm V496~Sct} \cr
&=& ((m - M)_I + \Delta I_C)_{\rm YY~Dor} - 2.5 \log 10.5 \cr
&=& 18.67 - 3.1\pm0.2 - 2.55 = 13.02\pm 0.2 \cr
&=& ((m - M)_I + \Delta I_C)_{\rm LMC~N~2009a} - 2.5 \log 6.6 \cr
&=& 18.67 - 3.6\pm0.2 - 2.05 = 13.02\pm 0.2 \cr 
&=& ((m - M)_I + \Delta I_C)_{\rm V382~Vel} - 2.5 \log 3.9 \cr
&=& 11.1 + 3.4\pm0.2 - 1.47 = 13.03\pm 0.2. 
\label{distance_modulus_i_v496_sct_v382_vel_yy_dor_lmcn2009a}
\end{eqnarray}
Thus, we obtain $(m-M)_{I, \rm V496~Sct}= 13.02\pm0.2$.

\clearpage
































































\end{document}